%% file: PhD.tex
\documentclass[10pt,a4paper,twoside]{book}

\input{english.tex}
\begin{document}


\setstretch{1.5}
\input{title.tex}
\newpage  
\blanc


\lhead[\textbf{\thepage}]{\textbf{\rightmark}}
\rhead[\textbf{\leftmark}]{\textbf{\thepage}}
\cfoot[]{}
\renewcommand{\chaptermark}[1]{\markboth{\thechapter\ \ #1}{}}
\renewcommand{\sectionmark}[1]{\markright{\thesection\ #1}}
\pagestyle{fancy}
\pagenumbering{roman}


\lhead[\textbf{\thepage}]{\textbf{Acknowledgements}}
\rhead[\textbf{Acknowledgments}]{\textbf{\thepage}}
\section*{Acknowledgments}
\addcontentsline{toc}{chapter}{{Acknowledgments}}
\input{acknowledgments.tex}

\newpage


\lhead[\textbf{\thepage}]{\textbf{Table of contents}}
\rhead[\textbf{Table of contents}]{\textbf{\thepage}}
\tableofcontents
\newpage


\lhead[\textbf{\thepage}]{\textbf{Notations and conventions}}
\rhead[\textbf{Notations and conventions}]{\textbf{\thepage}}
\section*{Notations and conventions}
\addcontentsline{toc}{chapter}{{Notations and conventions}}
\input{notations.tex}
\newpage


\lhead[\textbf{\thepage}]{\textbf{Introduction}}
\rhead[\textbf{Introduction}]{\textbf{\thepage}}
\section*{Introduction}
\addcontentsline{toc}{chapter}{{Introduction}}
\input{introduction.tex}

\newpage


\pagenumbering{arabic}
\lhead[\textbf{\thepage}]{\textbf{\rightmark}}
\rhead[\textbf{\leftmark}]{\textbf{\thepage}}
\renewcommand{\chaptermark}[1]{\markboth{\thechapter\ \ #1}{}}
\renewcommand{\sectionmark}[1]{\markright{\thesection\ #1}}


\chapter{Supersymmetry} 
\input{supersymmetry.tex}


\chapter{$\mathcal{N}=2$ Super Yang-Mills theory}
\input{yangmills.tex}


\chapter{Localization, deformation and equivariant integration}
\input{localization.tex}


\chapter{Finite dimensional reduction}
\input{finite.tex}


\chapter{Instanton corrections in the general case}
\input{towards.tex}


\chapter{Saddle point equations}
\input{saddlepoint.tex}


\chapter{Seiberg-Witten geometry}
\input{geometry.tex}


\chapter{Open questions and further directions}
\input{questions.tex}


\appendix


\chapter{Spinor properties}
\input{spinors.tex}


\chapter{Lie algebras}
\input{lie.tex}


\providecommand{\bysame}{\leavevmode\hbox to3em{\hrulefill}\thinspace}
\providecommand{\href}[2]{#2}

\end{document}

%% file: english.tex
\usepackage{amssymb}
\usepackage{amsfonts}
\usepackage{mathrsfs}
\usepackage{latexsym}
\usepackage{amsmath}
\usepackage{graphics}
\usepackage{graphicx}
\usepackage{hhline}                   
\usepackage{euscript}               
\usepackage{rotating}
\usepackage{setspace}                 
\usepackage{fancyheadings}            
\usepackage[dvips]{hyperref}

\input{definitions.tex}

\newenvironment{remark}{\noindent\underline{Remark.}}{$\Box$}
\newcommand{\microsection}[1]{\noindent\underline{#1}}
\newcommand{\Ref}[1]{(\ref{#1})}

\newcommand{\blanc}{\ifthenelse{\boolean{@twoside}}{\thispagestyle{empty}~\newpage}{} } 
\newcommand{\horline}{\hrule\smallskip}

%% file: definitions.tex
\DeclareSymbolFont{NumBold}{U}{bbold}{m}{n}
\DeclareSymbolFontAlphabet{\mathNumbb}{NumBold}
\DeclareMathSymbol{\Id}{\mathord}{NumBold}{`1}
\newcommand{\mbf}[1]{\mbox{\boldmath{$#1$}}}
\newcommand{\alg}{\mathfrak}
\newcommand{\mr}{\mathrm}

\newcommand{\A}{\mathcal{A}}
\newcommand{\B}{\mathcal{B}}
\newcommand{\C}{\mathcal{C}}
\newcommand{\D}{\mathcal{D}}
\newcommand{\E}{\mathcal{E}}

\newcommand{\G}{\mathcal{G}}
\renewcommand{\H}{\mathcal{H}}
\newcommand{\I}{\mathcal{I}}

\renewcommand{\L}{\mathcal{L}}

\newcommand{\N}{\mathcal{N}}
\renewcommand{\O}{\mathcal{O}}
\renewcommand{\P}{\mathcal{P}}

\newcommand{\R}{\mathcal{R}}
\renewcommand{\S}{\mathcal{S}}

\newcommand{\V}{\mathcal{V}}
\newcommand{\W}{\mathcal{W}}

\renewcommand{\a}{\alpha}
\renewcommand{\b}{\beta}
\newcommand{\g}{\gamma}
\renewcommand{\d}{\delta}
\newcommand{\eps}{\varepsilon}
\newcommand{\ep}{\epsilon}
\renewcommand{\o}{\omega}
\renewcommand{\r}{\rho}
\newcommand{\f}{\phi}
\newcommand{\vf}{\varphi}
\renewcommand{\l}{\lambda}
\newcommand{\s}{\sigma}
\renewcommand{\t}{\tau}
\renewcommand{\th}{\theta}
\newcommand{\vth}{\vartheta}

\newcommand{\z}{\zeta}
\newcommand{\vr}{\varrho}

\newcommand{\gG}{\Gamma}
\newcommand{\gD}{\Delta}
\newcommand{\gL}{\Lambda}
\newcommand{\gTh}{\Theta}
\newcommand{\gO}{\Omega}
\newcommand{\gS}{\Sigma}

\newcommand{\EuA}{\EuScript{A}}

\newcommand{\EuD}{\EuScript{D}}

\newcommand{\EuF}{\EuScript{F}}
\newcommand{\EuG}{\EuScript{G}}

\newcommand{\EuQ}{\EuScript{Q}}

\newcommand{\EuZ}{\EuScript{Z}}

\newcommand{\bA}{\mathbb{A}}

\newcommand{\bJ}{\mathbb{J}}

\newcommand{\e}{\mathop{\mathrm{e}}\nolimits}

\newcommand{\sn}{\mathop{\mathrm{sn}}\nolimits}
\newcommand{\logmod}[1]{\ln\left| #1 \right| }
\newcommand{\logmodL}[1]{\logmod{\frac{#1}{\gL}}}

\newcommand{\Heavi}{\mathscr{H}}

\newcommand{\Tr}{\mathop{\mathrm{Tr}}\nolimits}
\newcommand{\tr}{\mr{T}}
\newcommand{\vpint}{\displaystyle -\hskip -1.1em \int}
\newcommand{\rank}{\mathop{\mathrm{rank}}\nolimits}
\newcommand{\diag}{\mathop{\mathrm{diag}}\nolimits}
\newcommand{\Ch}{\mathop{\mathrm{Ch}}\nolimits}
\newcommand{\Td}{\mathop{\mathrm{Td}}\nolimits}
\newcommand{\Ind}{\mathrm{Ind}}

\newcommand{\sign}{\mathop{\mathrm{sign}}\nolimits}
\newcommand{\Spin}{\mathop{\mathrm{Spin}}\nolimits}
\newcommand{\dd}{\mathrm{d}}
\newcommand{\pd}{\partial}
\newcommand{\cD}{\nabla}
\newcommand{\Integers}{\mathbb{Z}}
\newcommand{\Real}{\mathbb{R}}
\newcommand{\Compl}{\mathbb{C}}
\newcommand{\Integer}{\mathbb{Z}}
\newcommand{\Sphere}{\mathbb{S}}

\let\IM=\Im
\let\RE=\Re
\let\Im=\undefined
\let\Re=\undefined
\newcommand{\Im}{\mathop{\IM\mathfrak{m}}\nolimits}
\newcommand{\Re}{\mathop{\RE\mathfrak{e}}\nolimits}
\newcommand{\Image}{\mathop{\mathrm{Im}}\nolimits}
\newcommand{\Ker}{\mathop{\mathrm{Ker}}\nolimits}
\renewcommand{\v}{\upsilon}
\newcommand{\Sym}{\mathop{\mathrm{Sym}}\nolimits}
\newcommand{\Ant}{\wedge}
\newcommand{\Tor}{\mathbb{T}}
\newcommand{\Mod}{\mathfrak{M}}
\newcommand{\Hom}{\mathop{\mathrm{Hom}}\nolimits}
\newcommand{\ds}{\displaystyle}
\newcommand{\cc}{\overline}
\newcommand{\bsl}{\backslash}
\newcommand{\tld}{\widetilde}
\newcommand{\isom}{\simeq}

\newcommand{\Lie}{\mathop{\mathrm{Lie}}\nolimits} 
\newcommand{\<}{\langle}
\renewcommand{\>}{\rangle}
\newcommand{\Fun}{\mathop{\mathrm{Fun}}\nolimits}
\newcommand{\Vol}{\mathop{\mathrm{Vol}}\nolimits}
\newcommand{\Eu}{\mathop{\mathrm{Eu}}\nolimits}
\newcommand{\Pf}{\mathop{\mathrm{Pf}}\nolimits}

\newcommand{\Prep}{\mathcal{F}}
\newcommand{\Pol}{\mathcal{P}}
\renewcommand{\k}{\mbf{k}}
\newcommand{\Moduli}{\mathfrak{M}}
\newcommand{\BundleD}{\mathfrak{D}}
\newcommand{\sD}{\EuD}
\newcommand{\sQ}{\EuQ}
\newcommand{\sCC}{\EuZ}
\newcommand{\sA}{\EuA}

%% file: title.tex

\thispagestyle{empty}

\makeatletter
\begin{titlepage}
\begin{flushright}
{hep-th/0502180}\\
\end{flushright}
\vskip 1cm
\begin{center}
{\LARGE\bf On certain aspects of string theory/gauge theory correspondence}\\
\vskip 2cm
{\Large PhD Thesis}
\vskip 2cm
{\bf Sergey Shadchin}
\end{center}
\vskip 1cm
\centerline{\em Laboratoire de Physique Th\'eorique, b\^atiment 210,}
\centerline{\em Universit\'e Paris-Sud, 91405, Orsay, France}
\centerline{\tt email: serezha@th.u-psud.fr}
\vskip 2cm
\input{abstract.tex}

\bigskip
\end{titlepage}
\makeatother

%% file: abstract.tex

$\N=2$ supersymmetric Yang-Mills theories for all classical gauge groups, that is, for $SU(N)$, $SO(N)$, and $Sp(N)$ is considered. The formal expression for almost all models accepted by the asymptotic freedom are obtained. The equations which define the Seiberg-Witten curve are proposed. In some cases they are solved. It is shown that for all considered the 1-instanton corrections which follows from these equations agree with the direct computations. Also they agree with the computations based on Seiberg-Witten curves which come from the $M$-theory consideration. It is shown that for a large class of models the $M$-theory predictions matches with the direct compuatations. It is done for all considered models at the 1-instanton level. For some models it is shown at the level of the Seiberg-Witten curves.

%% file: acknowledgments.tex

First of all I would like to express gratitude to scientific advisor, Nikita Nekrasov, who opened to me new domain of Theoretical Physics. His deep knowledge and creativity always maked our discussions very interesting and fruitful. The subject he prposed makes me to learn a lot of mathematics and physics which a did not know before. And this is thanks to his clear and lucid explainations that this work has succeded.

I am very grateful to all scientists with whom I had pleasure to discuss different questions and who helped me to find new ideas and get new vision of old ones. In particular I thank Sergey Alexandrov, Alexey Boyarsky, Alexey Gorinov, Denis Grebenkov, Ivan Kostov, Andrey Losev, Jeong-Hyuck Park, Oleg Ruchayski, Ksenia Rulik.

I am grateful to the Institut des Hautes \'Etudes Scientifiques, where my work has been launched. I am thankful to \'Ecole Poletechnque for the financial support. Let me also express my gratitude to scientists of the Laboratoire de Physique Th\'eorique d'Orsay. In particular I would like to thank Ulrich Ellwanger, Michel Fontanaz, Grigory Kortchemski and Samuel Wallon for numerous interesting and stimulating converstions.

I am very grateful to Pierre Vanhove who has read the draft of this manuscript and made very important comments, which helped me to make the text more readable.

Finally I am thankful to Constantin Bachas and Edouard Br\'ezin who accepted to be my reviewers and to Jean Iliopoulos, Ruben Minasian and Pierre Vanhove who agreed to be the members of my jury.

%% file: notations.tex
The following convention will be used through the paper:

\horline
Indices:
\begin{itemize}
\item Greek indices $\mu, \nu, \dots$ run over $0,1,2,3$, 
\item small latin indices $i,j,\dots$ run over $1,2,3$, 
\item capital latin indices $A, B, \dots$ run over $1,2$. They are supersymmetry indices,
\item small greek indices $\a,\b,\dots$ run over $1,2$. They are spinor indices,
\item capital latin indices $I, J, \dots$ run aver $0,1,2,3,4,5,6$. This is six dimensional indices.
\end{itemize}

\horline
\begin{itemize}
\item $\t_1$, $\t_2$ and $\t_3$ are the Pauli matrices defined in the standard way \Ref{Pauli},
\item The Euclidean $\s$-matrices are:
$$
\begin{aligned}
\s_{\mu,\a\dot{\a}} &= (\Id_2, - i \t_1, -i \t_2, -i \t_3), \\
\bar{\s}_\mu^{\dot{\a}\a} &= (\Id_2, + i \t_1, +i \t_2, +i \t_3) = {\left(\s_{\mu,\a\dot{\a}}\right)}^\dag, 
\end{aligned}
$$
\item in Minkowskian space two homomorphisms $SL(2,\Compl) \to SO(3,1)$ are governed by:
$$
\begin{aligned}
\s_{\mu,\a\dot{\a}} &= (\Id_2, -  \t_1, - \t_2, - \t_3), \\
\bar{\s}_\mu^{\dot{\a}\a} &= (\Id_2, +  \t_1, + \t_2, + \t_3). 
\end{aligned}
$$
(we apologize for the confusing notations -- we can only hope that every time it will be clear whether we work with Euclidean  or Minkowski signature).
\end{itemize}

\horline
\begin{itemize}
\item $\sD_\a$, $\bar{\sD}_{\dot{\a}}$ are covariant derivatives in superspace, see \Ref{SUSYCovariant},
\item $\sQ_\a$, $\bar{\sQ}_{\dot{\a}}$ are the supersymmetry operators, defined in \Ref{SUSYDiff}.
\end{itemize}

\horline
\begin{itemize}
\item $\d_{ab}$ is the Kronecker delta. By definition $\d_{ab} = 1$ when $a = b$ and $\d_{ab} = 0$ otherwise.
\item $\ep_{\mu_1,\dots,\mu_d}$ is the $d$-dimensional Levi-Civita tensor. $\ep_{12\dots d} = +1$,
\item the spinor metric is
$$\ep = \Vert \ep_{\a\b} \Vert = \left(
\begin{array}{cc}
0 & -1 \\
1 & 0
\end{array}
\right).
$$
\end{itemize}

\horline
\begin{itemize}
\item $\Id_n$ is $n \times n$ unit matrix,
\item the symplectic structure is denoted by
$$
\bJ_{2n} = \left(
\begin{array}{cc}
 0 & \Id_n \\
 -\Id_n  & 0
\end{array}
\right).
$$
\end{itemize}

\horline
The generators of the spinor representation of $SO(3,1)$ are
$$
\begin{aligned}
\s^{\mu\nu} &= \frac{1}{4} \Big( \s^\mu \bar{\s}^\nu - \s^\nu \bar{\s}^\mu \Big), \\
\bar{\s}^{\mu\nu} &= \frac{1}{4} \Big( \bar{\s}^\mu \s^\nu - \bar{\s}^\nu \s^\mu \Big),
\end{aligned}
$$
they satisfy
$$
\begin{aligned}
\s^{\mu\nu,\a\b}\s^{\r\s}_{\a\b} &= \frac{1}{2}\Big( g^{\mu\r}g^{\nu\s} - g^{\mu\s}g^{\nu\r} \Big) - \frac{i}{2}\ep^{\mu\nu\r\s}, \\
\bar{\s}^{\mu\nu,\dot{\a}\dot{\b}}\bar{\s}^{\r\s}_{\dot{\a}\dot{\b}} &= \frac{1}{2}\Big( g^{\mu\r}g^{\nu\s} - g^{\mu\s}g^{\nu\r} \Big) + \frac{i}{2}\ep^{\mu\nu\r\s}.
\end{aligned}
$$

\horline
In the Euclidean space the complex conjugation rises and lowers the spinor indices without changing their dottness. In the Minkowski space the height of the index is unchanged whereas its dottness does change.

\horline
\begin{itemize}
\item Mostly we denote by $G$ the gauge group. Its Lie algebra is denoted by $\alg{g} = \Lie(G)$. Sometimes when we identify the gauge group and the group of the rigid gauge transformations, which acts at the infinity, we denote it by $G_\infty$. Its maximal torus is denoted by $\Tor_\infty \subset G_\infty$.  $h^\vee$ is the dual Coxeter number. We use the notation $a$ for the elements of $\Lie(\Tor_\infty)$. The set of positive roots for the gauge group is denoted by $\gD^+$. The Dynkin index for a representation $\vr$ is $\ell_\vr$. The set of weights for a representation $\vr$ is denoted by $\mbf{w}_\vr$.
\item We denote by $G_D$ the dual (in the sense of \cite{SelfDualSolution}) group (see the definition at the end of section \ref{ADHMforSU(N)}). Its maximal torus is denoted by $\Tor_D \subset G_D$. The Cartan subalgebra is $\alg{t} = \Lie(\Tor_D)$. $W_D$ is its Weyl group.
\item The flavor group is denoted by $G_F$ (see the definition at the end of section \ref{Matter}). Its maximal torus is $\Tor_F \subset G_F$.
\item The Killing form on the Lie algebra of the gauge group is denoted as $\< \a,\b\>$. In the adjoint representation it is given by $\ds\< \a,\b\> = \frac{1}{h^\vee} \Tr_{\mr{adj}} \{\a\b\}$ where the trace is taken over the adjoint representation. 
\end{itemize}

\horline
In section \ref{O-background} we have introduces so-called $\gO$-background. The main object is the matrix of the Lorentz rotations $\gO_{\mu\nu}$ which we represent as follows
$$
\gO = \frac{1}{\sqrt{2}}\left( 
\begin{array}{cccc}
0 & 0 & 0 & -\eps_1 \\
0 & 0 & -\eps_2 & 0 \\
0 & \eps_2 & 0 & 0 \\
\eps_1 & 0 & 0 & 0
\end{array}
\right).
$$

It will be useful to introduce the following combinations of the parameters $\eps_1$ and $\eps_2$:
\begin{itemize}
\item $\ds\eps_\pm = \frac{\eps_1 \pm \eps_2}{2}$,
\item $\eps = \eps_1 + \eps_2 = 2\eps_+$.
\end{itemize}

\horline
If $\V$ is a vector space, then $\Pi \V$ is the vector space with changed statistics (bosons $\leftrightarrow$ fermions).

\horline
We study gauge theory on $\Real^4$. Sometimes it is convenient to compactify $\Real^4$ by adding a point $\infty$ at infinity, thus producing $\Sphere^4 = \Real^4 \cup \{\infty\}$.

\horline
We consider a principal $G$-bundle over $\Sphere^4$, with $G$ being one of the classical groups ($SU(N)$, $SO(N)$ or $Sp(N)$). To make ourselves perfectly clear we stress that $Sp(N)$ means in this paper the group of matrices $2N \times 2N$ preserving the symplectic structure, sometimes denoted in the literature as $USp(2N)$.

\horline
In our notations the gauge boson field (the connection) $A_\mu$ are real. Therefore the covariant derivative is defined as follows: $\cD_\mu  = \pd_\mu - i A_\mu$. The curvature (stress tensor) is defined by  \Ref{Strength}. Sometimes the connection $A_\mu$ is supposed to be antihermitian (especially in mathematical texts). In that case the field strength is defined by 
$$
F_{\mu\nu}^{\mr{m}} = \pd_\mu A_\nu^{\mr{m}} - \pd_\nu A_\mu^{\mr{m}} + [A_\mu^{\mr{m}},A_\nu^{\mr{m}}].
$$ 
We can establish the connection with the mathematical formalism as follows
$$
\begin{aligned}
A^{\mr{m}}_\mu &= -i A_\mu, & F^{\mr{m}}_{\mu\nu} &= -i F_{\mu\nu}. 
\end{aligned}
$$
In these notations we have the following definition of the cuvature tensor:
$$
[\cD_\mu,\cD_\nu] = - i F_{\mu\nu}.
$$

\horline
In section \ref{scn:TopTwist} we will introduce twisted fields $\bar{\psi}, \psi_\mu, \bar{\psi}_{\mu\nu}$. In order to make contact with the topological multiplet \cite{TQFT} $(A_\mu^{\rm{top}}, \phi^{\mr{top}}, \l^{\mr{top}}, \eta^{\mr{top}}, \psi_\mu^{\mr{top}}, \chi_{\mu\nu}^{\mr{top}})$ let us write the rule of correspondence \Ref{TopCorrespondence}:
$$
\begin{aligned}
A_\mu^{\mr{top}} &= A_\mu, & \psi_\mu^{\mr{top}} &= \psi_\mu, \\
\phi^{\mr{top}} &= -2\sqrt{2} H, & \l^{\mr{top}} &= -2\sqrt{2}H^\dag, \\
\eta^{\mr{top}} &= -4 \bar{\psi}, & \chi_{\mu\nu}^{\mr{top}} &= \bar{\psi}_{\mu\nu}.  
\end{aligned}
$$

\horline
\begin{itemize}
\item The vacuum expectation of the field $\f$ belonging to the topological multiplet will be denoted through the paper as $a$. 
\item The vacuum expectation of an observable $\O$ over the field configurations with the fixed value of $\f$ at infinity (which is equal to $a$) is denoted as
$$
\< \O \>_a = \int_{\ds\lim_{x\to\infty}\f(x) = a} \D\left\{\mr{fields}\right\}\e^{\mr{action}}\O
$$
\item The vacuum expectation of the Higgs field $H$ will differ to $a$ by the factor $\ds-\frac{1}{2\sqrt{2}}$:
$$
\< H\>_a = - \frac{1}{2\sqrt{2}}a
$$
\end{itemize}

\horline
We will use the complex coupling constant $\t$ which is related with the Yang-Mills coupling constant $g$ and with the instanton number in the following way
$$
\t = \frac{4\pi i}{g^2} + \frac{\gTh}{2\pi}.
$$
In section \ref{scn:SeibergWitten} we introduce the instanton counting parameter $q$ which is related to $\t$, $g$ and $\gTh$ as follows:
$$
q = \e^{2\pi i \t} = \e^{-\frac{8\pi^2}{g^2}} \e^{i\gTh}.
$$

%% file: introduction.tex

The duality between the gauge theories and the string theory is now of the great importance. The actual knowledge suggests that all the superstring theories in ten dimensions can be obtained as different limits of a unique eleven dimensional theory, known as $M$-theory \cite{Mtheory,PowerMtheory,Heterotic11D,11Drevisited}.

In spite of the existence of numerous arguments in favor of this approach, the $M$-theory is not yet built. Therefore one tries to find some non-direct evidences which confirm (or reject) this theory. The main strategy is to compare its prediction with results which can be obtained in a different (and independent of the $M$-theory) way.

Among other predictions which provides $M$-theory there are those which concern to the Wilsonian effective action \cite{WilsonianAction,WilsonianActionDine} along the Coulomb branch for $\N=2$ super Yang-Mills theory \cite{PrepFromM}. The leading part of the non-perturbative effective action for the gauge group $SU(2)$ which contains up to two derivatives and and four fermions  was computed by Seiberg and Witten \cite{SeibergWitten}. After its appearance the Seiberg-Witten solution was generalized in both directions: to other classical groups and to various matter content \cite{SimpleSing,VacuumStruct,QuantModSpace,CurveForSO(2r+1),HyEl,SeibergWittenII,NewCurves,SYMandIntegr}. 

Till recently while generalizing one established the expression for the algebraic curve and the meromorphic differential from the first principles and then computed the instanton corrections to the leading part of the  effective action. This part can be expressed with the help of a unique holomorphic function $\Prep(a)$, referred as prepotential \cite{PrepotGates,PrepotDeWitt,Prepotential,IntroToN=2SUSY}. With the help of the extended superfield formalism the Lagrangian for the effective theory can be written as an $\N=2$ $F$-term:
$$
S_{\mr{eff}} = \frac{1}{4\pi}\Im \left\{\frac{1}{2\pi i}\int \dd^4 x \dd^4 \th \Prep(\Psi)\right\}.
$$
The classical prepotential, which provides the microscopic action, is
$$
\Prep_{\mr{class}}(a) = \pi i \t_0 \langle a,a \rangle,
$$
where $\ds \t_0 = \frac{4\pi i}{g^2_0} + \frac{\gTh_0}{2\pi}$. Note that we use the normalization of the prepotential which differs from some other sources by the factor $2\pi i$.

The complete Wilsonian effective action does contain other terms, for example the next one, which contains four derivatives and eight fermions can be expressed with the help of a real function $\H(a,\cc{a})$ as the $\N=2$ $D$-term \cite{DtermHenn,DtermDeWitt,DtermWess,DtermRocek,DtermYung1,DtermYung2,DineSeiberg}:
$$
S_{4\mr{-deriv}} = \int \dd^4 x \dd^4 \th \dd^4 \bar{\th} \H(\Psi,\bar{\Psi}).
$$

In \cite{SWfromInst,SWandRP} a powerful technique was proposed to follow this way in the opposite direction: to compute first the instanton corrections and to extract from them the Seiberg-Witten geometry and the analytical properties of the prepotential.

In \cite{ABCD}  the solution of $\N = 2$ supersymmetric Yang-Mills theory for the classical groups other that $SU(N)$ using the method proposed in \cite{SWfromInst,SWandRP} was obtained. This method consists of the reducing functional integral expression for the vacuum expectation of an observable (in fact, this observable equals to 1, hence we actually compute the partition function as it defined in statistical physics) to the finite dimensional moduli space of zero modes of the theory. That is, to the instanton moduli space, the moduli space of the solutions of the self-dual equation
$$
F_{\mu\nu} - \star F_{\mu\nu} = 0
$$
with the fixed value of the instanton number
$$
k = -\frac{1}{16\pi h^\vee} \int \Tr_{\mr{adj}} F \wedge F. 
$$
Notation $\Tr_{\mr{adj}}$ means that the trace is taken over the adjoint representation.

In \cite{SPinSW} we continue to investigate the possibility to solve the $\N = 2$ supersymmetric Yang-Mills theory with various matter content (limited, of cause, by the asymptotic freedom condition).

Roughly speaking our task can be split into two parts. First part consists of the writing the expression for the finite dimensional integral to which vacuum expectation in question can be reduced. To accomplish this task in \cite{SWfromInst,ABCD} the explicit construction for the instanton moduli space was used. Already for the pure gauge theory its construction (the famous ADHM construction of instantons, \cite{ADHM}) is rather nontrivial (see for example \cite{MultInstMeasure,MultiInstCalc,MultiInstCalcII,ItoSasakuraSU,ItoSasakura,ItoSasakuraSUSYQCD}). In the presence of matter it becomes even more complicated. 

Fortunately there is another method which lets to skip the explicit description of the moduli space and to directly write the required integral. This method uses some algebraic facts about the universal bundle over the instanton moduli space. It will be explained in section \ref{UniversalBundle}. Using this method we will obtain the prepotential as a formal series over the dynamically generated scale.

The second part of the task is to extract the Seiberg-Witten geometry from obtained expressions. To do this we will use the technique proposed in \cite{SWandRP}. It is based on the fact that in the limit of large  instanton number the integral can be estimated by means of the saddle point approximation. This approximation can be effectively described by the Seiberg-Witten data --- the curve and the differential. One may wonder why the prescription obtained in this limit will provide the exact solution even in the region of finite $k$, where the saddle point approximation  certainly will not work. The answer is that the real reason why the Seiberg-Witten prescription works is the holomorphicity of the prepotential, pointed out in \cite{SeibergWitten}, whereas the saddle point approximation just makes it evident and easy to extract.

The paper is organized as follows: in chapter \ref{supersymmetry} we recall some aspects of $\N=1$ and $\N=2$ supersymmetry. In chapter \ref{Outline} we give an outline of the important facts about $\N=2$ super Yang-Mills theories: the Seiberg-Witten theory, topological twist, and its relation to the $M$-theory. Chapter \ref{Localization} is devoted to some aspects of the equivariant integration. Also we give a short introduction to the ADHM construction. In chapter \ref{finite} we use the ADHM construction to compute the instanton corrections for some cases. In chapter \ref{towards} we describe a method to write the formula for the instanton corrections. In chapter \ref{saddlepoint} we reduce the problem of the instanton correction computations to the problem of minimizing a functional. And finally in chapter \ref{geometry} we solve the saddle point equations for some models. Using relations between the saddle point equation for different models we establish the same relations between the prepotentials for these models and finally we find the {\em hyperelliptic approximation} for the Seiberg-Witten curves for all the models. This allows us to compute the 1-instanton corrections which comes from the algebraic curve and compare it with the direct computations result. In each case perfect agreement between results of two approaches is observed. 

\begin{figure}
\includegraphics[width=\textwidth]{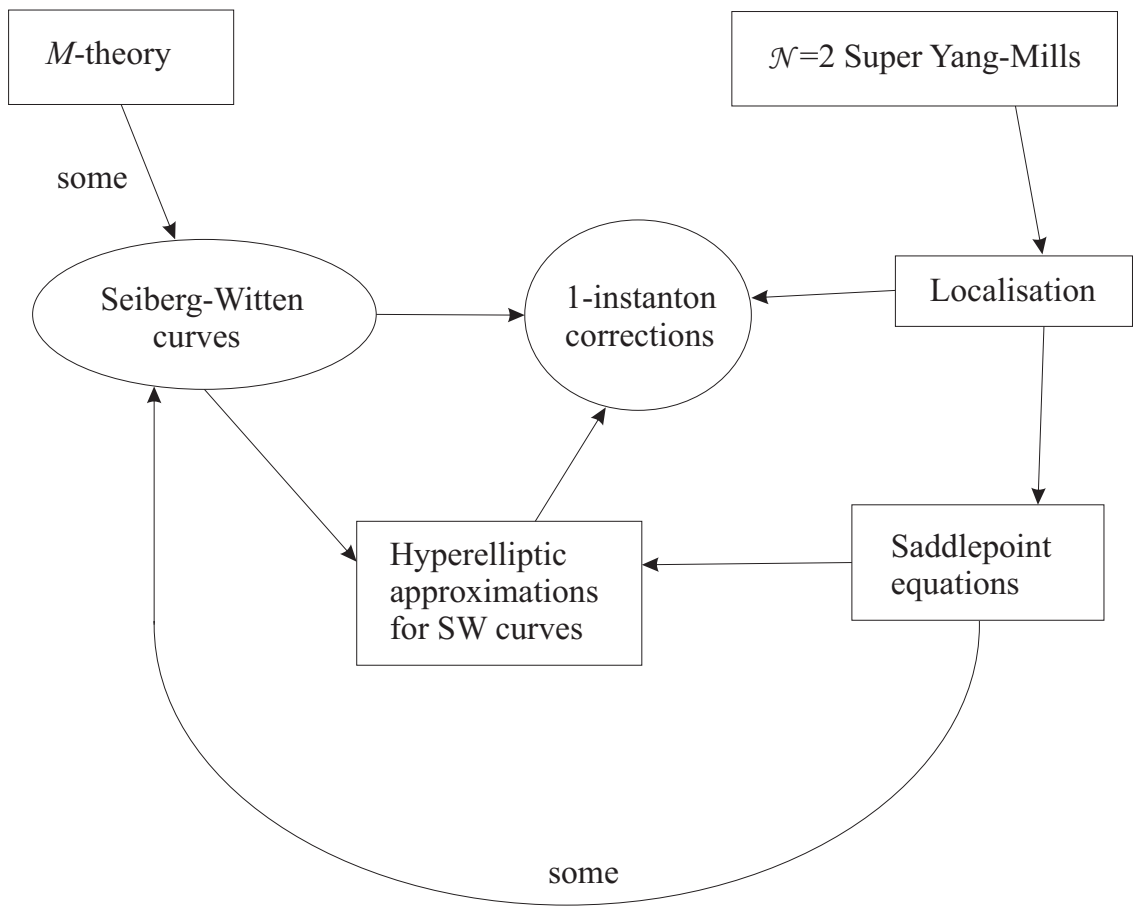}
\caption{The roadmap of the text}\label{Logic}
\end{figure}

The logic of the presentation is not always linear. In order to simplify the reading we have included a schematic roadmap of this text, figure \ref{Logic}. The word ``some'' near some arrow means that the passage is possible only for some models. 

%% file: supersymmetry.tex
\label{supersymmetry}

In this section we will shortly describe some properties of the superspace, which is necessary to consider super Yang-Mills theory. There are lot of well-written texts on supersymmetry \cite{SUSYandSUGRA,West,WestNieuwenhuizen,SUSYBilal,Intro2SUSYLykken,LecturesSWandIntegr}. Not even trying to describe the subject in all details, we have just pick some elements in order to make our story self-consistent. 


\section{Algebra of supersymmetry}

The Coleman and Mandula theorem \cite{ColemanMandula} states that the only allowed symmetry of the $S$-matrix is the Poincar\'e algebra plus maybe some internal symmetries which commute with it. This theorem concerns only transformation with commuting parameters. Therefore this statement is about the maximal allowed external symmetry Lie algebra. But if we include also some transformations with anticommuting parameters, that is, transform the Lie algebra to a superalgebra, we can obtain a supplementary symmetry in the theory. In this way the supersymmetry arises.

Let $P_\mu$ and $J_{\mu\nu}$ be the generators of the Poincar\'e algebra. Their commutation relations are the following
$$
\begin{aligned}
\left[P_\mu,P_\nu\right] &= 0, \\
[J_{\mu\nu},P_\r] &= i g_{\r\nu} P_\mu - i g_{\r\mu} P_\nu, \\
[J_{\mu\nu},J_{\r\s}] &= ig_{\nu\r} J_{\mu\s} - i g_{\mu\r} J_{\nu\s} -i g_{\nu\s}J_{\mu\r} + i g_{\mu\s}J_{\nu\r}.
\end{aligned}
$$

They can be represented by the following differential operators:
\begin{equation}
\label{PoincareDiff}
\begin{aligned}
P_\mu &= i \pd_\mu, \\
J_{\mu\nu} &= ix_\mu \pd_\nu - i x_\nu \pd_\mu + S_{\mu\nu}.
\end{aligned}
\end{equation}
These operators act on the argument of scalar functions and describe their transformation under rotations and translations of the Poincar\'e group. $S_{\mu\nu}$ is the spin operator. It describes the transformation of a function belonging to a higher spin representation of the Lorentz group. For example, if we consider a spinor function $\psi^\a(x)$ the spin operator takes the following form
$$
{\Big( S_{\mu\nu} \psi (x)\Big)}^\a = i \s_{\mu\nu}{}^\a{}_\b \psi^\b(x).
$$

The supersymmetry is realized as the largest supergroup of symmetry of the $S$-matrix \cite{HaagLopSohn}. It is described as follows. In addition to the operators \Ref{PoincareDiff}, which naturally have bosonic statistics, one introduces a supplementary set of operators $\sQ_\a^A$ and $\bar{\sQ}_{A,\dot{\a}} = {(\sQ_\a^A)}^\dag$, $A = 1,\dots,\N$, which are fermions. They have spinor indices. The (anti)commutation relations of the enlarged Poincar\'e algebra are the following (we use the standard normalization)
\begin{equation}
\label{SUSYalgebra}
\begin{aligned}
&{[}P_\mu,\sQ_\a^A{]}& &= 0,\\
&[P_\mu,\bar{\sQ}_{A,\dot{\a}}]& &= 0, \\
&[J_{\mu\nu},\sQ_\a^A]& &= i \s_{\mu\nu,\a}{}^\b \sQ_\b^A, \\
&[J_{\mu\nu},\bar{\sQ}^{\dot{\a}}_A]& &= i \bar{\s}_{\mu\nu}{}^{\dot{\a}}{}_{\dot{\b}} \bar{\sQ}^{\dot{\b}}_A, \\
&\{ \sQ_\a^A, \bar{\sQ}_{B,\dot{\b}}\}& &= 2 \s^\mu_{\a\dot{\b}}P_\mu \d^A_B, \\
&\{ \sQ_\a^A , \sQ_\b^B \}& &= \ep_{\a\b} Z^{AB} \sCC, \\
&\{ \bar{\sQ}_{A,\dot{\a}}, \bar{\sQ}_{B,\dot{\b}} \}& &= \ep_{\dot{\a}\dot{\b}} Z_{AB}^\ast \sCC.
\end{aligned}
\end{equation}
Here $Z^{AB}$ is an antisymmetric matrix. A new operator $\sCC$ is the central extension of the supersymmetry algebra. It is known as the \emph{central charge}. This operator commutes with all other generators of the super Poincar\'e algebra.

\begin{remark}
Note that we have adopted a rule according to which hermitian conjugation swaps upper and lower supersymmetry indices.
\end{remark}

\begin{remark} 
The dumb spinor indices will be omitted in general. To make formulae unambiguous we adopt the rule according to which undotted indices are summed from up-left to right-down, and dotted -- from down-left to right-up. For example $\psi \chi \equiv \psi^\a \chi_\a$, $\bar{\psi} \bar{\chi} \equiv \bar{\psi}_{\dot{\a}} \bar{\chi}^{\dot{\a}}$.
\end{remark}


\section{Superspace}

If we wish to represent operators $\sQ_\a^A$ and $\bar{\sQ}_{A,\dot{\a}}$ in the spirit of \Ref{PoincareDiff} we should introduce some additional coordinates. Namely, let us introduce $\N$ left handed spinor coordinates $\th^\a_A$ and $\N$ righthanded\footnote{for $\N > 2$ it does not have any practical value, since irreducible field multiplets will suffer too many constraints } $\bar{\th}^{A,\dot{\a}}$. these coordinates are anticommuting. Also introduce a boson real coordinate $z$ which corresponds to the central charge. The complete set of coordinates becomes therefore
$$
z^a = \left( x^\mu, \th^\a_A, \bar{\th}^{A,\dot{\a}}, z \right).
$$
The space with these coordinates will be referred as the \emph{superspace}. 

The following differential operators satisfy the supersymmetry algebra \Ref{SUSYalgebra}.
\begin{equation}
\label{SUSYDiff}
\begin{aligned}
\sCC &= i\frac{\pd}{\pd z}, \\
\sQ_\a^A &= \frac{\pd}{\pd \th_A^\a} + i \s^\mu_{\a\dot{\b}}\bar{\th}^{A,\dot{\b}}\pd_\mu + \frac{i}{2} \ep_{\a\b} Z^{AB}\th^\b_B \frac{\pd}{\pd z}, \\
\bar{\sQ}_{A,\dot{\a}} &= \frac{\pd}{\pd \bar{\th}^{A,\dot{\a}}} + i \th_A^\b\s^\mu_{\b\dot{\a}} \pd_\mu + \frac{i}{2}\ep_{\dot{\a}\dot{\b}} Z_{AB}^\ast\bar{\th}^{B,\dot{\b}} \frac{\pd}{\pd z}.  
\end{aligned}
\end{equation}
\begin{remark}
Our choice of the sign of the second summand in these formulae is closely related to our definition of the momentum operator $P_\mu$ \Ref{PoincareDiff}. The choice $P_\mu = + i \pd_\mu$ is, in its turn, fixed by our choice of the Minkowskian metric \Ref{MinkowskiMetric} and the corresponding formulae in Quantum Mechanics:
$$
\begin{aligned}
H &= +i \frac{\pd}{\pd t},  & \vec{P} &= -i \frac{\pd}{\pd \vec{x}}. 
\end{aligned}
$$
\end{remark}

\begin{remark}
In the opposition with the bosonic case the fermionic derivative is hermitian:
$$
{\left(\frac{\pd}{\pd \th_A^\a}\right)}^\dag = \frac{\pd}{\pd \bar{\th}^{A,\dot{\a}}}.
$$
\end{remark}

The general transformation of the super Poincar\'e algebra can be represented as follows:
$$
-ia^\mu P_\mu - \frac{i}{2}\o^{\mu\nu}J_{\mu\nu} + \z^\a_A \sQ_\a^A +  \bar{\z}^{B,\dot{\b}}\bar{\sQ}_{B,\dot{\b}} -i t \sCC.
$$
It corresponds to the following supercoordinate transformations:
\begin{equation}
\label{SUSYTransformCoord}
\begin{aligned}
x^\mu &\mapsto x^\mu + a^\mu + \o^{\mu\nu}x_\nu +  i\z^\a_A \s^\mu_{\a\dot{\b}}\bar{\th}^{A,\dot{\b}} - i \th^\a_B \s^\mu_{\a\dot{\b}} \bar{\z}^{B,\dot{\b}}, \\
\th^\a_A &\mapsto \th^\a_A + \z^\a_A + \frac{1}{2} \o^{\mu\nu} \s_{\mu\nu}{}^\a{}_\b \th^\b_A, \\
\bar{\th}^{A,\dot{\a}} &\mapsto \bar{\th}^{A,\dot{\a}} + \bar{\z}^{A,\dot{\a}} + \frac{1}{2}\o^{\mu\nu} \bar{\s}_{\mu\nu}{}^{\dot{\a}}{}_{\dot{\b}} \bar{\th}^{A,\dot{\b}}, \\
z &\mapsto z + t + \frac{i}{2}\z^\a_A \ep_{\a\b} Z^{AB} \th^\b_B + \frac{i}{2}\bar{\z}^{A,\dot{\a}} \ep_{\dot{\a}\dot{\b}}Z_{AB}^\ast \bar{\th}^{B,\dot{\b}}.
\end{aligned}
\end{equation}


\section{Geometry of the superspace}

In this section we consider some geometrical properties of the superspace. In particular, we recall how to derive the covariant derivative from the geometrical point of view. More details can be found, for example, in \cite{SUSYandSUGRA,West,WestNieuwenhuizen}.

Four dimensional Minkowski (Euclidean) space can be seen as a coset $ISO(3,1) / SO(3,1)$\footnote{``$I$'' stands for ``inhomogeneous''} ($ISO(4) / SO(4)$), where $ISO(3,1)$ ($ISO(4)$) is the Poincar\'e group. In the same way the superspace can be seen as a the super Poincar\'e group $SISO(3,1)$ ($SISO(4)$) factor Lorentz group. 

The geometrical properties of the superspace can be deduced from the fact that the Killing vectors of the super Poincar\'e symmetry of the space are obtained by the group multiplication. It allows to get the connection.

Any element of the super Poincar\'e group can be parametrized as follows
$$
g(z^a,\o^{\mu\nu}) = \exp\left\{-ix^\mu P_\mu + \th^\a_A \sQ_\a^A + \bar{\th}^{B,\dot{\b}} \bar{\sQ}_{B,\dot{\b}} -i z \sCC\right\} \exp\left\{-\frac{i}{2}\o^{\mu\nu} J_{\mu\nu}\right\}.
$$
A representative of a conjugacy class can be given by the first factor, that is, by
$$
\tld{g}(z^a) = \exp\left\{-ix^\mu P_\mu + \th^\a_A \sQ_\a^A + \bar{\th}^{B,\dot{\b}} \bar{\sQ}_{B,\dot{\b}}-i z\sCC\right\}.
$$

The vielbein $e_a{}^b$ and the spin connection $w_a^{\mu\nu}$ can be obtained in the following way:
$$
{\tld{g}}^{-1}(z^a) \dd \tld{g}(z^a) = \dd z^a e_a{}^\mu P_\mu + \dd z^a e_a{}^\a_A\sQ_\a^A + \dd z^a e_a{}^{B,\dot{\b}} \bar{\sQ}_{B,\dot{\b}} + \dd z^a e_a{}^z \sCC+ \frac{1}{2}\dd z^a w_a^{\mu\nu} J_{\mu\nu}.
$$ 

Computations give the following values for $e_a{}^b$:
$$
\begin{array}{c|cccc}
a\downarrow, b\rightarrow & P_\mu & \sQ^A_\a & \bar{\sQ}_{A,\dot{\a}} & \sCC \\
\hline
\dd x^\nu & -i \d_\nu^\mu & 0 & 0 & 0 \\
\dd \th^\b_B & \bar{\th}^{B,\dot{\g}}\s^\mu_{\b\dot{\g}} & \d^\a_\b \d^B_A & 0 & \frac{1}{2}\ep_{\b\g}Z^{BC}\th_C^\g \\
\dd \bar{\th}^{B,\dot{\b}} & \th^\a_B \s^\mu_{\a\dot{\b}} & 0 & \d^{\dot{\a}}_{\dot{\b}} \d^B_A & \frac{1}{2}\ep_{\dot{\b}\dot{\g}}Z^\ast_{BC}\bar{\th}^{C,\dot{\g}} \\
\dd z & 0 & 0 & 0 & -i 
\end{array}
$$
The spin connection $w_a^{\mu\nu}$ appears to be zero.

The covariant derivative can be obtained as follows:
$$
\sD_b = {e^{-1}}_b{}^a \left( \pd_a + \frac{1}{2}w_a^{\mu\nu} S_{\mu\nu} \right).
$$
Having inverted the vielbein matrix we get the following expressions (compare with \Ref{PoincareDiff} and \Ref{SUSYDiff}):
\begin{equation}
\label{SUSYCovariant}
\begin{aligned}
\sD_\mu &= i\pd_\mu, \\
\sD_\a^A &= \frac{\pd}{\pd \th^\a_A} - i \s^\mu_{\a\dot{\b}}\bar{\th}^{A,\dot{\b}}\pd_\mu - \frac{i}{2} \ep_{\a\b} Z^{AB}\th^\b_B \frac{\pd}{\pd z}, \\
\bar{\sD}_{A,\dot{\a}} &= \frac{\pd}{\pd \bar{\th}^{A,\dot{\a}}} - i \th^\b_A \s^\mu_{\b\dot{\a}}\pd_\mu - \frac{i}{2}\ep_{\dot{\a}\dot{\b}}Z^\ast_{AB}\bar{\th}^{B,\dot{\b}}\frac{\pd}{\pd z}, \\
\sD_z &= i\frac{\pd}{\pd z}.
\end{aligned}
\end{equation}

Since the supersymmetry transformation define Killing vectors with respect to this connection we conclude that the covariant derivatives commute with generators of the supersymmetry, that is, with the supercharges $\sQ_\a^A$ and $\bar{\sQ}_{A,\dot{\a}}$. Of cause, this statement can be checked straightforwardly.

\begin{remark}
There is another way to deduce \Ref{SUSYCovariant} which is simpler and closely related to the traditional way to introduce ``long'' derivatives. Taking into account \Ref{SUSYTransformCoord} we conclude that the derivative with respect to $\th^\a_A$ does not transforms covariantly:
$$
\begin{aligned}
\frac{\pd}{\pd \th^\a_A} &= \frac{\pd {\th^\b_B}'}{\pd \th^\a_A} \frac{\pd}{\pd {\th^\b_B}'} + \frac{\pd {x^\mu}'}{\pd \th^\a_A} \frac{\pd}{\pd {x^\mu}'} + \frac{\pd z'}{\pd \th^\a_A} \frac{\pd}{\pd z'} \\
&= \frac{\pd}{\pd {\th^\a_A}'} + \frac{1}{2}\o^{\mu\nu} \s_{\mu\nu}{}^\b{}_\a \frac{\pd}{\pd {\th^\b_A}'} \\
&- i \s^\mu_{\a\dot{\b}}\bar{\z}^{A,\dot{\b}}\frac{\pd}{\pd {x^\mu}'} - \frac{i}{2}\z^\b_B \ep_{\b\a}Z^{BA}\frac{\pd}{\pd z'}.
\end{aligned}
$$
The requirement that the last line in this expression is absent leads us directly to \Ref{SUSYCovariant}.
\end{remark}

The commutation rules for the covariant derivatives are the following:
\begin{equation}
\label{SUSYDerComm}
\begin{aligned}
\{ \sD^A_\a,  \bar{\sD}_{B,\dot{\b}} \} &= -2i \s^\mu_{\a\dot{\b}}\pd_\mu \d^A_B, \\
\{ \sD^A_\a, \sD^B_\b \} &= - i \ep_{\a\b} Z^{AB} \frac{\pd}{\pd z}, \\
\{ \bar{\sD}_{A,\dot{\a}}, \bar{\sD}_{B,\dot{\b}} \} &= - i \ep_{\dot{\a}\dot{\b}} Z^\ast_{AB} \frac{\pd}{\pd z}.
\end{aligned}
\end{equation}
All others are trivial. They could be used to reconstruct the curvature and the torsion of the superspace, but we will not need them.

Let us also introduce new coordinates which are covariantly constant in the $\bar{\th}^{A,\dot{\a}}$ and $z$ directions:
\begin{equation}
\label{SUSYCoord}
y^\mu = x^\mu - i \th_A \s^\mu\bar{\th}^A.
\end{equation}
It satisfies
$$
\bar{\sD}_{A,\dot{\a}} y^\mu = \sD_z y^\mu = 0.
$$


\section{Supermultiplets}

In this section we describe some supermultiplets which will be useful for the following.

In the spirit of field theory, where particles are seen as some irreducible representation of the Poincar\'e group, we would like to describe irreducible representations of the super Poincar\'e group. However, there is a difference. In the super case an irreducible multiplet contains more than one particle. At least, it contains bosons and fermions. Therefore, we will describe families of particles by means of irreducible representations.
  
As an supersymmetric extension of the Wigner theorem \cite{Wigner} we can say that all super multiplets can be described by means of families of function defined on the superspace, and which transform under an (irreducible) representation of the Lorentz group (the group we have factored out).


\subsection{$\N=1$ chiral multiplet.} 

Consider the simplest case: $\N=1$ (and therefore the central charge is absent) and the scalar representation of the Lorentz group. That is, we consider a scalar function $\Phi(x,\th,\bar{\th})$. Notice, however, that this function provides a reducible representation of the super Poincar\'e group, since we can impose the condition
$$
\bar{\sD}_{\dot{\a}} \Phi(x,\th,\bar{\th}) = 0
$$
which commute with the supersymmetry transformation, since the covariant derivative does.

This constraint can be solved using the coordinate \Ref{SUSYCoord}. The result is
$$
\begin{aligned}
\Phi(y,\th) &= H(y) + \sqrt{2} \th \psi(y) + \th\th f(y) \\
&= H(x) + i\th \s^\mu\bar{\th} \pd_\mu H(x) - \frac{1}{4} (\th \th)( \bar{\th} \bar{\th}) \pd_\mu \pd^\mu H(x) \\
&+ \sqrt{2}\th\psi(x)  - \frac{i}{\sqrt{2}} \th\th (\pd_\mu \psi (x)\s^\mu\bar{\th})+ \th \th f(x).
\end{aligned}
$$
Here $H(x)$ is a scalar field, $\psi^\a(x)$ is a Weyl spinor and $f(x)$ is an auxiliary field which does not have any dynamics (Lagrangian's do not contain any of its derivatives).


\subsection{$\N=1$ vector multiplet.} 

Now consider a general scalar function defined on the $\N=1$ superspace, which satisfies the reality condition:
$$
V(x,\th,\bar{\th}) = V^\dag(x,\th,\bar{\th}).
$$ 
Its component expansion is
$$
\begin{aligned}
V(x,\th,\bar{\th}) &= \vf(x) + \sqrt{2}\th\chi(x) + \sqrt{2}\bar{\th}\bar{\chi}(x) +\th \th g(x) + \bar{\th}\bar{\th} g^\dag(x) + \th \s^\mu\bar{\th} A_\mu(x) \\
&- i(\bar{\th}\bar{\th})\th \left( \l(x) + \frac{1}{\sqrt{2}} \s^\mu\pd_\mu \bar{\chi}(x)\right) + i (\th \th) \bar{\th}\left( \bar{\l}(x) + \frac{1}{\sqrt{2}} \bar{\s}^\mu\pd_\mu \chi(x) \right) \\
&+ \frac{1}{2}(\th\th)( \bar{\th}\bar{\th}) \left( D(x) - \frac{1}{2} \pd_\mu \pd^\mu \vf(x) \right). 
\end{aligned}
$$

The reality condition shows that $\vf^\dag(x) = \vf(x)$, $D^\dag(x) = D(x)$ and $A_\mu^\dag(x) = A_\mu(x)$. Real vector field is naturally associated with a vector boson, which is a gauge boson of a gauge theory. Since such bosons are in the adjoint representation of the gauge group, it is reasonable to take the vector superfield itself in the adjoint.

In fact, this supermultiplets is not irreducible, it contains a chiral multiplet (also in the adjoint representation). To gauge it out we can consider the following transformation:
\begin{equation}
\label{SuperGauge}
\e^{2V} \mapsto \e^{2V'} = \e^{-i\gL^\dag} \e^{2V} \e^{i\gL}.
\end{equation}
where 
$$
\gL(y,\th) = \a(y) + \dots
$$
is a chiral multiplet. Under such a transformation the vector component $A_\mu(x)$ transforms as follows:
$$
A_\mu(x) \mapsto A_\mu'(x) = A_\mu(x) - \cD_\mu(\Re\a(x)),
$$
where $\cD_\mu$ is the covariant derivative with the connection $A_\mu(x)$: 
$$\cD_\mu = \pd_\mu -i [A_\mu,\cdot].$$
This formula justifies the identification $A_\mu(x)$ as a gauge boson.

There is a specific gauge where the component expansion of the vector superfield becomes quite simple. It is the \emph{Wess-Zumino gauge}. In that gauge fields $\vf(x)$, $\chi^\a(x)$, $\bar{\chi}_{\dot{\a}}(x)$ and $g(x)$ are eliminated. Therefore we have the rest:
$$
V_{WZ}(x,\th,\bar{\th}) = \th \s^\mu\bar{\th}A_\mu(x) - i (\bar{\th}\bar{\th})(\th \l(x)) + i (\th\th) (\bar{\th}\bar{\l}(x)) + \frac{1}{2}(\th\th)(\bar{\th}\bar{\th})D(x).
$$
\begin{remark}
Even having fixed the Wess-Zumino gauge we still have a freedom to perform the gauge transformation  (and this is the only remaining freedom).
\end{remark}

\begin{remark}
The Wess-Zumino gauge does not commute with the supersymmetry transformation.
\end{remark}


\subsection{Supersymmetric field strength}

There is another way to represent the same field content. We can find an expression which remains unchanged under \Ref{SuperGauge}. It is given by 
$$
W_\a(x,\th,\bar{\th}) = -\frac{1}{8} \bar{\sD}_{\dot{\a}}\bar{\sD}^{\dot{\a}}\e^{-2V(x,\th,\bar{\th})} \sD_\a \e^{2V(x,\th,\bar{\th})}.
$$
Its component expansion is (we use $y^\mu = x^\mu - i \th \s^\mu \bar{\th}$):
$$
W_\a(y,\th) = -i\l_\a(y) + \th_\a D(y) - i\s^{\mu\nu}{}_\a{}^\b \th_\b F_{\mu\nu}(y) - \th^\b \th_\b \s^\mu_{\a\dot{\b}} \cD_\mu \bar{\l}^{\dot{\b}}(y).
$$
In this formula we see the appearance of the field strength 
\begin{equation}
\label{Strength}
F_{\mu\nu}(x) = \pd_\mu A_\nu(x) - \pd_\nu A_\mu(x) -i[ A_\mu(x), A_\nu(x)]
\end{equation}
which corresponds to the connection $A_\mu(x)$.

The superfield $W_\a(x)$ is chiral: $\bar{\sD}_{\dot{\a}} W_\a(x,\th,\bar{\th}) = 0$. In the abelian case it satisfies the following constraint (reality condition):
$$
\bar{\sD}_{\dot{\a}}\bar{W}^{\dot{\a}}(x,\th,\bar{\th}) = \sD^\a W_\a(x,\th,\bar{\th})
$$
which commute with the supersymmetry transformation. Therefore it can be seen as an another example of the Wigner theorem (now applied to a spinor function).

The reality condition assures that $D(x)$ is real field, and $F_{\mu\nu}(x)$ satisfied the Bianchi identity, which allows us to identify it with the curvature of a connection $A_\mu(x)$

In the non-abelian case these relations become more sophisticated. Namely, one should introduce the superconnection $\sA_\a^A$ and replace everywhere
$$
\begin{aligned}
\sD_\a^A \mapsto \tld{\sD}_\a^A &= \sD_\a^A -i \sA_\a^A, \\
\bar{\sD}_{A,\dot{\a}} \mapsto \tld{\bar{\sD}}_{A,\dot{\a}} &= \bar{\sD}_{A,\dot{\a}} +i \bar{\sA}_{A,\dot{\a}}, \\
\sD_\mu \mapsto \tld{\sD}_\mu &= \sD_\mu -i \sA_\mu.
\end{aligned}
$$
The relation with the the gauge field $A_\mu$ is established via
$$
{\left.\sA_\mu(x,\th,\bar{\th})\right|}_{\th=0,\bar{\th}=0} = A_\mu(x).
$$
Details can be found in \cite{SUSYandSUGRA,West}.


\subsection{$\N=2$ chiral multiplet \cite{GaugeMultiplet}.}

The most natural superfield representation for the $\N=2$ chiral multiplet is given in the extended superspace, which has the coordinates $x^\mu, \theta^\alpha_A, \bar{\theta}_{\dot{\alpha}}^A$, $A = 1,2$. The chirality condition for scalar superfield $\Psi(x,\th,\bar{\th},z)$ means that 
$$
\bar{\sD}_{A,\dot{\a}} \Psi(x,\th,\bar{\th},z) = 0.
$$
Using the algebra of covariant derivatives we see that it implies that this superfield does not depend on central charge coordinate $z$.

As usual when we consider chiral multiplets we introduce covariantly constant coordinate $y^\mu = x^\mu - i \th_A \s^\mu \bar{\th}^A$. The component expansion for the $\N=2$ chiral multiplet is the following:
\begin{equation}
\label{SUFI}
\begin{aligned}
\Psi(y,\th) &= H(y) + \sqrt{2}\th_A\psi^A(y) +\frac{1}{\sqrt{2}}\th_A\s^{\mu\nu}\th^A F_{\mu\nu}(y) + \frac{1}{\sqrt{2}}  \th_A L^A{}_B(y)\th^B \\
&- \frac{2i\sqrt{2}}{3} (\th^A\th^B)\left(\th_A \left\{\s^\mu \cD_\mu \bar{\psi}_B(y) + \frac{1}{\sqrt{2}} [H^\dag(y),\psi_B(y)]\right\}\right) \\
&- \frac{1}{3} (\th^A\th^B)(\th_A\th_B) \left( \cD^\mu \cD_\mu H^\dag(y) - [H^\dag(y),D(y)] - \frac{i}{\sqrt{2}} \{ \bar{\psi}^C(y),\bar{\psi}_C(y) \} \right).
\end{aligned}
\end{equation}
The matrix $L^A{}_B$ consists of auxiliary fields. This superfield is not an arbitrary chiral $\N=2$ superfield. It subjects to the following reality conditions (compare with \Ref{sigmaReality})
$$
\ep^{AC}L^\ast_C{}^D\ep_{DB} = L^A{}_B.
$$
This auxiliary field matrix can be expressed with the help of auxiliary fields for $\N=1$ chiral and vector multiplets as follows (we denote $f(y) = f'(y) + if''(y)$ and $f^\dag(y) = f'(y) - i f''(y)$)
$$
L^A{}_B(y) = \left(
\begin{array}{cc}
iD(y) & -\sqrt{2}f(y) \\
\sqrt{2}f^\dag(y) & -iD(y) 
\end{array}
\right) = -i\sqrt{2}f''(y) \t_1 - i \sqrt{2}f'(y) \t_2 +  iD(y) \t_3.
$$

Covariantly this restriction can be written (in the abelian case) as
$$
\sD^A \sD^B \Psi(x,\th,\bar{\th}) = \bar{\sD}^A \bar{\sD}^B {[\Psi(x,\th,\bar{\th})]}^\dag.
$$
In the non-abelian case we should introduce superconnection as in the case of the vector multiplet.

Using the language of the $\N=1$ supermultiplets one can re-express this superfield as follows:
$$
\Psi(y,\th) = \Phi(y,\th_1) + i\sqrt{2}\th_2 W(y,\th_1) + \th_2\th_2 \G(y,\th_1)
$$
where $\Phi(y,\th)$ and $\G(y,\th)$ are two $\N=1$ chiral multiplets. These two chiral supermultiplets are not independent. The second one can be obtained from the first one and the vector superfield in the following way:
$$
\G(y,\th) = - \frac{1}{2} \int \dd^2 \bar{\th} \Phi^\dag(y - 2i \th \s \bar{\th},\bar{\th}) \e^{2V(y, \th,\bar{\th}) } 
$$
While doing the integral in the righthand side $y^\mu$ is supposed to be fixed.


The supersymmetry transformation for $\N=2$ chiral multiplet is given by 
$$
\d_{\z,\bar{\z}} \Psi(x,\th,\bar{\th}) = (\z_A \sQ^A + \bar{\z}^A\bar{\sQ}_A) \Psi(x,\th,\bar{\th}).
$$
The component expansion for this equation gives
\begin{equation}
\label{SUSYTransform}
\begin{aligned}
\d_{\z,\bar{\z}} H &= \sqrt{2} \z_A \psi^A, \\
\d_{\z,\bar{\z}} H^\dag &= \sqrt{2} \bar{\z}^A \bar{\psi}_A, \\
\d_{\z,\bar{\z}} \psi^A_\a &=  \s^{\mu\nu}{}_\a{}^\b \z^A_\b F_{\mu\nu} +  i\z^A_\a [H,H^\dag] - i\sqrt{2} \s^\mu_{\a\dot{\b}} \bar{\z}^{A,\dot{\b}}\cD_\mu H , \\
\d_{\z,\bar{\z}} \bar{\psi}_A^{\dot{\a}} &= \bar{\s}^{\mu\nu,\dot{\a}}{}_{\dot{\b}}\bar{\z}^{\dot{\b}}_A F_{\mu\nu}-  i\bar{\z}_A^{\dot{\a}}[H,H^\dag] - i \sqrt{2} \bar{\s}^{\mu,\dot{\a}\b} \z_{A,\b}\cD_\mu H^\dag, \\
\d_{\z,\bar{\z}} A_\mu &= i \z^A \s_\mu \bar{\psi}_A - i \psi^A \s_\mu \bar{\z}_A.
\end{aligned}
\end{equation}
Here we have come slightly ahead and used the equations of motion which follow from the action \Ref{MicroActionComponent} of $\N=2$ super Yang-Mills theory:
$$
\begin{aligned}
f &= 0, & D &= [H,H^\dag]. \\
\end{aligned}
$$

Let us finally rewrite for further references the $\N=2$ superfield \Ref{SUFI} in less $SU(2)_I$ covariant and more tractable way. We have
$$
\begin{aligned}
\Psi(y,\th) &= H(y) + \sqrt{2}\th_1 \psi^1(y) + \sqrt{2}\th_2 \psi^2(y) \\
&+ \th_1\th_1 f(y) + \th_2\th_2 f^\dag(y) + i\sqrt{2}\th_1\th_2 D(y) + \sqrt{2} \th_2 \s^{\mu\nu} \th_1 F_{\mu\nu} \\
&-i\sqrt{2}(\th_1\th_1)(\th_2\s^\mu \cD_\mu \bar{\psi}_2(y)) + i (\th_1\th_1)(\th_2[H^\dag(y),\psi^1(y)]) \\
&-i\sqrt{2}(\th_2\th_2)(\th_1\s^\mu \cD_\mu \bar{\psi}_1(y)) - i (\th_2\th_2)(\th_1[H^\dag(y),\psi^2(y)]) \\
&-(\th_1\th_1)(\th_2\th_2)\left( \cD^\mu \cD_\mu H^\dag - [H^\dag,D]+i\sqrt{2}\{\bar{\psi}_1,\bar{\psi}_2\} \right).  
\end{aligned}
$$


\subsection{Hypermultiplet.}

The matter in $\N=2$ supersymmetric theory can be described with the help of the hypermultiplet \cite{FayetHyper,SohniusHyper}.

In a $SU(2)_I$ invariant way it can be described as follows. Consider an $SU(2)_I$ doublet of scalar superfields $Q^A(x,\th,\bar{\th},z)$. Its derivatives $\sD^A_\a Q^B$ and $\bar{\sD}^A_{\dot{\a}} Q^B$ belong to the reducible representation $\mbf{1}\oplus \mbf{3}$ of $SU(2)_I$. If we project out the three dimensional representation, the rest will be the hypermultiplet. That is, we impose the following condition
$$
\begin{aligned}
\sD^A_\a Q^B + \sD^B_\a Q^A &= 0 & &\Leftrightarrow & \sD^A_\a Q^B &= \frac{1}{2} \ep^{AB} \sD_{C,\a} Q^C, \\
\bar{\sD}^A_{\dot{\a}} Q^B + \bar{\sD}^B_{\dot{\a}} Q^A &= 0 & &\Leftrightarrow & \bar{\sD}^A_{\dot{\a}} Q^B &= \frac{1}{2} \ep^{AB} \bar{\sD}_{C,\dot{\a}} Q^C.
\end{aligned}
$$

\begin{remark}
The superfield $Q^A$ is not chiral. Therefore, it does depend on the central charge coordinate $z$. Since the matrix $Z^{AB}$ is antisymmetric, it is proportional to $\ep^{AB}$ when $\N=2$. After an appropriate rescaling of $z$ we can put simply $Z^{AB} = \ep^{AB}$  
\end{remark}

Consider the (infinite, thanks to the presence of the bosonic coordinate $z$) series which represents this superfield. Some first terms are given by the following formula
$$
Q^A(x,\th,\bar{\th},z) = q^A(x) + \sqrt{2}\th^A \chi(x) + \sqrt{2}\bar{\th}^A \bar{\tld{\chi}} - i z X^A(x) + \dots
$$ 
Here $(q^1,q^2)$ are an $SU(2)_I$ doublet of complex scalars, $\chi_\a$ and $\tld{\chi}_\a$ are two spinor singlets and $(X^1,X^2)$ are an doublet of auxiliary fields. Terms contained in ``$\dots$'' can be expressed as spacetime derivatives of these fields.

The on-shell supersymmetry transformations for the massive hypermultiplet coupled with the gauge multiplet are given by 
\begin{equation}
\label{SUSYHyp}
\begin{aligned}
\d_{\z,\bar{\z}} q^A &=  \sqrt{2}\z^A \chi + \sqrt{2}\bar{\z}^A \bar{\tld{\chi}}, \\
\d_{\z,\bar{\z}} \chi_\a &= i\sqrt{2} \s^\mu_{\a\dot{\a}} \bar{\z}^{\dot{\a}}_A\cD_\mu q^A - 2i \bar{\z}^A_{\dot{\a}} q^\dag_A H + \sqrt{2} m q^A \z_{A,\a},\\
\d_{\z,\bar{\z}} \bar{\tld{\chi}}_{\dot{\a}} &= i\sqrt{2} \z^{A,\a}\s^\mu_{\a\dot{\a}} \cD_\mu q_A - 2i \z^A_\a q^\dag_A H + \sqrt{2} m q^A \bar{\z}^{\dot{\a}}_A.
\end{aligned}
\end{equation}
where $H(x)$ is a Higgs field from the $\N=2$ chiral multiplet, $m$ is the massive hypermultiplet matter. In the covariant derivative $\cD_\mu = \pd_\mu - i A_\mu$ we use the connection which is also the part of the chiral multiplet. 

\begin{remark}
The multiplication $H q^A$ should be understood as follows: in the adjoint representation we have $H = H^a T^{\mr{adj}}_a$, $a=1,\dots,\dim G$ where $T^{\mr{adj}}_a$ are the generators of the gauge group (structure constants). Taking a representation $\vr$ of the gauge group one considers corresponding generators $T^{\vr}_a$. The superfield $Q^A$ is acted on by this representation. And $H q^A$ means $H^a T_a^{\vr} q^A$ which is well-defined. The same remark should be taken into account while considering $\cD_\mu q^A$.
\end{remark}

This field content can be repackaged into two $\N=1$ chiral superfield. Unfortunately, in non-$SU(2)_I$ invariant way. However, the practical computations with repackaged superfields are much simpler. These two chiral superfields have the following form:
$$
\begin{aligned}
Q(y,\th) &= q(y) + \sqrt{2}\th \chi(y) + \th \th X(y), \\
\tld{Q}(y,\th) &= \tld{q}^\dag(y) + \sqrt{2}\th\bar{\tld{\chi}}(y) + \th\th \tld{X}^\dag (y), 
\end{aligned}
$$
where $q(x) \equiv q^1(x)$, $\tld{q}^\dag(x) \equiv q^2(x)$, $X(x) \equiv  X^1(x)$ and $\tld{X}^\dag(x) \equiv X^2(x)$. Note that the hermitian conjugation in the last line does not affect on $y^\mu$. Also note that the $\N=1$ chiral multiplet $Q$ is acted on by the representation $\vr$ of the gauge group, whereas $\tld{Q}$ -- by the dual representation $\vr^\ast$.

%% file: yangmills.tex

\label{Outline}

In this chapter we give an outline of known facts about $\mathcal{N} = 2$ supersymmetric Yang-Mills theory: the action, the famous Seiberg-Witten theory, which allows to compute the non-perturbative corrections to the Green functions via the prepotential (see its definition is the section \ref{scn:WilsonAction}), and the stringy tools used in this theory. Also we discuss the twist which makes it a topological field theory and BV derivation of this topological field theory.


\section{The field content}

The field content of the pure $\N = 2$ super Yang-Mills theory is described by the $\N=2$ chiral superfield \Ref{SUFI}:
$$
\begin{aligned}
& & &A_\mu(x) & & \\
\psi_\alpha^1(x) &= \psi_\a(x) & &  & \psi_\alpha^2(x) &= \l_\a(x) \\
& & &H(x) & & 
\end{aligned}
$$
where 
\begin{itemize}
\item $A_\mu(x)$ is a gauge boson, 
\item $\psi_\a^A(x)$, $A = 1,2$ are two gluinos, represented by Weyl spinors, and 
\item $H(x)$ is the Higgs field, which is a complex scalar.
\end{itemize}

We have arranged these fields in this way in order to make explicit the $SU(2)_I$ symmetry. It acts on the rows. Accordingly $A_\mu(x)$ and $H(x)$ are singlets and $(\psi_\a^1(x),\psi_\a^2(x))$, are a doublet. 

Since vector bosons are usually associated with a gauge symmetry, $A_\mu(x)$ is supposed to be a gauge boson corresponding to a gauge group $G$. It follows that it transforms in the adjoint representation of $G$. To maintain the $\N = 2$ supersymmetry $\psi_\a^A(x)$ and $H(x)$ should also transform in the adjoint representation. Therefore, all the fields are supposed to be $\alg{g} = \Lie(G)$ valued functions. 

Let us also describe the matter hypermultiplet. The field content is the following: 
$$
\begin{aligned}
& & &\chi_\a(x) & & \\
q^1(x) &= q(x) & &  & q^2(x) &= \tld{q}^\dag(x) \\ 
& & &\bar{\tld{\chi}}_{\dot{\a}}(x) & & 
\end{aligned}
$$
where $\chi_\a$ and $\bar{\tld{\chi}}_{\dot{\a}}$ are two $SU(2)_I$ singlets Weyl spinors. $q$ and $\tld{q}^\dag$ form a doublet of complex bosons. To couple the matter fields with the gauge multiplet we should specify a representation $\vr$ of the gauge group. Then $q$ and $\chi$ are acted on by the gauge transformation in this representation, whereas $\tld{q}$ and $\tld{\chi}$ by the dual one $\vr^\ast$.


\section{The action}

Let us now write the action for $\N=2$ supersymmetric Yang-Mills theory. This action is uniquely defined by the following requirements (see, for example, \cite{BilalDuality,BilalSW,LecturesSWandIntegr})
\begin{itemize}
\item it contains only two derivative terms, and not higher,
\item it is renormalizable.
\end{itemize}

The action which satisfies these conditions is (after integration out all the auxiliary fields)
\begin{equation}
\label{MicroActionComponent}
\begin{aligned}
S_{\mr{YM}} &= \frac{\gTh_0}{32 \pi^2 h^\vee}\int \dd^4 x\Tr F_{\mu\nu} \star F^{\mu\nu} \\
&+  \frac{1}{g^2_0 h^\vee} \int \dd^4 x \Tr\left\{ - \frac{1}{4} F_{\mu\nu}F^{\mu\nu}  + \cD_\mu H^\dag \cD^\mu H - \frac{1}{2} {[H,H^\dag]}^2 \right. \\
&\left.+ i \psi^A \s^\mu\cD_\mu \bar{\psi}_A  - \frac{i}{\sqrt{2}}  \psi_A [H^\dag,\psi^A] + \frac{i}{\sqrt{2}}\bar{\psi}^A[H, \bar{\psi}_A] \right\}.
\end{aligned}
\end{equation}
Using $\N=1$ superfields one can rewrite this action as follows:
$$
S_{\mr{YM}} = \frac{1}{8\pi h^\vee} \Im \left\{ \t_0 \Tr\left(\int \dd^4 x \dd^2 \th  W^\a W_\a  + \int \dd^4 x \dd^2 \th \dd^2 \bar{\th} \Phi^\dag \e^{2V} \Phi \right) \right\}.
$$
Here $\displaystyle\tau_0 = \frac{4\pi i}{g^2_0} + \frac{\gTh_0}{2\pi}$, $g^2_0$ being the Yang-Mills coupling constant (and the Plank constant as well) and $\gTh_0$ is the instanton angle. Its contribution to the action is given by the topological term, $\gTh_0 k$ where $k\in \Integers$ is the instanton number:
\begin{equation}
\label{InstNumb}
k = - \frac{1}{32\pi^2 h^\vee} \int \dd^4 x \Tr F_{\mu\nu} \star F^{\mu\nu}.
\end{equation}
Here $h^\vee$ is the dual Coxeter number. Its values for different groups are collected in the Appendix \ref{LieAlg}.

The most natural form of this action can be obtained with the help of $\N=2$ chiral superfield \Ref{SUFI}:
\begin{equation}
\label{MicroAction}
S_{\mr{YM}} = \frac{1}{4\pi h^\vee} \Im \left\{ \int \dd^4x \dd^4\th \frac{\t}{2} \Tr \Psi^2  \right\}.
\end{equation}

The coupling constant $g$ is running in the Yang-Mills theories. At high energies it can go to infinity (Landau p\^ole) or to zero (or, in marginal cases, remain finite). The theories with the second and third type of behavior are referred as asymptotically free. Physically it means that the action \Ref{MicroActionComponent} better describes the model  at high energies. So, if we take the high energy limit, we will see the action becomes exact. 

Therefore, for asymptotically free theories the action \Ref{MicroActionComponent} is the exact or bare or microscopic one. However, when one goes from high to low energies, the bare action is getting dressed. The perturbative and non-perturbative correction should be taken into account and we arrive to the Wilsonian effective action.


\section{Wilsonian effective action}
\label{scn:WilsonAction}

By definition the Wilsonian effective action $S_{\rm{eff}}$ is defined in a similar way as a standard effective action, $\Gamma_{\rm{eff}}$. However there are some distinctions. The latter is defined as a generating functional of one-particle irreducible Feynman diagrams. It can be obtained from the generating functional of all Feynman diagrams $W$ by the Legendre transform. The former type of effective actions, the Wilsonian one, is defined in as $\Gamma_{\mr{eff}}$ except that one introduces explicitly an infra-red cut-off $\Lambda$ (often we will call it dynamically generated scale). Therefore, the Wilsonian effective action is cut-off dependent. There is no big difference between $S_{\mr{eff}}$ and $\Gamma_{\mr{eff}}$ when there are no massless particles in the theory. However, in the $\N=2$ super Yang-Mills theory there are such particles. The property that makes plausible to consider the Wilsonian effective action is that it is a holomorphic function of $\Lambda$, which is not the case for $\Gamma_{\mr{eff}}$. 

If one requires that $\N=2$ supersymmetry remains unbroken in low energy region, one can get very restrictive conditions to the form of the Wilsonian effective action. Namely when one goes to the low energies region, one observes that thanks to the term 
\begin{equation}
\label{FlatDirection}
- \frac{1}{2}{[H,H^\dag]}^2
\end{equation}
in the microscopic action massless Higgs fields satisfy the equation $[H,H^\dag] = 0$ and therefore belong to the Cartan subalgebra of the gauge group $G$. The same conclusion is also valid for the gauge field. The non-perturbative analysis shows that at low energies $\N=2$ supersymmetric Yang-Mills theory is alway in the Coulomb branch, where one finds $r = \rank G$ copies of the QED with photon fields being $A_{l,\mu}(x)$, $l=1,\dots,r$.

Having integrated out all the massive fields one gets the Wilsonian effective action, which describes the physics at low energies. The leading term of the effective action (containing up to two derivatives and four fermions terms) can be obtained by relaxing the renormalizability condition. The result is the following
$$
S_{\mr{eff}} = \frac{1}{8\pi} \Im \left\{\frac{1}{2\pi i}\int \dd^4 x \dd^2 \th \Prep^{lm}(\Phi) W^\a_lW_{m,\a} + \frac{1}{2\pi i}\int \dd^2 \th \dd^2 \bar{\th}{ \left[\Phi^\dag \e^{2V} \right]}_l \Prep^l(\Phi)\right\}.
$$
For this action to be $\N=2$ supersymmetric the following conditions should be satisfied :
$$
\begin{aligned}
\Prep^l(a) &= \frac{\pd \Prep(a)}{\pd a_l}, & \Prep^{lm}(a) &= \frac{\pd^2 \Prep(a)}{\pd a_l \pd a_m}.
\end{aligned}
$$
Here we have introduced a holomorphic function $\Prep(a)$ on $r$ variables $a_l$, which is called the \emph{prepotential}. 

As usual, the most compact form of the effective action can be obtained with the help of the $\N=2$ superfield \Ref{SUFI}:
$$
S_{\mr{eff}} = \frac{1}{4\pi}\Im \left\{  \frac{1}{2\pi i}\int \dd^4 x \dd^4 \th \Prep(\Psi )\right\} .
$$

The expression of the classical prepotential can be easily read from \Ref{MicroAction}:
\begin{equation}
\label{ClassPrep}
\Prep_{\mr{class}}(a) = \pi i \t_0 \sum_{l=1}^r{a_l}^2 = \pi i\t_0 \langle a, a\rangle.
\end{equation}
Note that we use the normalization of the prepotential which differs from some other sources by the factor $2\pi i$.

Further analysis \cite{Prepotential} shows that all perturbative contributions to the prepotential consist of the 1-loop term\footnote{this fact is closely related to the topological nature of the $\N=2$ super Yang-Mills theory, see section \ref{scn:TopTwist}}. The expression one gets is 
\begin{equation}
\label{pertPrep}
\begin{aligned}
\Prep_{\mr{pert}}(a,\gL) &= - \sum_{\a \in \gD^+} {\langle \a, a\rangle }^2 \left(\logmodL{\langle \a, a\rangle} - \frac{3}{2}\right) \\
&+ \frac{1}{2} \sum_{\vr\in\mr{reps}} \sum_{\l \in \mbf{w}_\vr} {\left( \langle a,\l\rangle + m_\vr \right)}^2 \left( \logmodL{ \langle a,\l\rangle + m_\vr} - \frac{3}{2} \right)
\end{aligned}
\end{equation}
where $\gL$ is the dynamically generated scale. This formula gives the prepotential for the Yang-Mills theories with matter multiplets which belong to  representations $\vr$ of the gauge group and have masses $m_\vr$. In this formula the highest root is supposed to have length 2.

\begin{remark}
Term $-\frac{3}{2}$ is not fixed by the perturbative computations. It describe the finite renormalization of the classical prepotential. Our choice is made for the simplicity of further formulae.
\end{remark}

The description of the positive root for classical Lie algebras are in the Appendix \ref{LieAlg}.


\section{Seiberg-Witten theory}
\label{scn:SeibergWitten}

Besides the classical \Ref{ClassPrep} and the perturbative \Ref{pertPrep} parts of the prepotential, there is also a third part, due to the non-perturbative effects and coming from the instanton corrections to the effective action.

The classical $\N=2$ syper Yang-Mills theory has internal $U(2) = SU(2)_I \times U(1)_\R$ symmetry. Thanks to ABJ anomaly, which appears on the quantum level, the second factor is broken down to $\Integer_\b$ where $\b$ is the leading (and unique thanks to topological nature of the theory) coefficient of the $\b$-function. $\b$ is an integer and for assymptotically free theories non-negative, therefore the object $\Integer_\b \equiv \Integer / \b \Integer$ does make sens. It is computed in Appendix \ref{LieAlg} According to this the general form of the non-perturbative contribution can be represented by the follwing series over $\Lambda$:
\begin{equation}
\label{instPrep}
\Prep_{\mr{inst}}(a,\Lambda) = \sum_{k=1}^\infty \Prep_k (a) \Lambda^{k\b},
\end{equation}

In order to make evident that this expansion is nothing but the nonperturbative expansion caused by contributions of different vacua let us consider the renormgroup flow for the coupling constant $\t$. It can be easily obtained from \Ref{pertPrep} and is given by 
$$
\t(\gL_1) = \t(\gL_2) + \frac{\b}{2\pi i} \ln \frac{\gL_1}{\gL_2}.
$$  

Let us choose the energy scale in such a way, that the renormalization group flow becomes $\t(\Lambda) = \t_0 + \dfrac{\b}{2\pi i} \ln \Lambda$. Introduce the instanton counting parameter
\begin{equation}
\label{InstantonParameter}
q = \e^{2\pi i \t} = \e^{-\frac{8\pi^2}{g^2}} \e^{i\gTh} = \e^{2\pi i \t_0} \Lambda^\b.
\end{equation}

\begin{remark}
When $\b\neq 0$ we can completely neglect $\t_0$ and in this case we have $q \mapsto \Lambda^\b$. For the conformal theories, that is, for the theories where $\beta = 0$, we have $q = \e^{2\pi i \t_0}$. In both cases we can replace $\gL^\b$ by $\e^{-\frac{8\pi^2}{g^2}} \e^{i\gTh}$.
\end{remark}

Taking into account the fact that the value of the Yang-Mills action on the instanton background with the instanton number $k$ is $\ds -\frac{8\pi^2 k}{g^2} + i \gTh k$ we conclude that the $\gL^\b$ expansion in the same as instanton expansion.

The non-perturbative constributions to the prepotential give rise to the instanton corrections to the Green functions (and therefore can be extracted from them \cite{ItoSasakura,ItoSasakuraSU,ItoSasakuraSUSYQCD}). However the direct calculation of their contribution is very complicated, thus making quite useful the Seiberg-Witten theory  \cite{SeibergWitten,SeibergWittenII}. In this section we will explain some basic aspects of this theory. More detailed explanation can be found, for example, in \cite{BilalSW,LecturesSWandIntegr}.

The key observation is that the kinetic term in the effective Wilson action is proportional to $\ds- \Im \frac{1}{2\pi i}\Prep^{lm}(H)$. Since this function is analytic, it can not be positive everywhere. Therefore such a description is valid only within a certain region of the moduli space. To find a universal description we involve the following geometrical fact: consider an algebraic curve, let $A_1,\dots,A_r$ and $B_1,\dots,B_r$ be its basic cycles which satisfy $A_l \# B_m = \d_{lm}$ and $\l_1,\dots,\l_r$ be holomorphic differentials such that
$$
\oint_{A_l} \l_m = \d_{lm}.
$$
Then the real part of the period matrix
$$
2\pi i B_{lm} = \oint_{B_l} \l_m
$$
is negatively defined.

Therefore, if find a meromorphic differential $\l$, depending on the quantum moduli space of the theory (set of vacuum expectations of the Higgs field $H(x)$), which we will denote $a_l$, such that
\begin{equation}
\label{SWprescr}
\begin{aligned}
\frac{\pd \l}{\pd a_l} &= \l_l, &
\oint_{A_l}\lambda  &= a_l, & 
&\mbox{and} & 
\oint_{B_l} \lambda &= \frac{1}{2\pi i}\frac{\pd \Prep(a)}{\pd a_l},
\end{aligned}
\end{equation}
we could assure the positivity of the kinetic term.

Another way to get the description of the prepotential in terms of an auxiliary algebraic curve is to account properly the monodromies of the vector 
$$
\vec{\v} = \left(
\begin{array}{c}
a_1 \\
a_D^1 \\
\vdots \\
a_r \\
a_D^r
\end{array}
\right)
$$
where $a^l_D = \dfrac{\pd \Prep (a)}{\pd a_l}$. It allows to write a differential (Schr\"odinger like) equation for $a_l, a^l_D$. Its solutions can be expressed with the help of hypergeometric functions, whose integral representations reproduce the prescription \Ref{SWprescr}.


\section{Topological twist}
\label{scn:TopTwist}

Another property of $\N=2$ supersymmetric Yang-Mills theory which will  be important in what follows is its relations to so-called topological (or cohomological) field theories \cite{TQFT,IntroToCohFT}.

Namely, the action \Ref{MicroAction}, up to a term, proportional to ${\Tr} F_{\mu\nu} \star F^{\mu\nu}$, which is purely topological itself, can be rewritten as a $\bar{\sQ}$-exact expression for a fermionic operator $\bar{\sQ}$. One can construct this operator by twisting the usual  supersymmetry generators $\bar{\sQ}_{A,\dot{\a}}$ in the following way:
$$
\bar{\sQ} = \ep^{A\dot{\a}}\bar{\sQ}_{A,\dot{\a}}.
$$

\begin{remark}
Note that in this expression we have mixed supersymmetry indices $A,B,\dots$ and space-time spinor indices $\dot{\a},\dot{\b},\dots$. Geometrically it corresponds to the redefinition of the Lorentz group of the theory. Indeed, the group of symmetries is\footnote{after the Wick rotation and passing form $SO(3,1)$ to $SO(4)$, whose cover is $SU(2)_L\times SU(2)_R$}
$$
SU(2)_L \times SU(2)_R \times SU(2)_I.
$$
Now we redefine the Lorentz group by taking $SU(2)_R^\prime = \diag SU(2)_R \times SU(2)_I$.
\end{remark}

Let us see in some details how does it work. According to this prescription we redefine the fields of the theory as follows:
$$
\begin{aligned}
\psi_{A,\a} &= \frac{1}{2}\s^\mu_{\a A} \psi_\mu, & \bar{\psi}^{A,\dot{\a}} = \frac{1}{2}\ep^{A\dot{\a}}\bar{\psi} + \frac{1}{2}\bar{\s}_{\mu\nu}{}^{A\dot{\a}}\bar{\psi}^{\mu\nu}.
\end{aligned}
$$
By definition field $\bar{\psi}^{\mu\nu}$ is anti-self-dual: 
$$
\bar{\psi}^{\mu\nu} = - i \star \bar{\psi}^{\mu\nu}.
$$
These expressions can be inverted as follows:
$$
\begin{aligned}
\psi^\mu &= \bar{\s}^{\mu,A\a} \psi_{A,\a}, & \bar{\psi} &= \ep_{\dot{\a}A} \bar{\psi}^{A,\dot{\a}} = \ep^{A\dot{\a}}\bar{\psi}_{A,\dot{\a}}, & \bar{\psi}^{\mu\nu} = \bar{\s}^{\mu\nu}{}_{\dot{\a}A}\bar{\psi}^{A,\dot{\a}}.
\end{aligned}
$$
\begin{remark}
Previously we had the following action of the hermitian conjugation: ${(\psi^A_\a)}^\dag = \bar{\psi}_{A,\dot{\a}}$. It corresponds to the fact that in the signature $SO(3,1)$ the complex conjugation swaps left and right spinors. Since we have redefined the Lorentz group it is naturally to expect that this map becomes more complicated. In particular, the action of hermitian conjugation should be accompanied by the charge conjugation matrix (which was trivial before).  
\end{remark}

The action \Ref{MicroActionComponent} becomes
\begin{equation}
\label{twistedAction}
\begin{aligned}
S_{\mr{YM}} &= \frac{\gTh_0}{32 \pi^2 h^\vee}\int \dd^4 x\Tr F_{\mu\nu} \star F^{\mu\nu} +  \frac{1}{g^2_0 h^\vee} \int \dd^4 x \Tr\left\{ - \frac{1}{4} F_{\mu\nu}F^{\mu\nu}  + \cD_\mu H^\dag \cD^\mu H - \frac{1}{2} {[H,H^\dag]}^2 \right. \\
&+ \left. \frac{i}{2} \psi^\mu \cD_\mu \bar{\psi} - \frac{i}{2}{(\cD_\mu \psi_\nu - \cD_\nu \psi_\mu)}^- \bar{\psi}^{\mu\nu}  + \frac{i}{2\sqrt{2}} \psi_\mu [H^\dag,\psi^\mu] - \frac{i}{2\sqrt{2}}\bar{\psi}[H, \bar{\psi}] - \frac{i}{2\sqrt{2}}\bar{\psi}^{\mu\nu}[H, \bar{\psi}_{\mu\nu}] \right\}.
\end{aligned}
\end{equation}

Now let us rewrite the supersymmetry transformations for these new fields. But before we introduce all set of the twisted supercharges:
$$
\begin{aligned}
\sQ_\mu &= \bar{\s}_\mu^{A\a} \sQ_{A,\a}, & \bar{\sQ}_{\mu\nu} &= \bar{\s}_{\mu\nu}{}^{A\dot{\a}} \bar{\sQ}_{A,\dot{\a}}.
\end{aligned}
$$




Having redefined the parameters of this transformation in the same way as the gluino fields $\z^\a_A, \bar{\z}^{A,\dot{\a}} \mapsto \z^\mu, \bar{\z}, \bar{\z}^{\mu\nu}$  we can easily deduce the action of operators $\bar{\sQ}, \sQ_\mu$ and $\bar{\sQ}_{\mu\nu}$ on the fields. We have
\begin{equation}
\label{Qtwisted}
\begin{aligned}
&\bar{\sQ} H &  &= 0,  &  &\sQ_\mu H & &= \sqrt{2} \psi_\mu, \\
&\bar{\sQ} H^\dag&  &= \sqrt{2}\bar{\psi}, & &\sQ_\mu H^\dag& &= 0, \\
&\bar{\sQ} \psi_\mu& &= 2i\sqrt{2} \cD_\mu H, & &\sQ_\mu \psi_\nu& &= -4 {\left(F_{\mu\nu}\right)}^+ +2ig_{\mu\nu} [H,H^\dag], \\
&\bar{\sQ} \bar{\psi}& &= 2i [H,H^\dag], & &\sQ_\mu \bar{\psi}& &= 2i\sqrt{2}\cD_\mu H^\dag, \\
&\bar{\sQ} \bar{\psi}_{\mu\nu}& &= -2 {\left(F_{\mu\nu}\right)}^-,  & &\sQ_\mu \bar{\psi}_{\r\t}& &= -2i\sqrt{2} {\left(g_{\mu\r}\cD_\t H^\dag - g_{\mu\t}\cD_\r H^\dag\right)}^-, \\
&\bar{\sQ} A_\mu& &=  -i \psi_\mu,& &\sQ_\mu A_\nu&  &= -i g_{\mu\nu}\bar{\psi} - 2i \bar{\psi}_{\mu\nu}, 
\end{aligned}
\end{equation}
$$
\begin{aligned}
&\bar{\sQ}_{\mu\nu} H& &= 0, \\
&\bar{\sQ}_{\mu\nu} H^\dag& &= \sqrt{2} \bar{\psi}_{\mu\nu}, \\
&\bar{\sQ}_{\mu\nu} \psi_\r& &= -2i \sqrt{2}{\left(g_{\mu\r}\cD_\nu H - g_{\nu\r} \cD_\mu H\right)}^-, \\
&\bar{\sQ}_{\mu\nu} \bar{\psi}& &= 2{\left(F_{\mu\nu}\right)}^-, \\
&\bar{\sQ}_{\mu\nu} \bar{\psi}_{\r\t}& &=-{\left(g_{\r\mu}{\left(F_{\t\nu}\right)}^- - g_{\t\mu} {\left(F_{\r\nu}\right)}^- +g_{\t\nu}{\left(F_{\r\mu}\right)}^- - g_{\r\nu} {\left(F_{\t\mu}\right)}^- \right)}^- \\
& & &+ i {\left( g_{\mu\r}g_{\nu\t} - g_{\mu\t}g_{\nu\r} \right)}^-[H,H^\dag], \\
&\bar{\sQ}_{\mu\nu} A_\r& &= -i {\left(g_{\mu\r} \psi_\nu - g_{\nu\r}\psi_\mu \right)}^-. 
\end{aligned}
$$

where we denote by
$$
{(F_{\mu\nu})}^\mp = \frac{1}{2}\left( F_{\mu\nu} \mp i \star F_{\mu\nu} \right) 
$$
the (anti)self-dual part of the antisymmetric tensor $F_{\mu\nu}$. It worth noting that $\bar{\psi}_{\mu\nu}$ and $\bar{\sQ}_{\mu\nu}$ are by definition anti-selfdual.

One should not be worried about the inconsistency, which appears at first sight in two first lines. Remember the remark before \Ref{twistedAction}.

The crucial observation about the action \Ref{twistedAction} (made for the first time by Witten \cite{TQFT} in the context of the Donaldson invariant theory) is that it is $\bar{\sQ}$ exact up to a topological term \Ref{InstNumb}. More precisely we see that
\begin{equation}
\label{ActionTwistedExact}
\begin{aligned}
S_{\mr{YM}} &=\Im \left[\bar{\sQ}\left\{ \frac{\t_0}{16\pi h^\vee} \int \dd^4 x \Tr \left( {(F_{\mu\nu})}^- \bar{\psi}^{\mu\nu} - i \sqrt{2} \psi^\mu \cD_\mu H^\dag +i \bar{\psi}[ H,H^\dag] \right) \right\}\right].
\end{aligned}
\end{equation}
In this computation we have used the equation of motion for $\bar{\psi}_{\mu\nu}$:
\begin{equation}
\label{ChiEqMotion}
{\left(\cD_\mu \psi_\nu - \cD_\nu \psi_\mu\right)}^- =  {\sqrt{2}}[H,\bar{\psi}_{\mu\nu}].
\end{equation}
This is an inevitable price to pay for the integration out auxiliary fields $f(x)$, $f^\dag(x)$ and $D(x)$ --- three degrees of freedom, therefore three equations of motion to use. 

The operator $\bar{\sQ}$ is nilpotent up to a gauge transformation (with the parameter $-2\sqrt{2}H$). To see this we should use the equation of motion for $\bar{\psi}_{\mu\nu}$ \Ref{ChiEqMotion}. Thanks to this property we can call it the BRST-like operator. As we shall see, the suffix ``like'' can be, actually, removed.

\begin{remark}
The topological term \Ref{InstNumb} is $\bar{\sQ}$ closed. Indeed
$$
\bar{\sQ} \int \dd^4 x F_{\mu\nu}\star F^{\mu\nu} = 2 i\int \dd^4 x \left( \cD_\mu \psi_\nu - \cD_\nu \psi_\mu \right)\star F^{\mu\nu} = - 4i \int \dd^4 x \psi_\nu \cD_\mu \star F^{\mu\nu}  = 0
$$
thanks to the Bianchi identity. 
\end{remark}


\section{BV quantization vs. twisting}
\label{BVvsTwist}

In previous section we have obtained topological action by appropriate twisting of $\N=2$ super Yang-Mills action \Ref{MicroActionComponent}. However, in order to perform some field theoretical computations we should do some extra work. 

First of all, as we have mentioned in passing by in the end of previous section the algebra of twisted fermionic operators is closed only on-shell. And, as usual in gauge theories, in order to be able to compute path integrals we should fix the gauge. This step requires to introduce a nilpotent (off-shell) BRST operator $\bar{\sQ}$.

An amazing property of the action \Ref{twistedAction} is that it can be obtained by an appropriate gauge fixing procedure for the topological action \cite{BaulieuSinger,GaugeFixingInTQFT,LabastidaPenici}. 
\begin{equation}
\label{Stop}
S_{\mr{top}} = \frac{\gTh_0}{32 \pi^2 h^\vee} \int \dd^4 x \Tr\Big\{ F_{\mu\nu} \star F^{\mu\nu} \Big\}.
\end{equation}
Therefore we can remove the suffix ``like'' and call $\bar{\sQ}$ the BRST operator. 

The topological action is invariant under the following transformation:
$$
A_\mu \mapsto A_\mu - \cD_\mu \a + \a_\mu,
$$
where $\a_\mu(x)$ is a $\alg{g}$ valued function constrained by the condition that $A_\mu(x) + \a_\mu(x)$ belong to the same gauge class that $A_\mu(x)$, whereas $\a(x)$ is an arbitrary $\alg{g}$ valued function. The invariance with respect to the last term is noting but the usual gauge invariance. The invariance with respect to the first transformation is guaranteed by the Bianchi identity for the curvature $F_{\mu\nu}$.

Following the standard BV procedure \cite{BVLinDep} one introduces the ghosts corresponding to each symmetry, $\psi_\mu$ and $c$. These fields are supposed to be fermions with associated ghost number $+1$. However, the direct implementation of the gauge fixing procedure leads to the singular Lagrangian. This is the consequence of the fact that $\a_\mu$ and $\a_\mu - \cD_\mu \b$ (where $\b$ is an arbitrary $\alg{g}$ valued function) produce the same transformation of $F_{\mu\nu}$. Therefore, further gauge fixing is needed. To this extent we introduce a ghost for ghosts $\phi$ which is boson with ghost number $+2$.

To fix the gauge we should impose the following conditions on fields (and ghosts):
$$
\begin{aligned}
\cD^\mu A_\mu &= 0, \\
{(F_{\mu\nu})}^- &= 0, \\
\cD^\mu \psi_\mu &= 0.
\end{aligned}
$$

To do this we will need some supplementary fields. Namely, for each gauge condition we introduce the Lagrange multiplier: bosons $b, H_{\mu\nu}$ and fermion $\eta$. Note that $H_{\mu\nu}$ is anti-selfdual. To them we associate the following ghost numbers: $(0,0,-1)$. Moreover, we will need a set of antighosts: $\bar{c}, \chi_{\mu\nu}$ and $\l$ with the following ghost numbers: $(-1,-1,-2)$. $\chi_{\mu\nu}$ is anti-selfdual. In order to simplify the references let us put the ghost number and the statistics of the introduced fields into the Table \ref{GhostNumbers}


\begin{table}[t]
\begin{center}
\begin{tabular}{||c||c||c|c|c||c|c|c||c|c|c||}
\hhline{|t:=:t:=:t:=:=:=:t:=:=:=:t:=:=:=:t|}
\textbf{Fields} & $\mbf{A_\mu}$ & $\mbf{c}$ & $\mbf{\psi_\mu}$ & $\mbf{\phi}$ & $\mbf{b}$ & $\mbf{H_{\mu\nu}}$ & $\mbf{\eta}$ & $\mbf{\bar{c}}$ & $\mbf{\chi_{\mu\nu}}$ & $\mbf{\l}$\\
\hhline{|:=::=::=:=:=::=:=:=::=:=:=:|}	
\textbf{Ghost number}  & $0$ &$+1$ & $+1$ & $+2$ & $0$ & $0$ & $-1$ & $-1$ & $-1$ & $-2$ \\
\hhline{||-||-||-|-|-||-|-|-||-|-|-||}
\textbf{Statistics} &  $B$ & $F$ & $F$ & $B$ & $B$ & $B$ & $F$ & $F$ & $F$ & $B$ \\
\hhline{|b:=:b:=:b:=:=:=:b:=:=:=:b:=:=:=:b|}
\end{tabular}
\end{center}
\caption{Ghost number and statistics}\label{GhostNumbers}
\end{table}


The BRST transformation for the ghosts which corresponds to this symmetry is the following:
\begin{equation}
\label{BRST}
\begin{aligned}
\bar{\sQ} A_\mu &= - \cD_\mu c -i \psi_\mu, \\
\bar{\sQ} c &= - \frac{i}{2}\{c,c\} - \phi, \\
\bar{\sQ} \psi_\mu &=  -i\cD_\mu \phi -i \{ c,\psi_\mu \}, \\
\bar{\sQ} \phi &= - i [c,\phi].
\end{aligned}
\end{equation}
For the Lagrange multipliers and antighosts we have the following expressions:
\begin{equation}
\label{BRST-Lagr}
\begin{aligned}
&\bar{\sQ}\bar{c} & &= b, & &\bar{\sQ} b & &= 0, \\
&\bar{\sQ}\chi_{\mu\nu} & &= H_{\mu\nu} - i \{c,\chi_{\mu\nu}\}, & &\bar{\sQ} H_{\mu\nu} & &= -i[\phi,\chi_{\mu\nu}] - i [c,H_{\mu\nu}], \\
&\bar{\sQ}\l & &= \eta - i[c,\l], & &\bar{\sQ}\eta & &= -i[\phi,\l] - i\{ c, \eta \}.
\end{aligned}
\end{equation}
One can see that the operator $\bar{\sQ}$ is nilpotent. Last two lines is rather unusual for the antighost-Lagrange multiplier transformation. However, one can check that the nilpotency condition is fulfilled \cite{LabastidaPenici}.

Now to construct a gauge fixed action we will need the last ingredient, the gauge fermion. This function has the ghost number $-1$. The appropriate choice is the following:
\begin{equation}
\label{VYM}
 V_{\mr{YM}} = \frac{1}{h^\vee g^2_0}\int \dd^4 x \Tr \left\{ \frac{1}{2}\chi^{\mu\nu} \left( {(F_{\mu\nu})}^- + \frac{1}{4} H_{\mu\nu}\right) + \frac{i}{8} \l \cD_\mu \psi^\mu + \bar{c}\left( \cD_\mu A^\mu + b\right)\right\}.
\end{equation}
The gauge fixed action can be written now as follows $S_{\mr{top}} + \bar{\sQ} V_{\mr{YM}}$.

In order to get the action \Ref{ActionTwistedExact} we add to the gauge fixed action another $\bar{\sQ}$-exact term $\bar{\sQ}V'$ where
$$
V' = -\frac{i}{128 h^\vee g^2_0} \int \dd^4 x \Tr \Big\{ \eta [\phi,\lambda ]\Big\}.
$$  
This term does not spoil the non-singularity of the kinetic term of the Lagrangian \cite{TQFT}. It is only responsible for the introduction of a potential.

In order to simplify further formulae we will  slightly change the notations. Namely, instead of using the $\N=2$ gauge multiplet we will  use the topological multiplet. Pragmatically it means that we redefine our fields as follows:
\begin{equation}
\label{TopCorrespondence}
\begin{aligned}
\phi& = - 2\sqrt{2} H, &\l &= -2\sqrt{2} H^\dag, \\
\chi_{\mu\nu} &= \bar{\psi}_{\mu\nu}, & \eta &= -4\bar{\psi}.
\end{aligned}
\end{equation}

\begin{remark}
Note that if we forget for a moment about the multiplet which is responsible for  gauge fixing, the  multiplet $(c,\bar{c},b)$,  then the action of the BRST operator coincides with \Ref{Qtwisted} if we use the introduced notations and use the equations of motion for $H_{\mu\nu}$: $H_{\mu\nu} = -2 {(F_{\mu\nu})}^-$. Moreover, the BRST operator $\bar{\sQ}$ becomes the same as the twisted supersymmetry operator \Ref{Qtwisted}. However in order to get the nilpotency of the BRST uperator up to a gaguge transformation we should use the equation of motion for $\chi_{\mu\nu}$ \Ref{ChiEqMotion}.
\end{remark}


\section{Dimensional reduction}
\label{DimensionalReduction}

In that follows it will be useful to keep in mind one more way to get $\N=2$ supersymmetric Yang-Mills action. 

Let us start with the six dimensional Minkowskian $\N=1$ super Yang-Mills theory. Suppose that the space is compactified in the following way: $\Real^{1,3} \times \Tor^2$ where $\Tor^2$ is a two dimensional torus described by coordinates $x^4$ and $x^5$:
$$
\begin{aligned}
x^4 &\equiv x^4 + 2\pi R_4, &  x^5 &\equiv x^5 + 2\pi R_5,
\end{aligned}
$$
where $R_4$ and $R_5$ are the radii of compactification.

Consider two six dimensional Weyl spinors which we denote as $\Psi^A$, $A=1,2$. We can buid from  them a single object, the symplectic Majorana spinor (see the Appendix \ref{AppendixSpinors} for some details) which is defined by the following condition:
\begin{equation}
\label{MajSyml}
\Psi^A = \ep^{AB}\C_6^+{\bar{\Psi}_B}^\tr = \ep^{AB}\C_6^+\gG^0 \Psi^\ast_B,
\end{equation}
where we have denoted $\gG^I = \g_6^I$, $I=0,1,2,3,4,5$. The matrices $\C_6^+$ and $\g_6^I$ are defined in the Appendix \ref{AppendixSpinors}.

The supersymmetric action can be written as follows:
\begin{equation}
\label{N=1,d=6}
S_{\N=1,d=6} =\frac{1}{g^2h^\vee}  \int \dd^4x \Tr\left\{ - \frac{1}{4}F_{IJ}F^{IJ} + \frac{i}{2} \bar{\Psi}_A \gG^I \cD_I \Psi^A\right\}.
\end{equation}

Now suppose that the radii of compactification of coordinates $x^4$ and $x^5$ is so small that all the fields can be considered as independent of them. It follows that $F_{\mu 4} = \cD_\mu A_4$ and $F_{\mu 5} = \cD_\mu A_5$. Therefore if we define 
\begin{equation}
\label{HiggsViaCompactification}
\begin{aligned}
H &= \frac{A_4 + iA_5}{\sqrt{2}}, & H^\dag &= \frac{A_4 - i A_5}{\sqrt{2}}
\end{aligned}
\end{equation}
we obtain $F_{45} = [H,H^\dag]$ and therefore
$$
-\frac{1}{4}F_{IJ}F^{IJ} = - \frac{1}{4}F_{\mu\nu}F^{\mu\nu} + \cD_\mu H \cD^\mu H^\dag - \frac{1}{2}{[H,H^\dag]}^2.
$$

Now let us represent Weyl spinors $\Psi^A$ in the following form
$$
\Psi^A = \left(
\begin{array}{c}
\psi^A_\a \\
\chi^{A,\dot{\a}} \\
0 \\
0 
\end{array}
\right).
$$
Then the symplectic Majorana condition can be recast as follows:
$$
\chi^{A,\dot{\a}} = \ep^{AB} \ep^{\dot{\a}\dot{\b}}\bar{\psi}_{B,\dot{\b}}.
$$
Recall that for four dimensional Weyl spinors bar means the complex conuugation: $\bar{\psi} = \psi^\ast$. 

Consequently we can write
$$
\frac{i}{2}\bar{\Psi}_A\gG^I\cD_I \Psi^A = i\psi^A \s^\mu \cD_\mu \bar{\psi}_A - \frac{i}{\sqrt{2}} \psi_A[H^\dag,\psi^A] + \frac{i}{\sqrt{2}} \bar{\psi}^A[H,\bar{\psi}_A\}.
$$
Therefore the $\N=1$, $d=6$ supersymmetric Yang-Mills action \Ref{N=1,d=6} becomes exactly the $\N=2$, $d=4$ action \Ref{MicroActionComponent}.


\section{Matter}
\label{Matter}

Let us finally describe the matter in the $\N=2$ super Yang-Mills theory \cite{LecturesTQFT,TopQCD,N=2SUSYand4Man}. The action for the hypermultiplet coupled with the gauge multiplet can be written in the $\N=1$ superfield language as follows (for the sake of simplicity we consider only one matter multiplet):
$$
S_{\rm{mat}} = \frac{1}{2h^\vee g^2_0 }\int \dd^4 x \Tr \left\{ \dd^2 \th \dd^2 \bar{\th} \left( Q^\dag \e^{2V} Q + \tld{Q}\e^{2V} \tld{Q}^\dag  \right) + 2 \Re\left(  \int\dd^2 \th \sqrt{2} \tld{Q}\Phi Q + m \tld{Q} Q\right) \right\}.
$$
where $m$ is the mass of the multiplet.

Consider first the massless case. In that situation after integration out the auxiliary fields $X$ and $\tld{X}$ we arrive to the following expression:
$$
\begin{aligned}
S_{\mr{mat}} &= \frac{1}{h^\vee g^2_0} \int \dd^4 x \Tr \left\{ \cD_\mu q_A^\dag \cD^\mu q^A + i \chi^\a \s^\mu_{\a\dot{\a}}\cD_\mu \bar{\chi}^{\dot{\a}} +  i \tld{\chi}^\a \s^\mu_{\a\dot{\a}}\cD_\mu \bar{\tld{\chi}}^{\dot{\a}} \right. \\
&+  \tld{\chi}^\a \phi \chi_\a -  \bar{\chi}_{\dot{\a}} \phi^\dag \bar{\tld{\chi}}^{\dot{\a}}+ \sqrt{2}q^\dag_A \psi^{A,\a} \chi_\a -  \sqrt{2}\bar{\chi}_{\dot{\a}} \psi^{\dot{\a}}_A q^A + \sqrt{2}q^\dag_A \bar{\psi}_{\dot{\a}}^A \bar{\tld{\chi}}^{\dot{\a}} - \sqrt{2}\tld{\chi}^\a q^A \psi_{A,\a}\\
&+\left.q^\dag_A \left( \phi \phi^\dag + \phi^\dag\phi \right) q^A - \frac{1}{2} \left( {q^\dag}^A {T^{\vr}}^a q^B + {q^\dag}^B {T^{\vr}}^a q^A \right)q^\dag_A T_a^{\vr}q_B\right\}.   
\end{aligned}
$$

For the matter multiplet the topological twist consists of the identification $q^A \mapsto q^{\dot{\a}}$. One can see that the twisted supersymmetry transformation \Ref{SUSYHyp} is not closed off-shell. It happens since we have already integrated out the auxiliary fields $X$ and $\tld{X}$. In order to close the transformation we introduce another set of auxiliary fields: $h_\a$ and $\tld{h}_\a$. As in the case of the pure Yang-Mills theory we see that their transformation properties differ from properties of the old ones. 

In order to simplify the formulae we introduce new fields $\bar{\mu}^{\dot{\a}}$, $\mu_{\dot{\a}}$, $\nu_\a$ and $\bar{\nu}^\a$ as follows:
$$
\begin{aligned}
\sqrt{2} \bar{\tld{\chi}}^{\dot{\a}} &= \mu^{\dot{\a}}, & \chi_\a &= \sqrt{2} \nu_\a, \\
\sqrt{2} \bar{\chi}_{\dot{\a}} &= \bar{\mu}_{\dot{\a}}, & \tld{\chi}^\a &= \sqrt{2} \bar{\nu}^\a.
\end{aligned}
$$

Closed off-shell (up to a gauge transformations) BRST operator $\bar{\sQ}$ is given by the following relations:
$$
\begin{aligned}
\bar{\sQ} q^{\dot{\a}} &= \mu^{\dot{\a}}, & \bar{\sQ} \mu^{\dot{\a}} &= \phi q^{\dot{\a}}, \\
\bar{\sQ} q^\dag_{\dot{\a}} &= \bar{\mu}_{\dot{\a}}, & \bar{\sQ} \bar{\mu}_{\dot{\a}} &= - q^\dag_{\dot{\a}}\phi, \\
\bar{\sQ} \bar{\nu}^\a &= \bar{h}^\a, & \bar{\sQ} \bar{h}^\a &= - \bar{\nu}^\a\phi, \\ 
\bar{\sQ} \nu_\a &= h_\a, & \bar{\sQ} h_\a &= \phi \nu_\a.
\end{aligned}
$$ 

\begin{remark}
The choice of the off-shell closed BRST transformation is not unique (see, for example, \cite{TopQCD}). However, this one makes the geometrical properties of the action clear. 
\end{remark}

Using these formulae one can check that the matter action can be rewritten as a $\bar{\sQ}$-exact expression: $S_{\mr{mat}} = \bar{\sQ} V_{\mr{mat}}$ where
\begin{equation}
\label{Vmat}
\begin{aligned}
 V_{\mr{mat}} &= \frac{1}{h^\vee g^2_0}\int \dd^4 x \Tr \left\{ - \frac{i}{2} \chi_{\mu\nu} q^\dag_{\dot{\a}}\bar{\s}^{\mu\nu,\dot{\a}}{}_{\dot{\b}}q^{\dot{\b}} - \frac{1}{4} \left( \bar{\mu}_{\dot{\a}} \l q^{\dot{\a}} - q^\dag_{\dot{\a}} \l \mu^{\dot{\a}} \right)\right.\\
&+ \left. 2\bar{\nu}^\a\left( \s^\mu_{\a\dot{\a}}\cD_\mu q^{\dot{\a}} - h_\a\right) - 2 \left( \cD_\mu q^\dag_{\dot{\a}}\bar{\s}^{\mu,\dot{\a}\a} - \bar{h}^\a\right) \nu_\a  \right\} .
\end{aligned}
\end{equation}

Now consider the general case, where the mass is not zero. After integration out all the auxiliary field in this case we obtain the following supplementary terms in the action:
$$
S_{\mr{mass}} = \frac{1}{h^\vee g^2_0} \int \dd^4 \Tr \left\{ -m^2 q^\dag_A q^A + \sqrt{2} m q^\dag_A H q^A + \sqrt{2} m  q_A^\dag H^\dag q^A - m \bar{\tld{\chi}}^{\dot{\a}}\bar{\chi}_{\dot{\a}} - m \tld{\chi}^\a \chi_\a  \right\}.
$$

The presence of the mass leads to the deformation of the supersymmetry transformation \Ref{SUSYHyp}. It turns to be that the proper version of the off-shell BRST transformation is given by 
\begin{equation}
\label{QtwistedMat}
\begin{aligned}
\bar{\sQ} q^{\dot{\a}} &= \mu^{\dot{\a}}, & \bar{\sQ} \mu^{\dot{\a}} &= \phi q^{\dot{\a}} + m q^{\dot{\a}}, \\
\bar{\sQ} q^\dag_{\dot{\a}} &= \bar{\mu}_{\dot{\a}}, & \bar{\sQ} \bar{\mu}_{\dot{\a}} &= - q^\dag_{\dot{\a}}\phi - m q^\dag_{\dot{\a}}, \\
\bar{\sQ} \bar{\nu}^\a &= \bar{h}^\a, & \bar{\sQ} \bar{h}^\a &= - \bar{\nu}^\a\phi - m \bar{\nu}_a, \\ 
\bar{\sQ} \nu_\a &= h_\a, & \bar{\sQ} h_\a &= \phi \nu_\a + m \nu_\a.
\end{aligned}
\end{equation}

Note that this deformation leads to a new property of the BRST operator. Before we had
$$
\bar{\sQ}^2 = \EuG(\phi)
$$
where $\EuG(\phi)$ is the gauge transformation with the parameter $\phi$. Now the new BRST operator satisfies the new relation:
$$
\bar{\sQ}^2 = \EuG(\phi) + \EuF(m).
$$
Here $\EuF(m)$ is an operator which does not affect on the gauge multiplet, but multiplies all the fields of the hypermultiplet by $\pm m$. This transformation can be seen as an infinitesimal version of the following transformation:
$$
\begin{aligned}
Q \mapsto Q' &= \e^m Q, & \tld{Q} \mapsto \tld{Q}' &= \e^{-m}\tld{Q}
\end{aligned}
$$
Therefore, this operator can be identified with the \emph{flavor group action}. In the case when we have only one hypermultiplet, the flavor group is $U(1)$. Note that usually one describes the $U(1)$ action as a multiplication by $\e^{i\th}$. It can be achieved after the redefinition $m\mapsto im$.

\begin{remark}
The deformation of the BRST operator described before provides only the part of the required mass term. However, the missed part can be restored after adding to to the action a BRST exact term $\bar{\sQ}V_{\mr{mass}}$ where
\begin{equation}
\label{Vmass}
V_{\mr{mass}} =\frac{1}{h^\vee g^2_0}\int \dd^4 x \Tr \left\{  -\frac{1}{4}m \left( q^\dag_{\dot{\a}} \mu^{\dot{\a}} + \bar{\mu}_{\dot{\a}} q^{\dot{\a}}\right) \right\} 
\end{equation}
\end{remark}

\begin{remark}
Since the operator $\bar{\sQ}$ is not nilpotent, the fact that the full action
$$
S = S_{\mr{top}} + \bar{\sQ}\left(V_{\mr{YM}} + V' + V_{\mr{mat}} + V_{\mr{mass}} \right) 
$$
is BRST invariant does not follow from the fact that it is (up to the topological term) BRST-exact. It follows from the invariance of $V_{\mr{YM}}$, $V'$, $V_{\mr{mat}}$ and $V_{\mr{mass}}$ with respect to the transformation generated by $\bar{\sQ}^2$. 
\end{remark}


\section{$M$-theory derivation of the prepotential}

In this section we will briefly describe some aspects of the relation between the $\N=2$ super Yang-Mills theory and string theory. Namely, we discuss how to get the curves which are essential element of the Seiberg-Witten theory using some stringy arguments. Also we describe the stringy interpretation of the auxiliary algebraic curve, which appears in the Seiberg-Witten theory. The reference is \cite{PrepFromM}, see also \cite{MTheoryTested,EnnesMasterFunc}.

We consider a gauge  theory described on the language of type IIA theory in $\Real^{10}$. The coordinates are denoted by $x^0,x^1,\dots x^9$.
 
\begin{figure}
\includegraphics[width=\textwidth]{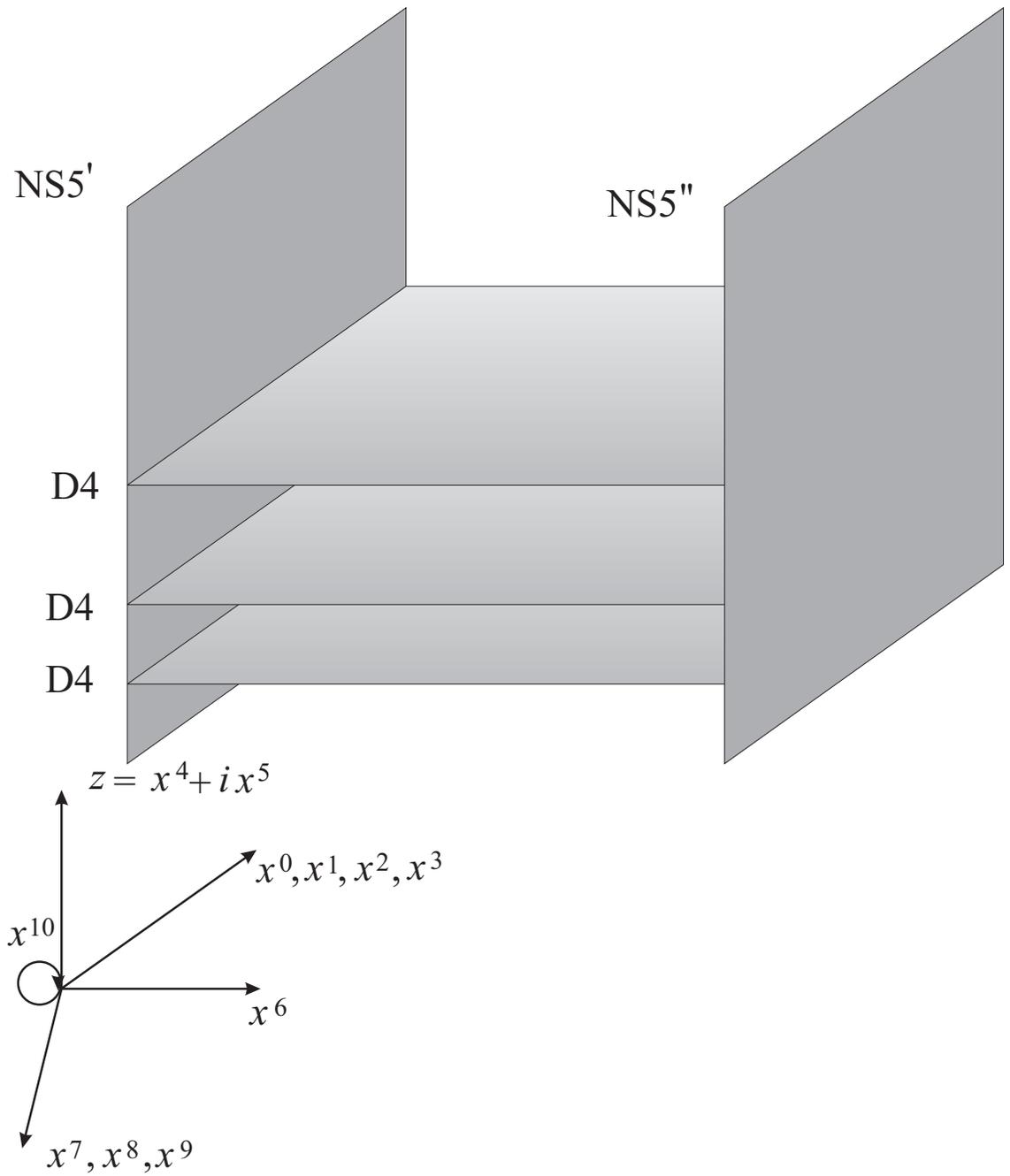}
\caption{$M$-theory setup for $SU(3)$ gauge theory}\label{Mtheory}
\end{figure}

We use the following setup (see figure \ref{Mtheory}): some NS5 branes with D4 branes suspended between them. The worldvolume of NS5 branes is along $x^0$, $x^1$, $x^2$, $x^3$, $x^4$ and $x^5$. Their positions correspond to different values of $x^6$. They have $x^7=x^8=x^9=0$. Their world volumes are described by $x^0$, $x^1$, $x^2$, $x^3$ and $x^6$. Since in the $x^6$ direction the world volume is finite, macroscopically it is described by $x^0,\dots x^3$, that is, the worldvolume is four dimensional. One considers the gauge theory on D4-branes.

The D$p$-brane action which generalizes the Nambu-Goto string action is the following
\begin{equation}
\label{DpBrane}
-\frac{T_p}{2}\int \dd^{p+1}\xi \sqrt{-\g}\left[ \g^{ab}\pd_a x^\mu \pd_b x^\nu g_{\mu\nu}(x) - \l_p(p-1)\right],
\end{equation}
where $g_{\mu\nu}(x)$ is the external metric, $\g_{ab}$ is the internal D$p$-brane matric, $\l_p$ is a constant, and $T_p$ is the D$p$-brane tension. In this section greek indices $\mu, \nu,\dots$ run over $0,1,\dots,9$ and for a D$p$-brane small latin indices $a,b,\dots$ run over $0,1,\dots,p$. 

This action implies the follwing equations of motion for the brane coordinates:
$$
\pd_a\sqrt{-\g}\g^{ab}\pd_b x^\mu + \sqrt{-\g}\g^{ab}\pd_a x^\r \pd_b x^\t \gG^\mu_{\r\t}(x) = 0
$$
where $\gG^\mu_{\r\t}(x)$ is the Cristoffel connection for the external metric.

The $x^6$ position of an NS5 brane depends only on $x^4$ and $x^5$. Let us ntroduce the complex coordinate $z = x^4 + i x^5$. The equation for $x^6$ for large $z$ becomes
$$
\pd_z \pd_{\bar{z}} x^6(z,\bar{z}) = 0.
$$ 
If we denote $z$-positions of D4 branes which are attached to an NS5 brane from the left by $a_i^{(L)}$ and positions of those D4 branes which are attached to it from the right by $a_i^{(R)}$ we can get the following solution for $x^6$:
$$
x^6 = \C \sum_{i=1}^{q_L} \logmod{z-a_i^{(L)}} - \C\sum_{i=1}^{q_R} \logmod{z-a_i^{(R)}},
$$
where $\C$ is a normalization constant.

If between two NS5 branes there are $N$ D4 branes, then the gauge theory will have $SU(N)$ as a gauge group (it can be shown that the $U(1)$ factor is frozen). To find the effective coupling constant let in the action \Ref{DpBrane} integrate out the internal D$p$-brane metric. In this way we get the induced volume action. In order to consider the gauge field which lives on this brane we should deform this action to the Born-Infeld one:
$$
\begin{aligned}
-T_4\l_4^{3/2}\int \dd^5 \xi \sqrt{-\det\left(\g^{\rm{induced}}_{ab} + F_{ab}\right)} &\approx \mr{Constant}  -T\l_4^{3/2}\int \dd^4 x \dd (x^6) F_{ab}F^{ab} \\
&= \mr{Constant} -T_4\l_4^{3/2}\left({x'}^6(z,\bar{z}) - {x''}^6(z,\bar{z}) \right)\int \dd^4 x F_{ab}F^{ab}
\end{aligned}
$$
where $F_{ab}$ is the field strength tensor, ${x'}^6(z,\bar{z})$ and ${x''}^6(z,\bar{z})$ are the $x^6$ positions of NS$5'$ and NS$5''$ branes respectively. Constant is proportional to the D4 brane volume. In this computation we have used the fact that $F_{\mu\nu}$ is independent of $x^6$. The coupling constant of this theory can be read from the last expression:
$$
\frac{1}{g^2(z,\cc{z})} = 4T_4\l_4^{3/2}\left({x'}^6(z,\bar{z}) - {x''}^6(z,\bar{z})\right),
$$
The logarithmic divergence in large $z$ is interpreted as a one loop $\b$-function contribution of the four dimensional theory.

Type IIA superstring theory can be reinterpreted as $M$-theory on $\Real^{10} \times \Sphere^1$. The eleventh coordinate $x^{10}$ is supposed to be compactified on a circle with radius $R$: $x^{10} \equiv x^{10} + 2\pi R_{10}$. Then the previous formula becomes
$$
s = \frac{x^6 + i x^{10}}{R_{10}} =  \sum_{i=1}^{q_L} \ln\left(z-a_i^{(L)}\right) - \sum_{i=1}^{q_R} \ln \left(z-a_i^{(R)}\right).
$$

If we then introduces the complex coupling constant $\ds\t = \frac{4\pi i}{g^2} + \frac{\gTh}{2\pi}$ we can write
$$
-i \t(z) = s'(z) - s''(z).
$$
Note that $\t(z)$ is a holomorphic function on $z$.

Type IIA NS5 brane can be interpreted as an M5 brane with a fixed value of $x^{10}$. Type IIA D4 brane can be seen as an M5 brane wrapped over $\Sphere^1$. Therefore we arrive at the crucial observation that NS5-D4 setup can be seen as a single M5 brane embedded in $\Real^{10}\times \Sphere^1$ in a complicated way. The worldvolume of this M5 brane can be described as follows: it fills the four dimensional space of the gauge theory: $x^0,\dots,x^3$, it is located at $x^7 = x^8 = x^9 = 0$. The intersection of the rest of 11-dimensional space and this M5 brane can be described as two dimensional subspace living in $\Real^3\times \Sphere^1$. Another viewpoint to this two dimensional subspace is the following: one introduces in the four dimensional space $\Real^3 \times \Sphere^1$ a complex structure, defined in such a way that $z = x^4 + i x^5$ and $\ds s = \frac{x^6 + i x^{10}}{R_{10}}$ are holomorphic. Then the two dimensional subspace in question is an algebraic curve. The point is that this curve is essentially the complex curve which appears as an auxiliary object in the Seiberg-Witten theory.

In order to find an explicit expression for the curve we introduce a single valued complex variable $y = \exp\left\{-s\right\}$. Then the curve is described by the equation
$$
F(y,z) = 0.
$$
The degree of $F$ as a polynomial on $y$ is the number of the NS5 branes. Therefore if one considers $SU(N)$ theory the only quadratic polynomials are needed. If one wish to consider the pure Yang-Mills theory this polynomial gains further restrictions and has the following form
$$
F(y,z) = y^2 + \prod_{l=1}^N (z-\a_l) y + 1.
$$
And this is exactly the Seiberg-Witten curve for the $SU(N)$ model.

One can go further and incorporate D6 branes in order to consider models with fundamental matter. To do this one should replace $\Real^3\times \Sphere^1$ by a non-trivial $\Sphere^1$ bundle over $\Real^3 \bsl \{\mr{singularities} \}$, known as multi-Taub-NUT space.

If one wishes to incorporate non-trivial matter multiplets in the theory, such as symmetric and antisymmetric, one should also introduce orientifold planes. 

Summarizing this discussion we can say, that the $M$-theory  provides the solutions for numerous models. Therefore the independent way to compute the effective action can be seen, in particular, as a test of the $M$-theory.

%% file: localization.tex

\label{Localization}

In this chapter we describe some essential tools which will be used to compute the prepotential for the low energy effective action. First of all we describe some aspects of the localization and find that the functional integral is localized on the instanton moduli space. When we describe the ADHM construction for this moduli space. After that we discuss some general properties of the equivariant integration: we introduce Thom and Euler classes, discuss the Duistermaat-Heckman formula. And finally we describe the deformation of the BRST charge, which will allows us to link the prepotential with some integrals over the instanton moduli space. 

Now we perform the Wick rotation and therefore lend to $\Real^4$. 


\section{Localization}
\label{localization}

In this section we describe how to reduce a functional integral, which represents a vacuum expectation for a quantity well chosen to a finite dimensional integral for the case of the topological field theory. 

Consider first a pure Yang-Mills theory, described by the action $S = S_{\mr{top}} + \bar{\sQ} (V_{\mr{YM}} + V')$. Let $\O$ be a $\bar{\sQ}$ closed observable: $\bar{\sQ} \O = 0$. For such a quantity we define its vacuum expectation as follows:
\begin{equation}
\label{vev}
\< \O \> = \int \D X \O \e^{S_{\mr{top}} + \bar{\sQ}(V_{\mr{YM}} + V')}
\end{equation}
where $\D X$ is the measure $\D X = \D A \D \phi \D \psi \D \eta \D \chi \D H \D \l$. Our computations will be based on the standard observation: if we add to the action a BRST exact term, the vacuum expectation value remain unchanged. The proof is simple taking into account the BRST closeness of both the observable $\O$ and the action itself we get
\begin{equation}
\label{BRSTinv}
\< \O \>' = \int \D X \O \e^{S + \bar{\sQ} \d V} = \< \O \> + \int \D X \O \e^{S} \bar{\sQ}\d V =\< \O \>  + \int \D X \bar{\sQ}\left( \O \e^{S}\d V\right) =  \< \O \>.
\end{equation}
Here we have used the Leibniz rule for the BRST operator and the fact the vacuum expectation of a BRST exact term equals zero.

Let us therefore modify the action in such a way that it becomes $S_{\mr{top}} + \bar{\sQ} \tld{V}$ where
\begin{equation}
\label{VYMmod}
\tld{V} = \int \dd^4 x \Tr \left\{ -\chi^{\mu\nu} \left( t {(F_{\mu\nu})}^- - \frac{1}{4}H_{\mu\nu}\right) + i \l \cD_\mu \psi^\mu  \right\} 
\end{equation}
(we suppose, that the measure $\D X$ is already divided by the volume of the gauge group, and we do not worry about the gauge fixing). Here $t$ is an arbitrary parameter. The whole integral does not depend on it provided it does not lead to new singularity of the Lagrangian.

If we integrate out the Lagrange multiplier $H_{\mu\nu}$ we arrive to the following expression for the action:
$$
S = S_{\mr{top}} + \int \dd^4 x \Tr \left\{ - t^2 {(F_{\mu\nu})}^-{(F^{\mu\nu})}^- + t \chi^{\mu\nu} {\left( \cD_\mu \psi_\nu - \cD_\nu \psi_\mu \right)}^- + i \eta \cD^\mu \psi_\mu + i \l \cD^\mu \cD_\mu \phi \right\}. 
$$

Since the integral does not depend on $t$ we can take $t\to\infty$ limit. We observe that in that case the integral localizes on the space of the solutions of the self-dual equation
\begin{equation}
\label{SDE}
F_{\mu\nu} = \star F_{\mu\nu}.
\end{equation}
\begin{remark}
Even though the second term seems to be negligible with respect to the first one, this is not the case. In fact, it serves to balance the Faddeev-Popov determinant, which comes from the first term.
\end{remark}


The space of the solutions of the selfdual equation is finite dimensional. Therefore the path integral can be reduced to a finite dimensional integral, which can be (in principle) computed exactly.


\section{ADHM construction}

Now it is a time to describe the moduli space of the solutions of the selfdual equation, the instanton moduli space. It is given by the ADHM construction \cite{ADHM} There are a number (see, for example, \cite{InstAndRec,SelfDualSolution,MultiInstCalc,MultiInstCalcII,CalcManyInst,MultInstMeasure}) of introduction to the subject. We pick some important details from them.

The ADHM construction is gauge group dependent. It exists only for the classical gauge groups, that is for $SU(N)$, $SO(N)$ and $Sp(N)$. Consider first the simplest case, the case of $SU(N)$.


\subsection{$SU(N)$ case}
\label{ADHMforSU(N)}

In order to construct the self-dual connection in the $SU(N)$ case we introduce a complex structure on $\Real^4$ with the help of the euclidean $\s$-matrices:
\begin{equation}
\label{ComplStruct}
x_{\a\dot{\a}} = \s^\mu_{\a\dot{\a}} x_\mu = \left(
\begin{array}{cc}
x^0 - i x^3 & -i x^1 - x^2 \\
-ix^1 + x^2 & x^0 + i x^3
\end{array}
\right) = \left(
\begin{array}{cc}
z_1 & -z_2^\ast \\
z_2 & z_1^\ast
\end{array}
\right).
\end{equation}

Moreover we need the following data: a $(N+2k)\times 2k $ complex matrix which depends linearly on the coordinates:
$$
\gD_{\dot{\a}} = \A_{\dot{\a}} + \B^\a x_{\a\dot{\a}}.
$$
We suppose that the matrix $\gD_{\dot{\a}}$ has maximal rang $2k$. The next ingredient is an annihilator of ${\gD^\dag}^{\dot{\a}}$ which we denote by $v(x)$:
\begin{equation}
\label{vCond}
{\gD^\dag}^{\dot{\a}} v = 0.
\end{equation}
$v$ is a matrix $N\times (2k+N)$ normalized as follows:
\begin{equation}
\label{vNorm}
v^\dag v = \Id_N.
\end{equation}

Having this data we can write the expression for the connection $A_\mu(x)$ as follows:
$$
A_\mu(x) = i v^\dag(x) \pd_\mu v(x).
$$
One can easily check that this connection is hermitian: $A_\mu^\dag = A_\mu$. Therefore it is a $U(N)$ connection in the fundamental representation.

Impose on $\gD_{\dot{\a}}$ the factorization condition:
\begin{equation}
\label{factorization}
{\gD^\dag}^{\dot{\a}}\gD_{\dot{\b}} = \d^{\dot{\a}}_{\dot{\b}}\R^{-1},
\end{equation}
where $\R(x)$ is an invertible $k\times k$ complex hermitian matrix.

Since the rang of the matrix $\gD_{\dot{\a}}$ is maximal and taking into account \Ref{vNorm} we conclude that
$$
\P = v v^\dag = \Id_{2k+N} - \gD_{\dot{\a}}\R {\gD^\dag}^{\dot{\a}}.
$$

It follows that the curvature is self-dual:
$$
F_{\mu\nu} = \pd_\mu A_\nu - \pd_\nu A_\mu -i [A_\mu,A_\nu] = -2i \eta^i_{\mu\nu} v^\dag \B^\a \R \B^\dag_\b v \t_{i,\a}{}^\b.
$$

\begin{remark} 
We have claimed before that $A_\mu$ is a $U(N)$ connection. However the trace part of this connection can be gauge out. Indeed, a solution of the self-dual equation satisfy also the Yang-Mills equation. Therefore we have $\cD_\mu F^{\mu\nu} = 0$. Taking the trace of this equation we get $\pd_\mu \Tr F^{\mu\nu} = 0$. It follows that 
$$
0 = \int \dd^4 x \Tr \left\{ A_\nu \pd_\mu \Tr F^{\mu\nu} \right\}= - \frac{N}{2}\int \dd^4 x \Tr F_{\mu\nu}\Tr F^{\mu\nu}.
$$
Therefore $\Tr F_{\mu\nu} = 0$ and $\Tr A_\mu  = \pd_\mu \a$. Thus we can say that $A_\mu(x)$ is, in fact, an $SU(N)$ connection.
\end{remark}

Let us express the factorization condition \Ref{factorization} in terms of $\A_{\dot{\a}}$ and $\B^\a$. Having develop on $x^\mu$ we get:
$$
\begin{aligned}
{\B^\dag}_\a \B^\b & = \frac{1}{2}\d_\a^\b {\B^\dag}_\g \B^\g, \\
{\B^\dag}_\a \A_{\dot{\a}} &= {\A^\dag}_{\dot{\a}}\B_\a, \\
{\A^\dag}^{\dot{\a}} \A_{\dot{\b}} &= \frac{1}{2}\d^{\dot{\a}}_{\dot{\b}} {\A^\dag}^{\dot{\g}}\A_{\dot{\g}}.
\end{aligned}
$$
Note that the first and second conditions can be packaged in the following one: ${\gD^\dag}_{\dot{\a}} \B_\a = \B^\dag_\a \gD_{\dot{\a}}$.

The meaning of the number $k$ can be clarified by means of the Osborn identity \cite{OsbornIdentity}
\begin{equation}
\label{Osborn}
\Tr_{\mr{fund}} F_{\mu\nu} \star F^{\mu\nu} = - {\left(\pd_\mu \pd^\mu\right)}^2  \ln \det \R.
\end{equation}

The factorization condition \Ref{factorization} implies
$$
2\R^{-1} = {\A^\dag}^{\dot{\a}}\A_{\dot{\a}} + {\A^\dag}^{\dot{\a}}\B^\b x_{\b\dot{\a}} + {x^\dag}^{\dot{\a}\b}\B^\dag_\b \A_{\dot{\a}} + {x^\dag}^{\dot{\a}\b}\B^\dag_\b \B^\g x_{\g\dot{\a}}.
$$
It follows that in the limit $x\to\infty$ we have the following assymptotics:
$$
\R^{-1} \to \frac{1}{2}x^2 \B^\dag_\a \B^\a. 
$$
Therefore exploiting the asymptotic expansion for $\R(x)$ we get
$$
\begin{aligned}
\pd_\mu \pd^\mu \Tr_{\mr{fund}} \ln \R &\to - \frac{4k}{x^2} & &\mbox{when} & x &\to \infty.
\end{aligned}
$$

Taking into account that for $SU(N)$ we have $2N \Tr_{\mr{fund}} F_{\mu\nu} \star F^{\mu\nu} = \Tr F_{\mu\nu} \star F^{\mu\nu}$, and using the formula \Ref{InstNumb} we conclude that $k$ is nothing but the instanton number.

Neither \Ref{factorization} nor \Ref{vNorm} changes under the
transformation
\begin{equation}
\label{transform}
\begin{aligned}
\Delta_{\dot{\alpha}} &\mapsto \Delta_{\dot{\alpha}}' =
U \Delta_{\dot{\alpha}} M & \ \ &\mbox{and} \ \
& v &\mapsto v' = Uv
\end{aligned}
\end{equation}
with $U$ being a $(N + 2k) \times (N + 2k)$ unitary matrix and $M$ being an invertible one. This freedom can be used to put the matrix $\B = (\B^1, \B^2)$ into the canonical form
$$
\B = \left(
\begin{array}{c}
0 \\
\Id_k \otimes \Id_2
\end{array}
\right).
$$
Then the relevant data is contained in the matrices $\A_{\dot{\a}}$ and $v$ which can be represented as follows:
$$
\begin{aligned}
\A &= (\A_{\dot{1}}, \A_{\dot{2}}) = \left(
\begin{array}{c}
S_{\dot{1}}\quad S_{\dot{2}} \\
X^\mu \otimes \sigma_\mu
\end{array}
\right), &
v &= \left(
\begin{array}{c}
T \\
Q_\a
\end{array}
\right).
\end{aligned}
$$
Matrices $S_{\dot{\alpha}}$ transform under the space-time rotations as righthand spinor, $X^\mu$ as a vector, $T$ is a scalar, and $Q_\a$ is a lefthanded spinor.

Having fixed the form of the matrix $\B^\a$ we still have a freedom to perform a transformation \Ref{transform} which can be read as
\begin{equation}
\label{residual}
\begin{aligned}
S_{\dot{\alpha}} &\mapsto S_{\dot{\alpha}}' = U_N S_{\dot{\alpha}} U_k^{-1}, & X^\mu &\mapsto {X^\mu}' = U_k X^\mu U^{-1}_k, \\
T &\mapsto T' = U_N T, & Q_\a &\mapsto Q_\a' = U_k Q_\a
\end{aligned}
\end{equation}
where $U_k \in U(k)$ and $U_N \in U(N)$. 

The factorization condition \Ref{factorization} requires the matrices $X^\mu$ to be hermitian: ${X^\mu}^\dag = X^\mu$ and also the following non-linear conditions to be satisfied:
$$\mu^i = \A^{\dag\dot{\alpha}} \tau_{i,\dot{\alpha}}{}^{\dot{\beta}}\A_{\dot{\beta}} = 0.$$

These conditions are known as the ADHM equations. They
are usually written in slightly different notations.
Namely, let
$$
\begin{aligned}
J &= S_{\dot{1}}, \ & I &= {S_{\dot{2}}}^\dag, \ & B_1 &= X^0 - i X^3 & \ \
&\mbox{and}\ \ &  B_2 &= -iX^1 + X^2.
\end{aligned}
$$
Then the ADHM equation are
\begin{equation}
\label{ADHM}
\begin{aligned}
\mu_\Real &= - \mu^3 = I I^\dag - J^\dag J  + [B_1,B_1^\dag] + [B_2,B_2^\dag] = 0,\\ 
\mu_\Compl &= \frac{1}{2}(\mu^1 -i \mu^2) = \mu^- =  IJ + [B_1,B_2] = 0. 
\end{aligned}
\end{equation}
If we consider two vector spaces $\V = \Compl^k$ and $\W = \Compl^N$ then $I, J, B_1$ and $B_2$ become linear operators acting as
$$
\begin{aligned}
I: \W &\to \V,& \ J: \V &\to \W, &   &\mbox{and}  & B_1, B_2: \V &\to \V.
\end{aligned}
$$

The space of such operators modulo transformations \Ref{residual} is the instanton moduli space.

The residual freedom \Ref{residual} corresponds to the freedom of the framing change in $\V$ and $\W$. Framing change in $\W$ corresponds to the rigid gauge transformation, which change, in particular, the gauge at infinity. Sometimes we will denote the group of the rigid gauge transformations as $G_\infty$. 

The change of frame in $\V$ becomes natural when one considers the instanton moduli space as a hyper-K\"ahler quotient. Indeed, the space of all (unconstrained) matrices  $\A_{\dot{\a}}$ has a natural metric $\dd\A^{\dag\dot{\a}}\dd\A_{\dot{\a}}$ and the hyper-K\"ahler structure which consists of the triplet of linear operators $\I^i$ which together with the identity operator is isomorphic to the quaternion algebra. These operators act as follows:
$$
\I^i \A_{\dot{\a}} = -i \t_{i,\dot{\a}}{}^{\dot{\b}}\A_{\dot{\b}}.
$$
The action of the unitary group $U(k)$ described by \Ref{residual} is Hamiltonian with respect to each symplectic structure. The Hamiltonian (moment), corresponding to the $i$-th symplectic form and the  algebra element $\xi \in \Lie(U(k))$ is
$$\mu^i_\xi = -i \t_{i,\dot{\a}}{}^{\b} \Tr (\xi \A^{\dag\dot{\a}}\A_{\dot{\b}}).$$
Hence the ADHM equations together with residual transformation can be interpreted as the hyper-K\"ahler quotient \cite{HyperKahler}:
$${\Moduli}_k = \mu^{-1}(0) / U(k).$$

We call the group which is responsible to the change of frame in $\V$ the dual group. In the case of $U(N)$ the dual group is $U(k)$.


\subsection{Solutions for the Weyl equations}

Before exploring other classical groups $SO(N)$ and $Sp(N)$ let us pause and consider the solutions for the Weyl equations in the instanton background. That is, consider the following equation:
\begin{equation}
\label{WEQ}
\cD^{\dot{\a}\a} \psi_\a = \bar{\s}^{\mu,\dot{\a}\a} \cD_\mu \psi_\a.
\end{equation}
For the fundamental representations of the gauge group we can get a simple formula for the $k$ independent solutions which  can be arranged to the $N \times k$ matrix \cite{OsbornDirac}
\begin{equation}
\label{WeylFund}
\psi^\alpha = v^\dag \B^\alpha \R = \bar{Q}^\a \R.
\end{equation}
One can show that thanks to the identity \cite{CorrGoddZeta}
$$
\bar{\psi}_\a \psi^\a = \R {\B^\dag}_\a v v^\dag B^\a \R = - \frac{1}{4}\pd_\mu \pd^\mu \R
$$
the following statements hold \cite{InstAndRec}:
\begin{equation}
\label{reciprocity}
\begin{aligned}
\int \dd^4x \bar{\psi}_\a\psi^\a & = \pi^2 \Id_k, & \int \dd^4x \bar{\psi}_\a\psi^\a x^\mu &= -\pi^2 X^\mu, & &\mbox{and} \\
\psi^\a x_{\alpha\dot{\a}} &\to - \frac{1}{x^2} S_{\dot{\a}} & \mbox{when}\;\;\;  x &\to \infty.
\end{aligned}
\end{equation}
Taking these equations as the definitions of $X^\mu$ and $S_{\dot{\alpha}}$ one recovers both  the ADHM constraints and the fact that the matrices $X^\mu$ are hermitian.

Let us look closely to the equations \Ref{vCond},\Ref{vNorm}. The first equation can be solved for $Q_\a$:
$$
Q_\a(x) = - {(X + x)}^{-2}{(X + x)}_{\a\dot{\a}} S^{\dag\dot{\a}}T(x).
$$
The second equation gives the following condition for $T(x)$:
\begin{equation}
\label{TCond}
T(x)^\dag \left(\Id_\W + S_{\dot{\a}}{(X + x)}^{-2}S^{\dag\dot{\a}}\right)T(x) = \Id_\W.
\end{equation}

The matrix in the brackets is positively defined and therefore there exists a matrix $M(x)$ such that 
\begin{equation}
\label{MCond}
M(x)^\dag M(x) = \Id_\W + S_{\dot{\a}}{(X + x)}^{-2}S^{\dag\dot{\a}}.
\end{equation}
It follows that $g(x) = M(x)T(x) \in U(N)$. Otherwise here we have found the explicit dependence on the gauge group.

\begin{remark}
When we consider group $SO(N)$ or $Sp(N)$ the equations \Ref{vCond}, \Ref{TCond} and \Ref{MCond} are still valid (modulo some minor changes) provided the following convention is accepted:
\begin{itemize}
\item for $SO(N)$ we replace ${(\cdot)}^\dag \mapsto {(\cdot)}^\tr$,
\item for $Sp(N)$ we replace ${(\cdot)}^\dag \mapsto {(\cdot)}^\dag \bJ_{2N}$.
\end{itemize} 
In particular the equation \Ref{MCond} implies $g(x) = M(x)T(x) \in G$.
\end{remark}

Let us also briefly describe the solutions for the Weyl equation in the adjoint representation. Let us use the following ansatz:
\begin{equation}
\label{WeylAdj}
\psi_\a = i \v^\dag\left( \C\R {\B^\dag}_\a  - \B_\a \R\C^\dag\right) v
\end{equation}
where $\C$ is a complex $k\times (N + 2k)$ matrix with constant coefficients. It follows by definition that $\ep_{\a\b} {\psi^\dag}^\b = \psi_\a$, therefore it belongs to the adjoint representation. 

Computation shows that $\psi_\a$ will  be solution of the Weyl equation if the matrix $\C$ satisfies the following condition ${\gD^\dag}^{\dot{\a}} \C + \C^\dag \gD^{\dot{\a}} = 0$, that is
\begin{equation}
\label{WeylAdjCondition}
\begin{aligned}
{\A^\dag}^{\dot{\a}} \C + \C^\dag \A^{\dot{\a}} &= 0, \\
{\B^\dag}_\a \C - \C^\dag \B_\a &= 0.  
\end{aligned}
\end{equation}

Lefthand sides of these equations are hermitian and anthihermitian matrices $k\times k$. Therefore they give $4k^2$ real conditions on $\C$. Matrix $\C$ has $2k(2k+N) = 4k^2 + 2kN$ real coefficients. Therefore the rest is $2kN$ solutions of the Weyl equation as it should be.


\subsection{$SO(N)$ case}

The extension to the $SO(N)$ case can be obtained with the help of the reciprocity construction \Ref{reciprocity}.

Note that according to the Table \ref{GroupTh} we have $\ell_{\mr{adj}} = 2h^\vee \ell_{\mr{fund}}$ for $SU(N)$ and $Sp(N)$ whereas $\ell_{\mr{adj}} = h^\vee \ell_{\mr{fund}}$ for $SO(N)$. Therefore formula \Ref{InstNumb} together with \Ref{Osborn} shows that in the case of $SO(N)$ to obtain the solution of the self-dual equation with the instanton number $k$ we should replace $k$ by $2k$ in the construction for $SU(N)$.

Let us choose the Darboux basis in $\V = \Compl^{2k}$, which corresponds to the split $\Compl^{2k} = \Compl^2 \otimes \Compl^k$. Correspondingly, we split the index which runs over $1,\dots,2k$ into two: the first, $A= 1,2$, and the second over $1,\dots,k$. Thus the solution for the Weyl equation can be written as the set of four $N \times k$ matrices $\psi_{\alpha A}$.
These matrices can be represented as follows:
$$\psi_{\alpha A} = \psi_\mu\sigma^\mu_{\alpha A}.$$
The twisted index $\mu$ that appears in the righthand side does not correspond to a Lorentz vector. The Weyl equation can be rewritten now as a set of four equations:
\begin{equation}
\label{psiEqn}
\begin{aligned}
\cD^\mu \psi_\mu &=0 & \ \ &\mbox{and} \ \ & {(\cD_\mu\psi_\nu)}^- = 0.
\end{aligned}
\end{equation}
It worth noting that these conditions mean that $\psi_\mu$ is orthogonal to the gauge transformations and that it satisfies the linearized self-dual equation.

The condition that $-iA_\mu$ belongs to the algebra of $SO(N)$ implies that it is real antisymmetric matrices. Hence the equation for $\psi_\mu$ has real coefficients and its solutions can be chosen real as well. The fact that $\psi_\mu$ are real means that $\psi_\mu \sigma^\mu_{\a A}$ can be considered as a quaternion. We recover here the quaternion construction introduced  in \cite{SelfDualSolution}. The possibility of this expansion with real coefficients implies that $S_{\dot{\alpha}}$ can also be expanded as $S_{A\dot{\alpha}} = S_\mu \sigma^\mu_{A\dot{\alpha}}$ where $S_\mu$ are real.

Using then the definition of $X^\mu$ \Ref{reciprocity} we derive the following statement:
$$\epsilon_{CA} X^{\mu,A}{}_B \epsilon^{BD} =  {({X^\mu}^\tr)}_C{}^D,$$
or, if we introduce the symplectic structure ${\bJ}_{2k}$ this can be written as
$${\bJ}_{2k} X^\mu {\bJ}_{2k}^\tr = {X^\mu}^\tr.$$

The dual group is a subgroup of $U(2k)$ which preserves this condition. It is the group $Sp(k) \subset U(2k)$.

The matrices $X^\mu$ and $S_{\dot{\alpha}}$
can be represented as follows:
\begin{equation}
\label{SOYZIJ}
\begin{aligned}
X^\mu &= \left(
\begin{array}{cc}
Y^\mu & {Z^\mu}^\dag \\
Z^\mu & {Y^\mu}^\tr
\end{array}
\right),\quad
& \quad S_{\dot{1}} = J = (K,K') &
& \ \ &\mbox{and}\ \  &
S_{\dot{2}} &= I^\dag = (-K^{' \ast},K^{\ast})
\end{aligned}
\end{equation}
where $Y^\mu$ is an hermitian matrix and $Z^\mu$ is an antisymmetric one.

Let
\begin{equation}
\label{SOBB}
B_{1,2} = \left(
\begin{array}{cc}
P_{1,2} & Q_{1,2}' \\
Q_{1,2} & P_{1,2}^\tr
\end{array}
\right),
\end{equation}
where $Q_{1,2}$ and $Q'_{1,2}$ are antisymmetric matrices. The ADHM equations for $SO(N)$ becomes:
\begin{equation}
\label{SOMN}
\begin{aligned}
\mu_\Compl &= \left(
\begin{array}{cc}
M_\Compl & N_\Compl' \\
N_\Compl & -M_\Compl^\tr
\end{array}
\right) = 0 & \ \
&\mbox{and} \ \ &
\mu_\Real &= \left(
\begin{array}{cc}
M_\Real & N_\Real' \\
N_\Real & -M_\Real^\tr
\end{array}
\right) = 0
\end{aligned}
\end{equation}
where
$$
\begin{aligned}
M_\Compl &= [P_1,P_2] + Q'_1 Q_2 - Q'_2 Q_1 - K^{' \tr} K, \\
N_\Compl &= Q_1 P_2 - P_2^\tr Q_1 + P_1^\tr Q_2 - Q_2 P_1 + K^\tr K,\\
N_\Compl' &=  Q_1' P_2^\tr- P_2 Q_1' + P_1 Q_2'- Q_2' P_1^\tr  - K^{\prime \tr} K',
\end{aligned}
$$
and
$$
\begin{aligned}
M_\Real &= \sum_{s = 1}^2 \left([P_s,{P_s}^\dag] + {Q_s}^\ast Q_s - {Q_s}' {Q_s}^{\prime \ast} \right) + K^{\prime \tr} K^{\prime \ast} -
K^\dag K, \\
N_\Real &= \sum_{s = 1}^2 \left(Q_s {P_s}^\dag - {P_s}^\ast Q_s  + {Q_s}^{\prime\ast} P_s- {P_s}^\tr {Q_s}^{\prime \ast}
\right)- K^\tr K^{\prime \ast} - K^{' \dag} K, \\
N_\Real' & = \sum_{s = 1}^2 \left( {Q_s}' {P_s}^\ast - {P_s}^\dag {Q_s}' + {Q_s}^\ast {P_s}^\tr - P_s {Q_s}^\ast \right) - K^{\prime \tr} K^\ast - K^\dag K'.
\end{aligned}
$$
Note that $N_\Compl, N'_\Compl, N_\Real$ and $N_\Real'$ are symmetric matrices.


\subsection{$Sp(N)$ case}

The group $Sp(N)$ is a subgroup of $U(2N)$ which preserves the symplectic structure ${\bJ}_{2N}$. The ADHM construction for $Sp(N)$ can be obtained by imposing some constraints on the ADHM construction for $SU(2N)$. A quick look at the Table \ref{GroupTh} shows that in this case there is no  doubling of the instanton charge.

Let us choose the Darboux basis in $\Compl^{2N}$, which corresponds to the split $\Compl^{2N} = \Compl^2 \otimes \Compl^N$, $\bJ_{2N} = \bJ_2 \otimes \bJ_{N}$.
Correspondingly, we split the index which runs over $1,\dots,2N$ into two: the first, $A= 1,2$, and the second: $l = 1,\dots,N$.

We can expand the solution of the Weyl equation as follows $\psi_{\alpha A} = \psi_\mu \sigma^\mu_{\alpha A}$. The fact that $-iA_\mu$ belongs to the Lie algebra of $Sp(N)$ imposes the following condition:
$${\bJ}_{2N} A_\mu^\ast {\bJ}_{2N} = - A_\mu.$$
The solutions $\psi_\mu$ can be chosen to be real. Thus the reciprocity formulae \Ref{reciprocity} show that in that case the matrices $X^\mu$ are not only hermitian, but also real and consequently symmetric. The dual group should preserve this condition and we arrive to the conclusion that this is $O(k) \subset U(k)$.

The reality of $\psi_\mu$ implies also that the matrices $S_{\dot{\alpha}}$ can be expanded as $S_{A\dot{\alpha}} = S_\mu \sigma^\mu_{A \dot{\alpha}}$ where $S_\mu$ are real. Hence for the matrices $I$ and $J$ we have
\begin{equation}
\label{IJSpN}
\begin{aligned}
J &= \left(
\begin{array}{c}
K \\
K'
\end{array}
\right) & &\mbox{and} & I^\dag &= \left(
\begin{array}{c}
- K^{' \ast} \\
K^\ast
\end{array}
\right).
\end{aligned}
\end{equation}

Hence the ADHM equation for $Sp(N)$ take the following form
$$
\begin{aligned}
\mu_\Compl &= K^\tr K' - K^{\prime\tr} K + [B_1,B_2], \\
\mu_\Real &= K^\tr K^\ast - K^\dag K + K^{\prime\tr} K^{\prime\ast} - K^{\prime \dag} K' + [B_1,B_1^\ast ] + [B_2,B_2^\ast].
\end{aligned}
$$
Here the matrices $B_{1,2}$ are symmetric. We see that $\mu_\Compl$ and $\mu_\Real$ are antisymmetric matrices.


\subsection{Spaces, matrices and so on}

To simplify further references we have put in the Table \Ref{ADHMTable} some relevant information about the ADHM data. In that follows we will  denote the dual group (in the sense of \cite{SelfDualSolution}) by $G_D$.


\begin{table}[t]
\begin{center}
\begin{tabular}{||c||c|c|c|c|c||}
\hhline{|t:=:t:=:=:=:=:=:t|}
$\mbf{G}$ & $\mbf{G_D}$ & \textbf{Size of $\mbf{\Delta_{\dot{\a}}}$} & \textbf{Size of $\mbf{v}$} & $\mbf{\V}$ & $\mbf{\W}$ \\
\hhline{|:=::=:=:=:=:=:|}	
$U(N)$ & $U(k)$ & $k \times N + 2k$ & $N \times N + 2k$ & $\Compl^k$ & $\Compl^N$ \\
\hhline{||-||-|-|-|-|-||}
$O(N)$ & $Sp(k)$ & $2k \times N + 4k$ & $N \times N + 4k$ & $\Compl^{2k}$ & $\Real^N$ \\
\hhline{||-||-|-|-|-|-||}
$Sp(N)$ & $O(k)$ & $k \times 2N + 2k$ & $2N \times 2N + 2k$ & $\Real^k$ & $\Compl^{2N}$ \\
\hhline{|b:=:b:=:=:=:=:=:b|}
\end{tabular}
\end{center}
\caption{Spaces, matrices, groups}\label{ADHMTable}
\end{table}



\section{Equivariant integration}
\label{equivInt}

In the previous section we have seen that the instanton moduli space, where the functional integral localizes to, can be seen as a space of linear operators $I$, $J$, $B_1$ and $B_2$ satisfying the ADHM equation \Ref{ADHM} and considered up to transformations generated by $G_\infty\times G_D$. The non-linear ADHM equations can not be solved for $k>3$. Therefore, we should find a way to perform required integration without introducing local coordinates on $\Moduli_k$.

This task can be accomplished with the help of the equivariant integration \cite{NikitaLectures,MooreYM2D}. Mathematically the problem can be formulated as follows. Let $X$ be a manifold. Let $G$ be a group which acts on this manifold. We denote the left action by $g \cdot x$, $g\in G$, $x\in X$. Let $M \subset X$ be a submanifold of $X$ on which the group $G$ acts freely. Then we wish to express the integral over the factor $N = M / G$ in terms of the integral over $X$.


\subsection{Integration over zero locus}
\label{IntZeroLoc}

Let us do it step-by-step. Suppose we have a closed form $\a$ defined on $M$. How to express $\ds \int_M \a$ as an integral over $X$? We will only need the case where $M = s^{-1}(0)$, where $s \in \gG(E)$ is a section of a vector bundle with a fiber $F$: $F \hookrightarrow E \stackrel{\pi}{\to} X$.

Let $\{x^\mu\}_{\mu=1}^{\dim X}$ be set of coordinates of $X$ in a local patch. In order to make our discussion sound field theoretically let us introduce an alternative notation for the base 1-forms: $\dd x^\mu = \psi^\mu$ and for de Rham differential $\dd = \bar{\sQ}$. Then we have:
$$
\begin{aligned}
\bar{\sQ} x^\mu &= \psi^\mu, & \bar{\sQ} \psi^\mu &= 0.
\end{aligned}
$$

Let $F$ be a vector space such that $\pi^{-1}(x) \isom F$  for a point $x \in M$. We should introduce a multiplet $(\chi,H) \in \Pi F^\ast \oplus F^\ast$ ($\chi$ is a fermion, therefore it belongs to $F^\ast$ with changed statistics, $\chi \in \Pi F^\ast$). In order to make the transformations for this multiplet covariant we should introduce a connection on the bundle $E$. Let us denote it $\gG_\mu$. Then we have
$$
\begin{aligned}
\bar{\sQ} \chi &= H - \gG_\mu \psi^\mu \chi, & \bar{\sQ} H &= - \gG_\mu \psi^\mu H + \frac{1}{2} R_{\mu\nu}\psi^\mu\psi^\nu \chi.
\end{aligned}
$$
where $R_{\mu\nu}$ is a curvature for the connection $\gG_\mu$. One can check that $\bar{\sQ}^2 = 0$. In order to see that $\bar{\sQ}^2 H = 0$ one should use the Bianchi identity for $R_{\mu\nu}$.

\begin{remark}
When the bundle $E$ is trivial (this is the case of the twisted $\N=2$ supersymmetric Yang-Mills theory) one has simply
$$
\begin{aligned}
\bar{\sQ} \chi &= H, & \bar{\sQ} H &= 0.
\end{aligned}
$$
\end{remark}

Then we required formula is
\begin{equation}
\label{local1}
\int_M \a = \int_X \D x \D \psi \D H \D \chi \iota^\ast \a \e^{i \bar{\sQ}\chi \left(  s(x) - \frac{1}{2 t} H \right)},
\end{equation}
where $\iota : M \to X$ is the inclusion map, $\D(\cdot)$ is a standard measure and we have used the fact that if we formally replace in a form $\a = \a(x,\dd x)$ all differentials $\dd x^\mu$ by Grassman variables $\psi^\mu$ we can write
$$
\int \a = \int \D x \D \psi \a(x,\psi).
$$
 
Taking into account the discussion in the section \ref{localization} we can see that the righthand side of \Ref{local1} does not depend on $t$. Therefore one can compute the integral in the large $t$ limit. It gives precisely the lefthand side.


\subsection{Integration over factor}
\label{intFactor}
Let $M$ be a manifold on which a group $G$ acts freely. We wish to  to express an integral over a factor $M/G$ as an integral over $M$. To do this we use the fact that de Rham cohomologies of $M / G$ are isomorph to so-called $G$-equivariant cohomologies of $M$ (which we denote by $H^\ast_G(M)$):
$$
H^\ast(M/G) \isom H^\ast_G (M).
$$ 

The latter can be described as follows. Let $\gO^\ast(M)$ be the de Rham complex of $M$. Denote by $\Fun(\alg{g})$ an algebra of function on $\alg{g} = \Lie(G)$. These function will be graded in such a way that $n$-th power homogeneous polynomial have the  degree $2n$.

\begin{remark}
Such an assignation is done in order to the Cartan differential (see few lines below) have a definite degree. It can be understood from the physical point of view if we consider the degree as the ghost number. Recall from the section \ref{BVvsTwist} that $\psi^\mu = \dd x^\mu$ has ghost number $+1$ and $\phi \in \alg{g}$ has the ghost number $+2$.
\end{remark}

Let  the group $G$ acts on the functions from $\Fun(\alg{g})$ by the adjoint representation, and on forms $G$-action be induced by left action on $M$. When one introduces another complex 
$$
\gO^\ast_G(M) = {\left( \gO^\ast \otimes \Fun(\alg{g})\right)}^G
$$
where  ${(\cdot)}^G$ means $G$-invariant part. Denote by $V(\phi) = \phi^a V_a$ a vector field on $M$ corresponding to $\phi \in \Fun(\alg{g})$ and introduce the Cartan differential
$$
\bar{\sQ} = \dd + i_{V(\phi)}.
$$
Its square is the Lie derivative with respect to $V(\phi)$. Hence $\bar{\sQ}^2 = 0$ on elements of $\gO^\ast_G(M)$. The cohomology of the Cartan differential $\bar{\sQ}$ are called $G$-equivariant cohomology of $M$:
$$
H^\ast_G(N) = \Image \bar{\sQ} / \Ker \bar{\sQ}.
$$  

Taking into account the isomorphism between $H^\ast(N)$ and $H^\ast_G(M)$ we can identify corresponding classes. Let $\a(\phi) = \a(\phi,x,\psi)$ be a representative of the class which contains $\tld{\a}$. Then the required formula can be obtained as follows. Let $(\cdot,\cdot)$ be a $G$-invariant metric on $M$. In coordinates we have $(v,w) = g_{\mu\nu}v^\mu w^\nu$. With the help of this metric we can raise and lower indices. Then the required formula is
\begin{equation}
\label{projection}
\int_{N = M/G} \tld{\a} = \int_M \frac{\D x\D \psi\D \phi\D \l \D\eta}{\Vol(G)} \e^{i \bar{\sQ} V_\mu(\l)\psi^\mu} \a(\phi,x,\psi).
\end{equation}
where we have introduced the projection multiplet $(\l,\eta)$. The Cartan differential acts on it and on $(x,\psi)$ as follows:
$$
\begin{aligned}
\bar{\sQ} x^\mu  &= \psi^\mu, & \bar{\sQ} \psi^\mu &= V^\mu(\phi), \\
\bar{\sQ} \l &= \eta, & \bar{\sQ} \eta &= [\phi,\l].
\end{aligned}
$$

Note that
$$
\bar{\sQ} V_\mu(\l) \psi^\mu = \eta^a V_{a,\mu} \psi^\mu + \l^a \left( (V_a,V_b) \phi^b + \pd_\mu V_{a,\nu} \psi^\mu \psi^\nu\right). 
$$
Therefore the $\l$ integral provides a delta function localized on 
$$
\phi^b = - {(V_a,V_b)}^{-1} \frac{1}{2} \left( \pd_\mu V_{a,\nu}  - \pd_\nu V_{a,\mu} \right) \psi^\mu \psi^\nu.
$$

Formula \Ref{projection} can be recast in more elegant form if we introduce the equivariant integration. Let us choose a Haar measure on $G$. And let $\dd \phi_1\dd\phi_2\dots\dd\phi_{\dim(G)}$ coincides with the Haar measure at the identity of $G$. Then we define a equivariant integration as follows:
$$
\oint_M \a = \frac{1}{\Vol(G)} \int_{\alg{g}} \prod_{a=1}^{\dim G} \frac{\dd \phi}{2\pi i} \int_M \a(\phi)
$$
\begin{remark}
In general, when the form $\a(\phi)$ is a polynomial on $\phi$, the integral does not converge. To cure this one introduces a convergence factor $\e^{-\frac{1}{2}\eps\< \phi,\phi\>}$ where $\<\cdot,\cdot\>$ is a Killing form on $\alg{g}$ and $\eps$ is a positive parameter. We will not need it since the form we wish to integrate is proportional to delta function on $\phi$.
\end{remark}

With this definition the formula \Ref{projection} takes the following form
$$
\int_{s^{-1}(0)/G} \tld{\a} = \oint_{s^{-1}(0)} \int \D \eta\D \l \e^{i\bar{\sQ} V_{\mu}(\l)\psi^\mu}\a.
$$
 

\subsection{Synthesis}

Now let us put things together. In the general case which are interested in here the solution exists when $s$ is an \emph{equivariant} section of $E$. It means that for any $g\in G$ we have $s(g \cdot x) = \vr(g) s(x)$ where $\vr(g)$ is the image of $g$ in the representation $\vr$ of $G$ which acts on $F$. This condition guarantees that $s^{-1}(0)$ is $G$-invariant.

We wish to express the integral of a closed form $\tld{\a} \in H^\ast(N)$ over $N = s^{-1}(0)/G$ as an integral over $X$. Now $\bar{\sQ}$ means the Cartan differential. Therefore, it acts as follows:
$$
\begin{aligned}
\bar{\sQ} x^\mu &= \psi^\mu, & \bar{\sQ} \psi^\mu &= V^\mu(\phi), \\
\bar{\sQ} \chi &= H - \gG_\mu \psi^\mu \chi, & \bar{\sQ} H &= \phi^a T^{\vr}_a \chi - \gG_\mu \psi^\mu H + \frac{1}{2} R_{\mu\nu} \psi^\mu \psi^\nu \chi, \\
\bar{\sQ} \l &= \eta, & \bar{\sQ} \eta &= [\l,\phi].
\end{aligned}
$$
If, as before $\a(\phi,x,\psi)$ belongs to the same class as $\tld{\a}$ then 
$$
\int_{s^{-1}(0)/G} \tld{\a} = \int_X \frac{\D x\D \psi\D H \D\chi \D\phi\D\l \D \eta}{\Vol(G)} \e^{i\bar{\sQ}\left(\chi s + \psi_\mu V^\mu(\l)\right)} \iota^\ast\a(\phi,x,\psi). 
$$ 
It can be rewritten with the help of the equivariant integration as follows:
$$
\int_{s^{-1}(0)/G} \tld{\a} = \oint_X \int \D \eta \D \l \D H \D \chi \e^{i\bar{\sQ}\left(\chi s + \psi_\mu V^\mu(\l) \right)}\iota^\ast\a (\phi,x,\psi).
$$


\subsection{Euler and Thom classes}

Consider again \Ref{local1}. Since the integral does not depend on $t$ we can set, for example,  $t = i$. Let us compute the exponent. We have
$$
i\bar{\sQ} \chi \left( s + \frac{i}{2} H \right) = - \frac{1}{2} {\left( H - is \right)}^2 - \frac{1}{2} s^2 + i \chi \cD_\mu s \psi^\mu + \frac{1}{4} \chi R_{\mu\nu} \psi^\mu \psi^\nu \chi + \frac{1}{2} \chi \phi^a T^\vr_a \chi
$$
where $\cD_\mu = \pd_\mu + \gG_\mu$ is the covariant derivative with the connection $\gG_\mu$. 

Let us now integrate out $H$. The integral is Gaussian and we arrive to
$$
i\bar{\sQ}\chi s \mapsto  - \frac{1}{2} s^2 + i \chi \cD_\mu s \psi^\mu + \frac{1}{4} \chi R_{\mu\nu} \psi^\mu \psi^\mu \chi + \frac{1}{2} \chi \phi^a T^\vr_a \chi. 
$$

Using the general arguments we can show that \Ref{local1} does not depend on $s(x)$ (see \Ref{BRSTinv}). Therefore we can simply set $s(x) = 0$. It leads to the following formula
\begin{equation}
\label{Euler}
\int_M \a = \int_X \D x \D \psi \D H \D \chi \iota^\ast \a \e^{i\bar{\sQ} \left( \chi s + \frac{i}{2} H \right)} = \int_X \iota^\ast \a \Eu_g(E)
\end{equation}
where $g = \e^\phi \in G$ and 
\begin{equation}
\label{EulerClass}
\Eu_g(E) = \frac{1}{{(2 \pi)}^{\dim F / 2}}\int_{\Pi F} \D \chi \e^{\frac{1}{4} \chi R_{\mu\nu} \psi^\mu \psi^\mu \chi + \frac{1}{2} \chi \phi^a T^\vr_a \chi}
\end{equation}
is the equivariant Euler class for a bundle $F \hookrightarrow E \stackrel{\pi}{\to} X$.

\begin{remark}
If $\bar{\sQ}$ is the de Rham differential when it becomes an ordinary Euler class 
$$
\Eu(E) = \frac{1}{{(2 \pi)}^{\dim F / 2}}\int_{\Pi F} \D \chi \e^{\frac{1}{4}\chi R_{\mu\nu}\psi^\mu\psi^\nu\chi}.
$$
If $E = TX$, and $\dim X = 2m$ then one can show using formulae for the Berezin integrals 
$$
\Eu(TX) = \frac{1}{{(2\pi)}^m} \Pf(R)
$$
where $\ds R = \frac{1}{2}R_{\mu\nu}\psi^\mu \psi^\nu$ is the curvature form. Then thanks to the Gauss-Bonnet-Hopf theorem
$$
\int_X \Eu(TX) = \chi_X,
$$
the Euler characteristic of $X$.
\end{remark}

The integral \Ref{Euler} does not depend on $s$. Therefore we can introduce another version of the Euler class:
$$
\Eu_g(E,s) = \frac{1}{{(2\pi)}^{\dim F /2}} \int_{\Pi F} \D \chi \e^{-\frac{1}{2}s^2 + i \chi \cD_\mu s \psi^\mu + \frac{1}{4} \chi R_{\mu\nu} \psi^\mu \psi^\mu \chi + \frac{1}{2} \chi \phi^a T^\vr_a \chi}.
$$
It can be seen as a pullback of $s : X \to E$ of a \emph{universal equivariant Thom class} $\Phi_g(E) \in \gO^\ast(E)$. The definition is the following. Denote by pair $(x^\mu,\xi^i)$ the local coordinates of $p \in E$. Let $\dd x^\mu = \psi^\mu$ and $\dd \xi^i$ be the basis of 1-forms on $E$. Define $\gG = \gG_\mu \psi^\mu$ and $\ds R = \frac{1}{2}R_{\mu\nu}\psi^\mu \psi^\nu$. Then
$$
\Phi_g(E) = \frac{1}{{(2\pi)}^{\dim F /2}} \int_{\Pi F} \D \chi \e^{-\frac{1}{2}\xi^2 + i \chi(\dd \xi + \gG\xi) + \frac{1}{2}\chi R \chi + \frac{1}{2}\chi \phi^a T_a^\vr \chi}
$$
It is clear that $\Eu_g(E,s) = s^\ast \Phi_g(E)$.

\begin{remark}
Usually in mathematical texts the Thom class is defined in a slightly different way. Consider the most explicit and simple example of the situation where $\dim F = 2$. In that case the general formula for the Thom class is the following \cite{Moore,BottTu}
$$
\Phi_g(E) = \r'(\xi^2)\left(\dd \xi + \gG \xi \right)\wedge\left(\dd \xi + \gG \xi \right) - \r(\xi^2) \Eu_g(E,0).
$$
where the function $\r(t)$ is such that $\ds \int_{\Real^2} \r'(\xi^2)\dd^2\xi = 1$. It is clear that our construction corresponds to the particular case 
$$
\r(t) = -\frac{1}{\pi}\exp\left\{-\frac{t}{2} \right\}.
$$
\end{remark}


\subsection{The Duistermaat-Heckman formula}

Another useful tool which we are going to exploit is the Duistermaat-Heckman formula. It allows us to express an integral over a symplectic manifold which is acted on by a torus $\Tor$ as a sum over the $\Tor$-stable points. Let us describe some relevant details.

Let $M$ be a $2n$ dimensional symplectic manifold, $\o$ be its symplectic form. Let $\Tor$ acts symplectically, and suppose that its action can be described by a Hamiltonian (momentum) map $\mu : M\to \alg{t}^\ast$, $\alg{t} = \Lie(\Tor)$. The choice of $\xi \in \alg{t}$ defines the Hamiltonian $h(\xi) = \<\mu,\xi\>$ and the action. It means that the $\dd h(\xi) = i_{V(\xi)} \o$. Let $x_f \in M$ be a fixed point of this action and $w_\a(x_f) \in \alg{t}^\ast$ a weight of this action on the tangent space to $x_f$. It means that on the tangent space to a fixed point $x_f$ the $\Tor$ action can be represented by a block diagonal matrix with blocks
$$
\left(
\begin{array}{cc}
\cos 2\pi w_\a(\xi) & \sin 2\pi w_\a(\xi) \\
-\sin 2\pi w_\a(\xi) & \cos 2\pi w_\a(\xi)
\end{array}
\right).
$$

Then the Duistermaat-Heckman formula states that
\begin{equation}
\label{DH}
\int_M \frac{\o^n}{n!}\e^{-\<\mu,\xi\>} = \sum_{x_f:\mr{fixed}} \frac{\e^{-\<\mu(x_f),\xi\>}}{\prod_{\a}\< w_\a(x_f),\xi\>}.
\end{equation}
In that follows we will basically use the shorthand notation $\< w_\a(x_f),\xi\> \equiv w_\a$.

To prove the formula we note that if we introduce the Cartan differential $\bar{\sQ} = \dd + i_{V(\xi)}$ then $\bar{\sQ} (\o - h(\xi)) = 0$, therefore $\o - h(\xi)$ is an equivariantly closed form. Note also that for any form 
$$
\int_M \bar{\sQ} \a = \int_M \dd \a + \int_M i_{V(\xi)} \a = 0
$$ 
(the second term vanished since it is not a top form). It follows that for any $\bar{\sQ}$ closed form $\a$ and for any $\Tor$ invariant form $\b$ we have
$$
\int_M \a = \int_M \a \e^{\bar{\sQ}\b}.
$$ 
If we choose $\b = - t V_\mu(\xi) \psi^\mu$ (cf \Ref{projection}) and $\a = \e^{\o - h(\xi)}$ then using 
$$
\bar{\sQ} \b = - t \pd_\mu V_\nu \psi^\mu\psi^\nu - t (V(\xi),V(\xi))
$$
and the standard localization arguments we arrive to \Ref{DH}.

\begin{remark}
When we deal with supermanifolds, which contain supercoordinates, the Duistermaat-Heckman formula should be modified as follows: $\prod_\a w_\a \mapsto \prod_\a {w_\a}^{\ep_\a}$ where $\ep_\a = \pm 1$ depends on the statistics of coordinate it comes from.
\end{remark}

It turns out to be easier to compute first the character of the torus element $q \in \Tor$:
$$
\Ind_q \equiv \sum_\a \ep_\a \e^{w_\a}.
$$
This can be done with the help of the equivariant analog of the Atiyah-Singer index theorem taking into account that the same quantity can be seen as the equivariant index of the Dirac operator. It worth noting that when $\Ind_q$ is derived equivariantely, the signs $\ep_\a$ comes from the alternated summation over cohomologies, and not from boson-fermion statistics.  

Once we have $\Ind_q$, the passage to the Duistermaat-Heckman formula can be done with the help of the following transformation (which can be seen as a proper time regularization, see section \ref{SU(N)PureYM}):
\begin{equation}
\label{TrWeights}
\sum_{\a} \ep_\a \e^{w_\a} \mapsto \prod_\a {w_\a}^{\ep_\a}.
\end{equation}
This transformation is performed in two steps: first we perform an integral transformation which converces $\e^{w_\a}$ to $\ln w_\a$. Then the exponent of the expression we have obtained $\ds\sum_\a \ep_\a \ln w_\a$ is precisely the rifgthand side of the announced formula.
 

\section{Back to Yang-Mills action}

Now it is time to look back at the action for super Yang-Mills. Consider first the pure Yang-Mills theory. Having compared \Ref{vev}, \Ref{VYM}, \Ref{VYMmod}, \Ref{local1}, \Ref{InstNumb} and  \Ref{Stop} we conclude that if $\O$ is a gauge invariant BRST closed operator, when $\< \O \>$ can be considered as an integral over the instanton moduli space of $\tld{\O}_k \in H^\ast(\Moduli_k)$, which belongs to the same cohomology class as $\O$. More precisely
\begin{equation}
\label{PYMvev}
{\< \O\>}_a = \sum_{k=0}^\infty \e^{2\pi i k\t} \int_{\Moduli_k} \tld{\O}_k.
\end{equation}
This is so since we have identified $s = {(F_{\mu\nu})}^-$ and the group which we factor by is the gauge group $\G = \{ g: \Real^4\to G: g(\infty) = \Id_G\}$. Note that the full gauge group is $\G_{\mr{full}} = \G \times G_\infty$, where $G_\infty$ is the group of the rigid gauge transformations, that is, the transformations at infinity.

Looking back to \Ref{Qtwisted} and \Ref{QtwistedMat} we see that $\bar{\sQ}^2$ produces the gauge transformation with the parameter $\phi$. From \Ref{FlatDirection} it follows that if the supersymmetry is unbroken then at infinity $\phi(x)\to a$, where $a \in \alg{g}$.  The notation ${\<\cdot\>}_a$ means that the vacuum expectation is taken with respect to such field configurations. Therefore, among others transformations, $\bar{\sQ}^2$ produces the rigid gauge transformations with parameters $a_l$, $l=1,\dots,r$. Taking into account the discussion in section \ref{intFactor} and the finite dimensional construction of the instanton moduli space we can schematically say that the full group of gauge transformations becomes the product $G_\infty \times G_D$.

The finite dimensional version of the Cartan differential squares, therefore, to 
\begin{equation}
\label{BRST^2}
\bar{\sQ}^2 = \EuG(a) + \EuF(m) + \EuD(\phi),
\end{equation}
where $\EuD(\phi)$ is a dual group transformation. 

Using the finite dimensional model for the instanton moduli space we can re-express the required vacuum expectation as a sum of the finite dimensional integrals. And therefore make the problem (in principle) doable.

In the presence of matter the situation is slightly different. First of all we note that if add to the pure Yang-Mills action terms which correspond to \Ref{Vmat} and \Ref{Vmass} then we can identify
$$
s = ({(F_{\mu\nu})}^- + i q^\dag_{\dot{\a}} \bar{\s}_{\mu\nu}{}^{\dot{\a}}{}_{\dot{\b}}q^{\dot{\b}}, \s^\mu_{\a\dot{\a}}\cD_\mu q^{\dot{\a}}),
$$
the multiplet $(\nu_\a,h_\a)$ with $(\chi,H)$ and $(q^{\dot{\a}},\mu^{\dot{\a}})$ with $(x,\psi)$. When the vacuum expectation can be localized to the moduli space of Seiberg-Witten monopoles, that is, to the solutions of the monopole equations
\begin{equation}
\label{SWmonopole}
\begin{aligned}
&{(F_{\mu\nu})}^- + i q^\dag_{\dot{\a}} \bar{\s}_{\mu\nu}{}^{\dot{\a}}{}_{\dot{\b}} q^{\dot{\b}} = 0, \\
&\s^\mu_{\a\dot{\a}}\cD_\mu q^{\dot{\a}} = 0
\end{aligned}
\end{equation}
up to a gauge transformation.

Another way to see the things is the following. First of all let us deform the action in such a way that the first equation becomes
$$
{(F_{\mu\nu})}^- + \frac{i}{t}q^\dag_{\dot{\a}} \bar{\s}_{\mu\nu}{}^{\dot{\a}}{}_{\dot{\b}} q^{\dot{\b}} = 0
$$
with an arbitrary $t$. In the $t\to\infty$ limit the equation reduces to the self-dual equation. Therefore the integral over the gauge multiplet localizes as before on the instanton moduli space. 

To deal with matter we observe that after integration out field $h^\a$ in \Ref{Vmat} the action becomes the equivariant Euler class \Ref{EulerClass} for a bundle $\mathfrak{D}_k$ over $\Moduli_k$ of the solutions for the Weyl equation. Indeed, the action which follows from \Ref{Vmat} forces fields to localize on the solutions of the Weyl equation
$$
\begin{aligned}
\s^\mu_{\a\dot{\a}} \cD_\mu q^{\dot{\a}} &= 0, \\
\s^\mu_{\a\dot{\a}} \cD_\mu \mu^{\dot{\a}} &= 0, \\
\bar{\s}^{\mu,\dot{\a}\a}\cD_\mu \nu_\a &= 0.
\end{aligned}
$$ 
There are no solutions for the first two equations. The solutions for the third are given by \Ref{WeylFund} for the fundamental representation and \Ref{WeylAdj} for the adjoint. The action on these solutions takes the following form:
\begin{equation}
\label{MatAct}
S_{\mr{mat}} =- \frac{1}{\pi^2} \int \dd^4 x \Tr \left\{ \bar{\nu}^\a \left( \phi+ m\right)\nu_\a   + \bar{h}^\a h_\a \right\}
\end{equation}

The equation \Ref{PYMvev} becomes
\begin{equation}
\label{YMmat}
{\< \O \>}_a = \sum_{k=0}^\infty \e^{2\pi i k\t} \int_{\Moduli_k} \tld{\O}_k \Eu_g(\mathfrak{D}_k)
\end{equation}
where $g  = (\e^m,\e^\phi,\e^a) \in G_F \times G_D \times G_\infty$.


\section{Lorentz deformation and prepotential}
\label{LorentzDef}

We have learned how to reduce the vacuum expectation to the finite dimensional integral. However in order to get access to the prepotential it is not sufficient. We should further deform our BRST operator $\bar{\sQ}$. It is already deformed in such a way that it squares induces $G_F\times G_D\times G_\infty$ transformation. We have another group with respect to which the action of the Yang-Mills theory is invariant. This is the Lorentz group. The deformed Yang-Mills action can be naturally described in the terms of so-called $\gO$-background.


\subsection{$\gO$-background}
\label{O-background}

In section \ref{DimensionalReduction} we have learned how to produce $\N=2$ super Yang-Mills action via dimensional reduction of $\N=1$, $d=6$ super Yang-Mills action. While compactifying we have used the following flat metric:
$$
\dd s^2_6 = g_{\mu\nu} \dd x^\mu \dd x^\nu - {(\dd x^4)}^2 - {(\dd x^5)}^2.
$$

Now let the torus $\Tor^2$ act on $\Real^{1,3}$ by Lorentz rotations. Its action is governed by the following vectors:
$$
\begin{aligned}
V_4^\mu &= \gO_4{}^\mu{}_\nu x^\nu, & V_5^\mu &= \gO_5{}^\mu{}_\nu x^\nu,
\end{aligned}
$$
where $\gO_a{}^\mu{}_\nu$, $a=4,5$ are matrices of Lorentz rotations. Since $\pi_1(\Tor^2)$ is commutative we conclude that the Lie bracket of $V_4^\mu$ and $V_5^\mu$ should vanish. It is equivalent to say that matrices $\gO_4$ and $\gO_5$ commute. Let us define the following metric \cite{SmallInst,SWandRP}:
$$
\begin{aligned}
\dd s_6^2 &= g_{\mu\nu}\left( \dd x^\mu + V^\mu_a \dd x^a\right)\left(\dd x^\nu + V^\nu_b \dd x^b\right) - {(\dd x^4)}^2 - {(\dd x^5)}^2 \\
&= G_{IJ} \dd x^I \dd x^J.
\end{aligned}
$$ 
We have
$$
\begin{aligned}
G_{\mu\nu} &= g_{\mu\nu}, & G^{\mu\nu} &= g^{\mu\nu} - V_a^\mu V_a^\nu, \\
G_{a\mu} &= V_{a,\mu}, & G^{a\mu} &= V_a^\mu, \\
G_{ab} &= - \d_{ab} + V_a^\mu V_{b,\mu}, & G^{ab} &= - \d^{ab}.
\end{aligned}
$$
One can also check that $G = \det G_{IJ} = -1$. Computation shows that this metric is flat when the matrices $\gO_4$ and $\gO_5$ commute.

In that follows we will use the six dimensional vielbein $e_I^{(J)}$ which satisfies
$$
\dd s_6^2 = g_{\mu\nu} e^{(\mu)}_I e^{(\nu)}_J \dd x^I \dd x^J - e^{(a)}_I e^{(a)}_J \dd x^I \dd x^J.
$$
It can be represented as follows:
$$
\begin{aligned}
e^{(\mu)}_\nu &= \d^\mu_\nu, & e^\mu_{(\nu)} &= \d^\mu_\nu, \\
e^{(\mu)}_a &= V^\mu_a, & e^a_{(\mu)} &= 0, \\
e^{(a)}_\mu &= 0, & e^\mu_{(a)} &= - V_a^\mu,\\
e^{(a)}_b &= \d^a_b & e^a_{(b)} &= \d^a_b.
\end{aligned}
$$

Let us write the action \Ref{N=1,d=6} in this background, keeping the compactification. Using the vielbein we get
$$
-\frac{1}{4}\sqrt{-G} F_{IJ}F_{KL}G^{IK}G^{JL} = -\frac{1}{4}F_{(I)(J)}F^{(I)(J)}.
$$
Computation shows that
$$
\begin{aligned}
F_{(\mu)(\nu)} &= F_{\mu\nu}, \\
F_{(a)(\mu)} &= F_{a\mu} - V_a^\r F_{\r\mu}, \\
F_{(a)(b)} &= V_a^\mu V_b^\nu F_{\mu\nu} - F_{a\nu} V_b^\nu - V_a^\mu F_{\mu b} + F_{ab}.
\end{aligned}
$$

Let us introduce the complex combination of $V_a^\mu$ and $\gO_a{}^\mu{}_\nu$ keeping in mind \Ref{HiggsViaCompactification}:
$$
\begin{aligned}
V^\mu &= \frac{1}{\sqrt{2}}\left( V_4^\mu + i V_5^\mu\right), & \bar{V}^\mu &= \frac{1}{\sqrt{2}}\left( V_4^\mu - i V_5^\mu\right), \\
\gO{}^\mu{}_\nu &= \frac{1}{\sqrt{2}}\left( \gO_4{}^\mu{}_\nu + i \gO_5{}^\mu{}_\nu\right), & \bar{\gO}^\mu{}_\nu &= \frac{1}{\sqrt{2}}\left( \gO_4{}^\mu{}_\nu - i \gO_5{}^\mu{}_\nu\right).
\end{aligned}
$$
The bosonic part of the action can be written as follows:
$$
\begin{aligned}
-\frac{1}{4}\sqrt{-G}F_{IJ}F_{KL}G^{IK}G^{JL} &= - \frac{1}{4}F_{\mu\nu}F^{\mu\nu} + \left( \cD_\mu H + V^\r F_{\r\mu}\right)\left( \cD^\mu H^\dag + \bar{V}^\r F_\r{}^\mu \right) \\
&- \frac{1}{2}{\left\{[H,H^\dag] - i\bar{V}^\mu V^\nu F_{\mu\nu} - i \left( V^\mu \cD_\mu H^\dag - \bar{V}^\mu \cD_\mu H \right)\right\}}^2.
\end{aligned}
$$

Note that when $\gO$ and $\bar{\gO}$ commute the last line can be rewritten as ${[\H,\H^\dag]}^2$ where
$$
\begin{aligned}
\H &= H - i V^\mu \cD_\mu, & \H^\dag &= H^\dag - i \bar{V}^\mu \cD_\mu.
\end{aligned}
$$
This shift can be explained as follows. Consider a function $\f$ belonging to the adjoint representation of the gauge group and to a representation of the Lorentz group. Let $S_{\mu\nu}$ be the spin operator for this representation. In the non-deformed case we had 
$$
\cD\f = \frac{1}{\sqrt{2}}\left( \cD_4 + i \cD_5\right) \f = -i[H,\f].
$$
In the $\gO$-background this expression is deformed as follows
$$
\begin{aligned}
\mbf{\cD} \f =& \frac{1}{\sqrt{2}}\left( \mbf{\cD}_{(4)} + i \mbf{\cD}_{(5)}\right) \f = \frac{1}{\sqrt{2}} \left( e_{(4)}^I \cD_I + ie_{(5)}^I \cD_I\right)\f \\
&= -i[H,\f] - V^\mu \cD_\mu \f + \frac{1}{2}\gO^{\mu\nu} S_{\mu\nu}\f= - i [\H,\f].
\end{aligned}
$$
Since $H$ itself is a scalar for this field the spin opertator is trivial. However it becomes non-trivial when acts on spinors. Therefore in the general case the shift is the following
$$
\begin{aligned}
H &\mapsto & \H &= H - i V^\mu \cD_\mu + \frac{i}{2}\gO^{\mu\nu} S_{\mu\nu}, \\
H^\dag &\mapsto & \H^\dag &= H^\dag - i \bar{V}^\mu \cD_\mu - \frac{i}{2}\bar{\gO}^{\mu\nu} \bar{S}_{\mu\nu}.
\end{aligned}
$$

Also note that when $\gO$ and $\bar{\gO}$ commute, the whole expression can be written up to a total derivative, which is irrelevan, as follows:

\begin{equation}
\label{BosonicPart}
\begin{aligned}
-\frac{1}{4}\sqrt{-G}F_{IJ}F_{KL}G^{IK}G^{JL} &= -\frac{1}{4}F_{\mu\nu}F^{\mu\nu} + \frac{1}{2}\left\{[\cD_\mu,\H],[\cD^\mu,\H^\dag]\right\} - \frac{1}{2}{[\H,\H^\dag]}^2 \\
&- \frac{1}{4}\bar{\gO}_{\r\mu}F^{\r\mu}\H - \frac{1}{4}\gO_{\r\mu}F^{\r\mu}\H^\dag
\end{aligned}
\end{equation}

The fermionic term can be written as follows:
$$
\frac{i}{2} \bar{\Psi}_A \gG^I e_{(I)}^J \mbf{\cD}_J \Psi^A,
$$
where in order to define the covariant derivative we should use the spin connection which can be written with the help of the Ricci coefficients:
$$
\g_{I,JK} = \frac{1}{2}e_{(I)}^M \left( e^L_{J} \mbf{\cD}_M e_{L(K)} - e^L_{(K)} \mbf{\cD}_M e_{L(J)}\right) = e_{(I)}^M e_{(J)}^L \mbf{\cD}_M e_{L(K)}.
$$
Computation shows that when $\gO_4$ and $\gO_5$ commute the only nonvanishing coefficient is 
$$
\g_{a,\mu\nu} = - \g_{a,\nu\mu} = - \gO_{a,\mu\nu}.
$$
 
For the covariant derivative we have the following expression:
$$
\mbf{\cD}_I \Psi = \pd_I \Psi - i [A_I,\Psi] + \frac{1}{2}\gS^{PQ}\g_{I,PQ} \Psi = \cD_I \Psi + \frac{1}{2}\gS^{PQ}\g_{I,PQ} \Psi,
$$
where $\ds \gS^{PQ} = \frac{1}{4} [ \gG^P,\gG^Q ]$ are the generators of the six dimensional Lorentz group in the spinor representation.

The calculation shows that the fermionic part of the Lagrangian can be represented as follows:
\begin{equation}
\label{FermionicPart}
\begin{aligned}
\frac{i}{2}\bar{\Psi}_A \gG^I e_{(I)}^J \mbf{\cD}_J \Psi^A &=  i \psi_A \s^\mu \cD_\mu \bar{\psi}^A  - \frac{i}{\sqrt{2}}\psi_A[\H^\dag, \psi^A] + \frac{i}{\sqrt{2}} \bar{\psi}^A[\H,\bar{\psi}_A] \\
&-\frac{1}{2\sqrt{2}}\gO_{\mu\nu} \bar{\psi}^A \bar{\s}^{\mu\nu}\bar{\psi}_A - \frac{1}{2\sqrt{2}} \bar{\gO}_{\mu\nu} \psi_A \frac{1}{2}\s^{\mu\nu}\psi^A.
\end{aligned}
\end{equation}

Having compared the initial action \Ref{MicroActionComponent} with the deformed one, which is the sum of \Ref{BosonicPart} and \Ref{FermionicPart}, we can note that there are only distingtion is coming from the formal shift $H \mapsto \H$ and from the additional terms which can be written as follows
$$
-\frac{1}{4}\bar{\gO}_{\r\mu}F^{\r\mu}\H - \frac{1}{4}\gO_{\r\mu}F^{\r\mu}\H^\dag-\frac{1}{2\sqrt{2}}\gO_{\mu\nu} \bar{\psi}^A \bar{\s}^{\mu\nu}\bar{\psi}_A - \frac{1}{2\sqrt{2}}\bar{\gO}_{\mu\nu}  \psi_A \s^{\mu\nu} \psi^A.
$$

Topological term should also be modified. Putting all things together we can see that the whole effect of the introducing of the $\gO$-background consists of the shift $H\mapsto \H$ and the following modification of the complex coupling constant:
$$
\t \mapsto \t(x,\th) = \t - \frac{1}{\sqrt{2}} {\left(\bar{\gO}_{\mu\nu}\right)}^+ \th^\mu\th^\nu,
$$
where we have used twisted supercoordinates: $\th^\mu = \bar{\s}^{\mu,A\a} \th_{A,\a}$.


\subsection{Getting the prepotential}

In order to use the powerful machinery of the equivariant integration we should be sure that the action in the $\gO$-background is still BRST exact with some BRST (or BRST-like) operator.

Being inspired by the formula \Ref{ActionTwistedExact} we perform some computations. First of all we have
$$
\sQ_\mu \left\{ \frac{1}{4}{\left(F_{\r\t}\right)}^- \bar{\psi}^{\r\t} -\frac{i}{2\sqrt{2}} \psi^\r \cD_\r H^\dag + \frac{i}{4}\bar{\psi}[H,H^\dag] \right\} = -\sqrt{2} \star F_{\mu\r}\cD^\r H^\dag - i \cD_\r \left( \bar{\psi}_{\mu\t}\bar{\psi}^{\r\t}\right).
$$  
When using the equations of motion for $\psi_\mu$ and $\bar{\psi}_{\mu\nu}$, which are modified in the $\gO$-background we can show that the Yang-Mills action in the $\gO$-background is $\bar{\sQ}_\gO$ exact where  
$$
\bar{\sQ}_\gO = \bar{\sQ} + \frac{1}{2\sqrt{2}}\gO^\mu{}_\nu x^\nu \sQ_\mu,
$$
provided we further shift the complex coupling constant:
$$
\t \mapsto \t(x,\th) = \t - \frac{1}{\sqrt{2}} \left[{\left(\bar{\gO}_{\mu\nu}\right)}^+ \th^\mu\th^\nu - \frac{1}{2\sqrt{2}}\bar{\gO}_{\mu\nu} \gO^\mu{}_\r x^\r x^\nu\right].
$$
This modification changes only the topologica term, which have not yet been discussed.

Note that thanks to the supersymmetry algebra which yields $\{ \bar{\sQ},\sQ_\mu\} = 4 i \cD_\mu$ up to a gauge transformation we get
$$
{\left(\bar{\sQ}_\gO\right)}^2 = i\sqrt{2}\gO^\mu{}_\nu x^\nu \pd_\mu.
$$

It worth noting that the superspace dependent complex coupling constant is annihilated by the following operator:
$$
\R_\gO = \th^\mu \frac{\pd}{\pd x^\mu} + \frac{1}{2\sqrt{2}} \gO^\mu{}_\nu x^\nu \frac{\pd}{\pd \th^\mu}. 
$$

This observation allows us to get access to the prepotential \cite{SWfromInst}. Indeed, taking into account the relation between the dynamically generated scale and the complex coupling constant \Ref{InstantonParameter} we conclude that in the $\gO$-background $\gL$ becomes effectively $x$ and $\th$ dependent. Moreover, this dependence is such that this new $\gL$ is annihilated this operator:
\begin{equation}
\label{LambdaAnnihil}
\R_\gO \gL(x,\th) = \R_\gO \e^{2\pi i \t(x,\th)} = 0.
\end{equation}

The second observation is that since the action is $\bar{\sQ}_\gO$ exact we can localize it on the zero-modes of the superfield $\ds\Psi(x,\th) =  H(x) +.\,.\,. = - \frac{1}{2\sqrt{2}} \f(x) + \dots$. Therefore the functional integral for the partition function of the theory on the $\gO$-background becomes
$$
Z(a;\gO) = {\< 1 \>}_a = \exp\Im \left\{\frac{1}{2\pi i}\int \dd^4 x \dd^4 \th\Prep(-\frac{1}{2\sqrt{2}} a,\gL(x,\th)) \right\}.
$$

The integral at the exponent on the righthand side can be computed using the localization arguments. The operator $\R_\gO$ can be seen as the Cartan differentail: $\ds\R_\gO = \dd + \frac{1}{2\sqrt{2}}i_V$. Therefore we can apply the Duistermaat-Heckman formula. 

Let us choose the coordinate system on $\Real^4$ where the matrix $\gO$ has the canonical form: 
$$
\gO = \frac{1}{\sqrt{2}}\left( 
\begin{array}{cccc}
0 & 0 & 0 & -\eps_1 \\
0 & 0 & -\eps_2 & 0 \\
0 & \eps_2 & 0 & 0 \\
\eps_1 & 0 & 0 & 0
\end{array}
\right).
$$
This choice corresponds to the complex structure introduced in \Ref{ComplStruct}.

The weights which correspond to the action of the operator at the lefthand side of \Ref{LambdaAnnihil} are
$$
\begin{aligned}
&-\frac{1}{4} \eps_1 & &\mbox{and} & &-\frac{1}{4}\eps_2.
\end{aligned}
$$
\begin{remark}
With this definition the weights of the action $\ds{\left(\bar{\sQ}_\gO\right)}^2$ are $\eps_1$ and $\eps_2$. This fact will be used in the next chapter. 
\end{remark}

Therefore using the localization and the fact that the prepotential is a homogenious function of the degree $2$ we get
\begin{equation}
\label{Z}
Z(a,m,\gL;\eps)=\exp\frac{1}{\eps_1\eps_2}\Prep(a,m,\gL;\eps),
\end{equation}
where the prepotential can be obtained from the function $\Prep(a,m,\gL;\eps)$ by taking $\eps_1, \eps_2\to 0$ limit: $\ds\Prep(a,m,\gL) = \lim_{\eps_1,\eps_2\to 0} \Prep(a,m,\gL;\eps)$. 

Now combining \Ref{PYMvev} (or in general case \Ref{YMmat}) with \Ref{Z} we get a way compute the prepotential:
$$
\exp\frac{1}{\eps_1\eps_2} \Prep(a,m,\gL;\eps) = \sum_{k=0}^\infty q^k \int_{\Moduli_k} \Eu_g(\mathfrak{D}_k),
$$
where $g = (\e^m,\e^\phi,\e^a,\e^{i\eps}) \in \Tor_F \times \Tor_D \times \Tor_\infty \times \Tor_L$.

Note that through the discussion of this chapter we have not say a word about the gauge fixing procedure, which should be performed in order to compute properly the functional integral. The proper accounting of this procedure leads to a supplementary factor for $Z^{\mr{pert}}(a,m,\gL;\eps)$. This factor leads, in particular, to the perturbative contribution to the prepotential \Ref{pertPrep}. See the discussion in sections \ref{PartSU} and \ref{SU(N)PureYM}.

Therefore the formula for the prepotential takes the following form:
$$
Z(a,m,\gL;\eps) = Z^{\mr{pert}}(a,m,\gL;\eps) \times \sum_{k=0}^\infty q^k Z_k(a,m;\eps) = \exp{\frac{1}{\eps_1\eps_2}\Prep(a,m,\gL;\eps)},
$$
where $Z^{\mr{pert}}(a,m,\gL;\eps)$ is the perturbative contribution to the partition function and
\begin{equation}
\label{Z_k}
Z_k(a,m;\eps) = \int_{\Moduli_k} \Eu_g(\BundleD_k).
\end{equation}

%% file: finite.tex
\label{finite}

In this chapter we derive the expression for the prepotential using the finite dimensional model for the instanton moduli space. We obtain formula which express $Z_k(a,m,\gL;\eps)$ in \Ref{Z_k} as a finite dimensional integral. After that we will rederive them in the spirit of the Duistermaat-Heckman formula.


\section{Direct computations: $SU(N)$ case}

Let us obtain the formulae for $SU(N)$ model without matter. We will  follow \cite{GeneralizedInst,DeformationInstanton,SWfromInst}. 

First of all let us introduce the finite dimensional analog of the BRST operator $\bar{\sQ}$. The instanton moduli space is the subset of the space of linear operators $B_1$, $B_2$, $I$ and $J$ factored by the dual group. This subset can be described by the ADHM equation \Ref{ADHM} $\mu_\Real = 0$ and $\mu_\Compl = 0$. The general discussion of the section \ref{equivInt} shows that in order to take into account this fact we should introduce the following supplementary multiplets:
$$
\begin{aligned}
&(\chi_\Real,H_\Real) & &\mbox{and} & &(\chi_\Compl,H_\Compl).
\end{aligned}
$$ 

The transformation properties of the matrices $B_1$, $B_2$, $I$, $J$ and the ADHM equation with respect to $\Tor_L$ are
$$
\begin{aligned}
B_1 &\mapsto \e^{i\eps_1}B_1, & B_2 &\mapsto \e^{i\eps_2}B_2, \\
I &\mapsto \e^{-i\eps_+} I, & J &\mapsto \e^{-i\eps_+} J, \\
\mu_\Real &\mapsto \mu_\Real, & \mu_\Compl &\mapsto \e^{i\eps} \mu_\Compl,
\end{aligned}
$$
where $\eps = \eps_1 + \eps_2$ and $\ds\eps_+ = \frac{1}{2}\eps = \frac{1}{2}(\eps_1 + \eps_2)$. Taking into account \Ref{Qtwisted}, \Ref{BRST^2} and the Lorentz deformation of the BRST operator we can write
\begin{equation}
\label{Qfinite}
\begin{aligned}
\bar{\sQ} B_{1,2} &= \psi_{1,2}, & \bar{\sQ} \psi_{1,2} &= [\phi,B_{1,2}] + i\eps_{1,2} B_{1,2}, \\
\bar{\sQ} I &= \psi_I, & \bar{\sQ}\psi_I &= \phi I - I a - i\eps_+ I , \\
\bar{\sQ} J &= \psi_J, & \bar{\sQ}\psi_J &= - J\phi + a J - i\eps_+ J, \\
\bar{\sQ} \chi_\Real &= H_\Real, & \bar{\sQ} H_\Real &= [\phi,\chi_\Real], \\
\bar{\sQ} \chi_\Compl &= H_\Compl, & \bar{\sQ} H_\Compl &= [\phi,\chi_\Compl] + i\eps \chi_\Compl, \\
\bar{\sQ} \eta &= \l, & \bar{\sQ} \l &= [\phi,\l].
\end{aligned}
\end{equation}

The finite dimensional version of \Ref{Z_k} is
\begin{equation}
\label{Z_KSU}
Z_k(a;\eps) = \int \frac{\D \phi}{\Vol(G_D)}\D \eta\D\l\D H \D\chi \D B_1\D B_2\D I \D J \D \psi \e^{i\bar{\sQ}\left( \chi\cdot\mu + t \chi \cdot H +  \psi \cdot V(\l) \right)}
\end{equation}
where 
$$
\begin{aligned}
\chi \cdot \mu &= \Tr \left\{ \chi_\Real \mu_\Real + \frac{1}{2}\left( \chi_\Compl^\dag \mu_\Compl + \chi_\Compl \mu_\Compl^\dag \right)\right\}, \\
\chi \cdot H &= \Tr \left\{ \chi_\Real H_\Real +\frac{1}{2}\left(  \chi_\Compl^\dag H_\Compl + \chi_\Compl H_\Compl^\dag \right) \right\},
\end{aligned}
$$
(note the torus action on $\chi_\Real$ and $\chi_\Compl$ is chosen in such a way that $\chi \cdot \mu$ be invariant) and $V(\l)$ is the dual group flow vector field: 
$$
\psi \cdot V(\l) = \Tr \left\{ \psi_1 [\l,B_1^\dag] + \psi_2 [\l,B_2^\dag] + \bar{\psi}_1 [\l,B_1] + \bar{\psi}_2 [\l,B_2] + \psi_I \l I - I^\dag \l \bar{\psi_I} - J \l \bar{\psi}_J + \psi_J \l J^\dag \right\}.
$$

Let us consider two way to do this integral: straightforwardly and using the Duistermaat-Heckman formula.


\subsection{Straightforward computation}
\label{straightforward}

To compute this integral we add to the exponent two BRST-exact terms, which should not change the integral:
\begin{equation}
\label{localizator}
i\bar{\sQ}t' \Tr\chi_\Real \l - \bar{\sQ} \frac{1}{2}t'' \Tr \left\{ \sum_{s=1}^2\left(B_s^\dag \psi_s - \bar{\psi}_s B_s \right) - I^\dag \psi_I + \bar{\psi}_I I - J^\dag \psi_J + \bar{\psi}_J J \right\}.
\end{equation}

The term, proportional to $t'$ produces
$$
t' \Tr H_\Real \l + t' \Tr \chi_\Real \eta.
$$
Therefore if we take first the limit $t'\to\infty$ we can first integrate out $H_\Real$ which gives the delta function of $t' \l$. Then we integrate out $\l$ which gives the factor $\ds\frac{1}{{t'}^{k^2 / 2}}$. And finally when we do the integral over $\chi_\Real$ and $\eta$ we remove this factor.

Now let us compute the contribution of terms which proportional to $t$. They can be written as follows:
$$
t \Tr H_\Real H_\Real + t H_\Compl H_\Compl^\dag + t \Tr \chi_\Real [\phi,\chi_\Real] + t \Tr \chi_\Compl^\dag \left( [\phi,\chi_\Compl] + i \eps \chi_\Compl\right). 
$$
Note that the first term in fact, does not present, since it is already integrated out (when $t' \gg t$ it can be neglected). The third term does not contribute neither since in the $t'\to\infty$ limit all terms which are proportional to a power $\Tr\chi_\Real[\phi,\chi_\Real]$ will be suppressed.

Now we take the $t\to\infty$ limit. $H_\Compl$ integral is Gaussian and produces the factor $\frac{1}{t^{k^2}}$. In order to compensate it we integrate out $\chi_\Compl$. But first let us reduce the $\phi$ integral from the Lie algebra of $G_D$ to the Lie algebra of its maximal torus. The price we pay is the Vandermond and the order of the Weyl group. In section \ref{HaarMeasures} we treat this question in the general framework. Let us here just cite the result for $G_D = U(k)$:
$$
\D \phi \mapsto \frac{1}{k!} \prod_{i=1}^k \frac{\dd \phi_i}{2\pi i} \prod_{i<j\leq k} \left( \phi_i - \phi_j\right)^2.
$$
The advantage is that now the matrix $\phi$ can be seen as the diagonal one with eigenvalues $i\phi_i$:
$$
\phi = \diag\{i\phi_1,\dots,i\phi_k \}.
$$

It follows that when we integrate out $\chi_{\Compl,ij}^\dag$ and $\chi_{\Compl,ij}$ we get factor $\phi_i - \phi_j + \eps$. Therefore when we integrate out $\chi_\Compl^\dag$ and $\chi_\Compl$ completely we get
$$
\eps^k \prod_{i< j \leq k} \left( {\left(\phi_i - \phi_j\right)}^2 + \eps^2 \right)
$$
\begin{remark}
Note that this contribution formally looks like the Vandermond if one set $\eps = 0$ (and remove the factor $\eps^k$). However, we cannot integrate out $\chi_\Real$ and $H_\Real$ before introducing $t'$ term. The reason is the presence of zero modes in the expression $\Tr\chi_\Real [\phi,\chi_\Real]$. Besides, even if we forget about these zero modes and integrate out $\chi_\Real$ we would get only the square root of the Vandermond. However, this similarity can be used to get the expression for the Vandermond using the transformation rules for $\chi_\Real$ (see below).
\end{remark}

Let us finally sent $t''\to \infty$ and integrate out the rest of ``fields'' (but $\phi$). The contribution to the exponent is the following:
\begin{multline}
\label{t''}
- t'' \Tr \left\{ \bar{\psi}_1 \psi_1 + \bar{\psi}_2 \psi_2 + \bar{\psi}_I \psi_I + \bar{\psi}_J \psi_J\right\} \\ 
- t'' \Tr \left\{ \sum_{s=1}^2B_s^\dag \left([\phi,B_s] + i\eps_s B_s \right) - I^\dag (\phi I - I a  - i\eps_+ I) - J^\dag(-J \phi + a J -  i \eps_+) J\right\}.
\end{multline}
 
When we integrate out $B_s$ (recall that it is complex) we get the following contributions (up to the factor $\ds \frac{1}{{t''}^{k^2}}$ which can be killed by $\psi_s$ integration):
$$
\frac{1}{\eps_s^k}\prod_{i<j\leq k}\frac{1}{\left( {\left(\phi_i - \phi_j\right)}^2 - \eps_s^2 \right)}.
$$
Let in now remember that $a \in \Tor_\infty$. That is, has the following form:
$$
a = \diag\{ ia_1,\dots,ia_N\}.
$$
Taking this observation into account one can see that the same procedure applied to $I$ and $J$ (as usual, accompanied by the integration over $\psi_I$ and $\psi_J$) leads to the factor
$$
\prod_{i=1}^k\prod_{l=1}^N \frac{1}{\left( {\left( \phi_i - a_l\right)}^2 - \eps_+^2\right)}.
$$

Now let us combine all pieces. In order to simplify formulae we introduce the following notations:
\begin{equation}
\label{PandDeltaSU}
\begin{aligned}
\Delta_\pm(x) &= \prod_{i<j\leq k} \Big( (\f_i \pm \f_j)^2 - x^2\Big) \\
\P(x) &= \prod_{l=1}^N (x-a_l).
\end{aligned}
\end{equation}
Then the integrand for the partition function is given by
\begin{equation}
\label{directSU}
z_k(a,\phi;\eps) = \frac{1}{k!}\frac{\eps^k}{\eps_1^k\eps_2^k}\frac{\Delta_-(0)\Delta_-(\eps)}{\Delta_-(\eps_1)\Delta_-(\eps_2)}\prod_{i=1}^k \frac{1}{\P(\f_i+\eps_+)\P(\f_i-\eps_+)}.
\end{equation}

The formula for $Z_k(a;\eps)$ is
$$
Z_k(a;\eps) = \int \prod_{i=1}^k \frac{\dd \phi_i}{2\pi i} z_k(a,\phi;\eps).
$$

\begin{remark}
The integral on $\phi$ seems to pass through the poles of the integrand. However this is not the case. To see this we should consider the expression \Ref{t''}. The integral over $B_s$ has schematically the form $\dd\int_\Real \dd x\e^{i\eps x^2}$. In order to make it convergent we should introduce the shift $\eps_s \mapsto \eps_s + i0$. Therefore this integral can be computed by residues. Note that another way to get this shift is to consider $-\eps_s$ as an infrared regularizer (the mass) for the $B_s$ integral (which is needed for zero modes of $\Tr B_s^\dag[\phi,B_s]$). The shift is then nothing but the Feynman rule of bypassing the poles.
\end{remark}


\subsection{Stable points computation}

Now let us describe another way to understand formula \Ref{directSU}. 

Suppose that we have already integrated out $\chi_\Real$, $H_\Real$ and the projection multiplet $(\l,\eta)$. Consider the superspace spanned by $B_1$, $B_2$, $I$, $J$, and $\chi_\Compl$. Formulae \Ref{Qfinite} suggests that $\psi_1$, $\psi_2$, $\psi_I$, $\psi_J$ and $H_\Compl$ should be their differentials. On this space the torus $\Tor_D \times \Tor_{\infty} \times \Tor_L$ acts as follows:
\begin{equation}
\label{Action}
\begin{aligned}
B_s &\mapsto \e^\phi B_s \e^{-\phi}\e^{i\eps_s}, \\
I &\mapsto  \e^{\phi}I\e^{-a}\e^{-i\eps_+}, \\
J &\mapsto \e^{a}J \e^{-\phi}\e^{-i\eps_+}, \\
\chi_\Compl &\mapsto \e^{\phi}\chi_\Compl \e^{-\phi}\e^{-i\eps}.
\end{aligned}
\end{equation}
 
This action has the only stable point, the origin. Therefore the integral over $B_s$, $I$, $J$ and $\chi_\Compl$ can be computed with the help of \Ref{DH} provided we know the weights $w_\a$. However, they can be easily read from the infinitesimal form of \Ref{Action}. We have:
$$
\begin{aligned}
&\phi_i - \phi_j + \eps_s & &\mbox{for $B_{s,ij}$}, \\
&\phi_i - a_l - \eps_+ & &\mbox{for $I_{il}$}, \\
&a_l - \phi_i - \eps_+ & &\mbox{for $J_{li}$}, \\
&\phi_i - \phi_j - \eps& &\mbox{for $\chi_{\Compl,ij}$}.
\end{aligned}
$$

Applying then the Duistermaat-Heckman formula \Ref{DH} and taking into account the statistics of the coordinates we get \Ref{directSU}.

\begin{remark}
Note that if one applies the same recipe for $\chi_\Real$ one gets the following weights 
$$
\begin{aligned}
&\phi_i - \phi_j, & i&\neq j.  
\end{aligned}
$$
If we include them into the Duistermaat-Heckman formula we get precisely the Vandermond.
\end{remark}


\section{Haar measures}
\label{HaarMeasures}

Even though the Haar measure can be obtained considering the weight for $\chi_\Real$ let us describe its  standard group theoretical derivation.

The general formula for the Haar measure reduced to the maximal torus of the group is given by:
$$
\dd \mu_G = \frac{1}{|W|} \prod_{i=1}^{\rank G} \frac{\dd \phi_i}{2\pi} {\left| \prod_{\alpha \in {\Delta}^+} \left(
\e^{\frac{i}{2}\langle \a, \phi\rangle} - \e^{-\frac{i}{2}\langle \a, \phi\rangle}\right)\right|}^2
$$ 

This measure gives a measure on the Lie algebra
(it corresponds to the limit of  small $\phi_i$)
$$
\dd\mu_{\alg{g}} = \frac{1}{|W|} \prod_{i=1}^{\rank G} \frac{\dd\phi_i}{2\pi} \prod_{\alpha \in
{\Delta}^+} {\langle\alpha, \phi\rangle}^2 .
$$

Using the root systems of algebras of type $B$, $C$ and $D$ we can write the measures:
$$
\begin{aligned}
\dd\mu_{B_k} &= \frac{1}{2^k k!} \prod_{i=1}^k \frac{\dd\phi_i}{2\pi} \prod_{i<j}{(\phi_i^2 - \phi_j^2)}^2 \prod_{i=1}^k \phi_i^2, \\
\dd\mu_{C_k} &= \frac{1}{2^k k!} \prod_{i=1}^k \frac{\dd\phi_i}{2\pi} \prod_{i<j}{(\phi_i^2 - \phi_j^2)}^2 \prod_{i=1}^k (2 \phi_i)^2, \\
\dd\mu_{D_k} &= \frac{1}{2^{k-1} k!} \prod_{i=1}^k \frac{\dd\phi_i}{2\pi} \prod_{i<j}{(\phi_i^2 - \phi_j^2)}^2 .
\end{aligned}
$$
The detailed investigation of the Haar measure can be found for example in  \cite{BourbakiLie}, and in \cite{KrStau1,KrStau2}.


\section{$SO(N)$ and $Sp(N)$ gauge groups}

Let us find the analog of \Ref{directSU} for pure Yang-Mills theories with the gauge groups $SO(N)$ and $Sp(N)$. Since we have already described the finite dimensional model for the instanton moduli space for these groups the only thing we need is to find the weights of the $\Tor_D \times \Tor_\infty \times \Tor_L$ action. 


\subsection{$SO(N)$ case}

Let us choose the matrices from the Cartan subalgebra of $SO(N)$ and $Sp(k)$ in the standard forms:
$$
\begin{aligned}
a &= \diag \left\{  \left(
\begin{array}{cc}
0 & -a_1 \\
+a_1 & 0
\end{array}
\right),\dots,\left(
\begin{array}{cc}
0 & -a_n \\
+a_n & 0
\end{array} \right),\diamondsuit \right\} \in \Lie(SO(N)), \\
\phi &= \left(
\begin{array}{cc}
\tilde{\phi} & 0 \\
0 & -\tilde{\phi}
\end{array} \right) \in \Lie(Sp(k))
\quad\mbox{where}\quad \tilde{\phi}  = \diag\{i\phi_1,\dots,i\phi_k\}.
\end{aligned}
$$
Here $\diamondsuit = 0$ for odd $N$ and is absent for even $N$.  The eigenvalues $a_1,\dots,a_n$ and $\phi_1,\dots,\phi_k$ are assumed real.

Now we have to rewrite the $\Tor_D\times \Tor_\infty\times\Tor_L$ action in terms of the building blocks for matrices $B_s$: \Ref{SOBB} $M_\Compl$ and $N_\Compl$ \Ref{SOMN}. We have ($s = 1,2$):
$$
\begin{aligned}
P_s &\mapsto \e^{\tilde{\phi}}P_s \e^{-\tilde{\phi}}\e^{i\eps_s}, & M_\Compl &\mapsto \e^{\tilde{\phi}}M_\Compl \e^{-\tilde{\phi}}\e^{i\eps_s}, \\
Q_s &\mapsto \e^{-\tilde{\phi}}Q_s \e^{-\tilde{\phi}}\e^{i\eps_s}, &  N_\Compl &\mapsto \e^{-\tilde{\phi}}N_\Compl \e^{-\tilde{\phi}}\e^{i\eps_s}, \\
Q_s' &\mapsto \e^{\tilde{\phi}}Q_s' \e^{\tilde{\phi}}\e^{i\eps_s}, &  N_\Compl' &\mapsto \e^{\tilde{\phi}}N_\Compl' \e^{\tilde{\phi}}\e^{i\eps_s}.
\end{aligned}
$$

In order to diagonalize $a$ let us introduce the following $N \times N$ matrix
\begin{equation}
\label{MatrixU}
U = \diag \left\{
\frac{1}{\sqrt{2}}\left(
\begin{array}{cc}
1 & 1 \\
-i & i
\end{array}
\right),\dots,\frac{1}{\sqrt{2}}\left(
\begin{array}{cc}
1 & 1 \\
-i & i
\end{array}
\right),\boxtimes \right\}
\end{equation}
where $\boxtimes = 1$ for odd $N$ and is absent for even $N$. One sees that
$$
\tilde{a} = U^\dag a U = \diag\left\{i a_1, - ia_1, ia_2,-ia_2,\dots,i a_n,-ia_n,\diamondsuit \right\}
$$

We have the following action for $\tilde{K} = U^\dag K$ and $\tilde{K}' = U^\dag K'$,
where $K$ and $K'$ are defined in \Ref{SOYZIJ}:
$$
\begin{aligned}
\tilde{K} &\mapsto \e^{\tilde{a}}\tilde{K}\e^{-\tilde{\phi}}\e^{-i\eps_+}, & \tilde{K'} &\mapsto \e^{\tilde{a}}\tilde{K'}\e^{\tilde{\phi}}\e^{-i\eps_+}.
\end{aligned}
$$

For even $N$ the weights for  $\tilde{K}_{li}$ and $\tilde{K}'_{li}$ are:
$$
\begin{aligned}
&a_l - \phi_i -\eps_+, & -&a_l - \phi_i - \eps_+, \\
&a_l + \phi_i -\eps_+, & -&a_l + \phi_i - \eps_+,
\end{aligned}
$$
where $l = 1,\dots,n$ and $i = 1,\dots,k$. For odd $N$ we have as well
$$
\begin{aligned}
&\phi_i -\eps_+, & -&\phi_i - \eps_+.
\end{aligned}
$$

The weights which correspond to $P_{s,ij}$, $Q_{s,ij}$, $Q_{s,ij}'$ for all $N$ are:
$$
\begin{aligned}
&\phi_i - \phi_j  +\eps_s, & i,j &= 1,\dots,k, \\
-&\phi_i - \phi_{j} + \eps_s, & i &< j, \\
&\phi_{i} + \phi_{j} + \eps_s, & i &< j.
\end{aligned}
$$
The same procedure applied to $\chi_{\Compl,ij}$ gives:
$$
\begin{aligned}
&\phi_i - \phi_j + \eps, & i,j &= 1,\dots,k, \\
-&\phi_i - \phi_j + \eps, & i &\leq j, \\
&\phi_i + \phi_j + \eps, &  i &\leq j.
\end{aligned}
$$
Here we have taken into account that the matrices $Q_s$ and $Q'_s$ are antisymmetric, whereas $\chi_\Compl$ is symmetric (since $\mu_\Compl$ is).


\subsection{$Sp(N)$ case}

Now let us consider the group $Sp(N)$. As in the previous section we choose the matrices from Cartan subalgebras of $G_\infty$ and $G_D$ in the standard form:
$$
\begin{aligned}
a &= \left(
\begin{array}{cc}
\tilde{a} & 0 \\
0 & \tilde{a}
\end{array}
\right) \in \Lie(Sp(N)) \quad \mbox{where} \quad \tilde{a} = \diag\{ ia_1,\dots,ia_N\}, \\
\phi &= \diag\left\{ \left( 
\begin{array}{cc}
0 & -\phi_1 \\
+\phi_1 & 0 
\end{array} \right), \dots, \left(
\begin{array}{cc}
0 & - \phi_n \\
+\phi_n & 0
\end{array}
\right), \diamondsuit \right\} \in \Lie(SO(k)),
\end{aligned}
$$
where $n = [k/2]$, and $\diamondsuit = 0$ if $k$ is odd and is absent if $k$ is even.  As before, $\phi_1,\dots\phi_n$ and $a_1,\dots,a_N$ are supposed to be real.

For matrices $K$ and $K'$ from \Ref{IJSpN} the $\Tor_D\times \Tor_\infty\times \Tor_L$ action becomes
$$
\begin{aligned}
K &\mapsto \e^{\tilde{a}} K\e^{-\phi}\e^{-i\eps_+} & K' &\mapsto \e^{-\tilde{a}}K'\e^{-\phi}\e^{-i\eps_+}.
\end{aligned}
$$

After introducing the $k \times k$ dimensional version of \Ref{MatrixU} for $\tilde{K} = K U$ and $\tilde{K}' = K' U$ we obtain
$$
\begin{aligned}
\tilde{K} &\mapsto \e^{\tilde{a}} \tilde{K}\e^{-\tilde{\phi}}\e^{-i\eps_+}, \\
\tilde{K}' &\mapsto \e^{-\tilde{a}}\tilde{K}'\e^{- \tilde{\phi}}\e^{-i\eps_+}.
\end{aligned}
$$
Hence the weight for the matrix elements of $K_{li}$ and $K'_{li}$ for even $k$ are
$$
\begin{aligned}
&a_l - \phi_i -\eps_+, & &a_l + \phi_i - \eps_+, \\
-&a_l - \phi_i - \eps_+, & -&a_l + \phi_i -\eps_+
\end{aligned}
$$
where $l = 1,\dots,n$ and $i = 1,\dots,N$. For odd $k$ we have a supplementary pair of weight:
$$
\begin{aligned}
&a_l -\eps_+ ,&  - &a_l - \eps_+.
\end{aligned}
$$

Now let us obtain weight for $B_s$, $s = 1,2$. Consider the case of even $k$. The $B_s$ transformation is the same as in the $SU(N)$ case. Therefore we arrive to the following weight for $i \neq j$:
$$
\begin{aligned}
&\eps_s + \phi_i -  \phi_j, &  &\eps_s + \phi_i +  \phi_j, \\
&\eps_s - \phi_i -  \phi_j, &  &\eps_s - \phi_i +  \phi_j. \\
\end{aligned}
$$
And for $i = j$
$$
\begin{aligned}
&\eps_s, & &\eps_s + 2\phi_i, & &\mbox{and} & &\eps_s - 2\phi_{i}.
\end{aligned}
$$
For odd $k$ we get additional wights:
$$
\begin{aligned}
&\eps_s, & &\eps_s + \phi_i, & &\mbox{and} & &\eps_s - \phi_i.
\end{aligned}
$$

The same procedure gives for $\chi_\Compl$ the following weights for even $k$ and $i\neq j$ :
$$
\begin{aligned}
&\eps + \phi_i -  \phi_j, &  &\eps + \phi_i +  \phi_j, \\
&\eps - \phi_i -  \phi_j, &  &\eps - \phi_i +  \phi_j. \\
\end{aligned}
$$
For $i = j$ the only weight we get is $\eps$ (we remember that $\mu_\Compl$ (and therefore $\chi_\Compl$) is antisymmetric). For odd $k$ the following weights appear:
$$
\begin{aligned}
&\eps, & &\eps + \phi_i, & &\mbox{and} & &\eps - \phi_i.
\end{aligned}
$$


\section{Expression for the partition function}

Let us finally combine all pieces and obtain expressions for $Z_k(a,\gL;\eps)$ which allows us to get the prepotential according to \Ref{Z}.


\subsection{$SO(N)$ case}

Let $N = 2n + \chi$ where $n = [N/2]$, $\chi \equiv N \pmod 2$. Denote
\begin{equation}
\label{PandDeltaSO}
\begin{aligned}
\Delta(x) &= \prod_{i<j\leq k} \Big( (\f_i + \f_j)^2 - x^2 \Big) \Big( (\f_i -\f_j)^2 - x^2\Big), \\
\P(x) &= x^\chi \prod_{l=1}^n (x^2 - a_l^2).
\end{aligned}
\end{equation}

When for the partition function integrand we have the following expression:
\begin{equation}
\label{directSO}
z_k(a,\phi;\eps) = \frac{\eps^k}{\eps^k_1\eps^k_2} \frac{\Delta(0)\Delta(\eps)}{\Delta(\eps_1)\Delta(\eps_2)}\prod_{i=1}^k \frac{4\phi_i^2 (4\f_i^2 - \eps^2)}{\P(\f_i + \eps_+)\P(\f_i - \eps_+)}.
\end{equation}


\subsection{$Sp(N)$ case}

Now consider the $Sp(N)$ case.  Let $k = 2n + \chi$, $n = [k/2]$, $\chi \equiv k \pmod 2$. Introduce
\begin{equation}
\label{PandDeltaSp}
\begin{aligned}
\Delta(x) &= \prod_{i<j\leq n} \Big( (\f_i + \f_j)^2 - x^2\Big)\Big( (\f_i - \f_j)^2 - x^2\Big), \\
\P(x) &= \prod_{l=1}^N (x^2 - a_l^2).
\end{aligned}
\end{equation}
Then
\begin{equation}
\label{directSp}
\begin{aligned}
z_k(a,\phi;\eps) &= \frac{\eps^n}{\eps_1^n\eps_2^n} {\left[ \frac{1}{\eps_1\eps_2 \prod_{l=1}^N (\eps_+^2 - a_l^2)} \prod_{i=1}^n \frac{\f_i^2(\f_i^2 - \eps^2)}{(\f_i^2 - \eps_1^2)(\f_i^2 -\eps_2^2)}\right]}^\chi \\
&\times \frac{\Delta(0)\Delta(\eps)}{\Delta(\eps_1)\Delta(\eps_2)}\prod_{i=1}^n\frac{1}{\P(\f_i - \eps_+)\P(\f_i + \eps_+)(4\f_i^2 - \eps_1^2)(4\f_i^2 - \eps_2^2)}
\end{aligned}
\end{equation}


\subsection{Matter}
\label{direct:mat}

Let us say some words about the matter. Using the fact that in the presence of matter fields we should consider the  equivariant integral of the equivariant Euler class of $\BundleD_k$ we can write corresponding contributions. Consider first the fundamental representation of $SU(N)$. 

As we have already seen the integral localizes on the solutions of the Weyl equation \Ref{MatAct}. All The solutions for the Weyl equation in the fundamental representation are given by \Ref{WeylFund}. In order to specify one solution we should introduce a vector $x \in \V = \Compl^k$. Then the solution for the Weyl equation which corresponds to this vector is
$$
\psi_x^\a = \bar{Q}^\a \R x.
$$
It gives the boson solution. If we wish to have a \emph{fermion} solution we should consider a fermion vector $\xi \in \Pi \V = \Pi \Compl^k$. Then taking into account \Ref{reciprocity} we can rewrite the contribution of \Ref{MatAct} as follows
\begin{equation}
\label{MatFundFinite}
\int \D x \D\xi \e^{- t \bar{x} x - t \bar{\xi}(\phi + m - i\eps_+) \xi}.
\end{equation}
Note that the integral does not depend on $t$. 

\begin{remark}
In that follows we redefine the mass: $m \mapsto im$.
\end{remark}

The integral is Gaussian, therefore we can compute it exactly and the result is the following supplementary factor of $z_k$:
\begin{equation}
\label{DirMatFund}
z_k^{\mr{fund}}(\phi,m;\eps) = \prod_{i=1}^k (\phi_k + m -\eps_+). 
\end{equation}

In the spirit of the Duistermaat-Heckman formula this expression can be understood as follows. We introduce the supplementary fermion coordinate $\xi \in \Pi \V$, $x$ being its differential. On the space $\Pi\V$ the torus $\Tor_D \times \Tor_F\times \Tor_L$ acts as follows:
$$
\xi \mapsto \e^{\phi}\e^{im}\xi\e^{-i\eps_+}. 
$$
The last factor will be explained in the next chapter. Anyway, one can consider it as a redefinition of the mass. Note that the physical value of $m$ is pure imaginary with $\Im m < 0$. If we add $-\eps_+$ this condition can not be violated since $\Im \eps_+ > 0$. 

The BRST operator acts on $\xi$ and $x$ as follows:
$$
\begin{aligned}
\bar{\sQ} \xi &= x, & \bar{\sQ} x &= [\phi,\xi] + im \xi - i\eps_+ \xi.
\end{aligned}
$$
Note that the exponent in \Ref{MatFundFinite} can be written as (cf \Ref{localizator}, $t''$ terms)
$$
-\frac{t}{2} \bar{\sQ}\left(\bar{\xi} x + x^\dag \xi\right). 
$$

The torus action has the unique stable point, the origin. It allows also to deduce the weights of the torus action. They are
$$
\begin{aligned}
&\phi_i + m - \eps_+, & i&=1,\dots,k.
\end{aligned}
$$ 
Thus we arrive to \Ref{DirMatFund}.

The matter in the adjoint representation can be considered in the same way. The only difference is that now for a solution for the Weyl equation we have supplementary condition to be satisfied \Ref{WeylAdjCondition}. One can do so in the spirit of the section \ref{IntZeroLoc}. The required ingredient is the pair of supplementary coordinates with the opposite statistics. The final answer will be obtained in the next chapter in more general context.


\section{Example: $Sp(N)$ instanton corrections}

In previous sections we have shown how to get the partition function. However, the thing we are looking for is the prepotential. Let us show how it can be extracted from the partition function. For the simplicity reason we consider the case of the pure $Sp(N)$ theory.

Looking at \Ref{directSp} we see that in the case of $Sp(N)$ to obtain $k$-th instanton correction we have to compute only $[k/2]$tiple integral. In particular to get $\Prep_1(a)$, $\Prep_2(a)$ and $\Prep_3(a)$ we should compute a single $\phi$ integral.

For $Z_1(a;\eps)$, $Z_2(a;\eps)$ and $Z_3(a;\eps)$ we have:
$$
Z_1(a;\eps) = - \frac{1}{2\eps_1\eps_2} \prod_{l=1}^N \frac{1}{a_l^2- \eps_+^2},
$$
$$
\begin{aligned}
Z_2(a;\eps) &= -\frac{1}{2\eps_1\eps_2} \sum_{l=1}^N \Big(S_l(a_l^-)D(a_l^-) + S_l(a_l^+)D(a_l^+) \Big) \\
&+ \frac{1}{8{(\eps_1\eps_2)}^2} \frac{\eps_2 T(\eps_1/2) - \eps_1 T(\eps_2/2)}{\eps_2 - \eps_1}, 
\end{aligned}
$$
$$
\begin{aligned}
Z_3(a;\eps) &=\frac{1}{4{(\eps_1\eps_2)}^2} \prod_{l=1}^N \frac{1}{a_l^2- \eps_+^2} \\
&\times\left\{ \sum_{l=1}^N \Big( S_l(a_l^-)D(a_l^-)E(a_l^-) +  S_l(a_l^+)D(a_l^+)E(a_l^+)\Big)\right. \\
&+\frac{1}{12\eps_1\eps_2}\left[ -\frac{T(\eps_1/2) + T(\eps_2/2)+2T(\eps_1)+2T(\eps_2)}{6} \right.\\
&+(\eps_1 + \eps_2)\left(-\frac{2\left(T(\eps_1)-T(\eps_2)\right)+5\left(T(\eps_1/2)-T(\eps_2/2)\right)}{\eps_1-\eps_2}\right.\\
&\left.\left.\left. -\frac{8\eps_1\left(T(\eps_2/2)-T(\eps_1)\right)}{3(\eps_1-2\eps_2)}-\frac{8\eps_2\left(T(\eps_1/2)-T(\eps_2)\right)}{3(\eps_2-2\eps_1)}\right)\right]\right\}
\end{aligned}
$$
where
$$
\begin{aligned}
a_l^\pm &= a_l \pm \eps_+, \\
D(t) &=\frac{1}{\left[4t^2 - \eps_1^2\right]\left[4t^2 - \eps_2^2\right]}, \\
E(t) &= \frac{t^2(t^2-(\eps_1+\eps_2)^2)}{(t^2-\eps_1^2)(t^2-\eps_2^2)},\\
S_l(t) &= \frac{1}{4a_lt} \prod_{k\neq l} \frac{1}{\left[{(t+\eps_+)}^2 - a_k^2\right]\left[{(t-\eps_+)}^2- a_k^2\right]},\\
T(t) &= \frac{1}{\Pol(\eps_+ + t)\Pol(\eps_+ - t)}.
\end{aligned}
$$
Using the definition \Ref{Z} we get the prepotential:
$$
\begin{aligned}
\Prep_1(a) &= -\frac{1}{2}\prod_{l=1}^N \frac{1}{a_l^2}, \\
\Prep_2(a) &= - \frac{1}{16}\left(\sum_{l=1}^N \frac{\tilde{S}_l(a_l)}{a_l^4} + \frac{1}{4}{\left.\frac{\pd^2 \tilde{T}(t)}{\pd t^2}\right|}_{t=0}\right), \\
\Prep_3(a) &= -\frac{1}{16}\prod_{l=1}^N\frac{1}{a_l^2} \left( \sum_{l=1}^N \frac{\tilde{S}_l(a_l)}{a_l^6} + \frac{1}{144}{\left.\frac{\pd^4\tilde{T}(t)}{\pd t^4}\right|}_{t=0}\right)
\end{aligned}
$$
where the tilde over $S$ and $T$ means that we set $\eps_1 = \eps_2 =0$ in the definition of these functions.

For the case $N=1$, that is, for the group $Sp(1) = SU(2)$ we have the following expression for the prepotential:
$$
\Prep_{\mr{inst}}(a,\Lambda) =  -\frac{\Lambda^4}{2a^2} - \frac{5\Lambda^8}{64a^6} -\frac{3\Lambda^{12}}{64a^{10}}+ O(\Lambda^{16})
$$
which coincides with both Seiberg-Witten  \cite{HKP-SU} and direct computations  \cite{SWfromInst} for $SU(2)$. $Z_k(a;\eps)$, $k=1,2,3$ can also be checked against the corresponding quantities for $SU(2)$  \cite{SWfromInst}.

%% file: towards.tex
\label{towards}

As we have seen, to write the expression for the partition function integrands we should have in hands a model for the instanton moduli space. In particular, to incorporate the matter we should find solutions for the Weyl equation for all interesting representations. However, we can avoid this work and find directly the weights which appear in the Duistermaat-Heckman formula. In this chapter we present some methods which will allow us to do that.


\section{Universal bundle}

\label{UniversalBundle}

It is well known \cite{Moore} that a manifold $M$ equipped by an almost complex structure and a hermitian metric allows to define a $\mbox{Spin}^\Compl$-structure. Moreover in that case the complexified tangent bundle can be view as $TM \otimes \Compl \simeq \Hom(S_+ \otimes L,S_-\otimes L)$. Here $S_\pm$ is the spinor bundles of positive and negative chiralities, and $L$ is the determinant bundle. Even if $S_\pm$ and $L$ do not exist separately their tensor product $S_\pm \otimes L$ does. That is why $S_\pm$ and $L$ is called sometimes virtual bundles.

On the sections of $S_+$ and $S_-$, that is, on dotted and undotted spinors, the maximal torus of the Lorentz group $\Tor_L$ acts as follows: 
$$
\begin{aligned}
\chi_{\dot{\a}} \mapsto {\chi_{\dot{\a}}}' &= {U_+}_{\dot{\a}}{}^{\dot{\b}} \chi^{\dot{\b}} &
&\mbox{and} &
\psi_\a \mapsto \psi_\a' &= {U_-}_\a{}^\b \psi_\b
\end{aligned}
$$
where
$$
\begin{aligned}
U_+ &= \left(\begin{array}{cc}
\e^{i\eps_+} & 0 \\
0 & \e^{-i\eps_+}
\end{array}\right), & 
U_- &= \left(\begin{array}{cc}
\e^{i\eps_-} & 0 \\
0 & \e^{-i\eps_-}
\end{array}\right), &
&\mbox{where} & \eps_\pm &= \frac{\eps_1 \pm \eps_2}{2}.
\end{aligned} 
$$
The complex coordinates transform as:
$$
\begin{aligned}
z_1\mapsto z_1' &= z_1 \e^{i\eps_1} & &\mbox{and} & z_2\mapsto z_2' &= z_2 \e^{i\eps_2},
\end{aligned}
$$
and the sections of $L$ as $s \mapsto s \e^{i\eps_+}$.

\begin{remark}
As an illustration we consider the four dimensional manifold. Then $(1,\dd z_1 \wedge \dd z_2)$ transforms as ``$s \times \mbox{dotted spinor}$'' and $(\dd z_1, \dd z_2)$ as ``$s \times \mbox{undotted spinor}$''.
\end{remark}

Taking into account these properties the ADHM construction can be represented by a following complex:
\begin{equation}
\label{ADHMcomplex}
\V\otimes \L^{-1} \stackrel{\tau}{\longrightarrow} \V \otimes \S_- \oplus \W \stackrel{\sigma}{\longrightarrow} \V\otimes \L
\end{equation}
where 
$$
\begin{aligned}
\tau &= \left(
\begin{array}{c}
B_1 \\
B_2 \\
I
\end{array}
\right), & \sigma &= (B_2,-B_1,J),
\end{aligned}
$$
where $\L$ and $\S_-$ can be viewed as fibers of $L$ and $S_-$ respectively. The ADHM equations \Ref{ADHM} assure that this is indeed a complex.

Now we recall the construction of the universal bundle. Let $\Mod$ be the instanton moduli space, given by the ADHM construction. Let us introduce local coordinates on $\Mod$: $\{m^I\}$, $I=1,\dots,\dim \Mod$. The tangent space to a point $m \in \Mod$ is spanned by solutions of the linearized self-dual equation \Ref{SDE}. Let us fix a basis of such a solutions: $\{ a^I_\mu(x,m) \}$. Consider now a family of instanton gauge fields parametrized by points of $\Mod$: $A_\mu(x,m)$. We can write
$$
\frac{\pd A_\mu}{\pd m^I} = h_{IJ} a^J_\mu + \cD_\mu \a_I
$$ 
where $\a_I$ is a compensating gauge transformation. We can combine it with the connection $A_\mu$ into a one form on $\Real^4\times \Mod $: $\bA = A_\mu \dd x^\mu + \a_I \dd m^I$ which can be seen as a connection of the vector bundle $\E$ over $\Real^4 \times \Mod $ with the fiber $\W$. This bundle is called the universal bundle.

Let $q$ be generic element of the torus $\Tor = \Tor_D \times \Tor_\infty \times \Tor_L \times \Tor_F$. The equivariant Chern character of $\E$ depending on $q$ can be computed as an alternating sum of traces over the cohomologies of the complex \Ref{ADHMcomplex} (see \cite{SWfromInst,SmallInst} for some details). Then we come to the formula
\begin{equation}
\label{ChernE}
\begin{aligned}
\Ch_q(\E) &\equiv \Tr_\E(q) = \Tr_\W(q) + \Tr_\V(q)\Big( \Tr_{\S_-}(q) - \Tr_{\L}(q) - \Tr_{\L^{-1}}(q)\Big) \\
&= \Tr_\W(q) - (\e^{i\eps_1} - 1)(\e^{i\eps_2} - 1)\e^{-i\eps_+} \Tr_\V(q)
\end{aligned}
\end{equation}
where $\Ch_q(\E)$ is the equivariant Chern character. 

The equivariant analog of the Atiyah-Singer theorem allows us to compute the equivariant index of the Dirac operator. It has the following form
$$
\Ind_q = \sum_{\a} \ep_\a \e^{w_\a} = \int_{\Compl^2} \Ch_q(\E) \Td_q(\Compl^2),
$$
where the sum is taken over all fixed points of the $\Tor$ action and all $\Tor$ action invariant subspaces of the tangent space to a fixed point, $w_\a$ being a weight of this action. In this formula $\Td_q(\Compl^2)$ is the equivariant analogue of the Todd class\footnote{the fact that we should use the Todd class, and not the $\hat{A}$-polynomial, as one could think, follows from the close relation between solutions of the Dirac equation and Dolbeaut cohomologies, discussed at the beginning of the section \ref{DiracDbar}.}, which for $\Compl^2 \simeq \Real^4$ has the simple form:
$$
\Td_q(\Compl^2) = \frac{\eps_1\eps_2}{(\e^{i\eps_1} - 1)(\e^{i\eps_2} - 1)}.
$$

The integration can be performed with the help of the Duistermaat-Heckman formula \Ref{DH}. The Hamiltonian of $\Tor_L$ action is $i\eps_1|z_1|^2 + i\eps_2|z_2|^2$. The only fixed point of this action on $\Compl^2$ is the origin. The weights are $i\eps_1$ and $i\eps_2$. Consequently we arrive at
\begin{equation}
\label{sumFund}
\Ind^{\rm{fund}}_q = \sum_{\a} \epsilon_\a \e^{w_\a} = \frac{{\left.\Ch_q(\E)\right|}_{z_1 = z_2 = 0}}{(\e^{i\eps_1} - 1)(\e^{i\eps_2} - 1)}.
\end{equation}

Let us denote the elements of $\Tor_\infty$, $\Tor_D$ and $\Tor_F$ as follows:
$$
\begin{aligned}
 q_G &= \diag \{ ia_1,\dots,ia_N \} \in \Tor_\infty \\
 q_D &= \diag \{ i\f_1,\dots,i\f_k \} \in \Tor_D\\
 q_F &= \diag \{ i m_1,\dots,i m_{N_f} \} \in \Tor_F
\end{aligned}
$$
where $a_1,\dots a_N, \f_1,\dots,\f_k, m_1,\dots,m_{N_f}$ are real, $N_f$ being the number of flavors. Then combining \Ref{ChernE} and \Ref{sumFund} we get for $N_f = 1, m = 0$
\begin{equation}
\label{IndSUFund}
\Ind^{\rm{fund}}_q = \sum_{\a} \ep_\a \e^{w_\a} = \frac{1}{(\e^{i\eps_1} - 1)(\e^{i\eps_2} - 1)}\sum_{l=1}^N \e^{ia_l} - \sum_{i=1}^k \e^{i\f_i - i\eps_+}.
\end{equation}

The generalization to $N_f>1$ is straightforward and we obtain:
$$
\Ind^{{\rm{fund}},N_f}_q = \frac{1}{(\e^{i\eps_1} - 1)(\e^{i\eps_2} - 1)}\sum_{f=1}^{N_f}\sum_{l=1}^N \e^{ia_l + im_f} - \sum_{f=1}^{N_f}\sum_{i=1}^k \e^{i\f_i - i\eps_+ + im_f}.
$$


\section{Alternative derivation for $\mathop{\mathrm{Ch}}\nolimits_q(\mathcal{E})$}
\label{DiracDbar}

The derivation of \Ref{IndSUFund} presented in previous section, yet quite general, may seem to be too abstract. Here we present an alternative way to get it. In particular this method allows us to see the origin of all terms which appears in the formula.

Before starting let us recall the relation between Dirac operator on complex manifolds and $\bar\pd$ operator. Define
$$
\bar\pd = \dd z_1 \cD_1 + \dd z_2 \cD_2.
$$
Thanks to the self-dual equation equations \Ref{SDE} this operator is nilpotent $\bar{\pd}^2 = 0$. Indeed, the self-dual equation is equivalent to
$$
\begin{aligned}
&[\cD_1,\cc{\cD}_{\bar{1}}] + [\cD_2,\cc{\cD}_{\bar{2}}] = 0,\\
&[\cD_1,\cD_2] = 0.
\end{aligned}
$$

The solutions of the Weyl equation \Ref{WEQ} can be naturally associated with Dolbeaut cohomology. The only thing that should be taken into account is the twist by the square root of the determinant bundle. 

Now let us recall the discussion of the section \ref{direct:mat}. A solution for the fermionic Dirac equation can be parametrized by $\xi \in \Pi \V$. Now we remember that we have a freedom to perform a gauge transformations which are trivial at infinity. Therefore we see that a solution of the Weyl equation are labeled by $\G \oplus \Pi\V$, where $\G = \{ g: \Sphere^4 \to G\}$.

We stress that it is not the moduli space, since we {\em do not} factor out the group of local gauge transformations $\G$.  

Now we have enough information to reconstruct the equivariant index of the Dirac operator. It is given by the sum of $\Tor$ action weights to fixed points.

Since the gauge transformation $g$ should be $\bar{\pd}$-closed, that is, holomorphic, we conclude that
$$g = \sum_{n_1,n_2\geq 0} g_{n_1,n_2} z_1^{n_1}z_2^{n_2}.$$
The weights of the $\Tor$ action on $g$ are 
$$
\begin{aligned}
w_\a &= ia_l + in_1\eps_1 + in_2\eps_2, &  l& =1,\dots,N, & n_1,n_2 &\geq 0.
\end{aligned}
$$
The $\Tor$ action on $\Pi\V$ is given by $\xi \mapsto \e^{-i\eps_+}q_D\xi$ where $q_D \in \Tor_D$. It follows that the weights are given by
$$
\begin{aligned}
w_\a &= i \f_i - i\eps_+, & i &= 1,\dots,k.
\end{aligned}
$$

Now we recall that the contribution of the fermionic variables comes with $\ep_\a = -1$ (see remark below \Ref{DH}). It implies that the equivariant index equals to
$$
\Ind_q^{\mr{fund}}  = \sum_{l=1}^N \sum_{n_1,n_2 \geq 0} \e^{ia_l + in_1\eps_1 + in_2\eps_2} - \sum_{i=1}^k \e^{i\f_i - i\eps_+}
$$
which is equivalent to \Ref{ChernE} after applying \Ref{sumFund}.


\section{Equivariant index for other groups}

In a similar way we can find the equivariant index for the fundamental representation of $SO(N)$ and $Sp(N)$.


\subsection{$SO(N)$ case} 

Let $N = 2n + \chi$ where $n = [N/2]$ and $\chi \equiv N \pmod 2$. Then 
\begin{equation}
\label{IndSOFund}
\Ind_q^{\mr{fund}} = \frac{1}{(\e^{i\eps_1}- 1)(\e^{i\eps_2} - 1)} \left( \chi + \sum_{l=1}^n \left(\e^{ia_l} + \e^{-ia_l}\right)\right) - \sum_{i=1}^k \left( \e^{i\f_i -i\eps_+} + \e^{-i\f_i - i\eps_+} \right).
\end{equation}


\subsection{$Sp(N)$ case} 

Let $k = 2n + \chi$ where $n = [k/2]$ and $\chi \equiv k \pmod 2$. Then
\begin{equation}
\label{IndSpFund}
\Ind_q^{\mr{fund}} = \frac{1}{(\e^{i\eps_1}- 1)(\e^{i\eps_2} - 1)} \sum_{l=1}^N \left( \e^{ia_l} + \e^{-ia_l} \right) - \sum_{i=1}^n \left( \e^{i\f_i} + \e^{-i\f_i} + \chi \right).
\end{equation}


\section{Equivariant index for other representations} 

Having computed the equivariant index for the fundamental representation let us turn to others representations. 


\subsection{$SU(N)$ case} 

As it was explained in \cite{SmallInst} the equivariant index for the adjoint representation of $SU(N)$ can be obtained as follows:
\begin{equation}
\label{IndAdjAtiyah}
\begin{aligned}
\Ind^{\rm{adj}}_q &= \sum_{\a}\ep_\a \e^{w_\a} = \int_{\Compl^2} \Ch_q(\E\otimes\E^\ast) \Td_q(\Compl^2) \\
&= \int_{\Compl^2}\Ch_q(\E)\Ch_q(\E^\ast)\Td_q(\Compl^2) = \frac{{\left.\Ch_q(\E)\Ch_q(\E^\ast)\right|}_{z_1=z_2=0}}{(\e^{i\eps_1}-1)(\e^{i\eps_2}-1)}.
\end{aligned}
\end{equation}
We can use the expression \Ref{ChernE} to compute this index. The result is
\begin{equation}
\label{IndSUAdj}
\begin{aligned}
\Ind^{\mr{adj}}_q &=  \frac{1}{(\e^{i\eps_1}-1)(\e^{i\eps_2} - 1)}\left[ N + \sum_{l\neq m}^N \e^{ia_l-ia_m} \right] \\
&- \sum_{i=1}^k\sum_{l=1}^N \left( \e^{i\f_i  - i\eps_+ - ia_l} + \e^{- i\f_i + ia_l - i\eps_+}\right) + k (1-\e^{-i\eps_1})(1-\e^{-i\eps_2})\\
&+ \sum_{i\neq j}^k \left( \e^{i\f_i - i\f_j} + \e^{i\f_i - i\f_j - i\eps_1 - i\eps_2} - \e^{i\f_i - i\f_j - i\eps_1} -  \e^{i\f_i - i\f_j - i\eps_2}\right).
\end{aligned}
\end{equation}

At the same way the indices for symmetric and antisymmetric representations can be obtained. Denote
$$
\begin{aligned}
\Ch_q^{\mr{sym}}(\E) &= \Ch_q(\Sym^2\E), \\
\Ch_q^{\mr{ant}}(\E) &= \Ch_q(\Ant^2\E).
\end{aligned}
$$
If $\Ch_q^{\mr{fund}}(\E) = \sum_\a \ep_\a \e^{w_\a}$ then
\begin{equation}
\label{IndSymAnt}
\Ch_q^{\mr{sym},\mr{ant}}(\E) = \frac{1}{2}\left[ {\left(\Ch_q^{\mr{fund}}\right)}^2 \pm \Ch^{\mr{fund}}_{q^2}\right] = \frac{1}{2}\left[ {\Big( \sum_\a \ep_\a \e^{w_a} \Big)}^2 \pm \sum_\a \ep_\a \e^{2w_\a}\right].
\end{equation}
We can now apply the analog of \Ref{IndAdjAtiyah} to compute the equivariant index for these representations. The result is the following:
\begin{equation}
\label{IndSUSym}
\begin{aligned}
\Ind_q^{\mr{sym}} &= \frac{1}{(\e^{i\eps_1} - 1)(\e^{i\eps_2} - 1)} \sum_{l\leq m \leq N} \e^{ia_l + ia_m} \\
&- \sum_{l=1}^N \sum_{i=1}^k \e^{ia_l + i\f_i - i\eps_+} - \sum_{i=1}^k \left( \e^{2i\f_i - i\eps_1} + \e^{2i\f_i - i\eps_2}\right) \\ 
&+ \sum_{i<j\leq k} \left( \e^{i\f_i + i\f_j} + \e^{i\f_i + i\f_j - i\eps_1 - i\eps_2} - \e^{i\f_i + i\f_j - i\eps_1} -  \e^{i\f_i + i\f_j - i\eps_2}\right),
\end{aligned}
\end{equation}
\begin{equation}
\label{IndSUAnt}
\begin{aligned}
\Ind_q^{\mr{ant}} &= \frac{1}{(\e^{i\eps_1} - 1)(\e^{i\eps_2} - 1)} \sum_{l <  m \leq N} \e^{ia_l + ia_m} \\
&- \sum_{l=1}^N \sum_{i=1}^k \e^{ia_l + i\f_i - i\eps_+} + \sum_{i=1}^k \left( \e^{2i\f_i} + \e^{2i\f_i - i\eps_1 - i\eps_2}\right) \\ 
&+ \sum_{i<j\leq k} \left( \e^{i\f_i + i\f_j} + \e^{i\f_i + i\f_j - i\eps_1 - i\eps_2} - \e^{i\f_i + i\f_j - i\eps_1} -  \e^{i\f_i + i\f_j - i\eps_2}\right).
\end{aligned}
\end{equation}


\subsection{$SO(N)$ case} 

Using Table \ref{GroupTh} we see that the adjoint representation of $SO(N)$ is the antisymmetric one. So using \Ref{IndSOFund} together with \Ref{IndSymAnt} we get 
\begin{equation}
\label{IndSOAdj}
\begin{aligned}
\Ind_q^{\mr{adj}} &= \frac{1}{(\e^{i\eps_1}-1)(\e^{\eps_2}-1)} \left[ \sum_{l<m\leq n} \left( \e^{ia_l + ia_m} + \e^{ia_l - ia_m} + \e^{- ia_l + ia_m} + \e^{- ia_l - ia_m} \right)  \right. \\
&+ \left. \chi \sum_{l=1}^n \left( \e^{ia_l} + \e^{-ia_l} \right) + n \right] + \chi \sum_{i=1}^k \left( \e^{i\f_i - i\eps_+} +\e^{-i\f_i - i\eps_+} \right) \\
&+ \sum_{i=1}^k\sum_{l=1}^n \left( \e^{i\f_i + ia_l -i\eps_+} +  \e^{-i\f_i + ia_l -i\eps_+} +  \e^{i\f_i - ia_l -i\eps_+} +  \e^{-i\f_i - ia_l -i\eps_+} \right)\\
&+ \sum_{i=1}^k \left( \e^{2i\f_i} +  \e^{-2i\f_i} +  \e^{2i\f_i-i\eps} +  \e^{-2i\f_i-i\eps} \right) + k (1-\e^{-i\eps_1})(1-\e^{-i\eps_2})\\
&+ (1-\e^{-i\eps_1})(1-\e^{-i\eps_2})\sum_{i<j\leq k}\left ( \e^{i\f_i + i\f_j} + \e^{i\f_i - i\f_j} + \e^{- i\f_i + i\f_j} + \e^{- i\f_i - i\f_j} \right)
\end{aligned}
\end{equation}


\subsection{$Sp(N)$ case} 

Table \ref{GroupTh} shows that the adjoint representation of $Sp(N)$ is the symmetric representation. Using \Ref{IndSpFund} and \Ref{IndSymAnt} we get:
\begin{equation}
\label{IndSpAdj}
\begin{aligned}
\Ind_q^{\mr{adj}} &= \frac{1}{(\e^{i\eps_1}-1)(\e^{i\eps_2}-1)} \left[ \sum_{l<m\leq N} \left( \e^{ia_l + ia_m} + \e^{ia_l - ia_m} + \e^{- ia_l + ia_m} + \e^{- ia_l - ia_m} \right)  \right.\\
&\left. + \sum_{l=1}^N \left( \e^{2ia_l} + \e^{-2ia_l} \right)+ N \right] -\chi \sum_{l=1}^N \left( \e^{ia_l - i\eps_+} + \e^{-ia_l-i\eps_+}\right) \\
&- \sum_{l=1}^N \sum_{i=1}^n \left( \e^{ia_l +i\f_i -i\eps_+} + \e^{ia_l -i\f_i -i\eps_+} + \e^{-ia_l +i\f_i -i\eps_+} + \e^{-ia_l -i\f_i -i\eps_+}\right) \\
&-\chi \left( \e^{-i\eps_1} + \e^{-i\eps_2} \right) + (1-\e^{-i\eps_1})(1-\e^{-i\eps_2}) \left[ n + \chi \sum_{i=1}^n \left( \e^{i\f_i} + \e^{-i\f_i}\right)\right] \\
&+ (1-\e^{-i\eps_1})(1-\e^{-i\eps_2}) \sum_{i<j\leq n} \left( \e^{i\f_i + i\f_j} + \e^{i\f_i - i\f_j} + \e^{-i\f_i + i\f_j} + \e^{-i\f_i - i\f_j}\right) \\
&- \left( \e^{-i\eps_1} + \e^{-i\eps_2} \right) \sum_{i=1}^n \left( \e^{2i\f_i} + \e^{-2i\f_i}\right). 
\end{aligned}
\end{equation}

Another case that we will  be interested in is the antisymmetric representation of $Sp(N)$. Using \Ref{IndSpFund} and \Ref{IndSymAnt} we get
\begin{equation}
\label{IndSpAnt}
\begin{aligned}
\Ind_q^{\mr{ant}} &= \frac{1}{(\e^{i\eps_1}-1)(\e^{i\eps_2}-1)} \left[ \sum_{l<m\leq N} \left( \e^{ia_l + ia_m} + \e^{ia_l - ia_m} + \e^{- ia_l + ia_m} \right.\right. \\
&\left.\left. + \e^{- ia_l - ia_m} \right)  + N \right] -\chi \sum_{l=1}^N \left( \e^{ia_l - i\eps_+} + \e^{-ia_l-i\eps_+}\right) + \chi \left( 1 + \e^{-i\eps} \right) \\
&- \sum_{l=1}^N \sum_{i=1}^n \left( \e^{ia_l +i\f_i -i\eps_+} + \e^{ia_l -i\f_i -i\eps_+} + \e^{-ia_l +i\f_i -i\eps_+} + \e^{-ia_l -i\f_i -i\eps_+}\right) \\
&+ (1-\e^{-i\eps_1})(1-\e^{-i\eps_2}) \left[ n + \chi \sum_{i=1}^n \left( \e^{i\f_i} + \e^{-i\f_i}\right)\right] \\
&+ (1-\e^{-i\eps_1})(1-\e^{-i\eps_2}) \sum_{i<j\leq n} \left( \e^{i\f_i + i\f_j} + \e^{i\f_i - i\f_j} + \e^{-i\f_i + i\f_j} + \e^{-i\f_i - i\f_j}\right) \\
&+ \left( 1 + \e^{-i\eps} \right) \sum_{i=1}^n \left( \e^{2i\f_i} + \e^{-2i\f_i}\right). 
\end{aligned}
\end{equation}


\section{Partition function}

Now we are ready to write the expression for the pertition function \Ref{Z}. First we note that since the tangent space to a point belonging to bundle is a direct sum of the tanget space to the point of the base and the tangent space to the point of the fiber. Taking into account the statistics of the fields (recall that the Yang-Mills connection $A_\mu$ belongs to the adjoint representation of the gauge group) we can write
\begin{equation}
\label{AddInd}
\Ind_q = \Ind_q^{\mr{adj,gauge}} - \sum_{\vr \in \mr{reps}} \Ind_q^{\vr,\mr{matter}}.
\end{equation}
The transformation \Ref{TrWeights} converts the sum to a product. In order to get the $k$-instanton partition function $Z_k(a,m;\eps)$ we should integrate over $\alg{t} = \Lie(T_D)$. We have to keep in mind the order of the Weyl group which we should divide the integral on.

Let us realize this program step-by-step. Compute first the weights products \Ref{TrWeights} for (almost) all cases allowed by the asymptotic freedom. We will consider all the matter representations contained in a tensor power of the fundamental one. For $SU(N)$ we can get all the representations in such a way. However, for other groups this is not the case. For example for $SO(N)$ we will miss some spinor representations. We should find all the solutions of the equation $\beta \geq 0$ where $\beta$ is defined by the righthand side of \Ref{betaDef}. Using Table \ref{GroupTh} we get the following list (Table \ref{Models}) of asymptotically free models.


\begin{table}
\begin{itemize}
\item \underline{$SU(N)$}:
\begin{itemize}
\item $N_f$ fundamental multiplets, $N_f \leq 2N$,
\item 1 antisymmetric multiplet and $N_f$ fundamental, $N_f \leq N + 2$,
\item 1 symmetric multiplet and $N_f$ fundamental, $N_f \leq N-2$,
\item 2 antisymmetric and $N_f$ fundamental, $N_f \leq 4$,
\item 1 symmetric and 1 antisymmetric multiplet,
\item 1 adjoint multiplet.
\end{itemize}
\item \underline{$SO(N)$}:
\begin{itemize}
\item $N_f$ fundamental multiplet, $N_f \leq N-2$,
\item 1 adjoint multiplet.
\end{itemize}
\item \underline{$Sp(N)$}:
\begin{itemize}
\item $N_f$ fundamental multiplet, $N_f \leq N+2$,
\item 1 antisymmetric multiplet and $N_f$ fundamental, $N_f \leq 4$,
\item 1 adjoint multiplet.
\end{itemize}
\end{itemize}
\caption{Models allowed by the asymptotic freedom}\label{Models}
\end{table}


Here we give the expression for the building blocks which are necessary to construct all the cases listed above.


\subsection{$SU(N)$ case}
\label{PartSU}
 
As it was noticed in the end of the section \Ref{LorentzDef} the partition function is the product of the perturbative part $Z^{\mr{pert}}(a,m,\gL;\eps)$ and 
$$
\sum_{k=0}^\infty q^k \int \prod_{i=1}^k \frac{\dd \phi_i}{2\pi i} z_k(a,\phi,m;\eps).
$$
\begin{remark}
The term $Z^{\mr{pert}}(a,m,\gL;\eps)$ comes from the first terms in \Ref{IndSUFund}, \Ref{IndSUAdj}, \Ref{IndSUSym}, \Ref{IndSUAnt} respectively. Under the transformation \Ref{TrWeights} these terms become the infinite products to be regularized. It can be shown \cite{SWfromInst,SWandRP} that after the proper time regularization they give precisely the perturbative contribution to the prepotential \Ref{pertPrep} (in the $\eps_1,\eps_2\to 0$ limit, see section \ref{SU(N)PureYM}). In that follows we will drop this term in all calculations and restore it, if ever, only in the final result. 
\end{remark}

We use the definition \Ref{PandDeltaSU}. Then
\begin{equation}
\label{zSUFund}
z_k^{\mr{fund}}(a,\phi,m;\eps) = \prod_{i=1}^k (\phi_i+m  -\eps_+), 
\end{equation}
\begin{equation}
\label{zSUAdjGauge}
z_k^{\mr{adj,gauge}}(a,\phi;\eps) = \frac{\eps^k}{\eps_1^k\eps_2^k}\frac{\Delta_-(0)\Delta_-(\eps)}{\Delta_-(\eps_1)\Delta_-(\eps_2)}\prod_{i=1}^k \frac{1}{\P(\f_i+\eps_+)\P(\f_i-\eps_+)}, 
\end{equation}
\begin{equation}
\label{zSUAdjMatter}
\begin{aligned}
z_k^{\mr{adj,matter}}(a,\phi,m;\eps) &= \frac{(m-\eps_1)^k(m-\eps_2)^k}{(m-\eps)^k m^k} \frac{\Delta_-(m-\eps_1)\Delta_-(m-\eps_2)}{\Delta_-(m)\Delta_-(m-\eps)} \\
&\times \prod_{k=1}^k \P(\f_i - m + \eps_+)\P(\f_i + m -\eps_+), 
\end{aligned}
\end{equation}
\begin{equation}
\label{zSUSym}
\begin{aligned}
z_k^{\mr{sym}}(a,\phi,m;\eps) &=  \frac{\Delta_+(m-\eps_1)\Delta_+(m-\eps_2)}{\Delta_+(m)\Delta_+(m-\eps)} \\
&\times \prod_{i=1}^k ( 2\f_i +m -\eps_1)(2\f_i + m -\eps_2)\P(-\f_i -m + \eps_+), 
\end{aligned}
\end{equation}
\begin{equation}
\label{zSUAnt}
z_k^{\mr{ant}}(a,\phi,m;\eps) = \frac{\Delta_+(m-\eps_1)\Delta_+(m-\eps_2)}{\Delta_+(m)\Delta_+(m-\eps)}\prod_{i=1}^k\frac{\P(-\f_i - m + \eps_+)}{(2\f_i + m)(2\f_i +m - \eps)}.
\end{equation}

To find similar expressions for $SO(N)$ and $Sp(N)$ we use \Ref{IndSOFund}, \Ref{IndSOAdj}, \Ref{IndSpFund}, \Ref{IndSpAnt} and \Ref{IndSpAdj}. The result is the following.


\subsection{$SO(N)$ case}

We will use the notations introduced in \Ref{PandDeltaSO}. Then
\begin{equation}
\label{zSOFund}
z_k^{\mr{fund}}(a,\phi,m;\eps) = \prod_{i=1}^k ((m-\eps_+)^2 - \f_i^2),
\end{equation}
\begin{equation}
\label{zSOAdjGauge}
z_k^{\mr{adj,gauge}}(a,\phi;\eps) = \frac{\eps^k}{\eps^k_1\eps^k_2} \frac{\Delta(0)\Delta(\eps)}{\Delta(\eps_1)\Delta(\eps_2)}\prod_{i=1}^k \frac{4\phi_i^2 (4\f_i^2 - \eps^2)}{\P(\f_i + \eps_+)\P(\f_i - \eps_+)}, 
\end{equation}
\begin{equation}
\label{zSOAdjMatter}
\begin{aligned}
z_k^{\mr{adj,matter}}(a,\phi,m;\eps) &= \frac{(m-\eps_1)^k(m-\eps_2)^k}{m^k(m-\eps)^k}\frac{\Delta(m-\eps_1)\Delta(m-\eps_2)}{\Delta(m)\Delta(m-\eps)} \\
&\times \prod_{i=1}^k \frac{\P(\f_i + m - \eps_+)\P(\f_i -m + \eps_+)}{(4\f_i^2 - m^2)(4\f_i^2 - (m-\eps)^2)}.
\end{aligned}
\end{equation}


\subsection{$Sp(N)$ case}

We use the definition \Ref{PandDeltaSp}. Then
\begin{equation}
\label{zSpFund}
z_k^{\mr{fund}}(a,\phi,m;\eps) = (m-\eps_+)^\chi\prod_{i=1}^n ( (m-\eps_+)^2 - \phi_i^2),
\end{equation}
\begin{equation}
\label{zSpAdjGauge}
\begin{aligned}
z_k^{\mr{adj,gauge}}(a,\phi;\eps) &= \frac{\eps^n}{\eps_1^n\eps_2^n} {\left[ \frac{1}{\eps_1\eps_2 \prod_{l=1}^N (\eps_+^2 - a_l^2)} \prod_{i=1}^n \frac{\f_i^2(\f_i^2 - \eps^2)}{(\f_i^2 - \eps_1^2)(\f_i^2 -\eps_2^2)}\right]}^\chi \\
&\times \frac{\Delta(0)\Delta(\eps)}{\Delta(\eps_1)\Delta(\eps_2)}\prod_{i=1}^n\frac{1}{\P(\f_i - \eps_+)\P(\f_i + \eps_+)(4\f_i^2 - \eps_1^2)(4\f_i^2 - \eps_2^2)}, 
\end{aligned}
\end{equation}
\begin{equation}
\label{zSpAdjMatter}
\begin{aligned}
z_k^{\mr{adj,matter}}(a,\phi,m;\eps) &= \frac{(m-\eps_1)^n(m-\eps_2)^n}{m^n(m-\eps)^n}\frac{\Delta(m-\eps_1)\Delta(m-\eps_2)}{\Delta(m)\Delta(m-\eps)}\\
&{\times\left[ (m-\eps_1)(m-\eps_2) \prod_{l=1}^N\Big((m-\eps_+)^2 - a_l^2\Big)\prod_{i=1}^n\frac{(\f_i^2 - (m-\eps_1)^2)(\f_i^2 - (m-\eps_2)^2)}{(\f_i^2 - m^2)(\f_i^2 - (m-\eps)^2)}\right]}^\chi \\
&\times \prod_{i=1}^n\P(\f_i +m - \eps_+)\P(\f_i -m + \eps_+)\prod_{s=1}^2(4\f_i^2 - (m - \eps_s)^2), 
\end{aligned}
\end{equation}
\begin{equation}
\label{zSpAnt}
\begin{aligned}
z_k^{\mr{ant}}(a,\phi,m;\eps) &= \frac{(m-\eps_1)^n(m-\eps_2)^n}{m^n(m-\eps)^n}\frac{\Delta(m-\eps_1)\Delta(m-\eps_2)}{\Delta(m)\Delta(m-\eps)}  \\
&{\times \left[ \frac{ \prod_{l=1}^N((m-\eps_+)^2 - a_l^2)}{m(m-\eps)}\prod_{i=1}^n\frac{(\f_i^2 - {(m-\eps_1)}^2)(\f_i^2 - {(m-\eps_2)}^2)}{(\f_i^2 - m^2)(\f_i^2 - {(m-\eps)}^2)}\right]}^\chi \\
&\times\prod_{i=1}^n \frac{\P(\f_i + m -\eps_+)\P(\f_i - m + \eps_+)}{(4\f_i^2 - m^2) (4\f_i^2 - {(m-\eps)}^2)}.
\end{aligned} 
\end{equation}
Now we should perform the integration over $\Tor_D$. The orders of the Weyl group of the dual group $|W_D|$ can be found in Table \ref{GroupTh}. Finally for a theory with some matter multiplets we get the following expression:
\begin{equation}
\label{partition}
Z_k(a,m;\eps) = \frac{1}{|W_D|} \oint \prod_{i=1}^k \frac{\dd \f_i}{2\pi i} z_k^{\mr{adj,gauge}} (a,\phi;\eps)\prod_{\vr \in \mr{reps}} z_k^{\vr,\mr{matter}}(a,m_\vr,\phi;\eps)
\end{equation}

\begin{remark}
The expressions for the adjoint representation integrand $z_k(q)$ for $SO(N)$ and $Sp(N)$ coincides with the expressions which can be obtained from the direct analysis of the instanton moduli space for these groups \Ref{directSU}, \Ref{directSO}, \Ref{directSp}, \Ref{DirMatFund} \cite{SWfromInst,ABCD}.
\end{remark}

To compute the contour integral we need a contour bypassing prescription. It can be obtained, as explained in \cite{ABCD}, by considering the four dimensional theory as a limit of a five dimensional theory, where the complexified torus $\Tor_\Compl$ acts on. As a result we obtain $(\eps_1,\eps_2,m) \mapsto (\eps_1 + i0,\eps_2 + i0, m - i0)$ prescription. It worth noting that the prescription for masses $m$ coincides with the Feynman prescription for bypassing the physical poles. See also the remark at the end of section \ref{straightforward}. The contour can be closed on the upper or lower complex halfplain. The choice is irrelevant since the residue at infinity of the integrand  vanishes. 


\section{1-instanton corrections and residue functions}
\label{1instComp}

Formula \Ref{partition}, yet far from the final result, allows, however, to perform various checks. In particular, we can check this formula against the known one instanton corrections.

After the work of Seiberg and Witten \cite{SeibergWitten} the 1-instanton corrections was computed for numerous combinations of (classical) groups and matter content. In particular, in references \cite{NaculichSUAntFund,EnnesSUSymFund,EnnesSUSymAnt,EnnesSU2AntFund,HKP-SU,SOandSp,CMandSW,EllipticMod} it was done for all cases allowed by asymptotic freedom.

In \cite{EnnesSU2AntFund,EnnesMasterFunc,EllipticMod,MTheoryTested} it was pointed out that in all cases the one instanton corrections can be described with the help of a rational function $S(x)$ referred as a {\em master function} or {\em residue function}. This function appears in the hyperelliptic truncation of the Seiberg-Witten curve as follows:
\begin{equation}
\label{HyperEllipticTrunc}
y(z) + \frac{1}{y(z)} = \frac{1}{\sqrt{S(z)\Lambda^\b}}.
\end{equation}

The rules to construct such a function was proposed in \cite{EllipticMod,MTheoryTested}. We have put them to the Table \ref{Srules}


\begin{table}
\begin{center}
\begin{tabular}{||c|c||c||}
\hhline{|t:=:=:t:=:t|}
\textbf{Group} & \textbf{Multiplet} & \textbf{Factor of $\mbf{S(x)}$}\\
\hhline{|:=:=::=:|}	 
& Adjoint, gauge & $\dfrac{1}{\prod_{l=1}^N {(x-a_l)}^2}$\\
\hhline{||~|-||-||}
& Fundamental & $x+m$\\
\hhline{||~|-||-||}
$SU(N)$ & Symmetric & ${(2x + m)}^2 \prod_{l=1}^N(x + a_l + m)$\\
\hhline{||~|-||-||}
& Antisymmetric & $\dfrac{1}{{(2x + m)}^2}\prod_{l=1}^N (x + a_l + m)$\\
\hhline{||~|-||-||}
& Adjoint, matter & $\prod_{l=1}^N ({(x - a_l)}^2 - m^2)$\\
\hhline{|:=:=::=:|}
& Adjoint, gauge & $\dfrac{x^{4-2\chi}}{\prod_{l=1}^n {(x^2 - a_l^2)}^2}$\\
\hhline{||~|-||-||}
$SO(2n + \chi)$ & Fundamental & $x^2 - m^2 $\\
\hhline{||~|-||-||}
$(\chi = 0,1)$& Adjoint, matter & $\dfrac{{(x^2 - m^2)}^\chi}{4x^2 - m^2} \prod_{l=1}^n ({(x + m)}^2 - a_l^2)({(x - m)}^2 -a_l^2)$\\
\hhline{|:=:=::=:|}
& Adjoint, gauge & $\dfrac{1}{x^4 \prod_{l=1}^N{(x^2 - a_l^2)}^2}$\\
\hhline{||~|-||-||}
$Sp(N)$ & Fundamental & $x^2 -m^2$\\
\hhline{||~|-||-||}
& Antisymmetric &  $\dfrac{\prod_{l=1}^N({(x+m)}^2 - a_l^2)({(x-m)}^2 - a_l^2)}{{(4x^2 - m^2)}^2}$\\
\hhline{||~|-||-||}
& Adjoint, matter & ${(4x^2 - m^2)}^2 \prod_{l=1}^N({(x+m)}^2 - a_l^2)({(x-m)}^2 - a_l^2)$\\ 
\hhline{|b:=:=:b:=:b|}
\end{tabular}
\end{center}
\caption{$S(x)$ building blocks}\label{Srules}
\end{table}


The residue function  has double and quadratic poles. Denote the corresponding ``residues'' as follows:
$$
\begin{aligned}
S(x) &\sim \frac{S_2(x_0)}{(x-x_0)^2}, & S(x) &\sim \frac{S_4(y_0)}{(x-y_0)^4}.
\end{aligned}
$$

Then in many cases the one instanton corrections are given by 
\begin{equation}
\label{Prep1}
\Prep_1(a,m) = \sum_{l=1}^N S_2(a_l).
\end{equation}
If the model contains  one antisymmetric representation of $SU(N)$ or the adjoint of $SO(N)$ one should add to \Ref{Prep1} term $-2S_2(-m/2)$, where $m$ is the mass of corresponding matter multiplet. For two antisymmetric multiplets of $SU(N)$ with masses $m_1$ and $m_2$ one adds $-2S_2(-m_1/2)-2S_2(-m_2/2)$.

Finally for the group $Sp(N)$ we have a quite different expression. One instanton corrections for all matter multiplets is given by
$$
\Prep_1(a,m) =\sqrt{S_4(0)}.
$$

The aim of this section is to show how the notion of the master function naturally appears in our approach. This analysis allows us to state that one instanton corrections computed by our method match with one instanton corrections computed from $M$-theory curves.

Put $k=1$. The 1-instanton contribution to the partition function \Ref{partition} is given by
\begin{equation}
\label{Z1}
Z_1(a,m;\eps) = \oint \frac{\dd\f}{2\pi i} z_1(a,m,\phi;\eps).
\end{equation}
The 1-instanton correction to the prepotential can be extracted from $Z_1(a,m,\eps)$ according to
\begin{equation}
\label{F1fromZ1}
Z_1(a,m;\eps) = \frac{1}{\eps_1\eps_2}\Prep_1(a,m)  + \dots,
\end{equation}
where ``$\dots$'' denotes all terms containing nonnegative powers of $\eps_1,\eps_2$. Combining these two formulae we get
$$
\Prep_1(a,m) = \lim_{\eps_1,\eps_2 \to 0} \eps_1\eps_2 \oint \frac{\dd\f}{2\pi i}z_1(a,m,\phi;\eps).
$$


\begin{table}
\begin{center}
\begin{tabular}{||c|c||c||}
\hhline{|t:=:=:t:=:t|}
\textbf{Group} & \textbf{Multiplet} & \textbf{Factor of $\mbf{z_1(a,m,\phi;\eps)}$}\\
\hhline{|:=:=::=:|}	 
& Adjoint, gauge & $\dfrac{\eps}{\eps_1\eps_2} \dfrac{1}{\prod_{l=1}^N ((\f-a_l)^2 - \eps_+^2)}$\\
\hhline{||~|-||-||}
& Fundamental & $\f + m -\eps_+$\\
\hhline{||~|-||-||}
$SU(N)$ & Symmetric & $(2\f + m -\eps_1)(2\f + m -\eps_2)\prod_{l=1}^N(\f + a_l + m - \eps_+)$\\
\hhline{||~|-||-||}
& Antisymmetric & $\dfrac{\prod_{l=1}^N (\f + a_l + m - \eps_+)}{(2\f + m)(2\f + m - \eps)}$\\
\hhline{||~|-||-||}
& Adjoint, matter & $\dfrac{(m-\eps_1)(m-\eps_2)}{(m-\eps)m} \prod_{l=1}^N ((\f - a_l)^2 - (m - \eps_+)^2)$\\
\hhline{|:=:=::=:|}
& Adjoint, gauge & $\dfrac{\eps}{\eps_1\eps_2}\dfrac{4\f^2 (4\f^2 - \eps^2)}{(\f^2 - \eps_+^2)^\chi\prod_{l=1}^n ((\f+\eps_+)^2 - a_l^2)((\f-\eps_+)^2 - a_l^2)}$\\
\hhline{||~|-||-||}
$SO(2n + \chi)$ & Fundamental & $(m- \eps_+)^2 - \f^2$\\
\hhline{||~|-||-||}
$(\chi = 0,1)$& Adjoint, matter & $\dfrac{(m-\eps_1)(m-\eps_2)}{m(m-\eps)}(\f^2 - (m-\eps_+^2)^2)^\chi$ \\
& & $\times \dfrac{\prod_{l=1}^n ((\f + m - \eps_+)^2 - a_l^2)((\f - m + \eps_+)^2 -a_l^2)}{(4\f^2 - m^2)(4\f^2 - (m-\eps)^2)}$\\
\hhline{|:=:=::=:|}
& Adjoint, gauge & $\dfrac{1}{\eps_1\eps_2\prod_{l=1}^N(\eps_+^2 - a_l^2)}$\\
\hhline{||~|-||-||}
$Sp(N)$ & Fundamental & $(m-\eps_+)$\\
\hhline{||~|-||-||}
& Antisymmetric & $\dfrac{\prod_{l=1}^N ((m-\eps_+)^2 - a_l^2)}{m(m-\eps)}$\\
\hhline{||~|-||-||}
& Adjoint, matter & $(m-\eps_1)(m-\eps_2)\prod_{l=1}^N ((m-\eps_+)^2 - a_l^2)$\\ 
\hhline{|b:=:=:b:=:b|}
\end{tabular}
\end{center}
\caption{$z_1(a,m,\phi;\eps)$ building blocks}\label{z1rules}
\end{table}


Analysis of \Ref{zSUFund}, \Ref{zSUAdjGauge}, \Ref{zSUAdjMatter}, \Ref{zSUAnt}, \Ref{zSUSym}, \Ref{zSOFund}, \Ref{zSOAdjGauge}, \Ref{zSOAdjMatter}, \Ref{zSpFund}, \Ref{zSpAdjGauge}, \Ref{zSpAdjMatter}, and \Ref{zSpAnt} together with \Ref{partition} shows that one can establish the rule to construct $z_1(a,m,\eps,\f)$ (see Table \ref{z1rules}). 

First observation is that that for $SU(N)$ and $SO(N)$ the following equality holds:
$$
\lim_{\eps_1,\eps_2\to0}\frac{\eps_1\eps_2}{\eps}z_1(a,m,\phi;\eps) = S(\f).
$$
Hence one can call $z_1(a,m,\phi;\eps)$ a deformed residue function. Using the properties of the contour integration
$$
\begin{aligned}
\oint \frac{\dd \f}{2\pi i} \frac{1}{(\f - x_0 - \eps_+)(\f - x_0 +\eps_+)} = \frac{1}{\eps}
\end{aligned}
$$
we arrive to the rule announce after \Ref{Prep1}.

\begin{remark}
For $Sp(N)$ the integrand does not depend on $\f$. It means that for $Sp(N)$ the one instanton corrections are given by
$$
\Prep_1(a,m) = \lim_{\eps_1,\eps_2\to 0} \eps_1\eps_2z_1(a,m;\eps).
$$
The rule for the residue function proposed in \cite{EllipticMod,MTheoryTested} are such that 
$$\sqrt{S_4(0)} = \lim_{\eps_1,\eps_2\to 0}\eps_1\eps_2 z_1(a,m;\eps).$$
This proves the validity of our formulae in the case of $Sp(N)$.
\end{remark}

The method of residue function, yet simple for $k=1$ case, seems to be difficult to generalize to other ($k>1$) cases. The reason is both the complexity of \Ref{Z1} and \Ref{F1fromZ1} when $k>1$. For example \Ref{Z1} generalizes as follows (for $SU(N)$ and $SO(N)$, the $Sp(N)$ case should be considered separetely):
$$
Z_k(a,m;\eps) = \oint \prod_{i=1}^k \frac{\dd \f_i}{2\pi i} \EuScript{R}(\f) \prod_{i=1}^k z_1(a,m,\phi_i;\eps)
$$
where $\EuScript{R}(\f)$ is a ratio of $\Delta$'s products. The integral can be computed by hands in low $k$ case. For example, it was done in \cite{WyllMar} for $k \leq 3$ for $SO(N)$ and $Sp(N)$ pure Yang-Mills theories and for $k\leq 2$ for symmetric and antisymmetric representations of $SU(N)$. Also these integrals can be computed for general $k$ in the case of $SU(N)$ (fundamental and adjoint representations, \cite{SWfromInst}). See the discussion in \cite{WyllMar} of what happens in the case of other classical groups.

%% file: saddlepoint.tex
\label{saddlepoint}

The formal expression \Ref{partition} allows, in principle, to compute all the instanton correction. However, there are two objection: first, for general group and representation this is not known how to rewrite this integral as a sum over the residues of the deformed residue function $z_1(a,m,\phi;\eps)$. Second objection comes from the fact, that the representation of the prepotential as of the formal series on $\Lambda$ makes its analytical properties obscure. In particular, it is not clear how the prepotential could be analytically continued beyond the convergence radius. 

Fortunately, the Seiberg-Witten theory \cite{SeibergWitten} can answer to the second question. Our goal in this section is to explain how the Seiberg-Witten data can be extracted from \Ref{partition}.


\section{Thermodynamic (classical) limit}

In \cite{SWandRP} the general method to extract the Seiberg-Witten data was proposed. The idea is the following. The prepotential can be obtained from the partition function $Z(a,m,\gL;\eps)$ in the limit $\eps_1,\eps_2\to 0$ (see \Ref{Z}). One can show that in this limit the main contribution to the partition function comes from $k \sim \dfrac{1}{\eps_1\eps_2}$. It follows that in order to extract Seiberg-Witten data we don't need to examine the whole series \Ref{Z}. It is sufficient to consider the expression \Ref{partition} taken in the limit $k \to\infty$.
 
In this limit the multiple integral on $\f_i$ becomes Feynman integral over the density of $\f_i$'s. Each $\f_i$ can be seen as a physical quantity which corresponds to a ``particle''. The instanton number $k$ plays the role of the number of such a ``particles''. Another point of view is to consider the inverse instanton number as a Plank constant in a quantum mechanical problem. The expression \Ref{partition} becomes the partition function of a system, described by a Hamiltonian, depending of the $\f_i$'s density. 

In the thermodynamic (classical) limit $k\to\infty$ this partition function can be computed by the saddle point approximation. It means that the main contribution is given by a classical configuration (we put aside the question of existence and uniqueness of such a configuration). The prepotential appears as the ``free energy'' in this context. As we shall see the Seiberg-Witten data appears naturally when we solve the equation of motion (saddle point equation).

After this short introduction let us pass to the concrete computations. First we note that the thermodynamic (or quantum mechanical) problem is formulated by means of the action (Hamiltonian). The integrand in the Feynman integral generically has the form $\e^{-\frac{1}{\eps_1\eps_2}H}$. Therefore we should convert the integrand of \Ref{partition} into the similar form. Keeping in mind the origin of this integrand (formula \Ref{TrWeights}) we can obtain a mnemonic rule to compute the Hamiltonian $H$ directly from the equivariant index of the Dirac operator:
$$
\Ind_q = \sum_{\a} \ep_\a \e^{w_\a} \mapsto \prod_{\a} {w_\a}^{\ep_\a}= \exp\left\{\sum_\a \ep_\a \ln w_\a\right\} \mapsto  H_{\eps_1,\eps_2} = - \eps_1\eps_2\sum_{\a} \ep_\a \ln |w_\a|.
$$

However, the Hamiltonian defined above contains much more information we need. Namely it can be represented as a series over the nonnegative powers of $\eps_1$ and $\eps_2$. The only contribution relevant in the thermodynamic limit comes from the terms independent of $\eps_1$ and $\eps_2$. Therefore the expression for the Hamiltonian can be rewritten as follows:
\begin{equation}
\label{HamClass}
\Ind_q = \sum_{\a}\ep_\a \e^{w_\a} \mapsto H = - \lim_{\eps_1,\eps_2\to 0} \eps_1\eps_2 \sum_\a \ep_\a \ln |w_\a|.
\end{equation}

Taking into account the additivity of the equivariant index \Ref{AddInd} we conclude that 
$$
H = H^{\mr{adj,gauge}} + \sum_{\vr\in \mr{reps}} H^{\vr,\mr{matter}}.
$$

\begin{remark}
We have just established a rule to represent $Z_k(a,m;\eps)$ given by \Ref{partition} as an exponent of a sum of $\ln|w_\a|$'s. We can ask now what will change if we multiply $Z_k(a,m;\eps)$ by $\Lambda^{k\b}$. The answer is that we should replace $\ln|w_\a|$ with $\ln\left|\dfrac{w_\a}{\Lambda}\right|$.
\end{remark}


\section{A trivial model example}

To illustrate the phenomenon, where the series is evaluated by the saddle point we take the following trivial example:
$$
Z (\gL,\eps) = \sum_{k=0}^\infty \frac{1}{k!} {\left(\frac{\gL}{\eps}\right)}^k.
$$
Suppose $\ds\frac{\gL}{\eps} \in \Real_+$ and $\gL \gg \eps$. Then the series is dominated by the single term, where $k  = k_\star \sim \gL/\eps$. Stirling's formula gives:
$$
Z(\gL,\eps)  \sim \e^{k_\star} \sim \exp\frac{\gL}{\eps}.
$$
Now this formula can be analytically continued to aritrary $\gL \in \Compl$, and by expanding the answer in powers of $\gL$ we get correctly the terms in the original series for small $k$.


\section{$SU(N)$ case, pure Yang-Mills theory}
\label{SU(N)PureYM}

Let us consider in some details the simplest case: the $SU(N)$ theory without matter multiplets. The weights are given by \Ref{IndSUAdj}. 

Let us show how the first term in \Ref{IndSUAdj} gives the perturbative correction to the prepotential \cite{SWandRP}.  

As we have already mentioned, the transformation \Ref{TrWeights} can be seen as the proper time regularization. It is given by the formula
$$
\e^{i\langle x,w_\a(p)\rangle} \mapsto {\left. \frac{\dd}{\dd s}\right|}_{s=0}\frac{\Lambda^s}{\Gamma(s)} \int_0^\infty \frac{\dd t}{t} t^s \e^{i\langle tx,w_\a(p)\rangle} = - \ln \left|\frac{\langle x,w_\a(p)\rangle}{\Lambda}\right|.
$$

It follows that the contribution of the first term of \Ref{IndSUAdj} to the Hamiltonian \Ref{HamClass} is given by
$$
\lim_{\eps_1,\eps_2\to 0}\eps_1\eps_2 \sum_{l,m=1}^N \g_{\eps_1,\eps_2}(a_l - a_m,\Lambda)
$$
where 
$$
\g_{\eps_1,\eps_2} (x,\Lambda) = {\left. \frac{\dd}{\dd s}\right|}_{s = 0} \frac{\Lambda^s}{\Gamma(s)}\int_0^\infty \frac{\dd t}{t} t^s \frac{\e^{itx}}{(1-\e^{i\eps_1 t})(1 - \e^{i\eps_2 t})}.
$$
The $\eps$ expansion of $\gamma_{\eps_1\eps_2}$ is given by 
$$
\g_{\eps_1,\eps_2}(x,\Lambda) = \frac{1}{\eps_1\eps_2} \k_\Lambda(x) + \dots,
$$
where ``$\dots$'' are terms finite in the thermodynamic limit and
$$
\k_\Lambda(x) = \frac{1}{2} x^2 \left(\ln \left|\frac{x}{\Lambda}\right| - \frac{3}{2}\right).
$$
For more properties of $\g_{\eps_1,\eps_2}(x,\Lambda)$ see Appendix A in \cite{SWandRP}.

Finally the contribution to the Hamiltonian of the first term is given by
$$
\sum_{l\neq m} \k_\Lambda(a_l - a_m) = \sum_{l\neq m} \frac{1}{2} {(a_l - a_m)}^2 \left(\ln \left|\frac{a_l - a_m}{\Lambda}\right| - \frac{3}{2}\right).
$$

In this expression we can recognize the perturbative part of the prepotential \Ref{pertPrep}. It explains the remark after \Ref{zSUAnt}.

To handle the last line in \Ref{IndSUAdj} we use the following identity:
$$
f(0) + f(\eps_1+\eps_2) - f(\eps_1) - f(\eps_2) = \eps_1\eps_2 f''(0)+ \dots,
$$
where ``$\dots$'' are the higher $\eps$-terms. It gives
$$
\ln(\f_i - \f_j) + \ln (\f_i - \f_j - \eps) -  \ln (\f_i - \f_j -\eps_2) - \ln (\f_i - \f_j - \eps_1) = -\eps_1\eps_2 \frac{1}{{(\f_i - \f_j)}^2} + \dots.
$$

Finally with the help of $\Ref{rhoSU}$ we have the following expression for the Hamiltonian:
$$
H = - \sum_{l\neq m} \k_\Lambda(a_l - a_m) + 2\eps_1\eps_2\sum_{i=1}^k \ln \left| \frac{\P(\f_i)}{\Lambda^N} \right| + {(\eps_1\eps_2)}^2\sum_{i\neq j} \frac{1}{{(\f_i - \f_j)}^2}
$$
In the thermodynamic limit $k\to\infty$ the number of $\f_i$'s becomes infinite. It is natural to introduce its density which is  normalizable in the limit. In order to keep the normalizability we define:
\begin{equation}
\label{rhoSU}
\r(x) = \eps_1\eps_2\sum_{i=1}^k \d(x-\f_i).
\end{equation}
In the thermodynamic limit this function becomes smooth. With the help of the density function the Hamiltonian can be rewritten as follows:
$$
H = - \sum_{l\neq m} \k_\Lambda(a_l - a_m) + 2\sum_{l=1}^N\int \dd x\r(x) \ln \left|\frac{x - a_l}{\Lambda}\right| + \vpint_{x\neq y} \dd x\dd y\frac{\r(x)\r(y)}{{(x-y)}^2}.
$$

The obtained expression is rather suggestive. After integration by parts and introducing the {\em profile function}\footnote{in the $SU(N)$ case this function is closely related to the profile of the Young tableaux, as defined in  \cite{SWandRP}}
\begin{equation}
\label{ProfSU}
f(x) = -2 \r(x) + \sum_{l=1}^N |x-a_l|
\end{equation}
the Hamiltonian can be rewritten in a nice form:
\begin{equation}
\label{HamSUAdjGauge}
H[f] = -\frac{1}{4} \int \dd x\dd y f''(x) f''(y) \k_\Lambda(x-y).
\end{equation}

The partition function \Ref{Z} can be represented as follows:
\begin{equation}
\label{varEqn}
Z(a,m,\gL;\eps) \sim \int \mathcal{D} f \e^{-\frac{1}{\eps_1\eps_2} H_{\eps_1,\eps_2}[f]}.
\end{equation}
We are interested in the classical approximation of this integral only.


\section{$SU(N)$, matter multiplets}

In this section we obtain expressions for the Hamiltonians similar to \Ref{HamSUAdjGauge} for the matter multiplets using the rule \Ref{HamClass} and formulae \Ref{zSUFund}, \Ref{zSUAdjMatter}, \Ref{zSUSym}, and \Ref{zSUAnt}.


\subsection{Matter in the fundamental representation.}
$$
\begin{aligned}
H_{\eps_1,\eps_2} &= \sum_{l=1}^N  \k_\Lambda(a_l + m) - \eps_1\eps_2 \sum_{i=1}^k \ln \left| \frac{\f_i + m}{\Lambda}\right|\\
&= \sum_{l=1}^N  \k_\Lambda(a_l + m) - \int \dd x \r(x) \ln \left| \frac{x+m}{\Lambda}\right|.
\end{aligned}
$$
With the help of the profile function we can rewrite the Hamiltonian as follows:
\begin{equation}
H[f] = \frac{1}{2} \int \dd x \dd y f''(x) \k_\Lambda(x+m).
\label{HamSUFund}
\end{equation}


\subsection{Matter in the symmetric representation} We have
$$
\begin{aligned}
H_{\eps_1,\eps_2} &= \sum_{l\leq m\leq N} \k_\Lambda(a_l + a_m + m) - \eps_1\eps_2\sum_{i=1}^k \ln \left| \frac{\P(-\f_i -m)}{\Lambda^N}\right| \\
&- 2\eps_1\eps_2 \sum_{i=1}^k \ln \left| \frac{\f_i + m/2}{\Lambda}\right| - {(\eps_1\eps_2)}^2 \sum_{l\leq m \leq N} \frac{1}{{(\f_i + \f_j + m)}^2}.
\end{aligned}
$$
The density function lets us to rewrite this expression as follows:
$$
\begin{aligned}
H &= \sum_{l\leq m \leq N} \k_\Lambda(a_l + a_m + m) - \int \dd x \r(x)\ln \left| \frac{\P(-x-m)}{\Lambda^N} \right| \\
&- 2 \int \dd x \r(x)\ln \left| \frac{x + m/2}{\Lambda} \right| - \frac{1}{2}\int \dd x \dd y \frac{\r(x)\r(y)}{{(x+y+m)}^2}.
\end{aligned}
$$
Using the profile function we get finally
\begin{equation}
\label{HamSUSym}
H[f]= \frac{1}{8} \int \dd x \dd y f''(x) f''(y) \k_\Lambda(x+y+m) + \int \dd x f''(x)\k_\Lambda(x+m/2).
\end{equation}


\subsection{Matter in the antisymmetric representation} 
$$
\begin{aligned}
H_{\eps_1,\eps_2} &= \sum_{l\leq m\leq N} \k_\Lambda(a_l + a_m + m) - \eps_1\eps_2\sum_{i=1}^k \ln \left| \frac{\P(-\f_i -m)}{\Lambda^N}\right| \\
&+ 2\eps_1\eps_2 \sum_{i=1}^k \ln \left| \frac{\f_i + m/2}{\Lambda}\right| - {(\eps_1\eps_2)}^2 \sum_{l\leq m \leq N} \frac{1}{{(\f_i + \f_j + m)}^2}.
\end{aligned}
$$
The density function lets us to rewrite this expression as follows:
$$
\begin{aligned}
H &= \sum_{l\leq m \leq N} \k_\Lambda(a_l + a_m + m) - \int \dd x \r(x)\ln \left| \frac{\P(-x-m)}{\Lambda^N} \right| \\
&+ 2 \int \dd x \r(x)\ln \left| \frac{x + m/2}{\Lambda} \right| - \frac{1}{2}\int \dd x \dd y \frac{\r(x)\r(y)}{{(x+y+m)}^2}.
\end{aligned}
$$
Using the profile function we get finally
\begin{equation}
\label{HamSUAnt}
H[f]= \frac{1}{8} \int \dd x \dd y f''(x) f''(y) \k_\Lambda(x+y+m) - \int \dd x f''(x)\k_\Lambda(x+m/2).
\end{equation}


\subsection{Matter in the adjoint representation}
$$
\begin{aligned}
H_{\eps_1,\eps_2} &= \sum_{l\neq m} \k_\Lambda(a_l - a_m + m) + N\k_\Lambda(m)\\
&- \eps_1\eps_2\sum_{i=1}^k \ln \left| \frac{\P(\f_i -m)\P(\f_i + m)}{\Lambda^{2N}}\right| - {(\eps_1\eps_2)}^2 \sum_{l\neq m} \frac{1}{{(\f_i - \f_j + m)}^2}.
\end{aligned}
$$
The density function lets us to rewrite this expression as follows:
$$
\begin{aligned}
H &= \sum_{l\neq m} \k_\Lambda(a_l - a_m + m) + N \k_\Lambda(m)\\
&- \int \dd x \r(x)\ln \left| \frac{\P(x-m)\P(x+m)}{\Lambda^{2N}} \right| - \int \dd x \dd y \frac{\r(x)\r(y)}{{(x-y+m)}^2}.
\end{aligned}
$$
Using the profile function we get
\begin{equation}
\label{HamSUAdjMatter}
H[f]= \frac{1}{4} \int \dd x \dd y f''(x) f''(y) \k_\Lambda(x-y+m).
\end{equation}


\section{$SO(N)$ case}

Now let us apply the rule \Ref{HamClass} to the orthogonal group. 


\subsection{Pure gauge theory} 

Formulae \Ref{zSOFund}, \Ref{zSOAdjGauge} and \Ref{zSOAdjMatter} lead to the following expression for the Hamiltonian
$$
\begin{aligned}
H_{\eps_1,\eps_2} &=  -2\sum_{l<m\leq n} \Big( \k_\Lambda(a_l - a_m) + \k_\Lambda(a_l + a_m)\Big) - 2\sum_{l=1}^n \k_\Lambda(a_l) \\
&+2{(\eps_1\eps_2)}^2 \sum_{i<j\leq k} \left( \frac{1}{{(\f_i-\f_j)}^2} + \frac{1}{{(\f_i + \f_j)}^2} \right) \\
&+ 2\eps_1\eps_2 \sum_{i=1}^k \ln \left| \frac{\P(\f_i)}{\Lambda^{2n+\chi}}\right| - 4\eps_1\eps_2 \sum_{i=1}^k \ln \left| \frac{\f_i}{\Lambda}\right|.
\end{aligned}
$$
As in the $SU(N)$ case we introduce the density function as follows:
\begin{equation}
\label{rhoSO}
\r(x) = \eps_1\eps_2\sum_{i=1}^k \Big(\d(x-\f_i) + \d(x+\f_i)\Big).
\end{equation}
Simple computation shows that
$$
\begin{aligned}
\frac{1}{2} \vpint \dd x \dd y \frac{\r(x)\r(y)}{{(x-y)}^2} &= 2{(\eps_1\eps_2)}^2 \sum_{i<j\leq k} \left( \frac{1}{{(\f_i-\f_j)}^2} + \frac{1}{{(\f_i + \f_j)}^2} \right), \\
\int \dd x \r(x) \ln \left|\frac{\P(x)}{\Lambda^{2n+\chi}} \right| &= 2\eps_1\eps_2 \sum_{i=1}^k \ln \left| \frac{\P(\f_i)}{\Lambda^{2n+\chi}}\right|.
\end{aligned}
$$
Using these formulae we get
$$
\begin{aligned}
H &= -2\sum_{l<m\leq n} \Big( \k_\Lambda(a_l - a_m) + \k_\Lambda(a_l + a_m)\Big) - 2\sum_{l=1}^n \k_\Lambda(a_l)\\
&+ \frac{1}{2}\vpint \dd x \dd y \frac{\r(x)\r(y)}{{(x-y)}^2} + \int \dd x \r(x) \ln \left| \frac{\P(x)}{\Lambda^{2n+\chi}}\right| - 2 \int \dd x \r(x) \ln \left| \frac{x}{\Lambda}\right|
\end{aligned}
$$
Introducing the profile function 
\begin{equation}
\label{ProfSO}
f(x) = -2\r(x) + \chi|x| + \sum_{l=1}^n \Big( |x-a_l| + |x+a_l|\Big)
\end{equation}
we rewrite the expression for the Hamiltonian as follows:
\begin{equation}
\label{HamSOAdjGauge}
H[f] = - \frac{1}{8} \int \dd x \dd y f''(x)f''(y) \k_\Lambda(x-y) + \int \dd x f''(x) \k_\Lambda(x).
\end{equation}


\subsection{Matter in the fundamental representation} 

Formula \Ref{zSOFund} gives 
$$
\begin{aligned}
H_{\eps_1,\eps_2} &=  \sum_{l=1}^n \Big( \k_\Lambda(a_l - m) + \k_\Lambda(a_l + m)\Big)-\eps_1\eps_2 \sum_{i=1}^k \ln \left|\frac{\f_i^2 - m^2}{\Lambda^2}\right| \\
&=  \sum_{l=1}^n \Big( \k_\Lambda(a_l - m) + \k_\Lambda(a_l + m)\Big)- \int \dd x \r(x) \ln\left| \frac{x-m}{\Lambda}\right|.
\end{aligned}
$$
With the help of the profile function \Ref{ProfSO} it can be rewritten as follows:
\begin{equation}
\label{HamSOFund}
H[f] = \frac{1}{4}\int \dd x f''(x) \k_\Lambda(x+m) + \frac{1}{4}\int \dd x f''(x) \k_\Lambda(x-m).
\end{equation} 


\subsection{Matter in the adjoint representation} 

Applying the rule \Ref{HamClass} to \Ref{zSOAdjMatter} we get (for the perturbatif terms we use directly \Ref{IndSOAdj}):
$$
\begin{aligned}
H_{\eps_1,\eps_2} &= \sum_{l<m\leq n} \Big( \k_\Lambda(a_l + a_m + m) + \k_\Lambda(a_l - a_m + m) + \k_\Lambda(-a_l + a_m + m) \\
&+ \k_\Lambda(-a_l - a_m + m) \Big) + \chi \sum_{l=1}^n \Big( \k_\Lambda(a_l + m) + \k_\Lambda(-a_l + m) \Big) + n\k_\Lambda(m)\\
&- 2{(\eps_1\eps_2)}^2 \sum_{i<j\leq k} \left( \frac{{(\f_i - \f_j)}^2 + m^2}{{({(\f_i-\f_j)}^2 - m^2)}^2}  + \frac{{(\f_i + \f_j)}^2 + m^2}{{({(\f_i+\f_j)}^2 - m^2)}^2}\right) \\
&-\eps_1\eps_2 \sum_{i=1}^k \ln \left| \frac{\P(\f_i + m)\P(\f_i - m)}{\Lambda^{4n + 2\chi}}\right| + 2 \eps_1\eps_2 \sum_{i=1}^k\ln \left| \frac{\f_i^2 - m^2/4}{\Lambda^2}\right|.
\end{aligned} 
$$
Using the following algebraic identity
\begin{equation}
\label{AlgId}
\frac{1}{{(a+b)}^2} + \frac{1}{{(a-b)}^2} = 2\frac{a^2+b^2}{{(a^2 - b^2)}^2}
\end{equation}
we can write using the density function \Ref{rhoSO}:
$$
\frac{1}{2}\int \dd x \dd y \frac{\r(x)\r(y)}{{(x+y+m)}^2}  = 2{(\eps_1\eps_2)}^2 \sum_{i<j\leq k} \left( \frac{{(\f_i+\f_j)}^2 + m^2}{{{(\f_i+\f_j)}^2 - m^2)}^2} + \frac{{(\f_i - \f_j)}^2 + m^2}{{({(\f_i-\f_j)}^2 - m^2)}^2}\right).
$$
Thus the Hamiltonian can be rewritten as follows:
$$
\begin{aligned}
H &=  \sum_{l<m\leq n} \Big( \k_\Lambda(a_l + a_m + m) + \k_\Lambda(a_l - a_m + m) + \k_\Lambda(-a_l + a_m + m) \\
&+ \k_\Lambda(-a_l - a_m + m) \Big) + \chi \sum_{l=1}^n \Big( \k_\Lambda(a_l + m) + \k_\Lambda(-a_l + m) \Big) + n\k_\Lambda(m)\\
&- \frac{1}{2} \int \dd x \dd y \frac{\r(x)\r(y)}{{(x+y+m)}^2} - \int \dd x \r(x) \ln \left| \frac{\P(x+m)}{\Lambda^{2n + \chi}}\right| + 2 \int \dd x \r(x) \ln \left| \frac{x+m/2}{\Lambda} \right|.
\end{aligned}
$$
Using the profile function we get 
\begin{equation}
\label{HamSOAdjMatter}
H[f] = \frac{1}{8} \int \dd x \dd y f''(x)f''(y) \k_\Lambda(x+y+m) - \int \dd x f''(x) \k_\Lambda(x + m/2).
\end{equation}


\section{$Sp(N)$ case}

Let us finally apply the rule \Ref{HamClass} to the theories with the symplectic gauge group. We start with the pure Yang-Mills theory.


\subsection{Pure gauge theory} 

Applying the rule \Ref{HamClass} to formulae \Ref{zSpFund}, \Ref{zSpAdjGauge}, \Ref{zSpAdjMatter}, and \Ref{zSpAnt} we obtain:
$$
\begin{aligned}
H_{\eps_1,\eps_2} &= - \sum_{l,m}^N \Big( \k_\Lambda(a_l -a_m) + \k_\Lambda(a_l + a_m) \Big) - \sum_{l=1}^N \k_\Lambda(2a_l)\\
&+2{(\eps_1\eps_2)}^2 \sum_{i<j\leq n} \left( \frac{1}{{(\f_i-\f_j)}^2} + \frac{1}{{(\f_i + \f_j)}^2} \right) + 2\eps_1\eps_2\sum_{i=1}^n \ln\left| \frac{\P(\f_i) \f_i^2}{\Lambda^{2N+2}}\right| \\
&+ 2\chi{(\eps_1\eps_2)}^2 \sum_{i=1}^n \frac{1}{\f_i^2}.
\end{aligned}
$$
In order to reduce the sum to the integral introduce the density of $\f_i$'s as follows:
\begin{equation}
\label{rhoSp}
\r(x) = \eps_1\eps_2 \sum_{i=1}^n \Big( \d(x-\f_i) + \d(x+\f_i) \Big).
\end{equation}
It follows that the relevant in the thermodynamic limit Hamiltonian is
$$
\begin{aligned}
H &= - \sum_{l,m}^N \Big( \k_\Lambda(a_l -a_m) + \k_\Lambda(a_l + a_m) \Big) - \sum_{l=1}^N \k_\Lambda(2a_l)\\
&+\frac{1}{2}\vpint \dd x\dd y \frac{\r(x)\r(y)}{{(x-y)}^2} + \int \dd x \r(x) \ln \left| \frac{x^2 \P(x)}{\Lambda^{2N + 2}}\right| \\
&=- \sum_{l,m}^N \Big( \k_\Lambda(a_l -a_m) + \k_\Lambda(a_l + a_m) \Big) - \sum_{l=1}^N \k_\Lambda(2a_l)\\
&+ \frac{1}{2}\vpint \dd x\dd y \frac{\r(x)\r(y)}{{(x-y)}^2} + \frac{1}{2}\int \dd x \r(x)  \sum_{l=1}^N \left( \ln \left| \frac{x-a_l}{\Lambda}\right| + \ln \left| \frac{x+a_l}{\Lambda}\right| \right) \\
&+ 2\int \dd x \r(x) \ln \left| \frac{x}{\Lambda} \right|.
\end{aligned}
$$
\begin{remark}
It worth noting that in the thermodynamic limit the last $\chi$-dependent term becomes irrelevant and therefore can be dropped. It seems naturals since in the $k\to\infty$ limit the difference between $k$-even and $k$-odd cases disappears. 
\end{remark}

Further simplification can be achieved after the integration twice by part and introducing the following analogue of the profile function:
\begin{equation}
\label{ProfSp}
f(x) = -2\r(x) +\sum_{l=1}^N \Big(|x-a_l| + |x+a_l|\Big).
\end{equation}
Note that this function (as well as \Ref{rhoSp}) is explicitly symmetric. Then
\begin{equation}
\label{HamSpAdjGauge}
H[f] = - \frac{1}{8} \int \dd x \dd y f''(x)f''(y)\k_\Lambda(x-y) - \int \dd x f''(x)\k_\Lambda(x).
\end{equation}


\subsection{Matter in the fundamental representation} 

Formula \Ref{zSpFund} give
$$
\begin{aligned}
H_{\eps_1,\eps_2} &= \sum_{l=1}^N \Big( \k_\Lambda(a_l - m) + \k_\Lambda(a_l + m)\Big)\\
&-\eps_1\eps_2 \sum_{i=1}^n \ln|\f_i^2 -m^2| \\
&= - \sum_{l=1}^N \Big( \k_\Lambda(a_l - m) + \k_\Lambda(a_l + m)\Big) \\
&= -\frac{1}{2}\int \dd x\r(x) \ln \left|\frac{x + m}{\Lambda}\right| - \frac{1}{2} \int \dd x\r(x) \ln \left|\frac{x - m}{\Lambda}\right|.
\end{aligned}
$$
With the help of the profile function \Ref{ProfSp} we obtain
\begin{equation}
\label{HamSpFund}
H[f] = \frac{1}{4} \int \dd x f''(x) \k_\Lambda(x-m) + \frac{1}{4} \int \dd x f''(x) \k_\Lambda(x+m).
\end{equation} 


\subsection{Matter in the antisymmetric representation} 

We have (for the perturbative term we use directly \Ref{IndSpAnt})
$$
\begin{aligned}
H_{\eps_1,\eps_2} &=  \sum_{l<m} \Big( \k_\Lambda(a_l + a_m + m) + \k_\Lambda(a_l - a_m + m) \\
&+ \k_\Lambda(-a_l + a_m + m) + \k_\Lambda(-a_l -a_m + m) \Big) + N\k_\Lambda(m)\\
&- 2{(\eps_1\eps_2)}^2 \sum_{i<j\leq n} \left( \frac{{(\f_i - \f_j)}^2 + m^2}{{({(\f_i-\f_j)}^2 - m^2)}^2}  + \frac{{(\f_i + \f_j)}^2 + m^2}{{({(\f_i+\f_j)}^2 - m^2)}^2}\right) \\
&- 2\chi{(\eps_1\eps_2)}^2 \sum_{i=1}^n \frac{\f_i^2 + m^2}{{(\f_i^2 - m^2)}^2} -  \eps_1\eps_2\sum_{i=1}^n \ln\left| \frac{\P(\f_i + m)\P(\f_i - m)}{\Lambda^{4N}}\right| \\
&+ 2\eps_1\eps_2\sum_{i=1}^n \ln\left| \frac{\f_i^2 - m^2/4}{\Lambda^2}\right|.
\end{aligned}
$$
Using the algebraic identity \Ref{AlgId} we can rewrite obtained expression with the help of the density function \Ref{rhoSp} as follows:
$$
\begin{aligned}
H &= \sum_{l<m} \Big( \k_\Lambda(a_l + a_m + m) + \k_\Lambda(a_l - a_m + m) \\
&+ \k_\Lambda(-a_l + a_m + m) + \k_\Lambda(-a_l -a_m + m) \Big) +N\k_\Lambda(m)\\
&-\frac{1}{2}\int \dd x\dd y \frac{\r(x)\r(y)}{{(x+y+m)}^2} + \int \dd x \r(x) \left( \ln\left| \frac{x-m/2}{\Lambda}\right|+ \ln\left| \frac{x+m/2}{\Lambda}\right| \right) \\
&- \frac{1}{2} \int \dd x \r(x) \sum_{l=1}^N \left(  \ln\left| \frac{{(x + m)}^2 -a_l^2}{\Lambda}\right| +  \ln\left| \frac{{(x - m)}^2  -a_l^2}{\Lambda}\right| \right).
\end{aligned}
$$
It can be rewritten using the profile function as well:
\begin{equation}
\label{HamSpAnt}
\begin{aligned}
H[f] &=\frac{1}{8} \int \dd x \dd yf''(x) f''(y) \k_\Lambda(x+y+m) -\int \dd x f''(x) \k_\Lambda(x + m/2).
\end{aligned} 
\end{equation}


\subsection{Matter in the adjoint representation} 

After examination \Ref{zSpAdjMatter} we get 
$$
\begin{aligned}
H_{\eps_1,\eps_2} &= \sum_{l<m\leq N} \Big( \k_\Lambda(a_l + a_m + m) + \k_\Lambda(a_l - a_m + m) + \k_\Lambda(-a_l + a_m + m) \\
&+ \k_\Lambda(-a_l -a_m + m) \Big) + \sum_{l=1}^N \Big( \k_\Lambda(2a_l + m) + \k_\Lambda(-2a_l + m) \Big) + N\k_\Lambda(m)\\
&- 2{(\eps_1\eps_2)}^2 \sum_{i<j\leq n} \left( \frac{{(\f_i - \f_j)}^2 + m^2}{{({(\f_i-\f_j)}^2 - m^2)}^2}  + \frac{{(\f_i + \f_j)}^2 + m^2}{{({(\f_i+\f_j)}^2 - m^2)}^2}\right) \\
&- 2\chi{(\eps_1\eps_2)}^2 \sum_{i=1}^n \frac{\f_i^2 + m^2}{{(\f_i^2 - m^2)}^2} -  \eps_1\eps_2\sum_{i=1}^n \ln\left| \frac{\P(\f_i + m)\P(\f_i - m)}{\Lambda^{4N}}\right| \\
&- 2\eps_1\eps_2\sum_{i=1}^n \ln\left| \frac{\f_i^2 - m^2/4}{\Lambda^2}\right|.
\end{aligned}
$$
With the help of the density function \Ref{rhoSp} it can be rewritten as
$$
\begin{aligned}
H &=  \sum_{l<m\leq N} \Big( \k_\Lambda(a_l + a_m + m) + \k_\Lambda(a_l - a_m + m) + \k_\Lambda(-a_l + a_m + m) \\
&+ \k_\Lambda(-a_l -a_m + m) \Big) + \sum_{l=1}^N \Big( \k_\Lambda(2a_l + m) + \k_\Lambda(-2a_l + m) \Big) + N\k_\Lambda(m)\\
&-\frac{1}{2}\int \dd x\dd y \frac{\r(x)\r(y)}{{(x+y+m)}^2} - \int \dd x \r(x) \left( \ln\left| \frac{x-m/2}{\Lambda}\right|+ \ln\left| \frac{x+m/2}{\Lambda}\right| \right) \\
&- \frac{1}{2} \int \dd x \r(x) \sum_{l=1}^N \left(  \ln\left| \frac{{(x + m)}^2 -a_l^2}{\Lambda}\right| +  \ln\left| \frac{{(x - m)}^2  -a_l^2}{\Lambda}\right| \right).
\end{aligned}
$$
In terms of $f(x)$ it becomes
\begin{equation}
\label{HamSpAdjMatter}
\begin{aligned}
H[f] &=\frac{1}{8} \int \dd x f''(x) f''(y) \k_\Lambda(x+y+m) +\int \dd x f''(x) \k_\Lambda(x + m/2).
\end{aligned}
\end{equation}


\section{Hamiltonians}


\begin{table}
\begin{center}
\begin{tabular}{||c|c||c||}
\hhline{|t:=:=:t:=:t|}
\textbf{Group} & \textbf{Multiplet} & \textbf{Contribution to  $\mbf{H[f]}$}\\
\hhline{|:=:=::=:|}	 
& Adjoint, gauge & $ -\dfrac{1}{4}\int \dd x\dd y f''(x)f''(y)\k_\Lambda(x-y)$\\
\hhline{||~|-||-||}
& Fundamental & $  \dfrac{1}{2} \int \dd x f''(x) \k_\Lambda(x+m)$\\
\hhline{||~|-||-||}
$SU(N)$ & Symmetric & $\dfrac{1}{8}\int \dd x\dd y f''(x)f''(y) \k_\Lambda(x+y + m) +  \int \dd x f''(x) \k_\Lambda(x + m/2)$ \\
\hhline{||~|-||-||}
& Antisymmetric & $\dfrac{1}{8} \int \dd x\dd y f''(x)f''(y)\k_\Lambda(x + y + m) -  \int \dd x f''(x) \k_\Lambda(x + m/2)$\\
\hhline{||~|-||-||}
& Adjoint, matter & $\dfrac{1}{4} \int \dd x\dd y f''(x)f''(y)\k_\Lambda(x - y + m)$\\
\hhline{|:=:=::=:|}
& Adjoint, gauge & $-\dfrac{1}{8} \int \dd x\dd y f''(x)f''(y)\k_\Lambda(x + y) + \int \dd x f''(x) \k_\Lambda(x)$\\
\hhline{||~|-||-||}
$SO(N)$ & Fundamental & $  \dfrac{1}{2} \int \dd x f''(x) \k_\Lambda(x+m)$\\
\hhline{||~|-||-||}
& Adjoint, matter & $\dfrac{1}{8} \int \dd x\dd y f''(x)f''(y)\k_\Lambda(x + y + m) -  \int \dd x f''(x) \k_\Lambda(x + m/2)$ \\
\hhline{|:=:=::=:|}
& Adjoint, gauge & $-\dfrac{1}{8} \int \dd x\dd y f''(x)f''(y)\k_\Lambda(x + y) - \int \dd x f''(x)\k_\Lambda(x)$\\
\hhline{||~|-||-||}
$Sp(N)$ & Fundamental & $ \dfrac{1}{2} \int \dd x f''(x) \k_\Lambda(x+m)$\\
\hhline{||~|-||-||}
& Antisymmetric & $\dfrac{1}{8} \int \dd x\dd y f''(x)f''(y)\k_\Lambda(x + y + m)  - \int \dd x f''(x) \k_\Lambda(x + m/2)$\\
\hhline{||~|-||-||}
& Adjoint, matter & $\dfrac{1}{8}\int \dd x\dd y f''(x)f''(y) \k_\Lambda(x+y + m)  + \int \dd x f''(x)\k_\Lambda(x+m/2)$\\ 
\hhline{|b:=:=:b:=:b|}
\end{tabular}
\end{center}
\caption{Hamiltonians}\label{Hams}
\end{table}


Let us put the result obtained above to the table. Recall the definitions \Ref{rhoSU}, \Ref{ProfSU}, \Ref{rhoSO}, \Ref{ProfSO}, \Ref{rhoSp} and \Ref{ProfSp}:

\begin{itemize}
\item \underline{$SU(N)$}:
$$
\begin{aligned}
\r(x) &= \eps_1\eps_2\sum_{i=1}^k \d(x-\f_i), \\
f(x) &= -2\r(x) + \sum_{l=1}^N |x-a_l|,
\end{aligned}
$$
\item \underline{$SO(2n + \chi)$}:
$$
\begin{aligned}
\r(x) &= \eps_1\eps_2 \sum_{i=1}^k  \Big(\d(x-\f_i) + \d(x+\f_i)\Big), \\
f(x) &= -2\r(x) + \sum_{l=1}^n \Big(|x-a_l| + |x+a_l|\Big) + \chi|x|,
\end{aligned}
$$
\item \underline{$Sp(N)$}: Let $k = 2n + \chi$, $\chi = 0,1$.
$$
\begin{aligned}
\r(x) &= \eps_1\eps_2 \sum_{i=1}^n \Big(\d(x-\f_i) + \d(x+\f_i)\Big), \\
f(x) &= -2\r(x) + \sum_{l=1}^N \Big(|x-a_l| + |x+a_l|\Big).
\end{aligned}
$$
\end{itemize}
Note that in the case of $SO(N)$ and $Sp(N)$ the density function and the profile function are symmetric.

The Table \ref{Hams} contains formulae \Ref{HamSUAdjGauge}, \Ref{HamSUFund}, \Ref{HamSUAnt}, \Ref{HamSUSym}, \Ref{HamSUAdjMatter}, \Ref{HamSOAdjGauge}, \Ref{HamSOFund}, \Ref{HamSOAdjMatter}, \Ref{HamSpAdjMatter}, \Ref{HamSpFund}, \Ref{HamSpAnt}, \Ref{HamSpAdjMatter}.


\section{Profile function properties}

Let us briefly discuss some properties of the profile function $f(x)$.

First of all we note that since $\r(x)$ has a compact support $f(x)$ behaves like $d |x|$ when $x \to \pm \infty$, where $d$ is the number of connected pieces of the support of $f(x)$. It equals to the dimension of the fundamental representation. 

In general when $|a_l - a_m| \gg \Lambda$ , $l\neq m$, the support of $\r(x)$ is a union of $d$ disjoint intervals. Each of them contains one of $a_l$'s. Let $[\a_l^-,\a_l^+]$ be such an interval: $a_l \in [\a_l^+,\a_l^-]$. Then
\begin{equation}
\label{locIntf''}
\int_{\a_l^-}^{\a_l^+} f''(x) \dd x  =  2\int_{\a_l^-}^{\a_l^+} \Big( \d(x - a_l) - \r''(x) \Big) \dd x =2.
\end{equation}
It follows that 
$$
\int_\Real f''(x)\dd x  = 2d.
$$
\begin{equation}
\label{f''(x)xdx}
\begin{aligned}
\int_{\a_l^-}^{\a_l^+} x f''(x) \dd x &= 2\int_{\a_l^-}^{\a_l^+} x \Big( \d(x - a_l) - \r''(x) \Big) \dd x \\
&= 2a_l - 2 {\left. \Big( x\r'(x) - \r(x) \Big)\right|}^{\a_l^+}_{\a_l^-} = 2a_l.
\end{aligned}
\end{equation}
The sum $\sum_{l=1}^d a_l$ equals zero for all group we consider and therefore we have
$$\int_\Real x f''(x) \dd x = 2 \sum_{l=1}^d a_l = 0.$$
Using the definition of  $\r(x)$ for $SU(N)$ \Ref{rhoSU} we have
\begin{equation}
\label{k=int}
\begin{aligned}
\int_\Real x^2 f''(x) \dd x &= 2\int_\Real x^2\left( \sum_{l=1}^N\d(x-a_l) - \r''(x) \right) \dd x \\
&= 2 \sum_{l=1}^N a_l^2 - 4 \int_\Real \r(x)\dd x = 2\sum_{l=1}^N {a_l}^2 - 4\eps_1\eps_2 k.
\end{aligned}
\end{equation}
It follows that this integral fixes the relation between the instanton number $k$ and $\dfrac{1}{\eps_1\eps_2}$.

The equation \Ref{k=int} can be used to represent the factor $q^k$ in the form similar to \Ref{varEqn}. Indeed, we have
\begin{equation}
\label{qasint}
\begin{aligned}
q^k &= \exp-\frac{1}{\eps_1\eps_2}\left\{ - \pi i \t\sum_{l=1}^N a_l^2 + \frac{\pi i \t}{2}\int_\Real x^2 f''(x)\dd x \right\} \\
&= \Lambda^{k\b}\exp-\frac{1}{\eps_1\eps_2}\left\{ - \pi i \t_0\langle a,a\rangle + \frac{\pi i \t_0}{2}\int_\Real x^2 f''(x)\dd x \right\}. 
\end{aligned} 
\end{equation}
The first term in the curly brackets can be identified with the classical prepotential \Ref{ClassPrep}. The second term in general should be added to the Hamiltonian. However, for the non-conformal theories, as it was already mentioned, $\t_0$ can be neglected, and so this term is irrelevant. It becomes relevant only in the conformal theories.


\section{Lagrange multipliers}

In \Ref{varEqn} the integration is taken only over the functions satisfying the condition \Ref{f''(x)xdx}. This condition is rather complicated to be considered as the definition of the domain of the functional integration.

However we can extend this domain to all the functions after introducing the Lagrange multipliers. The standard way is the following: let $\xi_1,...,\xi_d$ be the multipliers. Then instead of the Hamiltonian $H[f]$ we should minimize the following (Lagrange) functional:
\begin{equation}
\label{Lagrange}
\begin{aligned}
L[f,\xi] &= H[f]  +\sum_{l=1}^d \xi_l \left(  \frac{1}{2}\int_{\a^-_l}^{\a^+_l}x f''(x)\dd x - a_l\right)\\
&= S[f,\xi] - \sum_{l=1}^d \xi_l a_l.
\end{aligned}
\end{equation}
where
\begin{equation}
\label{LegendreF}
S[f,\xi] = H[f]  + \frac{1}{2}\sum_{l=1}^d \xi_l \int_{\a^-_l}^{\a^+_l}x f''(x)\dd x.
\end{equation}
Having found the minimizer $f_\star(x)$ of $L[f,\xi]$ we should also find the stationary point with respect to $\xi_l$. This provide the condition \Ref{f''(x)xdx}. In other words $S[f,\xi]$ should satisfy
\begin{equation}
\label{Legendre}
{\left.\frac{\partial S[f_\star,\xi]}{\partial \xi_l}\right|}_{f_\star = \mr{const}} = a_l.
\end{equation}
where the $\xi$-dependence of  $f_\star(x)$ can be neglected since the derivative of the functional with respect to function is zero at the minimizer. This equation determines $\xi_l$ as some functions of $a_l$. Plugging back these functions into \Ref{Lagrange} we obtain the value of the Hamiltonian at the critical point. That is, the (minus) prepotential. Otherwise the function $S[f_\star,\xi]$ is nothing but the Legendre transform of $-\Prep(a,m)$. 

Note that since $\sum_{l=1}^d a_l = 0$ the sum of $\xi_l$ is not fixed by this procedure.

The last term in \Ref{LegendreF} requires the knowledge of the support of the minimizer $f_\star(x)$ which itself is to be found. Hence the constraints can not be imposed in the form presented above. However another way exists \cite{SWandRP}. Note that $f'(-\infty) = -d$, $f'(+ \infty) = d$ and thanks to \Ref{locIntf''} 
$$f'(\alpha_l^+) - f'(\alpha_l^-) = \int_{\alpha^-_l}^{\alpha^+_l}f''(x)\dd x = 2.$$
Hence we can introduce a piecewise linear function (the {\em surface tension function}) $\sigma(t)$  such that $\sigma'(t) = \xi_l$ when $t = f'(x), x\in [\alpha_l^-,\alpha_l^+]$, that is, $t \in (-d + 2(l-1), -d + 2l )$. With the help of this function we can rewrite the last term in \Ref{LegendreF} as follows
\begin{equation}
\label{Larg}
\frac{1}{2}\sum_{l=1}^d \xi_l \int_{\alpha_l^-}^{\alpha_l^+} x f''(x)\dd x = - \frac{1}{2}\vpint_\Real \sigma(f'(x))\dd x
\end{equation}
provided  $\sigma(d) + \sigma(-d) = 0$. Together with the definition of $\sigma(t)$ it implies $\sum_{l=1}^d \xi_l = 0$ and all the $\xi_l$'s are now defined. 

The discussion presented above implies that in order to determine the prepotential we have proceed the following steps:
\begin{itemize}
\item find the minimizer $f_\star(x)$ of the Lagrange functional:
\begin{equation}
\label{LagrangeS}
S[f,\xi] = H[f] - \frac{1}{2}\vpint_\Real \sigma(f'(x))\dd x,
\end{equation}
where the Hamiltonian $H[f]$ is defined for each model with the help of Table \ref{Hams},
\item in order to obtain the prepotential we need to perform the Legendre transform with respect to $\xi$ of $S[f_\star,\xi]$.
\end{itemize}
As we shall see in the next section the Seiberg-Witten curves appear naturally while performing these steps.

%% file: geometry.tex
\label{geometry}

In this chapter we consider some examples of the saddle point equations and their solution. Let start with an example of $SU(N)$: pure Yang-Mills theory and matter in fundamental representation \cite{SWandRP}.


\section{Example: $SU(N)$, pure Yang-Mills and fundamental matter}
\label{Curve:SU(N)Fund}

Let $N_f$ be the number of flavors. With the help of the Table \ref{Hams} we can write the Hamiltonian of the model:
$$
H[f] = - \frac{1}{4} \int \dd x \dd y f''(x)f''(y)\k_\Lambda(x-y) + \sum_{f=1}^{N_f} \frac{1}{2}\int \dd x f''(x) \k_\Lambda(x+m_f).
$$

In order to minimize the functional \Ref{LagrangeS} we note, that it naturally depends not on $f(x)$, but rather on $f'(x)$. The saddle point (Euler-Lagrange) equations for $f'(x)$ are
\begin{equation}
\label{SPE}
2 \frac{\d S[f,\xi]}{\d f'(x)} =  \int \dd y f''(y) \k'_\Lambda(x-y) - \sum_{f=1}^{N_f}\k'_\Lambda(x+m_f) - \s'(f'(x)) = 0. 
\end{equation}
Using the definition of $\s(t)$ we conclude that $\s'(f'(x)) = \xi_l$ when $x\in [\a_l^-,\a_l^+]$. When $x$ is outside of the support of $f''(x)$, say $x\in(\a_l^+,\a_{l+1}^-)$, we can not determine $\s'(f'(x))$. The only thing we can say is that in this case $\xi_l \leq \s'(f'(x)) \leq \xi_{l+1}$. 

Taking the derivative we obtain:
\begin{equation}
\label{SPE:SU(N)Fund}
\begin{aligned}
\int \dd y f''(y)\ln\left| \frac{x-y}{\Lambda}\right| - \sum_{f=1}^{N_f} \ln \left| \frac{x+m_f}{\Lambda} \right| &= 0, & &x\in[\a_l^-,\a_l^+].
\end{aligned}
\end{equation}

In order to go further we exploit the primitive of the Sokhotski formula:
$$
\ln(x+i0) = \ln|x| - i\pi\Heavi(-x),
$$
where $\Heavi(x)$ is the Heaviside step function: 
$$
\Heavi(x) = \left\{ 
\begin{aligned}
&1, & x &> 0, \\
&0, & x &< 0.
\end{aligned}
\right.
$$
Define the primitive of the resolvent of $f''(x)$:
$$
F(z) = \frac{1}{4\pi i} \int_\Real \dd y f''(y) \ln \left( \frac{z-y}{\Lambda}\right).
$$


\begin{figure}
\includegraphics[width=\textwidth]{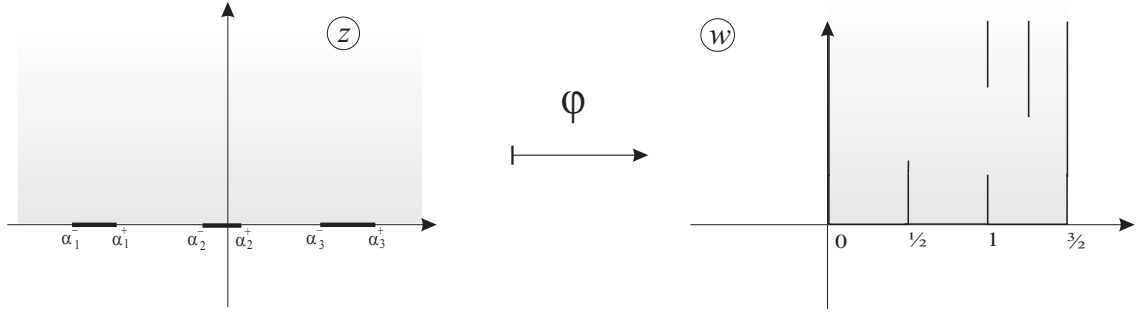}
\caption{Conformal map for $SU(3)$, $N_f=2$}\label{FigSUFund}
\end{figure}


For $F(x)$ we obtain the following equation:
\begin{equation}
\label{defOfF}
F(x) - \sum_{a=1}^{N_f}\frac{1}{4\pi i}\ln \left( \frac{x+m_f}{\Lambda}\right) = \vf(x),
\end{equation}
where the complex map $\vf(x)$ maps the real axis to boundary of the domain on the figure \ref{FigSUFund}. It is holomorphic (since the lefthand side is). It follows that $\vf(z)$ maps the upper half-plane to the domain. Suppose that $|a_l - a_m| \gg \Lambda$ if $l\neq m$ and $m_f \gg a_l$ for all $f$ and $l$. This information is sufficient to reconstruct this map. One gets (up to an additive constant):
\begin{equation}
\label{phiForPureSU}
\vf(z) = \frac{1}{2 \pi} \arccos\frac{P(z)}{2 \Lambda^{\b/2} \sqrt{Q(z)}},
\end{equation}
where according to \Ref{betaDef} $\b = 2N - N_f$ and
$$
\begin{aligned}
Q(x) &= \prod_{f=1}^{N_f} (x + m_f), & P(x) &= \prod_{l=1}^N (x - \a_l).
\end{aligned}
$$
We have introduces parameters $\a_l \in [\a_l^-,\a_l^+]$ which are the classical values of the Higgs vevs.

Define $y(z) = \exp{2 \pi i F(z)}$. Then the solution we have obtained can be written as an equation for $y(z)$:
\begin{equation}
\label{SUfundCurve}
y^2(z) -  P(z)y(z) + \Lambda^\b Q(z) = 0.
\end{equation}
The endpoints of $f''(x)$'s support satisfy the equation
$$
P^2(\a_l^\pm) - \Lambda^\b Q(\a_l^\pm) = 0.
$$

\begin{figure}
\includegraphics[width=\textwidth]{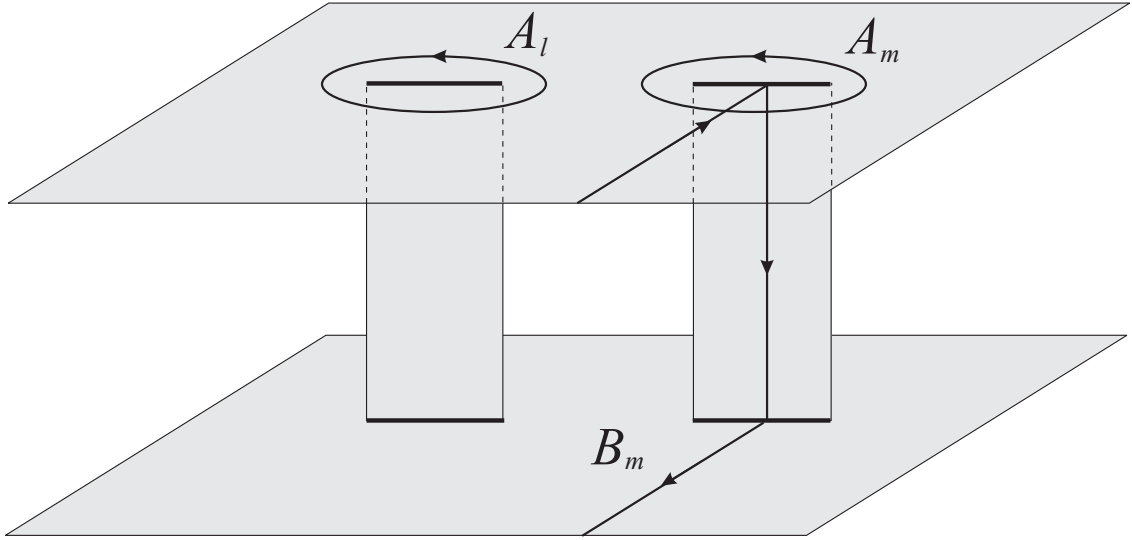}
\caption{Basic cycles}\label{Cycles}
\end{figure}

The Riemann surface of the function $y(z)$ is the two-fold covering of the Riemann sphere. It has cut which connect these two folds along the support of the profile function. Let us define the basic cycles of this Riemann surface (figure \ref{Cycles}). We see that the intersection number satisfies $A_l \# B_m = \d_{l,m}$.


Using some resolvent properties and \Ref{f''(x)xdx} one shows that
$$
\frac{1}{2} \int_{\a_l^-}^{\a_l^+} x f''(x)\dd x = \oint_{A_l} z \dd F(z) = \oint_{A_l} \frac{1}{2\pi i} z \frac{\dd y}{y} = a_l.
$$

Using the saddle point equation \Ref{SPE} we conclude that
$$
\begin{aligned}
\frac{\xi_{l+1}-\xi_l}{2\pi i} &= 2\int_{\a_l^+}^{\a_{l-1}^-} \left( F(z) - \sum_{f=1}^{N_f} \frac{1}{4\pi i} \ln \left(\frac{z+m_f}{\Lambda}\right)\right) \dd z \\
&= - 2 \int_{\a_l^+}^{\a_{l-1}^-} z \left( \dd F(z) - \sum_{f=1}^{N_f}\frac{1}{4\pi i} \frac{\dd z}{z + m_f} \right) = - \oint_{B_{l+1} - B_l} \frac{1}{2\pi i} z \frac{\dd y}{y}.
\end{aligned}
$$

Performing the Legendre transform inverse to \Ref{Legendre} we obtain
$$
\frac{\pd \Prep}{\pd a_l} = 2\pi i \oint_{B_l} \frac{1}{2\pi i} z \frac{\dd y}{y}.
$$

It follows that the prepotential for this theory can be reconstructed with the help of the Seiberg-Witten data: the curve \Ref{SUfundCurve} and the meromorphic differential 
\begin{equation}
\label{prepSU}
\l = \frac{1}{2\pi i} z \frac{\dd y}{y} =  z \dd F(z).
\end{equation}


\section{Fundamental matter for $SO(N)$ and $Sp(N)$}

In this section we extend the previous analysis for the matter in fundamental representation to other classical groups: $SO(N)$ and $Sp(N)$.


\subsection{$SO(N)$ case}
\label{scn:SOFund}
With the help of the Table \ref{Hams} we obtain the Hamiltonian. In order to obtain the saddle point equation we should take the variation with respect to the {\em symmetric} functions. The function $\s(t)$ is also supposed to be symmetric. The equation we get is
$$
\int \dd y f''(y) \k'_\Lambda(x-y) - 4\k_\Lambda'(x) - \sum_{f=1}^{N_f}\left(\k_\Lambda'(x+m_f) + \k_\Lambda'(x-m_f)\right) - 2 \s'(f'(x)) = 0.
$$
We see that this equation coincides with \Ref{SPE} for $2N_f + 4$ fundamental multiplets with masses $(0,0,0,0,m_1,-m_1,\dots,m_f,-m_f)$. It follows that the same should be true for the prepotential \cite{SOandSp}.

\begin{remark}
One could be worried about $N$-odd case, where one of the Higgs field vevs, which is equals to zero, matches with the zero mass coming form from the term $2\k'_\Lambda(x)$. However already from the expression \Ref{zSOAdjGauge} and \Ref{zSOFund} it is seen that they painless annihilate each other \cite{ABCD}.
\end{remark}

We define $F(z)$, $y(z)$ and $\l$ at the same way as in the $SU(N)$ case. We are able to write the Seiberg-Witten curve (as usual we define $N = 2n + \chi$, $\chi = 0,1$; according to \Ref{betaDef} $\b = 2N - 2N_f - 4$):
$$
y^2(z) + z^\chi \prod_{l=1}^n(z^2-\a_l^2) y(z) + \Lambda^{\b} z^4 \prod_{f=1}^{N_f}(z^2 - m_f^2) = 0.
$$


\subsection{$Sp(N)$ case}
\label{scn:SpFund}

In order to solve the saddle point equation for this model it is convenient to introduce another profile function defined as follows:
\begin{equation}
\label{ProfSpAltern}
\tilde{f}(x) = f(x) + 2 |x| = -2\r(x) + \sum_{l=1}^N\Big(|x-a_l| + |x+a_l|\Big) + 2|x|.
\end{equation}
The new profile function is also symmetric. We also should redefine the surface tension function $\s(t)$ as follows:
$$
\tilde{\s}'(t) = \left\{
\begin{aligned}
-&\xi_l, & t&\in (-2l,-2l-2), & l&=1,\dots,N \\
&0 & t&\in(-2,2) & \\
+&\xi_l, & t&\in (+2l,+2l+2), & l&=1,\dots,N \\
\end{aligned}
\right.
$$

\begin{figure}
\includegraphics[width=\textwidth]{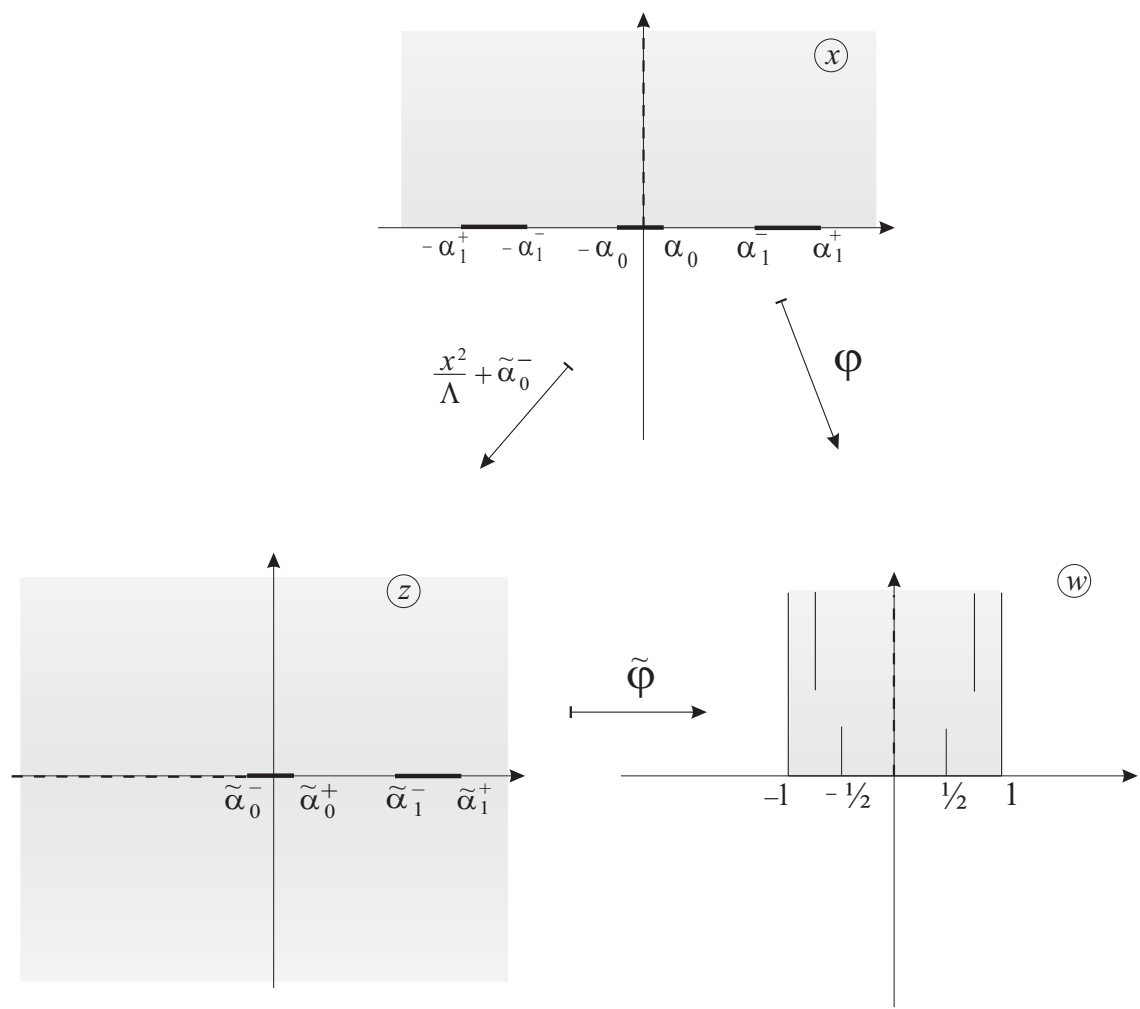}
\caption{Conformal map for $Sp(1)$, $N_f=1$}\label{FigSpFund}
\end{figure}

The Hamiltonian for the gauge multiplet is
$$
\tilde{H}[\tilde{f}] = H[f] = - \frac{1}{8}\int \dd x \dd y \tilde{f}''(x)\tilde{f}''(y) \k_\Lambda(x-y).
$$
And finally the saddle point equation for the model can be written as follows:
$$
\int \dd y \tilde{f}''(y)\k'_\Lambda(x-y) - \sum_{f=1}^{N_f} \left( \k'_\Lambda(x+m_f) + \k'_\Lambda(x-m_f)\right) - 2\tilde{\s}'(\tilde{f}'(x)) = 0.
$$
This equation looks like \Ref{SPE}. However we should remember that the support of $\tilde{f}''(x)$ contains the interval $[\a_o^-,\a_o^+] \ni 0$. Using the definitions \Ref{rhoSp} and \Ref{ProfSp} we get
\begin{equation}
\label{SpInt}
\int_{\a_o^-}^{\a_o^+} \tilde{f}''(x) \dd x = 4.
\end{equation}
It follows that for the primitive of the resolvent of $\tilde{f}(x)$ defined by \Ref{defOfF} we obtain the following equation
$$
F(z) - \sum_{f=1}^{N_f} \frac{1}{4\pi i} \left( \ln\left( \frac{z + m_f}{\Lambda} \right) + \ln \left( \frac{z-m_f}{\Lambda} \right) \right) = \vf(z),
$$
where $\vf(z)$ is a holomorphic function which maps the upper halfplain  to the domain on the figure \ref{FigSpFund}. In order to construct the map we use the reflection principle. Function $\vf(z)$ maps first quadrant to the half of the domain. It follows that together with the square map function $\vf(z)$ maps upper halfplain to the half of our domain. Hence we can use the result for $SU(N)$. The map $\tilde{\vf}(z)$ is given by
$$
\begin{aligned}
\tilde{\vf}(z) &= \frac{1}{2\pi} \arccos \frac{z \prod_{l=1}^N(z-\tilde{\a}_l)}{2\Lambda^{N+1-N_f/2} \sqrt{\prod_{f=1}^{N_f} (z + \tilde{m}_f) }}, & \tilde{m}_f = - \tilde{\a}_o^- - \frac{m_f^2}{\Lambda}.
\end{aligned}
$$


The endpoints of the intervals $[\tilde{\a}_l^-,\tilde{\a}_l^+]$ satisfy the equation:
$$
\tilde{\a}_l^\pm \prod_{l=1}^N (\tilde{\a}_l^\pm - \tilde{\a}) = \pm  2 \Lambda^{N+1-N_f/2} \prod_{f=1}^{N_f}\sqrt{\tilde{\a}_l^\pm + \tilde{m}_f}.
$$  
Using this condition we can rewrite the composition of $\tilde{\vf}(z)$ and $z \mapsto z^2 /\Lambda + \tilde{\a}_o^-$ as follows:
$$
\vf(z) = \frac{1}{2\pi}\arccos\frac{z^2\prod_{l=1}^N(z^2 -\a_l^2) + \Lambda^{\b/2} \prod_{f=1}^{N_f}im_f}{2 \Lambda^{\b/2}\sqrt{\prod_{f=1}^{N_f} (z^2 - m_f^2)}},
$$
where $\b = 4N + 4 - 2N_f$.

It follows that the curve can be written as 
$$
y^2(z) + \left[ z^2\prod_{l=1}^N(z^2 -\a_l^2) + \Lambda^{\b/2}\prod_{f=1}^{N_f}im_f \right]y(z) + \Lambda^{\beta} \prod_{f=1}^{N_f} (z^2 - m_f^2) = 0.
$$


\section{Symmetric and antisymmetric representations of $SU(N)$: equal masses}

Another model for which the curve can be obtained with the help of the  analysis of the saddle point equation is the $SU(N)$ gauge theory with symmetric and antisymmetric representations which have equal masses $m$. The Table \ref{Hams} shows that the same equation describes the $SU(N)$ gauge theory with two antisymmetric representations with the same masses $m$ and four fundamental multiplets with masses $m/2$. 

Taking into account the discussion after \Ref{qasint} we can write the Hamiltonian of the model as follows
$$
\begin{aligned}
H[f] &= -\frac{1}{4}\int \dd x \dd y f''(x)f''(y) \k_\Lambda(x-y) + \frac{1}{4}\int \dd x \dd y f''(x)f''(y) \k_\Lambda(x+y+m) \\
&- \frac{\pi i \t_0}{2}\int \dd x x^2 f''(x) - \frac{1}{2}\int \dd x \s(f'(x)).
\end{aligned}
$$
The saddle point equation is
$$
\int \dd y f''(y) \k'_\Lambda(x-y) - \int \dd y f''(y) \k'_\Lambda(x+y+m) = 2 \pi i \t_0 x + \s'(f'(x)).
$$ 
Taking the derivative we arrive to
\begin{equation}
\label{SU:SYM+ANT}
\begin{aligned}
\int \dd y f''(y) \ln |x-y| - \int \dd y f''(y) \ln |x+y + m| &= 2 \pi i \t_0,&  x &\in[\a_l^-,\a_l^+]. 
\end{aligned}
\end{equation}


\begin{figure}
\includegraphics[width=\textwidth]{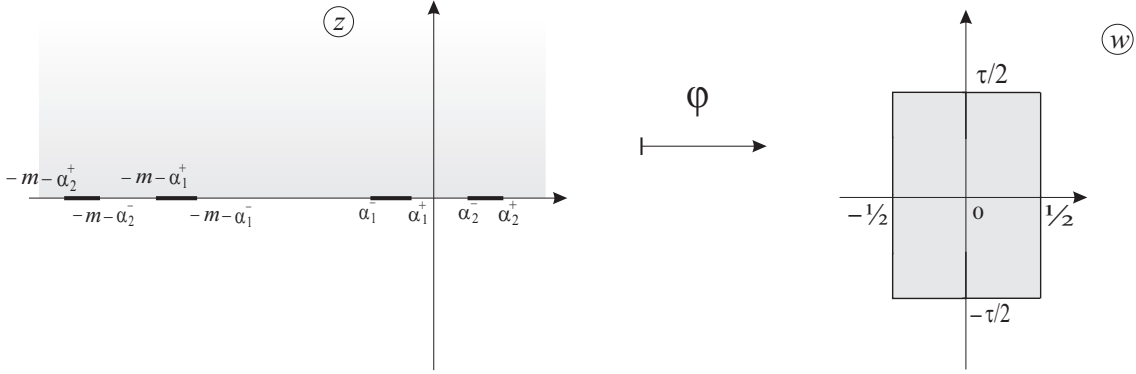}
\caption{Conformal map for $SU(2)$, and matter in symmetric and antisymmetric representation}\label{FigSUAntSym}
\end{figure}


The crucial observation is that the function on the lefthand side is antisymmetric under the reflection with respect to $-m/2$: $ x \mapsto - x - m$. So the righthand side is also antisymmetric. Hence the difference of the logarithms equals to $-2i\pi \t_0$ when $x\in[-\a_l^+ - m,-\a_l^- -m]$. Define
\begin{equation}
\label{defFSUSymAnt}
F(z) = \frac{1}{4\pi i} \int_\Real \dd x f''(x) \ln \left( \frac{z-x}{z+x+m}\right).
\end{equation}
The saddle point equation states that $F(z)$ maps the real axis to the boundary of the boundary of the domain on the figure \ref{FigSUAntSym}. So the upper halfplain is mapped to the whole domain. Such a map is known and given by the formula
\begin{equation}
\label{FforSymAnt}
F(z) = \frac{1}{2\o_1 } \sn^{-1}\frac{\vth_3(0)}{\vth_2(0)}\frac{\P(z)}{\P(-z-m)},
\end{equation}
where we have defined $\P(z) = \prod_{l=1}^N (z-\a_l)$. In this formula $\sn(x)$ is the Jacobi elliptic sinus, $2\o_1$ is its real period. It satisfies \cite{Gradstein}
$$
\begin{aligned}
\sn(\o_1 x) &= \frac{\vth_3(0)}{\vth_2(0)}\frac{\vth_1(x)}{\vth_4(x)}, \\
\vth_1(x+1/2) &= \vth_2(x) = \sum_{n\in\mr{odd}} q^{n^2/2} \e^{ i \pi n x}, \\
\vth_4(x+1/2) &= \vth_3(x) = \sum_{n\in\mr{even}} q^{n^2/2} \e^{i \pi n x}.
\end{aligned}
$$
The endpoints of the support of $f''(x)$ satisfy the equation 
$$
\P(\a_l^{\pm}) = \pm \frac{\vth_2(0)}{\vth_3(0)} \P(-m -\a_l^{\pm}).
$$

Using these formulae we can rewrite the expression for $F(z)$ as follows:
$$
\vth_4(2F) \P(z) - \vth_1(2F) \P(-z-m) = 0.
$$

This expression can be checked in various ways. First let us consider the limit $\t_0\to\infty$. In such a limit we have $\vth_3(0)/\vth_2(0) \sim q^{-1/2}$, $\sn(x) \approx \sin(x)$, and $\o_1 \approx \pi$. The expression \Ref{FforSymAnt} becomes
$$
F(z) \approx \frac{1}{2\pi }\sin^{-1} \frac{\P(z)}{q^{1/2}\P(-z-m)}.
$$ 
If we also take a limit $m\to\infty$ in such a way that $m^{2N} q = 4 \Lambda^{2N}$ stays finite we obtain the expression \Ref{phiForPureSU} for pure $SU(N)$ gauge theory, which is consistent with the fact that in this limit the massive representations decouple.

Another way to check this expression is to consider the hyperelliptic truncation of the curve, given by
$$
y(z) + \frac{1}{y(z)} = \frac{\P(z)}{q^{1/2}\P(-z-m)},
$$
where $y(z) = i \exp{2\pi i F(z)}$. Comparing this expression with \Ref{HyperEllipticTrunc} and referring to the Table \ref{z1rules} we see that the one instanton corrections are correctly described by this curve.


\section{Mapping to $SU(N)$ case}
\label{MappingToSU}

As another application, the saddle point equations help to establish the connection between different models. Some of them have been already found after examination the Seiberg-Witten curve \cite{SOandSp} and the 1-instanton corrections \cite{EllipticMod}. In this section we will examine the saddle point equations. If for two theories they match (after the appropriate identification the parameters of curves) it is natural to expect that the prepotentials will be the same.

As an example consider the $SU(N)$ theory with the symmetric and antisymmetric matter and some fundamental matter. We have the following saddle point equations:

\microsection{Antisymmetric matter.}
\begin{multline}
\label{SPE:SU(N)ANT}
\int \dd y f''(y) \logmodL {x-y} - \frac{1}{2} \int \dd y f''(y)\logmodL{x+y+m^{(a)}} + 2\logmodL{x+m^{(a)}/2} \\
-\sum_{f=1}^{N_f^{(a)}} \logmodL{x+m_f} = 2\pi i \t_0 + \s'(f'(x)), \;\;\;\;\; x\in[\a_l^-,\a_l^+]. 
\end{multline}
\microsection{Symmetric matter}
\begin{multline*}
\int \dd y f''(y) \logmodL {x-y} - \frac{1}{2} \int \dd y f''(y)\logmodL{x+y+m^{(s)}} - 2\logmodL{x+m^{(s)}/2} \\
-\sum_{f=1}^{N_f^{(s)}} \logmodL{x+m_f} = 2\pi i \t_0 + \s'(f'(x)), \;\;\;\;\; x\in[\a_l^-,\a_l^+]. 
\end{multline*}
The analysis of these two expressions leads us to the conclusion that the matter in the symmetric representation with mass $m$ is equivalent to the matter in antisymmetric representation with the same mass together with four fundamental multiplets with masses $m/2$.

In this section we establish such an equivalence between the models containing different groups and matter content. For each model we find its $SU(N)$ partner. We use the following notation: $\vec{a} = (a_1,\dots,a_n)$ for $SO(2n+\chi)$ models and $\vec{a} = (a_1,\dots,a_N)$ for $Sp(N)$ models.
 

\subsection{$SO(N)$, pure gauge} 

The saddle point equation is 
$$
\int \dd y f''(y) \logmodL{x-y} - 4 \logmodL{x} = 2\pi i \t_0.
$$
We conclude that this model is equivalent to the $SU(N)$ model with 4 massless fundamental multiplets. This fact was already used in the section \ref{scn:SOFund}.


\subsection{$SO(N)$, matter in fundamental representation} 

The contribution to the lefthand side of the saddle point equation is
$$
- \logmodL{x-m_f} - \logmodL{x+m_f}.
$$
It follows that each fundamental representation of $SO(N)$ is equivalent to two $SU(N)$ fundamental representations with masses $+m_f$ and $-m_f$.


\subsection{$SO(N)$, matter in adjoint representation} 

The contribution to the lefthand side of the saddle point equation is
\begin{multline*}
 -\frac{1}{2}\int \dd y f''(y) \logmodL{x+y+m} -\frac{1}{2}\int \dd y f''(y) \logmodL{x+y-m} \\
+ 2 \logmodL{x+m/2} + 2 \logmodL{x-m/2}.
\end{multline*}
Comparing with \Ref{SPE:SU(N)ANT} we see that the adjoint representation with mass $m$ in the $SO(N)$ case is equivalent to the two antisymmetric representation of $SU(N)$ with masses $+m$ and $-m$.


\subsection{$Sp(N)$, Pure gauge} 

The saddle point equation in this case is
$$
\int \dd y f''(y) \logmodL{x-y} + 4\logmodL{x}= 2\pi i \t_0.
$$
We conclude that $Sp(N)$ pure gauge theory with 4 massless fundamental matter multiplets is equivalent to the $SU(2N)$ theory with Higgs vevs $(\vec{a},-\vec{a})$.

There is another way to establish a map to the $SU(N)$ case which use the definition \Ref{ProfSpAltern} of the profile function. The  saddle point equation is
$$
\int \dd y \tilde{f}(y) \logmodL{x-y} = 2\pi i \t_0.
$$

We see that the model is equivalent to the $SU(2N+2)$ pure gauge model with the following values of the Higgs vevs: $(0,0,\vec{a},-\vec{a})$.

\begin{remark}
We should stress that in the case of embedding $Sp(N) \subset SU(2N + 2)$ the two of $2N+2$ Higgs vevs matches. This case should be treated carefully as shows the example of the section \ref{scn:SpFund}.
\end{remark}


\subsection{$Sp(N)$, matter in the fundamental representation} 

The contribution to the lefthand side of the saddle point equation is the same as in the $SO(N)$ case:
$$
- \logmodL{x-m_f} - \logmodL{x+m_f}.
$$
So the fundamental representation for $Sp(N)$ is equivalent to the fundamental representation for $SU(2N)$.


\subsection{$Sp(N)$, matter in the antisymmetric representation} 

With the help of the profile function \Ref{ProfSpAltern} the Hamiltonian can be rewritten as follows:
$$
\tilde{H}[\tilde{f}] = H[f] =\frac{1}{8} \int \dd x \dd y \tilde{f}''(x)\tilde{f}''(y)\k_\Lambda(x+y+m) - \int \dd x \tilde{f}''(x) \k_\Lambda(x + m/2).
$$
The lefthand side of the saddle point equation is
\begin{multline*}
- \frac{1}{2} \int \dd y \tilde{f}''(y) \logmodL{x+y+m} - \frac{1}{2} \int \dd y \tilde{f}''(y) \logmodL{x+y-m}  \\
+ 2 \logmodL{x+m/2} + 2\logmodL{x-m/2}.
\end{multline*}
It follows that the $Sp(N)$ model with antisymmetric matter representation with mass $m$ and two massless fundamental multiplets is equivalent to the $SU(2N)$ model with two antisymmetric representations with masses $+m/2$ and $-m/2$ and with moduli $(\vec{a},-\vec{a})$.


\subsection{$Sp(N)$, matter in the adjoint representation} 

Using the profile function \Ref{ProfSpAltern} one can rewrite the Hamiltonian of the model as follows:
$$
\tilde{H}[\tilde{f}] = H[f] =\frac{1}{8} \int \dd x \dd y \tilde{f}''(x)\tilde{f}''(y)\k_\Lambda(x+y+m) + \int \dd x \tilde{f}''(x) \k_\Lambda(x + m/2).
$$
The contribution to the lefthand side of the saddle point equation is
\begin{multline*}
- \frac{1}{2} \int \dd y \tilde{f}''(y) \logmodL{x+y+m} - \frac{1}{2} \int \dd y \tilde{f}''(y) \logmodL{x+y-m}  \\
- 2 \logmodL{x+m/2} - 2\logmodL{x-m/2}.
\end{multline*}
We conclude that the matter in the adjoint representation of $Sp(N)$ is equivalent to two symmetric multiplets with masses $+m$ and $-m$.

In the Table \ref{ToSU} we have collected these results. As usual, for the orthogonal group $SO(N)$ notation $\diamondsuit$ means 0 when $N$ is odd and it is absent when $N$ is even.


\begin{table}
\begin{center}
\begin{tabular}{||c|c||c|c|c||}
\hhline{|t:=:=:t:=:=:=:t|}
\textbf{Group} & \textbf{Multiplet} & \textbf{Higgs} & \textbf{Fund.} & \textbf{Anti.} \\
\hhline{|:=:=::=:=:=:|}
$SU(N)$ & Symmetric, $m$ & $\vec{a}$ &  $m/2$, $m/2$, $m/2$, $m/2$ & $m$ \\
\hhline{|:=:=::=:=:=:|}
& Adjoint, gauge & $(\diamondsuit,\vec{a},-\vec{a})$ &0, 0, 0, 0 & --- \\
\hhline{||~|-||-|-|-||}
$SO(N)$ & Fundamental, $m$ & $(\diamondsuit,\vec{a},-\vec{a})$ & $-m$, $+m$  & --- \\ 
\hhline{||~|-||-|-|-||}
& Adjoint, $m$ & $(\diamondsuit,\vec{a},-\vec{a})$ & --- & $+m$, $-m$ \\
\hhline{|:=:=::=:=:=:|}
& Adjoint, gauge & $(0,0,\vec{a},-\vec{a})$ & --- &  --- \\
\hhline{||~|-||-|-|-||}
& Adjoint, gauge &  &  & \\
$Sp(N)$& + 2 fund., $m=0$ & $(\vec{a},-\vec{a})$ & --- & --- \\
\hhline{||~|-||-|-|-||}
& Fundamental, $m$ & $(\vec{a},-\vec{a})$ & $+m$, $-m$ & --- \\ 
\hhline{||~|-||-|-|-||}
& Antisymmetric, $m$ & $(\vec{a},-\vec{a})$ & --- & $+m$, $-m$ \\
\hhline{||~|-||-|-|-||}
& Adjoint, $m$  &$(\vec{a},-\vec{a})$  & $+m/2$,$+m/2$,$-m/2$,$-m/2$ & $+m$, $-m$  \\
\hhline{|b:=:=:b:=:=:=:b|}
\end{tabular}
\end{center}
\caption{Mapping to $SU(N)$}\label{ToSU}
\end{table}



\section{Hyperelliptic approximations}
\label{HyperElliptic}

In this section we show how to extract the hyperelliptic approximation to the Seiberg-Witten curve from the saddle point equation. This allows us to prove that the 1-instanton correction which will be obtained from the curves matches with our computation presented in section \ref{1instComp} 

In \ref{1instComp} we have shown that our computations match with the algebraic curve computation provided the curve is given by \Ref{HyperEllipticTrunc} and the residue function have been constructed with the help of the Table \ref{Srules}. It follows that the only thing we should show is that when solving the saddle point equation in hyperelliptic approximation we obtain the correct rules for the residue function.

Note that for all (classical) groups and fundamental matter the hyperelliptic approximation is exact. It follows that the task is already accomplished for these models. 

Consider the first non-trivial case, the antisymmetric representation for $SU(N)$ model.


\subsection{$SU(N)$, antisymmetric matter and some fundamentals} 

The saddle point equation for this model is given by \Ref{SPE:SU(N)ANT}. In order to obtain the hyperelliptic approximation to the Seiberg-Witten curve we will simplify the second term. 

To do that we note that the approximation to the profile function which leads to the perturbative prepotential is the following (see \Ref{ProfSU}):
$$
f_{\mr{pert}}(x) = \sum_{l=1}^N |x-a_l|.
$$
The second derivative of this function has a pointwise support. The support of the exact solution is the union of intervals which has length of order $\Lambda \ll m$. Consider the primitive of the resolvent of $f''(x)$:
$$
F(z) = \frac{1}{4\pi i} \int_\Real \dd y f''(y) \ln \left(\frac{z-y}{\Lambda}\right).
$$
The primitive of $f_{\mr{pert}}$-resolvent is
$$
F_{\mr{pert}}(z) = \frac{1}{2\pi i }\sum_{l=1}^N \ln \left( \frac{z-a_l}{\Lambda}\right).
$$
The exact expression for $F(z)$ will be different. However, if $|z-a_l| \gg \Lambda$ for all $l=1,\dots,N$ we can still use this approximation. In particular when we compute integral over the cycles $A_l$ or $B_{l+1} - B_{l}$ we can use for $F(-z-m)$ the perturbative approximation. Coming back to the equation \Ref{SPE:SU(N)ANT} we conclude that in order to obtain 1-instanton correction we can put in the second term $f(x) = f_{\mr{pert}}(x)$. After this identification the equation becomes:
\begin{multline*}
\int \dd y f''(y) \logmodL{x-y} - \sum_{l=1}^N  \logmodL{x+a_l+m} \\
+ 2\logmodL{x+m/2} - \sum_{f=1}^{N_f} \logmodL{x+m_f} = 2\pi i \t_0, \;\;\;\;\; x\in [\a_l^-,\a_l^+].
\end{multline*}
To solve this equation let us define another profile function
$$
\tilde{f}(x) = f(x) + |x+m/2|.
$$
For this function we have the following saddle point equation:
$$
\begin{aligned}
&\int \dd y \tilde{f}''(y) \logmodL{x-y} - \sum_{l=1}^N  \logmodL{x+a_l+m} - \sum_{f=1}^{N_f} \logmodL{x+m_f} = 2\pi i \t_0, & x&\in [\a_l^-,\a_l^+].
\end{aligned}
$$
This equation looks like \Ref{SPE:SU(N)Fund} if we identify $\vec{a}\mapsto (-m/2,\vec{a})$, and $m_f \mapsto (-m-a_1,\dots,-m-a_N,m_1,\dots,m_{N_f})$. Using the result of the section \ref{Curve:SU(N)Fund} we can immediately write the solution \Ref{SUfundCurve}:
\begin{equation}
\label{HyperSU(N)ANT}
y(z) + \frac{1}{y(z)} = \frac{(2z + m)\prod_{l=1}^N (z-\a_l)}{\Lambda^{(N+2-N_f)/2}\sqrt{\prod_{l=1}^N(z+m+a_l)\prod_{f=1}^{N_f}(z+m_f)}}.
\end{equation}
\begin{remark}
Since we have identified the mass of the antisymmetric multiplet with one the Higgs vevs we should, in principle, write its contribution to the nominator as $(2z + \mu)$, where $\mu = m + O(\Lambda^{\b/2})$. However in order to compute the prepotential we will not need to compute any contour integral where contour passes near the point $-m/2$. It follows that the shift $\mu\mapsto m$ will take effect only in the higher instanton corrections which we are not interested in here.
\end{remark}

The equation \Ref{HyperSU(N)ANT} is the same as \Ref{HyperEllipticTrunc} provided we set$$
S(x) = \frac{\prod_{l=1}^N (x+m+a_l)\prod_{f=1}^{N_f}(x+m_f)}{{(2x+m)}^2\prod_{l=1}^{N_f}{(x-\a_l)}^2}.
$$
This expression matches with the value of the residue function which can be build with the help of the Table \ref{Srules}. The last observation proves that the solution of the saddle point equation \Ref{SPE:SU(N)ANT} gives the correct prediction for the 1-instanton correction.

The procedure presented above can be easily converted to the mnemonic rule to build the residue function which appears in \Ref{HyperEllipticTrunc}. It can be formulated as follows: any term of the form 
$$
\ep \logmodL{x-x_0}
$$ 
leads  to the ${(x-x_0)}^{-\ep}$ factor of the $S(x)$.


\subsection{$SU(N)$, matter in the symmetric representation} 

In order to obtain the hyperelliptic approximation for the case of symmetric representation we can either use the same technique as in the case of the antisymmetric multiplet or directly apply the result of the section \ref{MappingToSU}. Anyway the result for the simplified saddle point equation is
$$
\begin{aligned}
&\int \dd y f''(y) \logmodL{x-y} - \sum_{l=1}^N  \logmodL{x+a_l+m} -2 \logmodL{x+m/2}= 2\pi i \t_0, &  x&\in [\a_l^-,\a_l^+].
\end{aligned}
$$
Applying our rule we get the following contribution to the residue function:
$$
{(2z+m)}^2 \prod_{l=1}^N (z+m+a_l)
$$
which is in the agreement with the Table \ref{Srules}.


\subsection{$SU(N)$, matter in the adjoint representation} 

The contribution to the simplified saddle point equation is
$$
-\sum_{l=1}^N \logmodL{x - a_l + m}.
$$
It follows that the contribution to the residue function is
$$
\frac{1}{\prod_{l=1}^N (x-a_l + m)}.
$$
It agrees with the Table \ref{Srules}.


\subsection{$SO(N)$ models} 

In order to establish the same results for the $SO(N)$ models we can apply the result of the section \ref{MappingToSU}. The result for the adjoint gauge multiplet is:
$$
S(x) = \frac{x^{4-2\chi}}{\prod_{l=1}^N{(x-\a_l)}^2}.
$$
For the adjoint matter multiplet we get the following contribution to the residue function:
$$
\frac{{(x^2 - m^2)}^\chi}{4x^2 - m^2}\prod_{l=1}^N ({(x + m)}^2 - a_l^2)({(x - m)}^2 - a_l^2).
$$
These expression are in agreement with the Table \ref{Srules}.


\subsection{$Sp(N)$ models} 

Using the result of the section \ref{MappingToSU} we get the following residue function for the gauge multiplet:
$$
S(x) = \frac{1}{x^4\prod_{l=1}^N {(x^2 - \a_l^2)}^2}.
$$
The contribution which comes from the antisymmetric representation is defined by the following contribution to the saddle point equation
\begin{multline*}
-\frac{1}{2}\int \dd y f''(y) \logmodL{x+y+m}-\frac{1}{2}\int \dd y f''(y) \logmodL{x+y-m} \\
+ 2\logmodL{x+m} + 2 \logmodL{x-m} + 2\logmodL{x+m/2} + 2 \logmodL{x-m/2}.
\end{multline*}
Plugging into this expression the perturbative approximation of the profile function
$$
f_{\mr{pert}}(x) = \sum_{l=1}^N (|x-a_l| + |x+a_l|) + 2|x|
$$
we obtain the following contribution to the residue function:
$$
\frac{\prod_{l=1}^N {({(x+m)}^2 - a_l^2)({(x-m)}^2 - a_l^2)}^2}{(4x^2 - m^2)}.
$$
The contribution to the residue function which comes from the adjoint representation is
$$
(4x^2 - m^2)\prod_{l=1}^N {({(x+m)}^2 - a_l^2)({(x-m)}^2 - a_l^2)}^2.
$$
Obtained expressions is in agreement with the Table \ref{Srules}.

%% file: questions.tex
In this paper have derived the method which allows us to compute the low-energy effective action for $\N = 2$ supersymmetric Yang-Mills theories. We have considered almost all models allowed by the asymptotic freedom. Using the results of \cite{EnnesSU2AntFund,EnnesMasterFunc,EllipticMod,MTheoryTested} we have shown in section \ref{1instComp} that the equivariant deformation method provides the results which in the 1-instanton level agree with the previous computations. 

Also we have written the saddle point equation for each models and we have shown that in all cases when we can it solve obtained expressions for the Seiberg-Witten data agree with known results. 

We have shown that the saddle point equation technique is self-consistent: in spite of the fact that the curves and the differentials are obtained under rather strong condition $k\to\infty$, the final answer is nevertheless correct even if $k$ is low.

In section \ref{HyperElliptic} we obtained the hyperelliptic approximation to the Seiberg-Witten curve. Presumably, one can develop the method presented there and obtain the mathematically rigorous recursion scheme which will give all the instanton corrections. It would be interesting to establish its relation with other recursion schemes (such as, for example, \cite{Liouville}).   

Another direction would be the generalization of the moduli space singularities counting. Close relation between these singularities and Young tableaux allows us to compute the integral \Ref{partition} (see \cite{SWfromInst}). It would be interesting to generalize this approach to other models.

An important question which remains unsolved in the present paper is about the exact solutions for saddle point equations for all considered models. It seems, that the method of conformal map, used in \cite{SWandRP,ABCD,SPinSW} is limited by the models, considered in this paper. However, another methods exist. Their implementation whould be an interesting branch of investigations.

Another question which appears is the following. We have started from the exact microscopic action. Therefore the effective action we obtain should also be exact. We know how to extract from the integrals over the instanton moduli space the leading part of this action, the part, which can be reconstructited with the help of the prepotential. The question if it is possible and how to extract the subleading terms remains open.

And finally let us mention another direction for generalization. All the presented results are based on the ADHM construction which is known only for the classical groups. A way to perform the computations which does not use it would provide the effective action for all groups. It would be interesting to find it.

%% file: spinors.tex
\label{AppendixSpinors}

In this section we give a brief review of some properties of 4-dimensional spinors, related formulae and common notations. In order to build $\N=2$ super Yang-Mills theory we will need consider Minkowski space, that is, $\Real^{1,3}$. The choice of the metric is the following:
\begin{equation}
\label{MinkowskiMetric}
g_{\mu\nu} = \left(
\begin{array}{cccc}
+1 & 0 & 0 & 0 \\
0 & -1 & 0 & 0 \\
0 & 0 & -1 & 0 \\
0 & 0 & 0 & -1
\end{array}
\right).
\end{equation}
However to be capable to treat its instanton expansion we should preform a Wick rotation and therefore get in Euclidean space, $\Real^4$, where the metric equals to the Kronecker delta: $g_{\mu\nu} = \d_{\mu\nu}$. The spinor properties are different for these two spaces. We will mostly consider the euclidean spinors.


\section{Spinors in various dimensions}


\subsection{Clifford algebras}

We will start with some generalities about the Clifford algebras.

Let us consider the $d$ dimensional space with the diagonal metric with signature $(p,q)$, $p+q=d$. Otherwise 
$$
g_{\mu\nu} = \diag\{ \underbrace{+1,+1,\dots,+1}_{\mbox{$p$ times}},\underbrace{-1,-1,\dots,-1}_{\mbox{$q$ times}}\}.
$$

The Clifford algebra is generated by symbols ${\{\g^\mu_{p+q}\}}_{\mu=1}^d$ satisfying the following anticommutation relations:
\begin{equation}
\label{CliffordDef}
\{ \g^\mu,\g^\nu \} = g^{\mu\nu} \Id.
\end{equation}
Thanks to this relation an arbitrary element of the Clifford algebra can be written as follows
$$
\o = \o^{(0)} + \o^{(1)}_\mu \g^\mu + \frac{1}{2}\o^{(2)}_{\mu_1\mu_2}\g^{\mu_1\mu_2} + \dots + \frac{1}{k!}\o^{(k)}_{\mu_1\dots\mu_k}\g^{\mu_1\dots\mu_k}+ \dots,
$$
where $\g^{\mu_1\dots\mu_k}$ is the antisymmetric part of the product $\g^{\mu_1}\dots\g^{\mu_k}$. The dimension of the Clifford algebra is therefore $1 + C_d^2 + \dots = 2^d$.

Let us introduce the \emph{orientation operator}
$$
\gG_d = \frac{1}{d!} \ep_{\mu_1\dots\mu_d} \g^{\mu_1}\dots\g^{\mu_d}.
$$
Where are two possible signs in this definition, the choice corresponds to the sighn of the Levi-Civita tensor, which in its turn is defined by the choice of the orientation. The square of the orientation operator can be easily computed with the help of the defining relations \Ref{CliffordDef}. Indeed, since
$$
\gG_d^2 = \g^1\g^2\dots\g^{d-1}\g^d\g^1\g^2\dots\g^{d-1}\g^d = {(-1)}^{d-1} {(\g^d)}^2 \g^1\g^2\dots\g^{d-1}\g^1\g^2\dots\g^{d-1}
$$
It is easy to guess the following recurrent relation: $\gG_d^2 = {(-1)}^{d-1} {(\g^d)}^2 \gG_{d-1}^2$. It follows that 
$$
\gG_d^2 = {(-1)}^{(d-1)+(d-2)\dots+1}{(\g^1)}^2\dots{(\g^d)}^2 = {(-1)}^{\frac{d(d-1)}{2}+q}\Id.
$$

In the following we will be interested in the representations of the Clifford algebras. It is clear that if matrices $\g^\mu$ satisfy the basic relation \Ref{CliffordDef} then the conjugated matrices $\tld{\g}^\mu = U\g^\mu U^{-1}$ also satisfy these relations.

We will say that a matrix representation of the Clifford algebra is generated by the \emph{Dirac matrices}. The space on which these matrices act is the space of the Dirac spinors $\V_D$.

\microsection{Example.} Consider a trivial example $d=1$ and the signature of the matric is $(0,1)$. The Clifford algebra has only one generator satisfying $\g^2_1 = -1$. Therefore $\g_1=i$ and in this case the Clifford algebra is ismorphe to $\Compl$. Note that the representation of one dimensional Clifford algebra is also one dimensional.

\begin{remark}
It is clear that multiplying the generators by $i$ we can chnage the signature of the metric.
\end{remark}


\subsection{Recurrent relations}

Now suppose that we have constructed a representation of the Clifford algebra for the $d$ dimensional case. How to construct the representation for the dimension $d+1$? There are two possibilities.

If $d$ is even then it is easy to see that $\{ \gG,\g^\mu \} = 0$ $\forall \mu$. It follows that if we define 
$$
\begin{aligned}
\g^\mu_{d+1} &= \g^\mu_d, & \g^{d+1}_{d+1} &= \gG_d
\end{aligned}
$$ 
when it will be representation of the $d+1$ dimensional Clifford algebra. Note that in this case the dimension of the representation does not grow up.

If $d$ is odd then we can consider the following set of generators:
\begin{equation}
\label{CliffordMinus}
\begin{aligned}
\g^\mu_{d+1} &= \left(
\begin{array}{cc}
0 & -\g_d^\mu \\
\g_d^\mu & 0
\end{array}
\right) , & \g^{d+1}_{d+1} &=\left(
\begin{array}{cc}
0 & \Id \\
\Id & 0
\end{array}
\right).
\end{aligned}
\end{equation}
There are another realization of this construction. Indeed, we can define the following generators:
\begin{equation}
\label{CliffordPlus}
\begin{aligned}
\g^\mu_{d+1} &= \left(
\begin{array}{cc}
0 & \g_d^\mu \\
\g_d^\mu & 0
\end{array}
\right) , & \g^{d+1}_{d+1} &= \left(
\begin{array}{cc}
0 & \Id \\
-\Id & 0
\end{array}
\right).
\end{aligned}
\end{equation}
Both of this construction lead to the representation of $d+1$ dimensional Clifford algebra. Note that in this case the dimension of the metrices is doubled. These two constructions are conjugated by the matrix
$$
U = \left(
\begin{array}{cc}
0 & -q \Id \\
q^\ast\Id & 0
\end{array}
\right)
$$ 
where $\ds q = \e^{i\frac{\pi}{4}} = \frac{1+i}{\sqrt{2}}$.

It follows from this recurrent procedure that the dimension of the matrices of representation we have constructed is $2^{\left[\frac{d}{2}\right]}$. Therefore these matrices have $2^d$ components if $d$ is even and $2^{d-1}$ components if $d$ is odd. It suggests that in the even dimensional case we have a faithful representation, whereas in the odd dimensional case this is not true. The way out is to note that in the odd dimensional case the complex conjugated representation is not equiavalent to the initial one. These two representations together have $2\times 2^{d-1} = 2^d$ independent components, which is the dimension of the Clifford algebra.

Now let us apply the recurrent procedure to get some representation of the Clifford algebra for some $d$. The one dimensional representation is already considerred. Therefore we start with the case $d=2$.

\microsection{$d=2$.} Applying directly the prescription \Ref{CliffordMinus} we get
$$
\begin{aligned}
\g_2^1 &= \left(
\begin{array}{cc}
0 & 1 \\
1 & 0
\end{array}
\right), & \g_2^2 &= \left(
\begin{array}{cc}
0 & -\g_1 \\
\g_1 & 0 
\end{array}
\right) =  \left(
\begin{array}{cc}
0 & -i \\
i & 0 
\end{array}
\right).
\end{aligned}
$$
This choice corresponds to the signature $(2,0)$. If we wish to consider the Minkowski space (with sinature $(1,1)$) we ahould multiply the second matrix by a factor $i$. Therefore we get
$$
\begin{aligned}
\g_2^0 &= \left(
\begin{array}{cc}
0 & 1 \\
1 & 0
\end{array}
\right), & \g_2^1 &= \left(
\begin{array}{cc}
0 & 1 \\
-1 & 0 
\end{array}
\right).
\end{aligned}
$$

\microsection{$d=3$.} The orientation operator for the $d=2$, signature $(2,0)$ case is 
$$
\gG_2 = \g_2^1\g_2^2 = \left(
\begin{array}{cc}
i & 0 \\
0 & -i 
\end{array}
\right).
$$
Therefore the generators of the three dimensional Clifford algebra with the eucledian signature $(3,0)$ can be choosen as follows:
$$
\begin{aligned}
\g_3^1 &= \left(
\begin{array}{cc}
0 & 1 \\
1 & 0
\end{array}
\right), & \g_3^2 &= \left(
\begin{array}{cc}
0 & -i \\
i & 0 
\end{array}
\right), & \g_3^3 &= \left(
\begin{array}{cc}
1 & 0 \\
0 & -1 
\end{array}
\right).
\end{aligned}
$$
Note that this is nothing but the Pauli matrices \Ref{Pauli}.

\microsection{$d=4$.} Applying once again the prescription \Ref{CliffordMinus} we get for the Minkowkian signature $(1,3)$ the following representation:
$$
\begin{aligned}
\g_4^0 &= \left( 
\begin{array}{cc}
0 & \Id_2 \\
\Id_2 & 0 
\end{array}
\right), & \g_4^i &= \left(
\begin{array}{cc}
0 & -\t_i \\
\t_i & 0
\end{array}
\right) & &\Leftrightarrow & \g_4^\mu &= \left(
\begin{array}{cc}
0 & \s^\mu \\
\bar{\s}^\mu & 0 
\end{array}
\right),
\end{aligned}
$$
where $\s$-matrices are defined in \Ref{SigmaMinkowski}.

\microsection{$d=5$.} The orientation operator for the previous case is
$$
\gG_4 = \g_4^0\g_4^1\g_4^2\g_4^3 = \left(
\begin{array}{cc}
-i\Id_2 & 0 \\
0 & i\Id_2
\end{array}
\right).
$$
Since $\gG^2 = -\Id_4$ we conclude that the following set of generators
$$
\begin{aligned}
\g_5^\mu &= \g_4^\mu, & \g_5^4 &= \gG_4
\end{aligned}
$$
provides the representation of five dimensional Clifford algebra with the signature $(1,4)$.

\microsection{$d=6$.} Let us finally consider the six dimensional case. In order to get the representation of the Clifford algebra for the signature $(1,5)$ we use the prescription \Ref{CliffordPlus}. The result is the following:
$$
\begin{aligned}
\g_6^\mu &= \left(
\begin{array}{cc}
0 & \g_4^\mu \\
\g_4^\mu & 0 
\end{array}
\right), & \g_6^4 &= \left(
\begin{array}{cc}
0 & \gG_4 \\
\gG_4 & 0 
\end{array}
\right), & \g_6^5 &= \left(
\begin{array}{cc}
0 & \Id_4 \\
-\Id_4 & 0
\end{array}
\right).
\end{aligned}
$$

The schemas \Ref{CliffordPlus} and \Ref{CliffordMinus} show that in the Euledian signature the matrices $\g^\mu$, where $\mu$ is even odd can be choosen to be real whereas $\g^\nu$ for $\nu$ even can be choosen pure imaginary (recall that in the Eucledian case $\mu=1,2,\dots,d$ whereas in the Minkowskian $\mu = 0,1,\dots,d-1$). Moreover all the Dirac matrices are hermitian. Since multiplying them by $i$ we can change the signature we conclude that the following relation holds
$$
{(\g^\mu)}^\dag = \g_\mu = g_{\mu\nu}\g^\nu.
$$ 
Note that these two properties are stable under the conjugation by unitary matrices.


\subsection{Weyl and Majorana spinors}

Let us discuss the existence of the Weyl spinors. It is easy to see that if the dimension is even then the orientation operator $\gG_d$ (multiplied by $\pm i$, if necessary) has the following form:
$$
\gG_d = \left(
\begin{array}{cc}
\Id & 0 \\
0 & -\Id
\end{array}
\right).
$$
It allows us to define the projection operators to the space of the left and right handed spinors: $\ds \P^{\pm} = \frac{\Id \pm \gG_d}{2}$. Therefore in even dimensions the Weyl spinors exist. If $d$ is odd the orientation operator is proportional to the unit matrix and does not allow to define chirality.

Before discussing the Majorana spinors let us define the Dirac conjugation. It is easy to see that  the following matrices
$$
S_{\mu\nu} = \frac{1}{4i}\left[ \g_\mu,\g_\nu\right] = \frac{1}{4i}\left(\g_\mu\g_\nu - \g_\nu \g_\mu \right)
$$
satisfy the commutation relations for the Lorentz group $O(p,q)$:
$$
\left[ S_{\mu\nu}, S_{\r\s} \right] = i g_{\mu\r} S_{\nu\s} + i g_{\nu\s}S_{\mu\r} - i g_{\mu\s}S_{\nu\r} - i g_{\nu\r}S_{\mu\s}.
$$ 
The infinitesemal Lorentz rotation is defined on spinors as follows:
$$
\V_D\ni \psi \mapsto \left( \Id + \frac{i}{2}S_{\mu\nu}\o^{\mu\nu}\right) \psi 
$$
where $\o^{\mu\nu}$ are parameters of this rotation. The Dirac conjugation is defined as $\bar{\psi} = \psi^\dag \A$ where $\A$ is a unitary matrix (to assure $\bar{\bar{\psi}} = \psi$) choosen in such a way that $\bar{\psi}\psi$ is a scalar. This condition can be recast as follows:
$$
\A^{-1}\left[ \g_\mu, \g_\nu \right] \A = \left[ \g_\mu^\dag,\g_\nu^\dag \right]
$$
and indeed will be satisfied if $\A^{-1}\g_\mu \A = \pm \g_\mu^\dag = \pm \g^\mu$. Therefore we can take either $\A_+ = \g^1\g^2\dots\g^p$ or $\A_- = \g^{p+1}\g^{p+2}\dots\g^{p+q}$. In the Eucledian case we can take $\A = \A_- = \Id$, and in the Mikowskian case $\A = \A_+ = \g^0$. Note that since ${\g^\mu}^\dag = \g^\mu$ for $\mu = 1,\dots,p$ and ${\g^\mu}^\dag = - \g^\mu$ for $\mu = p+1, \dots, p+q$ we get $\A_+^\tr = \A_+$ and $\A_-^\tr = {(-1)}^q A_-$. 

Consider the Dirac equation for the massive spinors in the external gauge field $A_\mu$ which is supposed to be real:
$$
\left[i\cD_\mu \g^\mu - m \right]\psi = \left[\left(i\pd_\mu + A_\mu \right)\g^\mu - m\right]\psi = 0.
$$
The complex conjugated equation is
$$
\left[\left( - i \pd_\mu + A_\mu \right) {\g^\mu}^\ast - m \right]\psi^\ast = 0.
$$
If we wish the Dirac conjugated spinor $\bar{\psi}^\tr = \A^\tr \psi^\ast$ to satisfy the same equation as the initial one, but have the opposite charges ($A_\mu \mapsto -A_\mu$) we should identify $\psi$ and $\psi^\C = \C^- \bar{\psi}^\tr = \C^- \A^\tr\psi^\ast$ where $\C^-$ satisty the following condition:
\begin{equation}
\label{CMinus}
{(\C^-)}^{-1}\g^\mu \C^- = - \A^\tr{\g^\mu}^\ast\A^{-\tr}.
\end{equation}
This matrix together with the complex conjugation $\alg{c}$ defines an antilinear operation $\mbf{\C}^- = \C^-\A^\tr\alg{c}$. In the even dimensional case the projection operator $\gG$ anticommutes with the generators $\g^\mu$. It follows that the Clifford algebra is simple and threfore the representation generated by matrices $\g^\mu$ and the complex conjugated matrices should be equivalent.

Therefore the square of the antilinear operator is proportional to $\Id$ thanks to Schur lemma. Since $\mbf{\C}^- {(\mbf{\C}^-)}^2 = {(\mbf{\C}^-)}^2 \mbf{\C}^-$ we conclude that ${(\mbf{\C}^-)}^2 = \C^- \A^\tr{(\C^-\A^\tr)}^\ast = \a \Id$ where $\a \in \Real$. If we rescale $\mbf{\C}^-$ by a number $\l \in \Compl$ we get $\a \mapsto \a {|\l|}^2$. Therefore we can put either $\a = -1$ or $\a = +1$. In the last case it is possible to define the projection operators $\ds\P^\pm_{\C^-} = \frac{\Id \pm \mbf{\C}^-}{2}$ splitting the space of spinors into the selfconjugated and anti-selfconjugated. The space of self conjugated spinors, if exists, is called the space of Majorana spinors.

If the spinors are massless there is another option: we may find a matrix $\C^+$ satisfying 
\begin{equation}
\label{CPlus}
{(\C^+)}^{-1} \g^\mu \C^+ = + \A^\tr{\g^\mu}^\ast \A^{-\tr}.
\end{equation}
Combining this matrix with the copmplex conjugation we get the following antilinear operator $\mbf{\C}^+ = \C^+ \A^\tr\alg{c}$. When  ${(\mbf{\C}^+)}^2 = +\Id$ we can construct the projection operators $\ds\P_{\C^+}^\pm = \frac{\Id \pm \mbf{\C}^+}{2}$ which splits the space of Dirac spinors into two subspaces of Majorana spinors. 


Note that if ${(\mbf{\C}^\pm)}^2 = \C^\pm \A^\tr{(\C^\pm\A^\tr)}^\ast = - \Id$ there is still a way out to define a version of Majorana spinors, the symplectic Majorana spinors. To this extent we should enlarge the space of Dirac spinors $\V_D \mapsto \V_D \otimes \W$ where $\W$ is a vector space equipped by a symplectic form $\gO$. When we can define a projection operotor as follows
$$
\P^\pm_S = \frac{\Id_{\V_D}\otimes \Id_{\W} \pm \mbf{\C}\otimes \gO}{2}.
$$ 
Pragmatically it means that instead of one Dirac (or Weyl) spinor $\psi$ we consider a set of such a spinors: $\psi_1,\psi_2,\dots$. And the symplectic Majorana spinors are those which satisfy the following condition:
$$
\psi_i = \gO_{ij} \psi^\C_j = \pm \gO_{ij} \C \A^\tr \psi^\ast_j.
$$

Consider some examples of the matrices $\C_d^\pm$ in the Mikowski spaces (spaces with the signature $(1,d-1)$). To this extent we note that according to schemas \Ref{CliffordPlus} and \Ref{CliffordMinus} in this signature the matrices $\g^0, \g^1, \g^3, \dots$ are real whereas $\g^2,\g^4,\dots$ are imaginary. Recall that $\A = \g^0$. Therefore the conditions \Ref{CPlus} and \Ref{CMinus} are satisfied by the following matrices
$$
\begin{aligned}
\C_d^- &= \g^1_d\g^3_d\dots\g_d^{d-1} & &\mbox{and} & \C_d^+ &= \g^0_d\g^2_d\dots\g_d^d & &\mbox{if} & d &\equiv 2 \pmod 4,\\
\C_d^- &= \g^0_d\g^2_d\dots\g_d^d & &\mbox{and} & \C_d^+ &= \g^0_d\g^2_d\dots\g_d^d & &\mbox{if} & d &\equiv 0 \pmod 4.
\end{aligned}
$$ 
It is easy to write corresponding matrices for all other types of signature.

Consider some examples of these matrices.

\microsection{$d=2$.} Applying the rules we get
$$
\begin{aligned}
\C_2^- &= \g_2^1 = \left(
\begin{array}{cc}
0 & 1 \\
-1 & 0
\end{array}
\right), & \C_2^+ &= \g_2^0 =  \left(
\begin{array}{cc}
0 & 1 \\
1 & 0
\end{array}
\right).
\end{aligned}
$$
We see that ${(\mbf{\C}^\pm_2)}^2 = + \Id_2$. Therefore in $1+1$ there are spinors of Majorana of two types.

\microsection{$d=4$.} We get
$$
\begin{aligned}
\C_4^- &= \g_4^0\g_4^2 = \left(
\begin{array}{cc}
i\ep & 0 \\
0 & -i\ep
\end{array}
\right), & \C_4^+ &= \g_4^1\g_4^3 = \left(
\begin{array}{cc}
-\ep & 0 \\
0 & -\ep
\end{array}
\right),
\end{aligned}
$$
where the $2\times 2$ matrix $\ep$ is defined in \Ref{SpinorMetric}. In this case ${(\mbf{\C}_4^-)}^2 = +\Id_4$ whereas ${(\mbf{\C}_4^+)}^2 = - \Id_4$. Therefore in the four dimensional case there are only spinors of Majorana of type ``$-$'' (which respect the mass) can exist.

\microsection{$d=6$.} We obtain
$$
\begin{aligned}
\C_6^- &= \g_6^1\g_6^3\g_6^5 = \left(
\begin{array}{cc}
0 & \C_4^+ \\
-\C_4^+ & 0 
\end{array}
\right), & \C_6^+ &= \g_6^0\g_6^2\g_6^4 = \left(
\begin{array}{cc}
0 & -\C_4^+ \\
-\C_4^+ & 0 
\end{array}
\right)
\end{aligned}
$$
In both cases ${(\mbf{\C}^\pm_6)}^2 = -\Id_8$. Therefore the Majorana spinors can not exist in $1+5$ dimensions. However it is possible to define a symplectic Majoarana spinors. 


\section{Pauli matrices}

Define the \emph{Pauli matrices} in the standard way:
\begin{equation}
\begin{aligned}
\label{Pauli}
\t_1 &= \left(
\begin{array}{cc}
0 & 1 \\
1 & 0
\end{array}
\right), &
\t_2 &= \left(
\begin{array}{cc}
0 & -i \\
i & 0\end{array}
\right), &
\tau_3 &= \left(
\begin{array}{cc}
1 & 0 \\
0 & -1 
\end{array}
\right).
\end{aligned}
\end{equation}

They have naturally one upper and one lower spinor index: $\tau_{i,\alpha}{}^\beta$. This convention makes possible to multiply them. We have
\begin{equation}
\label{PauliMultiplication}
\begin{aligned}
\t_{i,\a}{}^\b\t_{j,\b}{}^\g &= \d_{ij}\d^\g_\a + i \ep_{ijk} \t_{k,\a}{}^\g &
\Big(\t_i \t_j &= \d_{ij}\Id_2 + i\ep_{ijk}\t_k\Big).
\end{aligned}
\end{equation}
Together with the unit matrix they form a basis of all $2\times 2$ complex matrices. This fact can be expressed as the \emph{Fierz identity}:
\begin{equation}
\label{FierzIdentity}
\begin{aligned}
\d_\a^\b \d_\g^\d + \t_{i,\a}{}^\b\t_{i,\g}{}^\d &= 2 \d^\b_\g \d^\d_\a &
\Big(\Id_2 \otimes \Id_2 + \t_i \otimes \t_i = 2 \Id_{2\times 2} \Big),
\end{aligned}
\end{equation}
where $\Id_{2\times 2}$ is the unit matrix in the vector space of all $2\times 2$ matrices. The Fierz identity is nothing but the completeness condition for the Pauli matrices.

The Pauli matrices are all hermitian:
$${\t_i}^\dag = \t_i.$$

Consider the matrix:
\begin{equation}
\label{SpinorMetric}
\begin{aligned}
\ep_{\a\b} &= \left(
\begin{array}{cc}
0 & -1 \\
1 & 0
\end{array}
\right), &
\ep^{\a\b} &= {\left(\ep^{-1}\right)}^{\a\b} = \left(
\begin{array}{cc}
0 & 1 \\
-1 & 0
\end{array}
\right).
\end{aligned}
\end{equation}
One can check that the Pauli matrices satisfy the equation:
\begin{equation}
\label{PauliReality}
\begin{aligned}
\t_i^\ast{}^\a{}_\b &= - \ep^{\a\g}\t_{i,\g}{}^\d\ep_{\d\b} & \Big( \t_i^\ast &= - \ep^{-1} \t_i \ep\Big).
\end{aligned}
\end{equation}

The meaning of this relation can be discovered as follows. Consider a matrix $X_\a{}^\b$. It can be developed as $X_\a{}^\b = X_0 \d^\b_\a + iX_k  \t_{k,\a}{}^\b$. The reality of $X_0$ and $X_i$ is equivalent to 
$$X^\ast = \ep^{-1} X \ep.$$
This equation is called the \emph{reality condition}.

For any $U \in SU(2)$ we have $U = \e^{i\phi^i\tau_i}$ where $\phi^i$ are real. Thus \Ref{PauliReality} yields
$$
\begin{aligned}
{\left(U^\ast\right)}^\a{}_\b &= \ep^{\a\g} U_\g{}^\d \ep_{\d\g} & \Big(U^\ast &= \ep^{-1} U \ep\Big).
\end{aligned}
$$
It follows that $\eps$ is stable under the $SU(2)$ transformations. Indeed
$$
\ep'_{\a\b} = U_\a{}^\g U_\b{}^\d \ep_{\g\d}  = \ep_{\a\b}.
$$
Hence the ``metric'' $\ep$ can be used to rise and lower the spinor indices.


\section{'t Hooft symbols}

In this section we consider selfdual and anti-selfdual forms in the Euclidean space. According to this we can make no difference between upper and lower spatial indices. The standard reference is \cite{tHooftSymbols}, however some details can be found in \cite{LandavshitzII}.

Any antisymmetric tensor in four dimensions $F_{\mu\nu}$ can be represented by means of two three dimensional vectors $a_i$ and $b_i$:
$$
F_{\mu\nu} = \left(
\begin{array}{cccc}
0 & a_1 & a_2 & a_3 \\
-a_1 & 0 & b_3 & -b_2 \\
-a_2 & -b_3 & 0 & b_1 \\
-a_3 & b_2 & -b_1 & 0 
\end{array}
\right) \equiv {(a,b)}_{\mu\nu}.
$$ 
\begin{remark}
The triples $a_i$ and $b_i$ will transform as vectors with respect to the subgroup $SO(3)$ of $SO(4)$ which preserves the vector $(1,0,0,0)$. However if we extend this subgroup to $O(3)$ by including the reflections $x^i \mapsto -x^i$ we find that $a_i$ is a vector whereas $b_i$ is a pseudo (or axial) vector.
\end{remark}

Using the Hodge star one can define for this tensor the \emph{dual} tensor as follows:
$$
\star F_{\mu\nu} \equiv {(\star F)}_{\mu\nu} = \frac{1}{2}\ep_{\mu\nu\r\s}F_{\r\s},
$$
where $\ep_{\mu\nu\r\s}$ is four dimensional Levi-Civita tensor defined as $\ep_{0123} = +1$. Calculation shows that the following identity holds:
$$\star{(a,b)}_{\mu\nu} = {(b,a)}_{\mu\nu}.$$
We see, that all tensors of form ${(a,a)}_{\mu\nu}$ satisfy the \emph{selfdual equation}:
$$
\star {(a,a)}_{\mu\nu} = {(a,a)}_{\mu\nu},
$$
and all the tensors which can be written as ${(-a,a)}_{\mu\nu}$ satisfy the \emph{anti-selfdual equation}:
$$
\star {(-a,a)}_{\mu\nu} = {(a,-a)}_{\mu\nu} = - {(-a,a)}_{\mu\nu}.
$$
Note that applying $\star$ twice we get the same tensor: $\star^2 F_{\mu\nu} = F_{\mu\nu}$.

\begin{remark}
In the general case when we apply $\star$ to an antisymmetric tensor with $m$ lower indices in $d$ dimensions we get: 
$$\star^2 = {(-1)}^{m(d-m)}\sign(\det g).$$
It follows that in Minkowski space the self-dual and anti-self-dual equations can have only trivial solution. Indeed if $\star F_{\mu\nu} = \pm F_{\mu\nu}$ when applying $\star$ and using $\star^2= - 1$ we get 
$$
- F_{\mu\nu} = \star^2 F_{\mu\nu} = \pm \star F_{\mu\nu} = F_{\mu\nu}.
$$
We can however introduce the imaginary unit $i$ in the definition of $\star$. This enables us to obtain nontrivial solution of the self-dual and anti-self-dual equations in the Minkowski space. This solution will be in general complex.
\end{remark}

One says that a tensor satisfying selfdual equation is \emph{selfdual}, and the tensor satisfying anti-selfdual equation is \emph{anti-selfdual}. We can find a basis of selfdual and anti-selfdual tensors as follows:
$$
\begin{aligned}
{(a,a)}_{\mu\nu} &= a_i \eta^i_{\mu\nu}, &
{(-a,a)}_{\mu\nu} &= a_i \bar{\eta}^i_{\mu\nu}.
\end{aligned}
$$

One says that $\eta^i_{\mu\nu}$ are \emph{selfdual t'Hooft symbols} and $\bar{\eta}^i_{\mu\nu}$ are \emph{anti-selfdual t'Hooft symbols} \cite{tHooftSymbols}. These symbols can be represented by six $4\times 4$ matrices as follows:
\begin{equation}
\label{tHooftSymbol}
\begin{aligned}
\eta^1_{\mu\nu} &= \left(
\begin{array}{cccc}
0 & 1 & 0 & 0 \\
-1 & 0 & 0 & 0 \\
0 & 0 & 0 & 1 \\
0 & 0 & -1 & 0 
\end{array}
\right), & 
\eta^2_{\mu\nu} &= \left(
\begin{array}{cccc}
0 & 0 & 1 & 0 \\
0 & 0 & 0 & -1 \\
-1 & 0 & 0 & 0 \\
0 & 1 & 0 & 0 
\end{array}
\right), &
\eta^3_{\mu\nu} &= \left(
\begin{array}{cccc}
0 & 0 & 0 & 1 \\
0 & 0 & 1 & 0 \\
0 & -1 & 0 & 0 \\
-1 & 0 & 0 & 0 
\end{array}
\right),
\end{aligned}
\end{equation}
and
\begin{equation}
\label{anti-tHooftSymbol} 
\begin{aligned}
\bar{\eta}^1_{\mu\nu} &= \left(
\begin{array}{cccc}
0 & -1 & 0 & 0 \\
1 & 0 & 0 & 0 \\
0 & 0 & 0 & 1 \\
0 & 0 & -1 & 0 
\end{array}
\right), &
\bar{\eta}^2_{\mu\nu} &= \left(
\begin{array}{cccc}
0 & 0 & -1 & 0 \\
0 & 0 & 0 & -1 \\
1 & 0 & 0 & 0 \\
0 & 1 & 0 & 0 
\end{array}
\right), &
\bar{\eta}^3_{\mu\nu} &= \left(
\begin{array}{cccc}
0 & 0 & 0 & -1 \\
0 & 0 & 1 & 0 \\
0 & -1 & 0 & 0 \\
1 & 0 & 0 & 0 
\end{array}
\right). 
\end{aligned}
\end{equation}

One can check that the t'Hooft symbols satisfy the following multiplication rule:
\begin{equation}
\label{tHooftMultiplication}
\begin{aligned}
\eta^i \eta^j &= - \delta^{ij}\Id_4 - \epsilon^{ijk}\eta^k  &
\Big(\eta^i_{\mu\nu}\eta^j_{\nu\rho} &= - \delta^{ij}\delta_{\mu\rho} - \epsilon^{ijk}\eta^k_{\mu\rho}\Big), \\
\bar{\eta}^i \bar{\eta}^j &= - \delta^{ij}\Id_4 - \epsilon^{ijk}\bar{\eta}^k  &
\Big(\bar{\eta}^i_{\mu\nu}\bar{\eta}^j_{\nu\rho} &= - \delta^{ij}\delta_{\mu\rho} - \epsilon^{ijk}\bar{\eta}^k_{\mu\rho}\Big).
\end{aligned}
\end{equation}
It follows that they form the representation of the $SU(2)$ group. Indeed from \Ref{tHooftMultiplication} we get
$$
\begin{aligned}
\left[-\frac{\eta^i}{2},-\frac{\eta^j}{2}\right] &= \epsilon^{ijk} \left( -\frac{\eta^k}{2} \right), \\
\left[-\frac{\bar{\eta}^i}{2},-\frac{\bar{\eta}^j}{2}\right] &= \epsilon^{ijk} \left( -\frac{\bar{\eta}^k}{2} \right).
\end{aligned}
$$

This fact can be easily seen from the representation theory point of view: the vector representation of $SO(4) \isom SU(2) \times SU(2) / \Integers_2$ is $(\frac{1}{2},\frac{1}{2})$. It follows from the properties of Clebcsh-Gordon coefficients for $SU(2)$ that 
$$
\left(\frac{1}{2},\frac{1}{2}\right) \otimes \left(\frac{1}{2},\frac{1}{2}\right) = (0,0) \oplus (1,1) \oplus (1,0) \oplus (0,1).
$$
The first term in the righthand side is the trace of a rang-2 tensor, the second is its symmetric traceless part and third and fourth are the decomposition of its antisymmetric part onto self-dual and anti-self-dual components.


\section{Euclidean spinors}

In this section we will mostly speak about spinors in the Euclidean space. Sometimes we will stress differences with the Minkowski space.

The double covering group for $SO(4)$ is $\Spin(4) \isom SU(2)\times SU(2)$. Thus we have two independent spinor representation each of them is isomorph to $SU(2)$. According to this the spinors in four dimensional euclidean space can have one \emph{doted} or one \emph{undotted} spinor index running over $1,2$ and $\dot{1},\dot{2}$ respectively. For the combinations $\psi^\alpha \chi_\alpha$ and $\psi_{\dot{\alpha}} \chi^{\dot{\alpha}}$ are supposed to be invariants we conclude that ${\left(\chi_\alpha\right)}^*$ transforms as $\psi^\alpha$. And the same rule for a doted index. 

\begin{remark}
Here and below the following rule will be held: the undotted indices follow form left-up to right-down and the doted -- from left-down to right-up.
\end{remark}

Three Pauli matrices and the unit one can be arranged to one four dimensional vector defined as
\begin{equation}
\label{sigmaDefinition}
\s_{\mu,\a\dot{\a}} = (\Id_2,-i\tau_1,-i\tau_2,-i\tau_3)
\end{equation}
The homomorphism from $SU(2)\times SU(2)$ to $SO(4)$ can be constructed as follows: consider a four vector $x^\mu$. We can build a matrix
\begin{equation}
\label{spinorCoordinate}
x_{\a\dot{\b}} = x^\mu \s_{\mu,\a\dot{\b}} = \left(
\begin{array}{cc}
x^0 - ix^3 & - x^2 - i x^1 \\
x^2 - ix^1 & - x^0 + ix^3
\end{array}
\right)
\end{equation}
satisfying $x^\dag x = x^2 \Id_2$ where $x^2 = x_\mu x_\mu$. Thus if we take two $SU(2)$ matrices $U_1$ and $U_2$ and transform $x \mapsto x' = U_1 x {U_2}^\dag$ we get
\begin{equation}
\label{SU2xSU2->SO4}
{x'}^2\Id_2 = {x'}^\dag x' = U_2 x^\dag {U_1}^\dag U_1 x {U_2}^\dag  = x^2 \Id_2.
\end{equation}

Hence this transformation generates an $SO(4)$ transformation of $x^\mu$ (since the group $SU(2)$ is simply connected we see that the determinant of $x^\mu$ transformation should be equal to 1). According to this four $\s$-matrices have one undotted and one doted index. Our convention is that they are both lower.

\begin{remark}
We see that the following rule holds: when complex conjugated, the indices rise and low without changing their dotness. Mention the difference with the Minkowski case: there the indices rise and low together with the changing of their dotness. This can be explained as follows: though in the euclidean case the both $SU(2)$ are independent, in the Minkowski case they are related by means of the complex conjugation. 
\end{remark}

The $\s$-matrices are not all hermitian, but rather satisfy the following conjugation rule:
$$
\begin{aligned}
{\s_0}^\dag &= \s_0, &  {\s_i}^\dag &= - \s_i.
\end{aligned}
$$
The reality condition for them can be expressed as follows:
\begin{equation}
\label{sigmaReality}
\begin{aligned}
{\s_\mu^\ast }^{\a\dot{\a}} &= \ep^{\a\b}\ep^{\dot{\a}\dot{\b}}\s_{\mu,\b\dot{\b}} & \Big({\s_\mu}^\ast &= \ep^{-1} \sigma_\mu \ep\Big).
\end{aligned}
\end{equation}
For any matrix which can be developed as $X_{\a\dot{\b}} = X^\mu \s_{\mu,\alpha\dot{\b}}$ the reality condition $X^\ast = \ep^{-1}X\ep$ means that the coefficients $X^\mu$ are real.

\begin{remark}
In Minkowski space the definition of $\s$-matrices misses $i$:
\begin{equation}
\label{SigmaMinkowski}
\begin{aligned}
\s_{\mu,\a\dot{\a}} &= (\Id_2,-\t_1,-\t_2,-\t_3), \\
\bar{\s}_\mu^{\dot{\a}\a} &= (\Id_2,+\t_1,+\t_2,+\t_3).
\end{aligned}
\end{equation}
This set of matrices governess an isomorphism $SL(2,\Compl) \to SO(3,1)$. There is another set of such matrices:
$$
\bar{\s}^{\mu,\dot{\a}\a} =  (\Id_2,+\t_1,+\t_2,+\t_3).
$$
Its analogue in the Euclidean space is just the hermitian conjugate of Euclidean $\s$-matrices: $\bar{\s}^{\mu,\dot{\a}\a} = (\s^\dag)^{\mu,\dot{\a}\a}$.
\end{remark}

The $\s$-matrices also satisfy a version of the Fierz identity \Ref{FierzIdentity} which has the following form:
\begin{equation}
\label{FierzForSigma}
\begin{aligned}
\s_{\mu,\a\dot{\b}}\s_{\mu,\g\dot{\d}} &= 2 \ep_{\a\g}\ep_{\dot{\b}\dot{\d}}, \\
\s_{\mu,\a\dot{\b}}\left(\s_{\mu}^\dag\right)^{\dot{\g}\d} &= 2 \d_\a^\d \d_{\dot{\b}}^{\dot{\g}}.
\end{aligned}
\end{equation}

Note that thanks to the doted-undotted  convention we can not directly multiply $\s_\mu$ and $\s_\nu$. However we can multiply it by $\bar{\s}_\nu$ (or ${\s_\nu}^\dag$). We get:
\begin{equation}
\label{sigmaMultiplication}
\begin{aligned}
\s_{\mu,\a\dot{\a}}\bar{\s}_\nu^{\dot{\a}\b} &= \d_{\mu\nu}\d_\a^\b + i \t_{i,\a}{}^\b \eta^i_{\mu\nu} & \Big(\s_\mu \bar{\s}_\nu &= \d_{\mu\nu} \Id_2 + i\t_i \eta^i_{\mu\nu}\Big), \\
\bar{\s}_\mu^{\dot{\a}\a}\s_{\nu,\a\dot{\b}} &= \d_{\mu\nu}\d^{\dot{\a}}_{\dot{\b}} + i {\t_i}^{\dot{\a}}{}_{\dot{\b}} \bar{\eta}^i_{\mu\nu} & \Big( \bar{\s}_\mu \s_\nu &= \d_{\mu\nu}\Id_2 + i \t_i \bar{\eta}^i_{\mu\nu}\Big).
\end{aligned}
\end{equation}
Here we see the appearance of selfdual \Ref{tHooftSymbol} and anti-selfdual \Ref{anti-tHooftSymbol} t'Hooft symbols .

\begin{remark}
If we swap $\s$-matrices and $\bar{\s}$-matrices we get the equations \Ref{sigmaMultiplication} but with the selfdual symbols replaced by anti-selfdual and vice versa. Notice going ahead that in this way we can construct the anti-instantons instead of the instantons.
\end{remark}

Let us introduce the Clebsch-Gordon coefficients which govern the spinor transformation with respect to the space rotation. Using \Ref{sigmaMultiplication} we get
\begin{equation}
\label{tHooftProjector}
\begin{aligned}
\s_{\mu\nu,\a}{}^\b &\equiv \frac{1}{4} \left( \s_{\mu,\a\dot{\g}}\bar{\s}_\nu^{\dot{\g}\b} - \s_{\nu,\a\dot{\g}}\bar{\s}_\mu^{\dot{\g}\b}\right) &= \frac{i}{2}\t_{i,\a}{}^\b \eta^i_{\mu\nu}, \\
\bar{\s}_{\mu\nu}^{\dot{\a}}{}_{\dot{\b}} &\equiv \frac{1}{4}\left(\bar{\s}_{\mu}^{\dot{\a}\g}\s_{\nu,\g\dot{\b}} - \bar{\s}_{\nu}^{\dot{\a}\g}\s_{\mu,\g\dot{\b}}\right) &= \frac{i}{2}\t_{i}{}^{\dot{\a}}{}_{\dot{\b}} \bar{\eta}^i_{\mu\nu}.
\end{aligned} 
\end{equation}
The appearance of 't Hooft symbols on the lefthand side allows us to call $\s_{\mu\nu}$ and $\bar{\s}_{\mu\nu}$ \emph{'t Hooft projectors}.

\begin{remark}
In Minkowski space they satisfy
$$
\s^{\mu\nu,\a\b}\s^{\r\s}{}_{\a\b} = \frac{1}{2}\left( g^{\mu\r}g^{\nu\s} - g^{\mu\s}g^{\nu\r} \right) - \frac{i}{2}\ep^{\mu\nu\r\s}
$$
\end{remark}

%% file: lie.tex

\label{LieAlg}

Here we cite some group theoretical data which is used (implicitly or explicitly) in our discussion of the derivation of prepotential. Since the derivation of the prepotential is based on the ADHM construction, which is known only for the classical groups, we consider only Lie algebras for $SU(N)$, $SO(N)$ and $Sp(N)$ groups. All details can be found, for example, in \cite{BourbakiLie}.

Apart from the standard group theoretical data, such as a root system, or the Weyl group description we also give the Dynkin indices for various representations and the coefficient $\b$ which appears in the $\gL$ expansion of the prepotential. 

Recall the Dynkin index definition. Since for the simple groups the Killing metric is unique up to multiplicative factor we conclude that for all representations $\ell_{\mr{adj}}\Tr_{\mr{adj}} = \ell_\vr \Tr_\vr$ where $\ell_\vr$ is the Dynkin index of this representation. Through the paper we normalize indices in such a way that $\ell_{\mr{fund}} = 1$ for all groups. 

The coefficient $\b$ is equal to 
\begin{equation}
\label{betaDef}
\b = \z \left( \ell_{\mr{adj}} - \sum_{\vr\in\mr{reps}} \ell_\vr \right).
\end{equation}
\begin{remark}
One could, of cause, renormalize the Dynkin index in order to absorb the parameter $\z$. To do this one can simply pose $\ell_{\mr{fund}} = \z$.
\end{remark}

We denote by $\gD^+$ the set of all positive roots. $h$ and $h^\vee$ the Coxeter and dual Coxeter number. We have collected some data in the Table \ref{GroupTh}. 


\begin{table}[t]
\begin{center}
\begin{tabular}{||c||c|c|c|c|c|c|c|c||}
\hhline{|t:=:t:=:=:=:=:=:=:=:=:t|}
\textbf{Algebra} & $\mbf{h}$ & $\mbf{h^\vee}$ & $\mbf{|W|}$ & \textbf{Adjoint}  & $\mbf{\ell_{\mr{adj}}}$ & $\mbf{\ell_{\mr{sym}}}$ & $\mbf{\ell_{\mr{ant}}}$ & $\mbf{\z}$ \\
\hhline{|:=::=:=:=:=:=:=:=:=:|}	
$A_n$  & $n+1$ &$n+1$ & $(n+1)!$ & $\mbox{fund}\otimes\mbox{fund}^\ast$ & $2n+2$ & $2n+4$ & $2n$ & 1 \\
\hhline{||-||-|-|-|-|-|-|-|-||}
$B_n$  & $2n$ & $2n-1$ & $2^n n!$ & $\Ant^2\mbox{fund}$ & $2n-1$  & $2n + 3$& $2n - 1$ & 2 \\
\hhline{||-||-|-|-|-|-|-|-|-||}
$C_n$ & $2n$ & $n+1$ & $2^n n!$ & $\Sym^2\mbox{fund}$ & $2n+2$ & $2n+2$ & $2n-2$ & 2\\
\hhline{||-||-|-|-|-|-|-|-|-||}
$D_n$ & $2n-2$ & $2n-2$ & $2^{n-1}n!$ & $\Ant^2\mbox{fund}$ & $2n-2$  & $2n + 2$& $2n - 2$ & 2\\
\hhline{|b:=:b:=:=:=:=:=:=:=:=:b|}
\end{tabular}
\end{center}
\caption{Group theoretical data}\label{GroupTh}
\end{table}



\section{Algebra $A_n$}

The algebra $A_n$ is the Lie algebra for the group $SU(n+1)$, $n\geq 1$. The root system can be describes as follows. Denote by $\{e_i\}$, $i=1,\dots,n+1$ an orthonormal base of the $\Real^{n+1}$. The set of positive roots is
$$
\begin{aligned}
\gD^+ &= \{ e_i - e_j \} , & &1\leq i < j \leq n+1.
\end{aligned}
$$
The adjoint representation lies in the tensor product of the fundamental and antifundamental representations.


\section{Algebra $B_n$}

This is the Lie algebra of the group $SO(2n+1)$, $n \geq 2$. We denote by $\{e_i\}$, $i = 1,\dots, n$ the base of $\Real^n$. The set of positive roots is
$$
\gD^+ = \left\{ 
\begin{aligned}
&e_i, & &1\leq i \leq n, \\
&e_i - e_j, & &1\leq i < j \leq n, \\
&e_i + e_j, & &1\leq i < j \leq n
\end{aligned}
\right.
$$
The adjoint representations is the antisymmetric one.


\section{Algebra $C_n$}

The Lie algebra of the group $Sp(n)$ is called $C_n$, $n\geq 2$. $\{ e_i \}$, $i=1,\dots,n$ is the base of $\Real^n$. The set of positive roots is
$$
\gD^+ = \left\{ 
\begin{aligned}
&e_i - e_j, & &1\leq i<j\leq n, \\
&e_i + e_j, & &1\leq i < j \leq n, \\
&2e_i, & &1\leq i \leq n.
\end{aligned}
\right.
$$
The adjoint representation is the symmetric one.


\section{Algebra $D_n$}

This is the Lie algebra of the group $SO(2n)$, $n\geq 3$. $\{ e_i \}$, $i=1,\dots,n$ is the base of $\Real^n$. The set of positive root is
$$
\gD^+ = \left\{
\begin{aligned}
&e_i - e_j, & &1\leq i < j \leq n, \\
&e_i + e_n, & &1\leq i < n, \\
&e_i + e_j, & &1\leq i < j < n.
\end{aligned}
\right.
$$
The adjoint representations is antisymmetric one.